\documentstyle[epsfig,a4,12pt]{report}

\font\tenbbb=msbm10 \font\sevenbbb=msbm7

\newfam\bbbfam \newfam\amsfam \newfam\bmfam     
\textfont\bbbfam=\tenbbb 
\scriptfont\bbbfam=\sevenbbb
\def\dbl {\fam\bbbfam}     

\begin{document}

        \onecolumn
        \setcounter{tocdepth}{2}
        


\baselineskip=16pt plus 0.01pt

\begin{titlepage}
	\centerline{\Large University of Wales Swansea}
	\centerline{\large Department of Physics}
	\centerline{June 14, 1999}
	\vspace{5cm}
	\centerline{\LARGE{\bf {\sc Five Dimensional Dynamical}}} 
	\vspace{0.2cm}
	\centerline{\LARGE{\bf {\sc Triangulations}}}
	\vspace{1.5cm}
	\centerline{by} 
	\vspace{1.5cm} 
	\centerline{\Large Alun George} 
	\vspace{5cm} 
	\centerline{A thesis submitted for the degree of}
	\centerline{Doctor of Philosophy.}
\end{titlepage}

\begin{titlepage} 
\begin{quote}
\raggedleft Er cof am fy annwyl Nain, \\ 
{\it Ellen Winifred Morris} \\
1911 -- 1998
\end{quote} 
\end{titlepage} 

\newpage \pagenumbering{roman} 
\pagestyle{plain} \setcounter{page}{2}
\addcontentsline{toc}{section}{Abstract}



\begin{center} {\bf Abstract} \end{center}

\vspace{3cm}

\noindent
The dynamical triangulations approach to quantum gravity is investigated in detail for the first time in five dimensions. In this case, the most general action that is linear in components of the $f$-vector has three terms. It was suspected that the corresponding space of couplings would yield a rich phase structure. This work is primarily motivated by the hope that this new viewpoint will lead to a deeper understanding of dynamical triangulations in general. Ultimately, this research programme may give a better insight into the potential application of dynamical triangulations to quantum gravity. This thesis serves as an exploratory study of this uncharted territory.

The five dimensional $(k,l)$ moves used in the Monte Carlo algorithm are proven to be ergodic in the space of combinatorially equivalent simplicial 5-manifolds. A statement is reached regarding the possible existence of an exponential upper bound on the number of combinatorially equivalent triangulations of the 5-sphere. Monte Carlo simulations reveal non-trivial phase structure which is analysed in some detail. Further investigations deal with the geometric and fractal nature of triangulations. This is followed by a characterisation of the weak coupling limit in terms of stacked spheres. Simple graph theory arguments are used to reproduce and generalise a well-known result in combinatorial topology. Finally, a comprehensive study of singular structures in dynamical triangulations is presented. It includes a new understanding of their existence, which appears to be consistent with the non-existence of singular vertices in three dimensions. The thesis is concluded with an overview of results, general discussion and suggestions for future work.


\newpage \pagenumbering{roman} 
\pagestyle{plain} \setcounter{page}{3}
\addcontentsline{toc}{section}{Crynodeb (Abstract in Welsh)}



\begin{center} {\bf Crynodeb} \end{center}

\vspace{3cm}

\noindent
Mae dull triongliannau dynamegol i ddisgyrchiant cwantwm yn cael ei ymchwilio'n fanwl mewn pum dimensiwm am y tro cyntaf. Yn yr achos yma, mae gan yr effaith mwyaf cyffredinol, sydd yn llinol yng nghydrannau y $f$-fector, dri term. Roedd yn debygol y byddai'r bwlch cyplysau cyfatebol yn ildio strwythur gwedd cyfoethog. Prif gymhelliant y gwaith yw'r gobaith fod y safbwynt newydd yma'n arwain at ddealltwriaeth dyfnach o driongliannau dynamegol yn gyffredinol. Yn y pen draw, gall y rhaglen ymchwil hon roi gwell mewnwelediad o ddefnydd potensial triongliannau dynamegol mewn disgyrchiant cwantwm. Cynigir y thesis yma fel astudiaeth chwilogol i'r maes newydd yma.

Mae'r symudiadau pum dimensiynol $(k,l)$ a ddefnyddir yn algorithm Monte Carlo wedi'u profi i fod yn ergodig yn y bwlch o 5-maniffoldiau symhlygol cyfartal cyfuniadol. Gwneir gosodiad ynglyn \^{a} bodolaeth posib o ff\^{i}n uchaf esbonnyddol ar y nifer o driongliannau o'r 5-sff\^{e}r sy'n gyfartal gyfuniadol. Mae efelychiad Monte Carlo yn datgelu strwythur gwedd annistadl ac fe'u dadansoddir yn fanwl. Mae ymchwiliadau pellach yn ymdrin \^{a} natur geometrig a ffractal y triongliannau. Dilynnir hyn trwy ddangos nodweddion ff\^{i}n y cyplyn gwan yn nhermau sfferau wedi'u stacio. Defnyddir ymresymiadau damcaniaeth graff syml i atgenhedlu a chyffredinoli canlyniad tra chyfarwydd yn nhopoleg cyfuniadol. Yn olaf cyflwynir astudiaeth cynhwysfawr o strwythurau hynod yn nhriongliannau dynamegol. Mae'n cynnwys dealltwriaeth newydd o'u bodolaeth sydd yn ymddangos i fod yn gyson ag anfodolaeth o fertigau hynod mewn tri dimensiwn. Wrth gloi mae'r thesis yn rhoi golwg cyflawn o'r canlyniadau, dadl gyffredinol ac awgrymiadau o waith i ddyfod.


\newpage \pagenumbering{roman} 
\pagestyle{plain} \setcounter{page}{4}
\addcontentsline{toc}{section}{Declarations}



\begin{center} {\bf Declarations} \end{center}
\vspace{3cm}

\noindent
This work has not previously been accepted in substance for any degree and is not being concurrently submitted in candidature for any degree.
\vspace{0.5cm}

\noindent 
Signed ...................................................\hspace{4.2cm}
Date ..............................

\vspace{2cm}

\noindent
This thesis is the result of my own investigations, except where otherwise stated. Sources are acknowledged by reference to a bibliography.
\vspace{0.5cm}

\noindent 
Signed ...................................................\hspace{4.2cm}
Date ..............................

\vspace{2cm}

\noindent
I hereby give consent for my thesis, if accepted, to be available for photocopying and for inter-library loan, and for the title and summary to be made available to outside organisations.
\vspace{0.5cm}

\noindent 
Signed ...................................................\hspace{4.2cm}
Date ..............................


\newpage \pagenumbering{roman} 
\pagestyle{plain} \setcounter{page}{5}
\addcontentsline{toc}{section}{Acknowledgements}



\begin{center} {\bf Acknowledgements} \end{center}
\vskip 30mm

\noindent
Many people have helped to make this thesis a reality\ldots

\vskip 5mm

\noindent
Firstly, I wish to thank Ray Renken, who first sparked my interest in dynamical triangulations during his visit to Swansea in 1996. I am greatly indebted to Simon Catterall who gave me a head start in my quest by allowing me to use his computer program. Thanks are also due to my supervisor Simon Hands for his advice and guidance, and to PPARC for funding my research.

\vskip 5mm

\noindent
I would like to thank the staff and postgraduate students of the Theory Group for their support over the last few years, particularly Martin Groves, Rhys Mainwaring, Evangelos Mavrikis, Sanjeev Shukla and David Tong. Fe hoffwn ddiolch i Elsie Prytherch am gywiro fy Nghymraeg. Finally, special thanks are due to Brett Taylor for many helpful discussions and proof-reading the manuscript.

\vskip 5mm

\noindent
\begin{quote}
Yn olaf hoffwn ddiolch i fy mam, fy nhad a'm chwaer, Ann, am eu holl gymorth yn ystod fy amser yn y brifysgol.
\end{quote}

\vskip 28mm

\begin{quote}
{\scriptsize ``All men dream: but not equally. Those who dream by night in the dusty recesses of their minds wake in the day to find that it was vanity: but the dreamers of the day are dangerous men, for they may act out their dream with open eyes, to make it possible. This I did.''\\
--- T.~E.~Lawrence, {\it Seven Pillars of Wisdom}.}
\end{quote} 

\newpage \pagenumbering{roman} 
\pagestyle{plain} \setcounter{page}{6}
\newpage \addcontentsline{toc}{section}{Contents}
\tableofcontents 

	\newpage
        \pagestyle{headings} 
        \pagenumbering{arabic}
        \setcounter{page}{1}
	\baselineskip=16pt plus 0.01pt

        \input epsf.tex
        

\input epsf.tex

\chapter{Introduction} \label{chap:chap1}

\section{Quantum Gravity} \label{sec:qg}

\begin{figure}[ht]
\leavevmode
\hbox{\epsfxsize=2.8cm \epsfbox{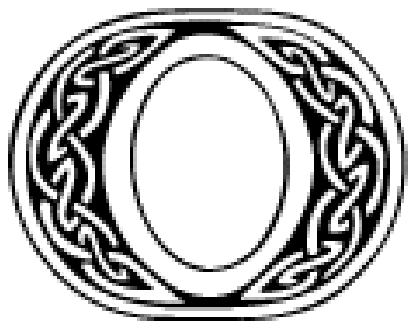}}
\parbox{13.1cm}{\baselineskip=16pt plus 0.01pt \vspace{-21mm}UR current understanding of gravity is based upon Einstein's general theory of relativity. General relativity has proved to be one of the great achievements in theoretical physics this century. It has predicted a number of gravitational $\,$phenomena $\,$which $\,$have $\,$subsequently $\,$been $\,$verified} 
\end{figure}
\vspace{-7mm}
experimentally to high degrees of precision. These effects include the advancement of Mercury's perihelion, gravitational redshift and gravitational lensing. Another effect currently under investigation is gravitational waves. The success of general relativity is merited by perfect agreement between theory and experiment at cosmological scales. However, general relativity is flawed in that it fails to describe the physics of spacetime singularities, such as those found in black holes and at the birth of the universe. In other words, we have a theory of gravity which is perfectly consistent with Nature at cosmological scales, yet fails for short distance scales. From this, one may cautiously infer that general relativity is a low energy limit of a more fundamental theory of gravity.

Perhaps the most outstanding scientific achievement of the century is the development of quantum theory. The effects of quantum mechanics and quantum field theory have also been tested in numerous experiments to very high degrees of precision at atomic and sub-atomic scales. These tests include the photoelectric effect, electron diffraction, the Stern-Gerlach experiment and calculation of the Lamb shift, to name but a few. It is this overwhelming success that has led physicists to believe that Nature is ruled by quantum theory. Presently, the only fundamental force which still eludes a satisfactory quantum description is gravity. 

One of the most important advancements in the understanding of fundamental forces was made by Maxwell. He proved that two seemingly disparate forces (electricity and magnetism) may be described by a single theory -- electromagentism. More recently, it has been shown that the electromagnetic and weak forces may be {\it unified} into a single electroweak force. This was made possible by realising that such theories can be understood in terms of symmetry groups. It is hoped that unification may be taken a step further by including the strong force, resulting in a grand unified theory (GUT). Unlike the electroweak theory, a GUT is characterised by a single coupling constant. Of course, the final and ultimate unification would be of all four known fundamental forces, incorporating a quantum theory of gravity. Such a theory is commonly referred to as a theory-of-everything.

For these reasons it is generally believed that there exists a quantum theory of gravity. Before going any further, we first ask ourselves an important question: what features do we expect quantum gravity to possess? First of all, gravitons, the excitations of the gravitational field, must be massless -- corresponding to long range interactions. Secondly, the theory must reduce to general relativity in the low energy limit and to Newtonian gravity in the weak coupling limit. Finally, the graviton should be a spin 2 particle since the spacetime metric is a rank 2 tensor.

\vskip 5mm

\noindent
Quantum gravity has been studied extensively by many research groups over the latter half of this century. Despite this fact, no satisfactory theory has yet been formulated. One of the main problems facing theorists is the fact that no experimental data from graviton scattering is (or will ever be) available, since the quantum effects of gravity are expected to emerge near the Planck scale, i.e. $\sim 10^{19}$\,GeV. This situation is in stark contrast to the early days of quantum field theory, when theory was led by experiment.

On that note, let us briefly mention some of the current avenues of research in the field of quantum gravity. For a comprehensive survey of the research programme as a whole, the reader is referred to a recent review by Rovelli~\cite{rovelli}. Perhaps the most promising and popular candidate theory to date is superstrings; a theory in which particles (including gravitons/gravitinos) are described by the excitations of one dimensional objects in ten dimensional spacetime. Loop quantum gravity, which takes the canonical quantisation approach, has also been studied intensively over the last decade with encouraging results. 

Theories of discretised spacetime have become increasingly popular of late~\cite{loll}. A subset of these, generically called {\it simplicial quantum gravity}, was founded upon the Euclidean quantum gravity programme~\cite{euclidean} and the pioneering work of Regge~\cite{regge}. Simplicial quantum gravity encompasses two closely related and complementary theories, namely, {\it quantum Regge calculus} and {\it dynamical triangulations} -- the latter being the subject of this thesis. Other candidate theories include non-commutative geometry, topological field theory, supergravity and twistors. 

The reconciliation of quantum theory and relativity has proved to be a formidable challenge. Indeed, their seemingly incompatible philosophical underpinnings may only be overcome by the formulation of a radical new theory, rather than a quantisation prescription {\it per se}. It is possible (if not likely) that a consistent formulation of quantum gravity will force us to abandon our current understanding of the small scale structure of spacetime. These difficult issues have been addressed by Isham~\cite{isham1,isham2}, amongst many others. 

At this point, it is expedient to describe the many problems associated with the (na\"{\i}ve) conventional field theoretic approach to quantum gravity. This will lead us naturally to a review of simplicial quantum gravity.

\subsection{Euclidean Quantum Gravity} \label{subsec:eqg}

We shall now consider the path integral approach to quantum gravity and highlight the associated problems. In this formalism one defines a functional integral over field configurations of a spacetime manifold $\cal M$ with metric $g_{\mu\nu}$. If $\cal M$ has a Lorentzian signature then the path integral is oscillatory and hence divergent, due to complex Boltzmann weights. Fortunately, it is possible to overcome this problem by Wick rotating time $t$ clockwise by $90^{\circ}$ in the complex plane to give the following transformation $t \rightarrow -it$. The resulting Euclidean path integral $\cal Z$$_{E}$ is written as
\begin{equation}
{\cal Z}_{E} = \int_{\cal M}\!\frac{{\cal D}g_{\mu\nu}}{\mbox{Vol(Diff)}}\exp(-S[g_{\mu\nu},\Lambda,G]). \label{eq:pathintegral}
\end{equation}
The measure includes a factor which prevents the overcounting of diffeomorphic field configurations. (Of course, to be precise, $\cal Z$$_{E}$ is a {\it Riemannian} path integral as spacetime is not necessarily flat by any means.) In $d$-dimensional Euclidean space the Einstein-Hilbert action $S[g_{\mu\nu},\Lambda,G]$ is defined (in natural units) as
\begin{equation}
S[g_{\mu\nu},\Lambda,G] = \frac{1}{16 \pi G}\int_{\cal M}\mbox{d}^{d}\xi\,(\det g_{\mu\nu})^{\frac{1}{2}}(2\Lambda- R), \label{eq:action}
\end{equation}
where $R$ is the scalar curvature, $\Lambda$ is the cosmological constant and $G$ is Newton's gravitational constant.

Having defined the Euclidean path integral $\cal Z$$_{E}$, we immediately encounter some serious set-backs. Firstly and perhaps most importantly, one can only {\it assume} that the signature may be rotated back to Lorentzian spacetime. Secondly, one expects the topology of spacetime to be dynamical in a quantum theory of gravity. Unfortunately, the topological classification of smooth (differentiable) 4-manifolds is still an outstanding problem in mathematics. The problems do not end here: the Einstein-Hilbert action $S[g_{\mu\nu},\Lambda,G]$ is known to be unbounded from below. Perturbation theory, in this framework, leads to yet more complications. The theory is perturbatively non-renormalisable since $G$ has a negative mass dimension. It {\it is} possible to overcome this problem by introducing higher derivative terms into the action, but only at the cost of unitarity~\cite{stelle}. One may strive to avoid many of these problems by considering the Euclidean path integral approach in a non-perturbative framework -- namely simplicial quantum gravity.

\section{Simplicial Quantum Gravity} \label{sec:sqg}

Simplicial quantum gravity is, in essence, a non-perturbative realisation of the Euclidean path integral over spacetime geometries. The key ingredient of this approach is lattice discretisation. This scheme brings the theory into the realm of statistical physics, where it can be studied numerically. Contact with the continuum may then be made at a critical point where the system becomes scale invariant. Traditionally, in such lattice field theories, the spacetime lattice serves as an inert background upon which the fields are defined. However, in this particular case, the lattice {\it itself}\, is dynamical, which furthermore must support a definition of curvature. As we shall see, this compels us to abandon the use of hypercubic lattices in favour of {\it simplicial} spaces. 

\vskip 5mm

\noindent
For the time being we shall offer a short r\'{e}sum\'{e} of simplicial quantum gravity; details and precise definitions are to be found later. Simplicial spaces are constructed by connecting together a finite set of `elementary building blocks' called {\it simplices}. A $d$-dimensional simplex (or $d$-simplex) is, in a certain geometrical sense, the most fundamental bounded unit of Euclidean space $\dbl R$$^{d}$. For example, a 0-simplex is a {\it vertex}, a 1-simplex is a {\it link}, a 2-simplex is a {\it triangle} and a 3-simplex is a {\it tetrahedron}, and so on. In general, a simplex is formed by the product of its constituent {\it subsimplices}. For example, a triangle is formed by the product of three vertices. As a general rule, $(d-1)$-subsimplices are called {\it faces} and $(d-2)$-subsimplices are called {\it hinges}.

In this thesis, we shall only consider a special subset of simplicial spaces, known as {\it simplicial manifolds}. As with smooth varieties, simplicial manifolds also support a definition of curvature in terms of the rotation of parallel transported vectors around hinges; a crucial fact discovered by Regge~\cite{regge}. Using this result, one can formulate a discretised counterpart of the Einstein-Hilbert action. These constructs enable us to `discretise' the Euclidean path integral. This discretisation scheme, in effect, reduces $\cal Z$$_{E}$ to a {\it partition function} $Z$
\begin{equation}
Z = \sum_{T\,\in\,{\cal T}}\exp(-S[T]),
\end{equation}
where $S[T]$ is the discretised Einstein-Hilbert action. The partition function is a summation over all triangulations $T$ of a space of triangulations $\cal T$, each weighted by its corresponding Boltzmann factor.

\vskip 5mm

\noindent
There are two main variants of simplicial quantum gravity which correspond to the two `degrees of freedom' of simplicial manifolds. These are the link lengths and the connectivity. The connectivity of a simplicial manifold is an abstract notion which refers to the different possible ways of interconnecting its constituent simplices. If the geometry is governed only by the link lengths, then the resulting theory is called quantum Regge calculus, the quantum extension of Regge calculus. Conversely, geometries that are controlled solely by the connectivity correspond to a theory known as dynamical triangulations. In dynamical triangulations, simplices are treated as being equilateral and hence link lengths play no role in the dynamics. Unlike quantum Regge calculus, it is a regularisation since the discretisation of spacetime introduces an intrinsic cut-off $a$, corresponding to the link length.

The goal of simplicial quantum gravity is to make contact with a physical {\it continuum limit}. This can only be done if the statistical ensemble loses scale dependence; a characteristic of continuous phase transitions. These critical points are searched for in the space of parameters using {\it Monte Carlo simulation} methods, which are discussed later in section~\ref{sec:montecarlo}. Before looking at Regge calculus, quantum Regge calculus and dynamical triangulations in more detail, let us first review some background mathematics.

\subsection{Combinatorial Topology} \label{subsec:ct}

The study of simplicial spaces and manifolds is known in mathematics as {\it combinatorial topology}\, (or {\it piecewise linear topology})~\cite{grunbaum,hudson,glaser,rourke}. This well-established branch of mathematics partly forms the theoretical basis of dynamical triangulations. The following sections are devoted to clarifying our definitions and conventions on the subjects of topology and combinatorial topology, which will be upheld throughout the remainder of this thesis. Fortunately, these basic constructs are easily defined and explained without the need for unnecessary mathematical rigour.

\subsubsection{Topology and Manifolds} \label{subsubsec:topman}

Two topological spaces $\cal A$ and $\cal B$ are said to be {\it homeomorphic} if there exist a continuous map $f\!\!:\!\!\!{\cal~A}\mapsto\!\!{\cal~B}$ and a continuous inverse map $f^{-1}\!\!:\!\!{\cal~B}\mapsto\!\!{\cal~A}$ between points on $\cal A$ and points on $\cal B$. A continuous map $g\!:\!\!{\cal~C}\mapsto\!\!{\cal~D}$ between topological spaces is called an {\it embedding} if it is a homeomorphism onto a subspace of $\cal D$. There are many kinds of topological spaces. We are concerned only with those that are locally homeomorphic to Euclidean space, i.e. {\it manifolds}. Throughout this thesis all manifolds are understood to be closed, compact, connected and orientable.

One goal of mathematics is to classify smooth manifolds (of a given dimension) according to their topology in terms of a finite set of topological invariants. This has been achieved in all dimensions except three and four~\cite{acmgeometry}. The topology of 2-manifolds are classified by a single integer invariant known as the {\it genus} $h$, which is essentially the number of handles. Alternatively, 2-manifolds may be classed in terms of the {\it Euler characteristic} $\chi$, which is related to the genus by $\chi = 2-2h$. In this thesis we shall only encounter the three most fundamental topologies: {\it $d$-balls} $B^{d}$, {\it $d$-spheres} $S^{d}$ and {\it $d$-tori} $T^{d}$,\,\footnote{A $d$-sphere is homeomorphic to the boundary of a $(d+1)$-ball, and a $d$-torus is homeomorphic to the product of $d$ 1-spheres.}.

\subsubsection{Simplices and Triangulations} \label{subsubsec:simptri}

In this section, the reader may assume that the subscript $i$ takes integer values from 0 to $d$ inclusively. Let ${\dbl R}^{n}$ be an $n$-dimensional Euclidean space. Formally, a $d$-simplex $\sigma^{d}$ may be defined as the convex hull formed by a set of $d+1$ affinely independent points in ${\dbl R}^{n}$, where $n \geq d$. From here on, a `simplex' is assumed to be $d$-dimensional. A simplex is composed of ${}^{d+1}$C$_{j+1}$ constituent $j$-dimensional subsimplices, for $j=0,\ldots,d-1$. An $i$-(sub)simplex $\sigma^{i}$ of a simplex $\sigma^{d}$ is defined as the $i$-simplex formed from a subset of $i+1$ points of the $d+1$ points of $\sigma^{d}$. As a general convention in this thesis, a `(sub)simplex' is taken to mean either a subsimplex or a simplex. 

A {\it simplicial complex} $K$ is a topological space constructed from a finite number of $k$-simplices with the property that any two simplices either intersect at a common subsimplex or are disjoint. The dimension of $K$ is that of its highest dimensional constituent $k$-simplex. Consider an $i$-(sub)simplex $\sigma^{i}$ of a simplicial complex $K$. The {\it star} of $\sigma^{i}$, written $St(\sigma^{i})$, is defined as the set of $k$-simplices in $K$ which intersect $\sigma^{i}$. 

In this thesis, we deal exclusively with {\it simplicial manifolds}, which form a special subset of simplicial complexes. By this we mean that all manifolds are complexes, but not all complexes are manifolds. A $d$-dimensional simplicial manifold (or {\it triangulation}) $T$ is a simplicial complex which satisfies the added contraints that demand all faces to be {\it pairwise} connected and that every star of $T$ is homeomorphic to $B^{d}$. For a simplicial manifold, the {\it local volume} $n(\sigma^{i})$ of $\sigma^{i}$ is defined as the number of simplices in $St(\sigma^{i})$. When referring to the local volumes of $i$-(sub)simplices {\it in general}, we shall use the symbol $n_{i}$. To any given simplicial manifold $T$, one may associate an {\it $f$-vector} 
\begin{equation}
f=(N_{0}\,,\ldots,N_{d}),
\end{equation}
where $N_{i}$ is the total number of $i$-(sub)simplices of $T$. In almost all cases the $f$-vector cannot uniquely identify a triangulation. This can only be done with knowledge of the connectivity. Given that simplicial manifolds are combinatorial in nature, one may expect certain relations between the components of an $f$-vector. This is indeed the case. It turns out that there are two sets of equations known as the Euler-Poincar\'{e} and Dehn-Sommerville relations.

\subsubsection{Euler-Poincar\'{e} and Dehn-Sommerville Relations}

Euler's theorem is a classic result of combinatorial topology. It relates the Euler characteristic $\chi$ to the total number of vertices $V$, links $L$ and faces $F$ of polyhedra. 
\begin{equation}
\chi = V - L + F = 2 \label{eq:euler}
\end{equation}
The generalisation of this result to $d$-dimensional simplicial manifolds is known as the Euler-Poincar\'{e} relation. 
\begin{equation}
\chi = \sum_{i=0}^{d} (-1)^{i} N_{i} \label{eq:eulerpoincare}
\end{equation}
The Euler characteristic $\chi$ is zero for odd dimensional closed manifolds and equal to two for even dimensional $d$-spheres.

The Dehn-Sommerville relations are a set of $d+1$ equations linear in the number $N_{i}$ of $i$-(sub)simplices of $d$-dimensional simplicial manifolds~\cite{sommerville}. They reduce to $\frac{d+2}{2}$ $\left(\!\frac{d+1}{2}\!\right)$ linearly independent relations in even (odd) dimensions. Collectively, they are written as
\begin{equation}
N_{i} = \sum_{j=i}^{d} (-1)^{d-j}\;\;{}^{j+1}\mbox{C}_{i+1}\;N_{j}, \label{eq:ds}
\end{equation}
where $i=0,\ldots,d$. These equations are necessary but {\it not sufficient} for a simplicial complex to be a simplicial manifold. In other words, all simplicial manifolds satisfy equations~(\ref{eq:ds}), {\it but} not all solutions of (\ref{eq:ds}) correspond to simplicial manifolds. 

\subsection{Regge Calculus} \label{subsec:regge}

Regge calculus was formulated by Regge in 1961 as a coordinate independent discretisation of general relativity~\cite{regge}. He showed that one can approximate smooth (spacetime) manifolds with piecewise-linear (simplicial) manifolds. His methods were used as an approximation of general relativity and have since been used for numerical and computational work~\cite{twbib,mtw}.

\begin{figure}[htp]
\centering{\input{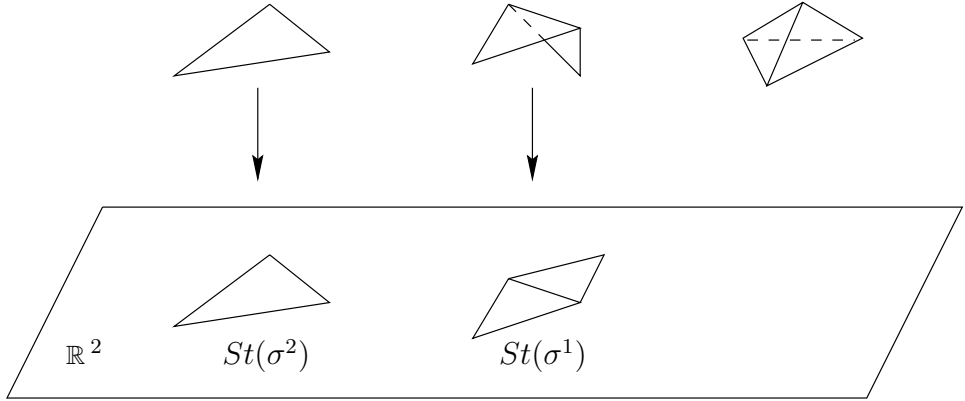}\par}
\caption{Embeddings of two dimensional stars in ${\dbl R}^{2}$.}
\label{fig:embedding}
\end{figure}

As mentioned earlier, curvature has a natural definition on simplicial manifolds in terms of parallel transportation of a vector around a hinge. For the moment we shall give a heuristic explanation of why curvature is only defined at hinges. Consider the embedding of an arbitrary $d$-surface $\cal S$ in $\dbl R$$^{n}$, where $n\geq d$. If $\cal S$ cannot be embedded in $\dbl R$$^{d}$, then it may be deemed to be curved. For simplicity, let us apply this notion to two dimensional simplicial manifolds. Clearly, the stars of triangles and links can always be embedded in $\dbl R$$^{2}$ and are therefore flat. These structures are illustrated in figure~\ref{fig:embedding}. 

\noindent
From figure~\ref{fig:embedding}, it is clear that a vertex star cannot {\it in general}$\,$ be embedded in $\dbl R$$^{2}$. This shows that curvature is defined at vertices (hinges). Of course, the embedding of a vertex star in ${\dbl R}^{2}$ {\it is} possible provided the link lengths satisfy certain conditions. Although difficult to visualise, these ideas can be generalised for hinges of higher dimensional simplicial manifolds. This brings us to the concept of {\it deficit angles}.

In its simplest form, the Einstein-Hilbert action $S_{R}$ of a $d$-manifold with metric $g_{\mu\nu}$ and scalar curvature $R$ is given by\,\footnote{Here, for simplicity, the coefficient has been rescaled `away' and the cosmological constant $\Lambda$ is set to zero.}
\begin{equation}
S_{R} = \int\mbox{d}^{d}\xi\,(\det g_{\mu\nu})^{\frac{1}{2}}R. \label{eq:Raction}
\end{equation}
In his seminal paper, Regge showed that the curvature at a hinge $h$ is proportional to its associated deficit angle $\delta_{h}$
\begin{equation}
\delta_{h} = 2\pi-\sum_{\sigma^{d}\,\in\,St(h)}\theta(\;\!\!\sigma^{d}), \label{eq:deficit}
\end{equation}
where $\theta(\;\!\!\sigma^{d})$ is the dihedral angle subtended by simplex $\sigma^{d}$ at $h$. The deficit angle $\delta_{h}$ is the rotation angle of a vector parallel transported around $h$. By summing over all $N_{d-2}$ hinges of a simplicial manifold $T$ we obtain an expression for the `integrated' scalar curvature. This gives us the discretised Regge action $S_{Regge}$
\begin{equation}
S_{R} \longrightarrow S_{Regge} = \sum_{h\,\in\, T}\delta_{h}V_{h}, \label{eq:reggeaction}
\end{equation}
where $V_{h}$ is the volume of $h$. An attractive feature of $S_{Regge}$ is that it can be shown to be equivalent to the continuum action $S_{R}$ in the limit where the link lengths are reduced to zero and their number is taken to infinity. In fact, $S_{Regge}$ can also be derived {\it from} the continuum action~\cite{flregge}.

\vskip 5mm

\begin{figure}[htp]
\centering{\input{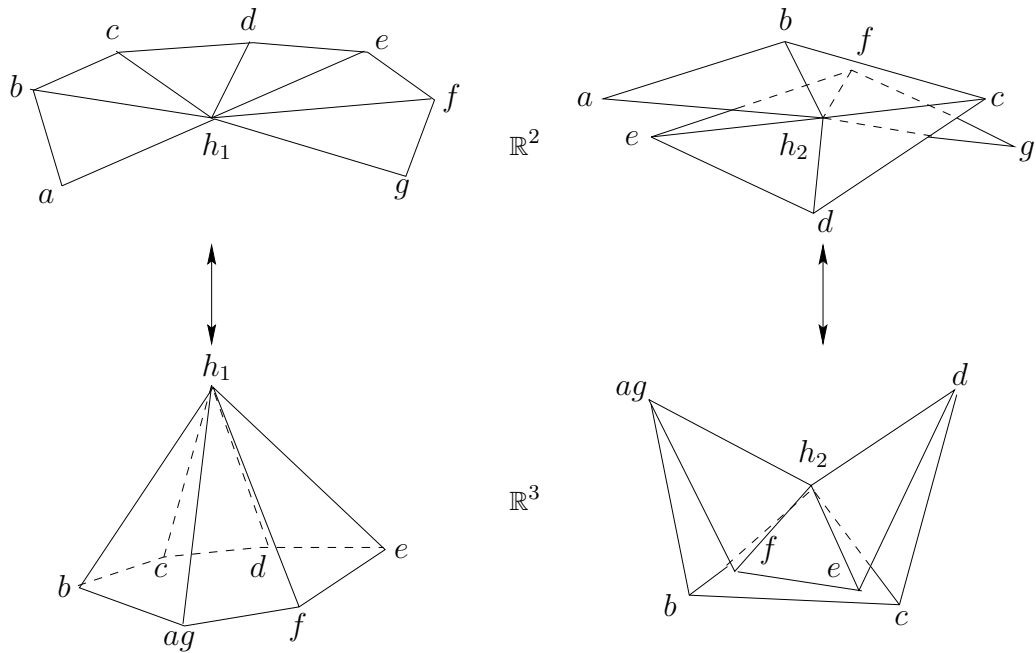}\par}
\caption{Curvature and deficit angles of two dimensional hinge stars.}
\label{fig:regge}
\end{figure}

\noindent
Clearly, all the dihedral angles and hinge volumes of a simplicial manifold are determined purely by the link lengths, which are subject to the triangle inequality or its higher dimensional generalisations. In other words, the curvature of the simplicial manifold is governed solely by the link lengths. It is useful at this point to illustrate the connection between deficit angles and curvature. Again, for convenience, we consider two dimensional simplices. In figure~\ref{fig:regge} the top pair of pictures show the stars of hinges $h_{1}$ and $h_{2}$ projected onto the plane $\dbl R$$^{2}$. 

\noindent
The top-left picture shows hinge $h_{1}$ with positive deficit angle. By conjoining vertices $a$ and $g$ and embedding the structure in $\dbl R$$^{3}$ we see that the surface has positive curvature (bottom-left). Conversely, the top-right picture shows hinge $h_{2}$ with negative deficit angle. This configuration forms a saddle-type structure with negative curvature when embedded in $\dbl R$$^{3}$ (bottom-right). If the deficit angle is zero then the structure is flat.

Classical problems in general relativity may be solved by varying the link lengths of a simplicial manifold to find the extremum of the action. This correponds to the solution of Einstein's equation in the limit of infinitesimally fine triangulations. We now move on to briefly mention the quantum extension of Regge Calculus. For a comprehensive (though dated) bibliography of the subject, the reader is referred to an article by Tuckey and Williams~\cite{twbib}.

\subsection{Quantum Regge Calculus} \label{subsec:qrc}

Quantum Regge calculus is a realisation of the Euclidean path integral $\cal Z$$_{E}$ in terms of discretised spacetime~\cite{rwqrc1}. As with Regge calculus, the dynamical variables are the link lengths of a simplicial manifold with fixed connectivity. The partition function is defined such that the measure includes a summation over all possible simplicial manifolds which satisfy the triangular inequality, or its higher dimensional generalisations. The continuum measure transforms as follows.
\begin{equation}
\int_{\cal M}\!\frac{{\cal D}g_{\mu\nu}}{\mbox{Vol(Diff)}} \longrightarrow \int_{0}^{\infty}\prod_{i=1}^{N_{1}}\mbox{d}l_{i}\,J(l_{1},\ldots,l_{N_{1}}) 
\end{equation}
So, the continuum measure of $\cal Z$$_{E}$ is replaced by a product of $N_{1}$ integrals over all possible link lengths $l_{i}$ from 0 to $\infty$. Hence, quantum Regge calculus is not a regularisation scheme; the triangulated lattice has no cut-off. Included in the measure is a non-trivial Jacobian $J(l_{1},\ldots,l_{N_{1}})$, whose form is not generally known. The partition function $Z_{Regge}$ is thus defined as
\begin{equation}
{\cal Z}_{E} \longrightarrow Z_{Regge} = \int_{0}^{\infty}\prod_{i=1}^{N_{1}}\mbox{d}l_{i}\,J(l_{1},\ldots,l_{N_{1}})\exp(-S_{Regge}).
\end{equation}
Numerical computation of $Z_{Regge}$ is possible only by introducing long and short length cut-offs to the integral. Contact with the continuum may be reached at a suitable scaling limit where the volume of the simplicial manifold is taken to infinity. One hopes, on the grounds of universality, that such a limit would be insensitive to discretisation details, such as the choice of triangulation connectivity. 

Quantum Regge calculus is not without its problems. Perhaps the most pressing difficulty relates to an ambiguity in the choice of measure. In addition to this, the integration over link lengths is known to overcount configurations, which may possibly be circumvented by introducing a Fadeev-Popov factor into the measure. In fact, a recent study has concluded that the theory `fails' in four dimensions~\cite{rolf}.

\subsection{Random Surfaces and Dynamical Triangulations} \label{sec:rsdt}
 
The origin of dynamical triangulations can be traced back to the application of random surfaces to string theory during the early 1980s and the seminal work of Weingarten~\cite{weingarten}. String theory may be viewed as being geometrical in nature since propagating strings sweep out world-sheets in their embedding space. It was Polyakov who first proposed the quantisation of (closed orientable) bosonic strings by defining a path integral over surfaces with internal metric $g_{ab}$, embedded in $D$-dimensional space~\cite{polyakov}. One may use this formalism as a quantisation of gravity, where the world-sheets are interpreted as spacetime manifolds and the $D$-dimensional embedding space become $D$ scalar fields $\phi$. In this framework, the Polyakov action is written as
\begin{equation}
S[g_{ab},\phi] = \sum_{i=1}^{D}\int\mbox{d}^{2}\xi\;(\det g_{ab})^{\frac{1}{2}}g^{ab}\partial_{a}\phi^{i}\partial_{\,b}\phi^{i}+\eta\int\mbox{d}^{2}\xi,
\end{equation} 
where $\eta$ is a constant. This is equivalent to pure two dimensional quantum gravity when $D=0$. Here, $D$ is interpreted as the central charge of the conformal matter coupled to gravity. This continuum formulation of two dimensional quantum gravity was solved analytically by Knizhnik {\it et al.} using Liouville theory~\cite{kpz}.

One can solve dynamical triangulations analytically using matrix models~\cite{bkkm}. In this case, one defines a path integral $\cal Z$$_{M}$ over Hermitian matrix fields $M_{ij}$ of dimension $N$.
\begin{equation}
{\cal Z}_{M} = \int\mbox{d}M_{ij}\exp(-S[M_{ij}])
\end{equation}
The action $S[M_{ij},g]$ is given by
\begin{equation}
S[M_{ij},g] = \frac{1}{2}M_{ij}M^{\ast}_{ij}+\frac{g}{\sqrt{N}}M_{ij}M_{jk}M_{ki},
\end{equation}
where $g$ is a coupling constant. A perturbative expansion in powers of $g$ yields all possible $\phi^{3}$ graphs. There exists a direct equivalence between dual $\phi^{3}$ graphs and two dimensional triangulations, as triangles have {\it three} faces. In other words, there exists a direct correspondence between $\cal Z$$_{M}$ and the partition function of two dimensional dynamical triangulations. The analytic results of this formalism has been shown to agree with those of Liouville theory for $c\leq 1$, where $c$ is the central charge\,\footnote{Liouville theory with $c>1$ matter `breaks down'.}. It is this success of dynamical triangulations which has encouraged physicists to consider higher dimensional generalisations.

\subsubsection{Two Dimensional Dynamical Triangulations} \label{subsubsec:2ddt}

Let us now consider dynamical triangulations in more detail by first deriving the action in two dimensions. From section~\ref{subsec:regge}, we know that the deficit angle of a hinge $h$ is given by
\begin{equation}
\delta_{h} = 2\pi-\sum_{\sigma^{2}\,\in\,St(h)}\theta(\;\!\!\sigma^{2}). \label{eq:def}
\end{equation}
For dynamical triangulations, equation~(\ref{eq:def}) requires modification since all simplices are equilateral. Specifically, the dihedral angle between two faces of a triangle is a {\it constant} $\Theta_{2}$. As a result, the deficit angle of a hinge $h$ depends solely on its local volume $n(h)$. Therefore, we have
\begin{equation}
\delta_{h} = 2\pi-n(h)\Theta_{2}. 
\end{equation}
Given that $\Theta_{2}=\pi/3$, it is clear that $h$ is flat if $n(h)=6$. As we shall now see, deficit angles are not particularly relevant to the two dimensional case. The Gauss-Bonnet theorem states that the integrated scalar curvature of a closed two dimensional manifold $\cal M$ is a topological invariant, related to the Euler characteristic $\chi$.
\begin{equation}
\int_{\cal M}\mbox{d}^{2}\xi\;R = 4\pi\chi \label{eq:gaussbonnet}
\end{equation}
Using this theorem, one can show that the two dimensional Einstein-Hilbert action is given by
\begin{equation}
S[g_{\mu\nu},\mu,G] = \mu A-\frac{\chi}{G},
\end{equation}
where $A$ is the area of $\cal M$ and $\mu$ plays the role of the cosmological constant. This may be discretised by realising that the area of a {\it simplicial} 2-manifold is proportional to $N_{2}$.
\begin{equation}
S[T] = \mu N_{2}-\frac{\chi}{G}
\end{equation}

The partition function of dynamical triangulations is a summation over all triangulations $T$ of a particular space of triangulations $\cal T$.
\begin{equation}
Z = \sum_{T\,\in\,{\cal T}}\exp(-S[T])
\end{equation}
As one might expect, there exist triangulations of $\cal T$ related by symmetry. The measure is a sum over physically inequivalent triangulations, therefore those related by symmetry must not be overcounted. The issues relating to the meaning of $\cal T$ are discussed in depth in section~\ref{subsec:combequiv}.

\section{Monte Carlo Simulations} \label{sec:montecarlo}

The numerical study of dynamical triangulations basically entails the detection of critical points in the space of couplings. Ultimately, this process involves calculating averages of thermodynamic variables. In dynamical triangulations the average value of an observable $O$ is given by
\begin{equation}
\langle O\rangle = \frac{1}{Z}\sum_{T\,\in\,{\cal T}}O\exp(-S[T]).
\end{equation}
The (grand canonical) partition function $Z$ obviously cannot be evaluated in practice. Instead, we consider the {\it canonical partition function} $Z_{c}$
\begin{equation}
Z_{c} = \sum_{T\,\in\,{\cal T}(N_{d})}\exp(-S[T]),
\end{equation}
where $T\in{\cal T}(N_{d})$ represents the subspace of triangulations $T$ of $\cal T$ with (fixed) volume $N_{d}$. (This sum can only be evaluated by making certain modifications. These are discussed later in chapters~\ref{chap:chap2} and~\ref{chap:chap3}.) Although $Z_{c}$ is a finite sum, the number of terms is, in general, very large. For example, even a $16^{2}$ Ising lattice has $2^{256} \approx 10^{77}$ possible spin configurations! This scale of calculation is obviously {\it far} beyond current computational capabilities. For most cases, the exact calculation of the canonical partition function is therefore effectively impossible. One can, however, calculate an {\it approximation} of $Z_{c}$ by sampling only a small proportion of the total number of configurations. Random sampling of configurations leads to very poor approximations because, typically, very few configurations contribute significantly to $Z_{c}$. One can overcome this problem using the {\it importance sampling} methods of Monte Carlo simulations. 

\subsubsection{Importance Sampling and Markov Chains}

The sampling of `important' configurations is achieved by generating a {\it Markov chain}. This is a sequence of configurations, propagated by performing a series of successive {\it local moves} on the ensemble. This means that any configuration in the Markov chain depends {\it only} on the preceding one. Markov chains are therefore, by their very nature, highly correlated. This fact must be taken into account when generating statistically independent data. The average value of an observable will converge to the thermal average as the Markov chain grows in length.

Local moves refer to algorithms which can transform one configuration into another. They are `local' in the sense that only a small portion of the configuration is altered, whose size is independent of the ensemble volume. For example, in the Ising model, the local move is a {\it single} spin flip. In the case of dynamical triangulations, the local moves are much more complicated. Unlike the Ising model, the transformation algorithm is dimension dependent. These moves will be reviewed later (see chapter~\ref{chap:chap2} for details). To ensure an accurate approximation of a thermal average, the local moves must be {\it ergodic} in the space of configurations. In other words, all configurations must be accessible to the Markov chain. Given that ${\cal P}(T \rightarrow T')$ is the normalised probability of transforming configuration $T$ into $T'$, then we insist that
\begin{equation}
\sum_{T'}{\cal P}(T\rightarrow T') = 1.
\end{equation}
This condition ensures that the Markov chain cannot come to an end. The Markov chain must also satisfy another condition known as {\it detailed balance}.
\begin{equation}
\exp(-S[T])\,{\cal P}(T\rightarrow T') = \exp(-S[T'])\,{\cal P}(T'\rightarrow T) \label{eq:detbal}
\end{equation}
This ensures that configurations are chosen with the correct probability distribution of $\exp(-S[T])$, provided the local moves are ergodic.

\subsubsection{Metropolis Algorithm}

A configuration may only be added to the Markov chain if accepted by an update acceptance algorithm. The Metropolis algorithm is most often used. Consider the transformation of configuration $T$ into $T'$. If the change in action is given by
\begin{equation}
\Delta S = S[T']-S[T],
\end{equation}
then the probability $\cal P$ of acceptance is as follows.
\[ {\cal P}(T\rightarrow T') = \left\{\begin{array}{ll}
\exp(-\Delta S) & \mbox{if $\Delta S >$ 0} \\
1 & \mbox{if $\Delta S \leq$ 0}
\end{array} \right. \] 
In other words, the update is automatically accepted providing the change in action is not positive. In practice, acceptance is determined in the following manner. If a pseudo-random number is found to be greater than $\exp(-\Delta S)$ then the update is accepted, otherwise it is rejected.

\vskip 5mm

\noindent
More often than not, the first few configurations of a Markov chain will be statistically unimportant. Therefore, in general, an ensemble requires a finite {\it thermalisation time} $\tau$ to evolve towards statistically important configurations. Any measurements taken prior to time $\tau$ will result in inaccurate estimates of thermal averages. Of course, successive configurations of Markov chains are highly correlated. Hence, in order to generate statistically independent data, one must allow a finite time for the ensemble to decorrelate. This is known as the {\it autocorrelation time}. The statistical treatment of independent and correlated data is explained in appendices~\ref{app:indep} and~\ref{app:binning} respectively. 

A well-known drawback of the Monte Carlo method is an effect known as {\it critical slowing down}. Near a critical point, one finds long range correlations in the ensemble. Under these circumstances, the number of local moves required to generate a statistically independent configuration grows very rapidly. In other words, the algorithm loses efficiency. 

\section{Dynamical Triangulations} \label{sec:dt}

So far, we have reviewed the historical development and certain technical aspects of simplicial quantum gravity and dynamical triangulations. The research of dynamical triangulations in `higher dimensions' ($d>2$) stems from the success of the theory in two dimensions. Let us now consider these extensions by first deriving the discretised Einstein-Hilbert action for arbitrary dimension and writing down the partition function. 

\subsection{Discretisation} \label{subsec:discretisation}

The continuum Einstein-Hilbert action $S[g_{\mu\nu},\Lambda,G]$ for a $d$-dimensional manifold $\cal M$ is defined as follows.
\begin{equation}
S[g_{\mu\nu},\Lambda,G] = \frac{1}{16 \pi G}\int_{\cal M}\mbox{d}^{d}\xi\,(\det g_{\mu\nu})^{\frac{1}{2}}\,(2\Lambda - R)
\end{equation}
We shall now formulate its discrete analogue $S[T]$, by considering each term of $S[g_{\mu\nu},\Lambda,G]$ individually. The cosmological constant term is obviously proportional to the volume of $\cal M$ and discretises to give
\begin{equation}
\int_{\cal M}\mbox{d}^{d}\xi\,(\det g_{\mu\nu})^{\frac{1}{2}} \longrightarrow V_{d}N_{d}, \label{eq:cosmo}
\end{equation}
where $V_{d}$ is the volume of a simplex. The volume of an equilateral simplex with link length $a$ is given by~\cite{kendall}
\begin{equation}
V_{d} = \frac{a^{\,d}}{d\,!}\!\left(\frac{d+1}{2^{\,d}}\right)^{\!\frac{1}{2}}\!\!.
\end{equation}

Given a $d$-dimensional simplicial manifold $T$, the deficit angle of a hinge $h$ in $T$ is defined as
\begin{equation}
\delta_{h} = 2\pi-n(h)\Theta_{d},
\end{equation}
where $\Theta_{d}$ is the dihedral angle subtended between two faces of a simplex. In arbitrary dimension~\cite{krasnod}, this is given by 
\begin{equation}
\Theta_{d} = \cos^{-1}\!\left(\frac{1}{d}\right)\!. \label{eq:dihedral}
\end{equation}
Therefore, the integrated curvature at a single hinge $h$ is
\begin{eqnarray}
\int_{h}\mbox{d}^{d}\xi\,(\det g_{\mu\nu})^{\frac{1}{2}}\,R & \longrightarrow & 2V_{d-2}\,\delta_{h} \label{eq:factor2} \\
& \longrightarrow & 2V_{d-2}\,(2\pi-n(h)\Theta_{d}).
\end{eqnarray}
By summing over all $N_{d-2}$ hinges of $T$ and using the fact that each simplex has ${}^{d+1}$C$_{d-1}$ hinges:
\begin{eqnarray}
\sum_{h\,\in\, T}n(h) & = & {}^{d+1}\mbox{C}_{d-1}\,N_{d} \\
& = & \frac{1}{2}\,d\;\!(d+1)N_{d}, 
\end{eqnarray}
we deduce that
\begin{eqnarray}
\int_{\cal M}\mbox{d}^{d}\xi\,(\det g_{\mu\nu})^{\frac{1}{2}}\,R & \longrightarrow & 2V_{d-2}\sum_{h\,\in\, T}\,(2\pi-n(h)\Theta_{d}) \\
& \longrightarrow & 2V_{d-2}\left(2\pi N_{d-2}-\frac{1}{2}\,d\,(d+1)N_{d}\Theta_{d}\right)\!. \label{eq:curv}
\end{eqnarray}
By combining and expanding (\ref{eq:cosmo}) and (\ref{eq:curv}), we get 
\begin{equation}
S[g_{\mu\nu},\Lambda,G] \longrightarrow S[T] = \frac{1}{16\pi G}\left(2\Lambda V_{d}N_{d}-4\pi V_{d-2}N_{d-2}+d\,(d+1)V_{d-2}N_{d}\Theta_{d}\right).
\end{equation}
This may be rewritten simply as
\begin{equation}
S[T,\kappa_{d-2}] = \kappa_{d}N_{d}-\kappa_{d-2}N_{d-2}, \label{eq:dicreteaction}
\end{equation}
where the constants $\kappa_{d}$ and $\kappa_{d-2}$ are given by
\begin{eqnarray}
\kappa_{d} & = & \frac{2\Lambda V_{d}+d\,(d+1)V_{d-2}\Theta_{d}}{16\pi G} \\
\kappa_{d-2} & = & \frac{V_{d-2}}{4G}. 
\end{eqnarray}
The exact form of $\kappa_{d}$ can vary slightly depending on how the cosmological constant is defined in the continuum Einstein-Hilbert action\,\footnote{The factor of 2 in~(\ref{eq:factor2}) is occasionally omitted, as in~\cite{acmgeometry}.}. This thesis is primarily concerned with dynamical triangulations of five dimensional simplicial manifolds. In this case the discretised Einstein-Hilbert action is obviously given by
\begin{equation}
S[T,\kappa_{3}] = \kappa_{5}N_{5}-\kappa_{3}N_{3}.
\end{equation}
Given the discretised Einstein-Hilbert action, one can formally write down the grand canonical partition function as
\begin{eqnarray}
{\cal Z}_{E} \longrightarrow Z(\kappa_{d-2},\kappa_{d}) & = & \sum_{T\,\in\,{\cal T}}\exp(-S[T,\kappa_{d-2}]) \\
& = & \sum_{ T\,\in\,{\cal T}}\exp(\kappa_{d-2}N_{d-2}-\kappa_{d}N_{d}).
\end{eqnarray}
The {\it canonical} partition function is defined as
\begin{equation}
Z_{c}(\kappa_{d-2},N_{d}) = \sum_{ T\,\in\,{\cal T}(N_{d})}\exp(\kappa_{d-2}N_{d-2}-\kappa_{d}N_{d}).
\end{equation}

Having explained the discretisation scheme in some detail, let us now move on to outline the general picture of dynamical triangulations in dimensions greater than two, which was first considered by Godfrey and Gross~\cite{godfreygross}.

\subsection{$d>2$ Dynamical Triangulations}

The first extension to three dimensional dynamical triangulations was made in the early 1990s~\cite{3dmatrixmodels,agishtein3d}. This work was soon followed by preliminary investigations of the four dimensional theory~\cite{agishtein4d,ambjorn4d}. Over the years, dynamical triangulations has become relatively well-established; a fact reflected by a number of review articles~\cite{david,ambjorncqg,johnston,burda} and books~\cite{acmgeometry,quantgeo}.

Three and four dimensional dynamical triangulations have two distinct phases which are fundamentally very similar in nature. The weak coupling phases (small $G$) correspond to elongated triangulations, which resemble branched polymers, and the strong coupling phases (large $G$) are characterised by `crumpled' triangulations with large (possibly infinite) fractal dimension. The phase transitions in both cases appear to be of first order, which is indicated by bistable behaviour, and hysteresis effects in three dimensions. This effectively eliminates the possibility of defining a continuum limit.

A fundamental distinction between three and four dimensions relates to the phenomenon of singular vertices. These vertices with unusually large local volumes are present in the strong coupling phase of four dimensions, but not in three dimensions. Studies have shown that they are significant in terms of the phase transition in four dimensions, though their role is not yet fully understood~\cite{hinptp}. This subject is dealt with in chapter~\ref{chap:chap6}. Other advances include the analytic study of stacked spheres, degenerate triangulations and simplicial manifolds with boundaries.

In recent years, much effort has been devoted to developing `modified models' of dynamical triangulations. Modifications include matter fields or new terms in the action. These advances are motivated by the hope that an expanded phase diagram may reveal extra phase structure which could lead to a continuum limit. It is in this spirit that we consider dynamical triangulations in five dimensions, as the most general action (linear in the components of the $f$-vector) has two independent coupling constants. It is hoped that this new perspective will yield a deeper understanding of dynamical triangulations in general. Before exploring the model in any detail, one must first show that it is well-defined. This involves proving the ergodicity of the five dimensional $(k,l)$ moves and the existence of an exponential bound (see chapters~\ref{chap:chap2} and~\ref{chap:chap3} respectively).

        	
\input epsf.tex

\chapter{Local Moves and Ergodicity} \label{chap:chap2}

\section{Local Moves} \label{sec:localmoves}

\begin{figure}[ht]
\leavevmode
\hbox{\epsfxsize=3.5cm \epsfbox{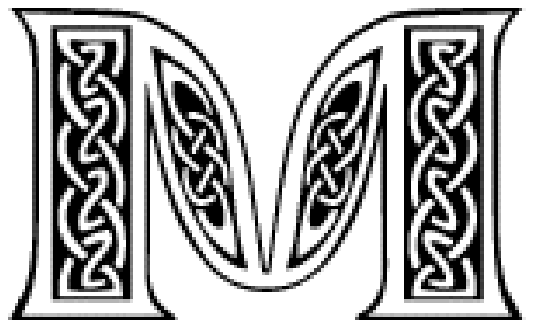}}
\parbox{12.4cm}{\baselineskip=16pt plus 0.01pt \vspace{-20mm}ONTE Carlo simulations have a wide scope of application in physics. These methods have so far been central to the research of dynamical triangulations. This chapter deals with certain technical aspects of the $\,$update $\,$algorithm in $\,$the $\,$context of five $\,$dimensional $\,$dynamical}
\end{figure}
\vspace{-7mm}
triangulations. To be precise, we shall clarify the meaning of the space of triangulations $\cal T$, describe the known local moves and attempt to prove their ergodicity in $\cal T$. 

The canonical partition function $Z_{c}(\kappa_{3},N_{5})$ cannot be computed in practice, as all known local moves change the triangulation volume $N_{5}$ (see later). To ensure ergodicity, one must therefore allow $N_{5}$ to fluctuate around a chosen {\it target volume} $N_{5}^{t}$. These fluctuations may be controlled by adding an extra Gaussian term to the action, thus redefining $Z_{c}(\kappa_{3},N_{5})$ as
\begin{equation}
Z'(\kappa_{3},N_{5}^{t},\gamma) = \sum_{T\,\in\,{\cal T}}\exp(\kappa_{3}N_{3}-\kappa_{5}N_{5}-\gamma(N_{5}-N_{5}^{t})^{2}),
\end{equation}
where $\gamma$ controls the degree of fluctuations in $N_{5}$. Let us now consider the subject of ergodicity in more detail. 
\vspace{1mm}

\subsection{Ergodicity} \label{subsec:ergodicity}

Local moves are ergodic if every configuration of a statistical ensemble is accessible in principle. This property is necessary for the accurate approximation of thermal averages. The matter of proving ergodicity is trivial in many statistical models. For example, in the Ising model the local move is a simple spin flip from $\uparrow$ to $\downarrow$ or vice versa. It is obvious that all configurations are eventually accessible, providing that spins are chosen at random. The flip local move is therefore manifestly ergodic.

Now consider the issue of ergodicity with regard to dynamical triangulations, beginning with the trivial case of one dimension, in which the only closed unbounded simplicial manifold is the circle $S^{1}$. The space of triangulations of $S^{1}$ is, of course, identical to the set of polygons. For a given volume $N_{1}$ there is only one possible triangulation, up to an automorphic relabelling of vertices (or links). The simplest local move which transforms $S^{1}$ triangulations entails exchanging one link for two, or vice versa. It is not even necessary to perform the moves on random simplices due to the ${\dbl Z}_{N_{1}}$ rotational symmetry. The moves are obviously ergodic in the space of $S^{1}$ triangulations. The one dimensional case is trivial because each simplex (link) has only {\it two} faces. Given that all faces of simplicial manifolds are {\it pairwise} connected, this results in a unique triangulation of any given volume. The trivial one dimensional case may be generalised to higher dimensions, in which the local moves become much more complicated. In general, a local move performed on a $d$-dimensional simplicial manifold $T$ involves the {\it retriangulation} of a small $B^{d}$ submanifold of $T$. Clearly, proving the ergodicity of local moves is a non-trivial matter for dimensions greater than one.

This brings us to an important aspect of the whole issue of ergodicity in dynamical triangulations. That is, over what space of triangulations are the moves required to be ergodic? For an Ising lattice of $N^{2}$ spins, the space of possible configurations is an unambiguous concept. It is simply the $2^{N^{2}}$ possible combinations of spins. There is no overcounting of symmetrical configurations because each configuration corresponds to a real physical state of the Ising lattice. 

\subsection{Combinatorial Equivalence} \label{subsec:combequiv}

For dynamical triangulations, the space of triangulations $\cal T$ requires precise and careful definition. Ideally, one hopes that 
\begin{equation}
\sum_{T\,\in\,{\cal T}} \longrightarrow \int_{\cal M}\frac{{\cal D}g_{\mu\nu}}{\mbox{Vol(Diff)}}
\end{equation}
is true in the continuum limit. As in the Ising model, there are triangulations which are related by automorphisms. These are equivalent by the relabelling of (sub)simplices and correspond to the same metric in the continuum. One can prevent the overcounting of symmetrical triangulations by including a factor that is equal to the inverse of the order of the automorphism group $C_{T}$ (of triangulation $T$). It is expected that this factor becomes less important for larger triangulations, as fewer are related by symmetry. Let us now formally define $\cal T$ based on this physical reasoning. 

Two simplicial manifolds $T$ and $T'$ are isomorphic if there exists a one-to-one correspondence between their subsimplices which preserves their connectivity. Such a mapping is denoted as $T\sim T'$. If no such map exists between $T$ and $T'$, then the triangulations are said to be {\it distinct}. Triangulation $T'$ is a {\it subdivision} of $T$ if every simplex of $T'$ is a simplex (or part of a simplex) of $T$. Two simplicial manifolds are {\it combinatorially equivalent} if they have isomorphic subdivisions\,\footnote{In the mathematical literature the term `combinatorial equivalent' often has the same meaning as `isomorphic'. This unfortunate situation can cause confusion!}. This is written as $T\approx T'$. Combinatorial equivalence is sometimes called a {\it piecewise linear homeomorphism}. Let $C$ represent the space of combinatorially equivalent triangulations. 

The partition function is (in our case) defined for fixed topology, therefore the measure is a summation over {\it distinct homeomorphic triangulations}. From our definitions, it is evident that $T\sim T'$ then $T\approx T'$. In other words, isomorphic triangulations are combinatorially equivalent. However, the opposite is not true in general. Combinatorially equivalent triangulations are homeomorphic. The conjecture that homeomorphic triangulations are combinatorially equivalent is known as the {\it Hauptvermutung}\,\footnote{German for `fundamental conjecture'.}. In other words, it states that homeomorphic triangulations have isomorphic subdivisions. It has been shown to be true for simplicial manifolds of dimension two~\cite{rado} and three~\cite{moise}, but false for $d>4$~\cite{milnor,kirby}. Let the space of homeomorphic triangulations be denoted by $H$. When the Hauptvermutung is true, then $C\subseteq H$ and the measure is equivalent to a summation over combinatorially equivalent triangulations modulo automorphisms. In general $C$ is comprised of a number of distinct automorphism classes. However, in the five dimensional case, the Hauptvermutung is false, therefore $C\subset H$. In other words, the measure is not equivalent to a summation over combinatorially equivalent triangulations modulo automorphisms.

\section{Alexander Moves} \label{sec:alexandermoves}

The Alexander moves were developed by Alexander in 1929, who called them {\it simple transformations}~\cite{alexander}. He showed that any {\it complex} could be transformed into any other {\it equivalent} complex using these topology preserving transformations. Alexander defined two complexes $K$ and $L$ to be equivalent if there exist {\it partitions} $K'$ and $L'$ respectively such that $K'$ and $L'$ are {\it congruent}. In his now classic paper, Alexander used the `algebra of complexes' to formalise his simple transformations. We shall now review the background mathematics leading to explanations of the above italicised terms.

\subsection{Algebra of Complexes}

A simplex $\sigma^{d}$ can be expressed as the product\footnote{Multiplication is associative, commutative and distributive.} of its $d+1$ constituent vertices $\sigma_{i}^{0}$, for $i=1,\ldots,d+1$. The set of vertices define the {\it point-set} of a simplex. Algebraically, $\sigma^{d}$ is defined as follows.
\begin{equation}
\sigma^{d} = \sigma_{1}^{0}\ldots\sigma_{d+1}^{0}
\end{equation}
The product of an $i$-simplex $\sigma^{i}$ by a $j$-simplex $\sigma^{j}$ is an $(i+j+1)$-simplex $\sigma^{i}\sigma^{j}$, providing that $\sigma^{i}$ and $\sigma^{j}$ do not intersect, i.e. $\sigma^{i}\cap\sigma^{j}=\emptyset$. A complex $K$ is a simplicial complex with the added constraint that each simplex in $K$ is of the same dimension. In algebraic terms, one can express a complex $K$ of $n$ simplices as the sum of its constituent simplices.
\begin{equation}
K = \sum_{i=1}^{n}\sigma^{d}_{i}
\end{equation}
In general, $K$ can be decomposed into a number of {\it subcomplexes}. For example, $K$ may be expressed as the sum of two non-intersecting subcomplexes $P$ and $Q$, such that $K = P+Q$. In general, the boundary of a complex $K$ is denoted by $\overline{K}$. Every subsimplex of a complex has an associated {\it complement}. The complement $J$ of an $i$-subsimplex $\sigma^{i}$ of a complex $K$ can be defined as the set of subsimplices such that $J\sigma^{i}$ is a subcomplex of $K$. By this definition, it is clear that $St(\sigma^{i})\equiv J\sigma^{i}$. One may then define the {\it remainder} subcomplex $Q$ of $K$ as $K-J\sigma^{i}$. 

A complex $K'$ is a partition of a complex $K$ if their point-sets coincide and if each simplex of $K'$ is a simplex or part of a simplex of $K$. Two complexes $K$ and $L$ (of the same dimension) are congruent if there is a one-to-one correspondence between the vertices of $K$ and the vertices of $L$. Following the convention set by Gross and Varsted~\cite{grossvarsted}, equivalent complexes will be renamed {\it Alexander equivalent}. We are now ready to review the simple transformations, or {\it Alexander moves} as they are now commonly called.

\subsubsection{Simple Transformations}

\begin{figure}[htp] 
\centering{\input{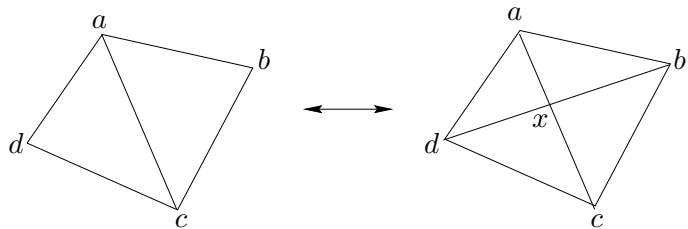}\par}
\caption{Two dimensional Alexander move.}
\label{fig:2dalex}
\end{figure}

Consider an $i$-(sub)simplex $\sigma^{i}$ of a complex $K$ with complement $P_{i}$ such that the remainder subcomplex is $Q$.
\begin{equation}
K = \sigma^{i}P_{i}+Q
\end{equation}
The set of $d+1$ Alexander moves can be written collectively as
\begin{equation}
K \longrightarrow K' = \sigma^{0}_{\ast}\,\overline{\sigma^{i}}P_{i} + Q,
\end{equation}
where $i=0,\ldots,d$, vertex $\sigma^{0}_{\ast}\notin K$ and $\overline{\sigma^{i}}P_{i}$ is the complement of $\sigma^{0}_{\ast}$. One may trivially define the inverse Alexander moves as 
\begin{equation}
K' \longrightarrow K = \sigma^{i}P_{i} + Q.
\end{equation}
Alexander proved that his set of $d+1$ transformations are in fact all equivalent to the move with $i=1$. This corresponds to a link, say the product of vertices $\sigma^{0}_{1}$ and $\sigma^{0}_{2}$, with complement $P$.
\begin{equation}
K = (\sigma^{0}_{1}\sigma^{0}_{2})P + Q \longleftrightarrow K' = \sigma^{0}_{\ast}(\sigma^{0}_{1}+\sigma^{0}_{2})P + Q \label{eq:alexmove}
\end{equation}
In other words, the remaining $d$ moves do not generate any additional complexes. Figure~\ref{fig:2dalex} illustrates the two dimensional Alexander move, where the link $ac$ has complement $b+d$ and $x$ is a new vertex. For this particular case one may rewrite the transformation explicitly as
\begin{equation}
ac(b+d) \longleftrightarrow x(a+c)(b+d).
\end{equation}

\section{$(k,l)$ Moves and Dual Graphs} \label{sec:kldual}

The $(k,l)$ moves were discovered more recently by Pachner~\cite{pachner1,pachner2}, who called them {\it elementary shellings}\,\footnote{They are sometimes referred to as the {\it Pachner moves}~\cite{lickorish} or {\it stellar subdivisions}~\cite{viro,barrett}.}. These topology preserving local moves transform $k$ simplices into $l$ simplices. The methods used by Pachner to develop his transformations are based on advanced principles of combinatorial topology which are beyond the scope of this thesis. In this section we merely explain the transformation algorithm in general terms.

\subsubsection{General Description}

\begin{figure}[htp] 
\centering{\input{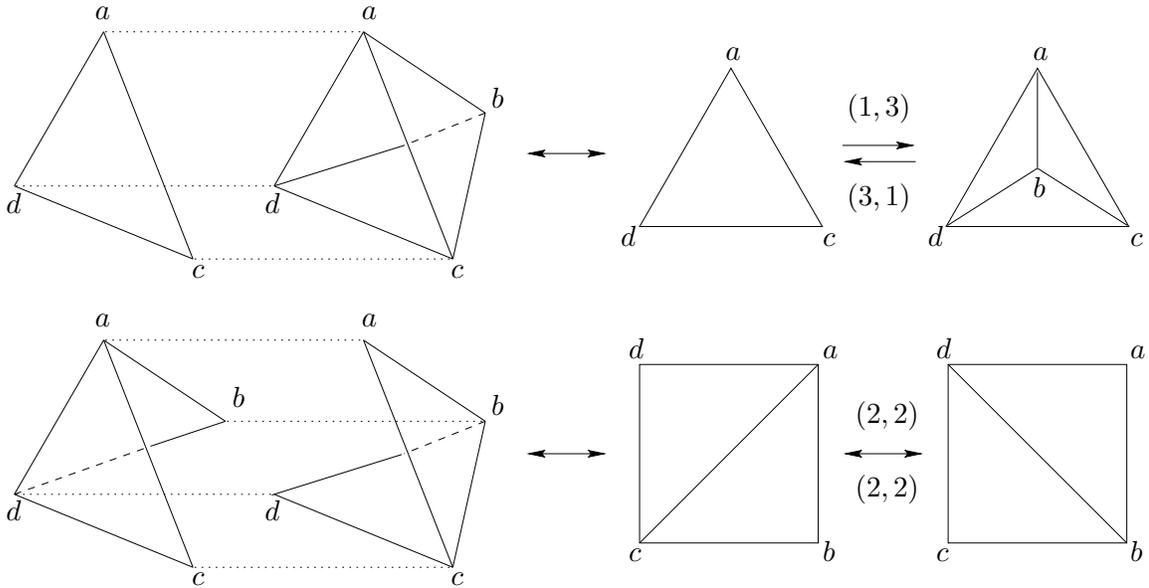}\par}
\caption{Two dimensional $(k,l)$ moves as the decompositions of a minimal 2-sphere.}
\label{fig:kl}
\end{figure}

The set of $(k,l)$ moves in $d$-dimensions may be regarded as the set of all possible decompositions of a minimal $d$-sphere into two disconnected $d$-balls. By this very definition, it is obvious that the moves preserve topology. In $d$-dimensions, the $(k,l)$ moves must satisfy
\begin{equation}
k+l = d+2, 
\end{equation}
since minimal $d$-spheres are composed of $d+2$ simplices. As an illustrative example, the two dimensional $(k,l)$ moves are represented by all the possible decompositions of a minimal 2-sphere, which is equivalent to the surface of a tetrahedron, see figure~\ref{fig:kl}. 

\subsection{Algebraic Formulation} \label{subsec:algform}

Let us now formalise the $(k,l)$ moves in terms of Alexander's algebra of complexes. A $d$-dimensional $(k,l)$ move may be equivalently considered as the replacement of an $i$-(sub)simplex of a complex $K$ with its {\it dual} $(d-i)$-(sub)simplex, where $i=0,\ldots,d$.  

One can define the $(k,l)$ moves in mathematical form for arbitrary $k$ and $d$. Let us first consider the case where $k=1$ and the complex $K$ is written in terms of a simplex $\sigma^{d}$ and the remainder complex $Q$.
\begin{equation}
K = \sigma^{d}+Q
\end{equation}
The $(1,d+1)$ move is defined as
\begin{equation}
K \longrightarrow K' = \sigma_{\ast}^{0}P_{0}+Q,
\end{equation}
where $\sigma_{\ast}^{0}\notin K$ and $P_{0}$ is the complement of $\sigma_{\ast}^{0}$. The complement $P_{0}$ is the $S^{d-1}$ subcomplex surface of $\sigma^{d}$. This is commonly known as the {\it vertex insertion} move. Its inverse, the {\it vertex deletion} move, is trivially defined as
\begin{equation}
K' = \sigma_{\ast}^{0}P_{0}+Q \longrightarrow K.
\end{equation}
Now consider the remaining moves with $k=2,\ldots,d$. Consider a $(d-k+1)$-subsimplex $\sigma^{d-k+1}$ of a complex $K$ with complement $P_{d-k+1}$.
\begin{equation}
K = \sigma^{d-k+1}P_{d-k+1}+Q
\end{equation}
A $(k,l)$ transformation is given by
\begin{equation}
K \longrightarrow K' = \sigma^{d-l+1}P_{d-l+1}+Q,
\end{equation}
where $\sigma^{d-l+1}$ is a $(d-l+1)$-subsimplex of $K$ with complement $P_{d-l+1}$. As an example, the transformations corresponding to the two dimensional $(k,l)$ moves pictured in figure~\ref{fig:kl} are given by 
\begin{eqnarray}
acd & \longleftrightarrow & (ac+ad+cd)b \\
ac(b+d) & \longleftrightarrow & (a+c)bd.
\end{eqnarray}

\subsection{Dual Graphs}

\begin{figure} 
\centering{\input{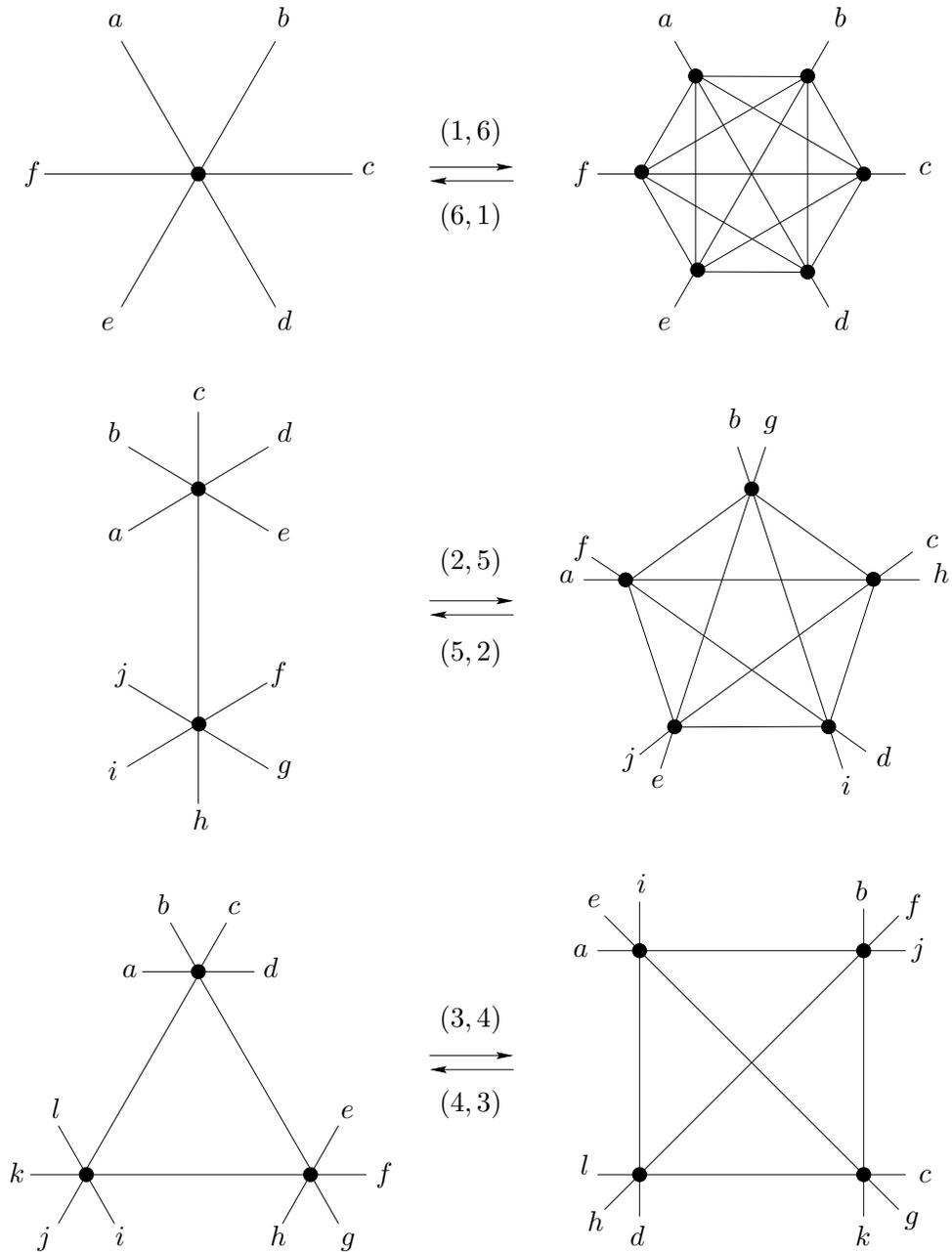}\par}
\caption{Five dimensional dual $(k,l)$ graphs. The free edges are labelled by letters $a$ to $l$.}
\label{fig:5ddual}
\end{figure}

As we have seen, the two dimensional $(k,l)$ moves are easily drawn and visualised. For higher dimensions, it is more convenient to transform the moves into their {\it dual} counterparts. This allows us to draw moves of dimension greater than two on paper. A dual $(k,l)$ {\it graph} is created by replacing each simplex with a {\it node} and each ($d-1$)-dimensional face by a {\it edge} between two nodes. A {\it free edge} (see figure~\ref{fig:5ddual}) is the dual of a $(d-1)$-sphere face. We shall now outline the general prescription for constructing dual graphs for arbitrary dimension. 
\begin{itemize}
\item Determine the initial $(k)$ and final $(l)$ number of simplices. Then draw their representative {\it nodes} -- say in the shape of a polygon.
\item Draw {\it edges} connecting every pair of nodes.
\item Draw extra free edges such that each node has $d+1$ attached edges. This corresponds to each simplex having $d+1$ faces.
\item Check that no two nodes are connected by more than one edge. This relates to the manifold condition that impose simplices to be pairwise connected. The five dimensional dual $(k,l)$ moves are shown in figure~\ref{fig:5ddual}.
\end{itemize}

\vskip 5mm

\noindent
From this duality, it is clear that an $i$-(sub)simplex of a $d$-dimensional triangulation is represented by a set of $d-i+1$ interconnected nodes of a dual graph. This relationship allows us to count the change in $N_{i}$ due to a $(k,l)$ move by calculating the change in $N_{d-i}^{dual}$, for $i=0,\ldots,d$.
\begin{equation}
\Delta N_{i} \equiv \Delta N_{d-i}^{dual} 
\end{equation}
$N_{d-i}^{dual}$ is the number of different combinations of $d-i+1$ nodes in the dual graph. The initial (final) dual graphs will have a total $k$ ($l$) nodes respectively. 
\begin{equation}
\Delta N_{i} = N_{d-i}^{dual,l} - N_{d-i}^{dual,k} 
\end{equation}
In other words we want to know the number of combinations of $d-i+1$ from $k$ and from $l$. This allows us to write down a very useful equation which gives $\Delta N_{i}$ for any $(k,l)$ move in any dimension $d$.
\begin{equation}
\Delta N_{i} = {}^{l}\mbox{C}_{d-i+1}-{}^{k}\mbox{C}_{d-i+1} \label{eq:deltani}
\end{equation}
Note that the equation is antisymmetric under the exchange of $k$ and $l$ as one would expect -- the $(l,k)$ move is the inverse of the $(k,l)$ move. Equation~(\ref{eq:deltani}) is used in later chapters. Having formally defined the local moves, let us now consider their ergodicity.

\subsection{Ergodicity of $(k,l)$ Moves} \label{sec:proofs}

In this chapter we are concerned only with proving the ergodicity of the {\it five} dimensional $(k,l)$ moves. It is felt that a general proof for arbitrary dimension may be rather ambitious and unnecessary for our purposes. Ideally, we want to show that the moves are ergodic in the space of combinatorially equivalent simplicial 5-manifolds. 

Alexander proved that his simple transformations are ergodic in the space of combinatorially equivalent simplicial manifolds. He achieved this by showing that two simplicial manifolds are Alexander equivalent if and only if they are combinatorially equivalent~\cite{alexander,glaser}. We shall attempt to construct a proof of ergodicity by following the same strategy as that taken by Gross and Varsted in their now widely cited paper~\cite{grossvarsted}. They showed that $(k,l)$ moves are ergodic in the space of combinatorially equivalent simplicial manifolds by proving that the $(k,l)$ moves are {\it equivalent} to the Alexander move. In other words, they proved that one can reproduce the effect of an Alexander move by a finite series of $(k,l)$ moves.

In two dimensions, it is easy to show that the Alexander move may be reproduced by a $(1,3)$ move followed by a $(2,2)$ move~\cite{grossvarsted}. In three and four dimensions the equivalence of the moves is much less obvious. This is seen only by first proving that 3-spheres and 4-spheres are {\it locally constructive}. A $d$-sphere may be transformed into a $d$-ball simply by removing a single simplex. For the purpose of this explanation, we call these structures {\it punctured} $d$-spheres. If a single simplex can be grown into a punctured $d$-sphere by the successive connection of simplices along $i$ faces (where $i=1,\ldots,d$) whilst preserving its topology, then the corresponding $d$-sphere is said to be locally constructive\,\footnote{A more precise mathematical definition of local constructivity is given by Gross and Varsted~\cite{grossvarsted}.}. In actual fact, we are interested in the locally constructivity of $S^{d-2}$ submanifolds. The relevance of this concept will become apparent later in section~\ref{sec:equivalence}.

\vskip 5mm

\noindent
For simplicity, let us first consider local constructivity of $S^{1}$ submanifolds of simplicial 3-manifolds. We begin with a single link, and then connect new links to either vertex of the 1-ball (line). Of course, any 1-ball with volume $N_{1}\geq 2$ is a punctured 1-sphere, since one can connect the two vertex boundaries with a link to form a 1-sphere (circle). Given that a link can be grown into a punctured 1-ball without a change in its topology, the 1-sphere is locally constructive. In the case of simplicial 4-manifolds, the submanifold in question is the 2-sphere. It has been shown that a single triangle {\it can} be grown into a punctured 2-ball (with volume $N_{2}\geq 3$) without a change in topology~\cite{grossvarsted}. For five dimensional triangulations, we are interested in proving that 3-spheres are locally constructive.

Local constructivity may seem an obvious property of $d$-spheres, but to {\it prove} this is a non-trivial matter for $d>1$. Gross and Varsted attempted to prove that the $(k,l)$ and Alexander moves are equivalent in arbitrary dimension $d$, but were unfortunately unable to prove the local constructivity of all $d$-spheres. In section~\ref{sec:equivalence} we prove that 3-spheres are locally constructive and use the result to show that the Alexander and $(k,l)$ moves are equivalent.

\section{Equivalence of Alexander and $(k,l)$ Moves} \label{sec:equivalence}

\begin{figure}[htp] 
\centering{\input{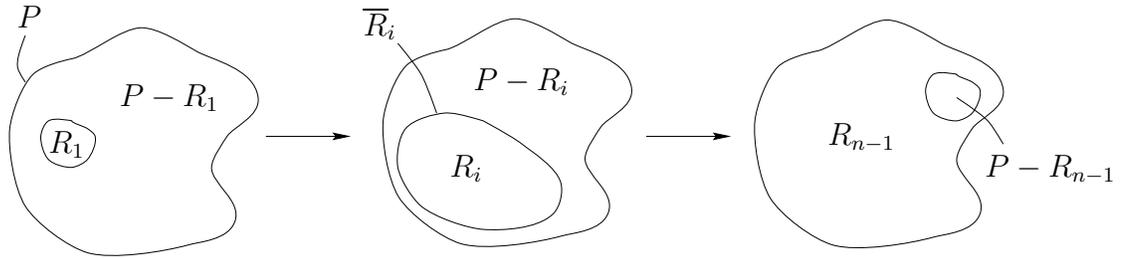}\par}
\caption{This illustration represents the two 3-balls $P$ and $P-R_{i}$ of the 3-sphere $P$, which share a 2-sphere boundary $\overline{R}_{i}$. In the first (last) picture $P-R_{1}$ ($R_{n-1}$) is a punctured 3-sphere.}
\label{fig:pr}
\end{figure}

Consider a closed five dimensional simplicial manifold $T$ which can be expressed as $T=(ab)P+Q$, where $P$ is the complement of the link $ab$ and $Q$ is the remainder submanifold of $T$. An Alexander move transforms $T$ to give
\begin{equation}
T \longrightarrow  T_{A} = x(a+b)P+Q, \label{eq:taualex}
\end{equation}
where $x$ is a vertex such that $x\not\in T$. The submanifold $P$ is a 3-sphere comprised of, say, $n$ tetrahedra. Let $ab$ be a link in $T$ and $pqrs$ be a tetrahedron in $P$. One can now express $T$ as follows. 
\begin{equation}
T = abpqrs+ab(P-pqrs)+Q
\end{equation}
We now perform a $(1,6)$ vertex insertion move on the simplex $abpqrs$, where $x\not\in T$.
\begin{equation}
abpqrs \longrightarrow (abpqr+abpqs+abprs+abqrs+apqrs+bpqrs)x
\end{equation}
This transforms $T$ into a new triangulation $T$$_{1}$.
\begin{equation} 
T \longrightarrow T_{1} = abx(pqr+pqs+prs+qrs)+x(a+b)pqrs+ab(P-pqrs)+Q
 \end{equation}
Now $T$$_{1}$ is of the form
\begin{equation}
T_{i} = abx\overline{R}_{i}+x(a+b)R_{i}+ab\;\!(P-R_{i})+Q, \label{eq:t16}
\end{equation}
where $\overline{R}_{1}=pqr+pqs+prs+qrs$ and $R_{1}=pqrs$. One could consider $R_{1}$ as a minimal 3-ball with $S^{2}$ boundary $\overline{R}_{1}$. In general, $R_{i}$ is a 3-ball submanifold of $P$, with 2-sphere boundary $\overline{R}_{i}$, see figure~\ref{fig:pr}.

The strategy of the proof involves subtracting tetrahedra from $P-R_{i}$ which are then added to $R_{i}$ by performing a series of $(k,l)$ moves. After $n$ moves we expect to have $R_{n}=P$ and $\overline{R}_{n}=0$. In this instance, (\ref{eq:t16}) becomes
\begin{equation}
T_{n} = x(a+b)P+Q = T_{A}.
\end{equation}
The resultant triangulation $T_{n}$ is identical to that generated by the Alexander move, given by transformation~(\ref{eq:taualex}). This process proves that the effect of an Alexander move may be reproduced by a finite series of $(k,l)$ moves. In other words, it is a proof of their equivalence.

\subsection{Local Constructivity}

\begin{figure}[htp] 
\centering{\input{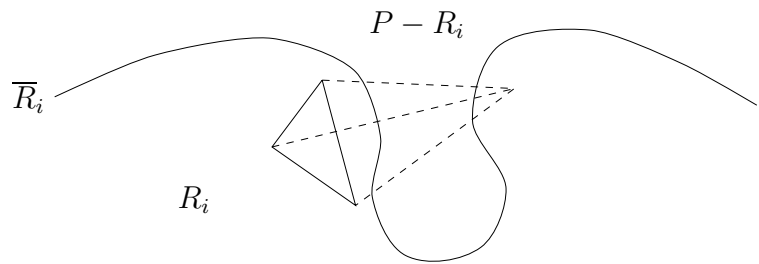}\par}
\caption{This illustration shows a tetrahedron in $P-R_{i}$ which shares one triangular face and four vertices with the 2-sphere $\overline{R}_{i}$.}
\label{fig:local}
\end{figure}

Every tetrahedron that is removed from $P-R_{i}$ and added to $R_{i}$ must satisfy the three conditions. These conditions ensure that the minimal 3-ball ${R}_{1}$ may be grown into a punctured 3-sphere ${R}_{n-1}$ without a change in its topology. The conditions insist that each tetrahedron of $P-R_{i}$ added to $R_{i}$ must:
\begin{itemize}
\item share at least one triangular face with the 2-sphere boundary $\overline{R}_{i}$. 
\item not share exactly one triangular face and four vertices with $\overline{R}_{i}$.
\item leave $T$$_{i}$ in the form of equation~(\ref{eq:t16}).
\end{itemize}
For the moment, let us concentrate on the first two conditions, which relate to the {\it local constructivity} of triangulations. Clearly, given that $T$ is a manifold, one can always find a tetrahedron of $P-R_{i}$ which shares at least one triangular face with $\overline{R}_{i}$. Therefore, the first condition is always satisfied. The importance of the second condition is illustrated by figure~\ref{fig:local}. One can see that a change in topology occurs if a tetrahedron of $P-R_{i}$ which shares one triangular face and four vertices with $\overline{R}_{i}$ is added to $R_{i}$. This ensures that the topology of $\overline{R}_{i}$ remains spherical, at least for $i=1,\ldots,n-1$.

\subsubsection{Proof of Local Constructivity}

\begin{figure}[htp] 
\centering{\input{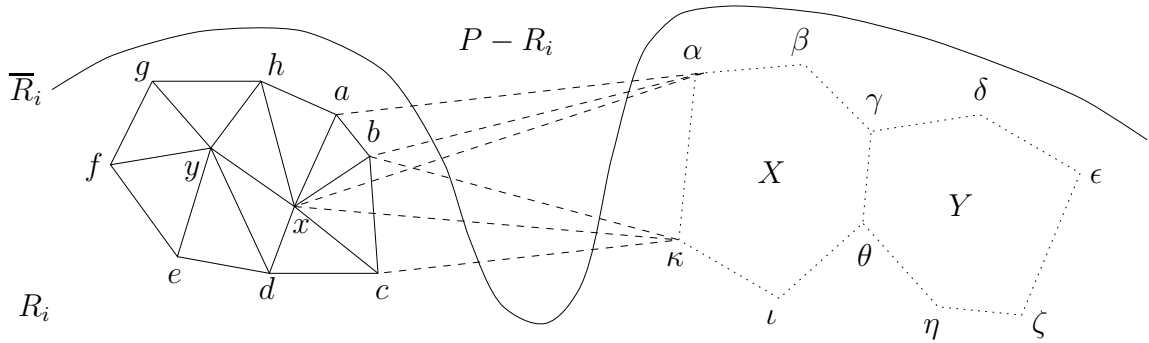}\par}
\caption{Tetrahedra of $P-R_{i}$ which share one face and four vertices with $\overline{R}_{i}$.}
\label{fig:localcon}
\end{figure}

One may show that the second condition can always be satisfied by proving that not {\it all} tetrahedra of $P-R_{i}$ can share one triangular face and four vertices with $\overline{R}_{i}$. This is done by assuming the opposite and showing that the resulting structure is not a manifold ({\it reductio ad absurdum}). This would prove the local constructivity of 3-spheres. Consider a tetrahedron $xab\alpha$ of $P-R_{i}$ formed by the product of triangle $xab$ and vertex $\alpha$ in $\overline{R}_{i}$. These structures are shown in figure~\ref{fig:localcon}. 

Given that $T$ is a manifold, one can always identify three triangular neighbours of $xab$ in $\overline{R}_{i}$. One of its neighbours, $xbc$ forms a tetrahedron $xbc\kappa$ with vertex $\kappa$. Clearly, $\alpha$ and $\kappa$ are connected by series of links in $\overline{R}_{i}$. The product of these links with $xb$ form a set of tetrahedra in $P-R_{i}$. One can then repeat this identification for every triangle that is common to vertex $x$. The corresponding tetrahedra are $xab\alpha$, $xbc\kappa$, $xcd\iota$, $xdy\theta$, $xyh\gamma$ and $xha\beta$. Vertices $\alpha$, $\beta$, $\gamma$, $\theta$, $\iota$ and $\kappa$ form a 1-sphere subcomplex of $\overline{R}_{i}$, which is the boundary of a 2-ball $X$. Again, if $T$ is a manifold, then the product of the triangles of $X$ with vertex $x$ {\it must} form a set of tetrahedra in $P-R_{i}$. These also have one face and four vertices in common with $\overline{R}_{i}$. The tetrahedra defined so far form a solid 3-ball handle attached to $\overline{R}_{i}$. 

One may then consider the triangles common to vertex $y$ and apply the same process. In this instance the product of triangles $yhx$, $yxd$, $yde$, $yef$, $yfg$ and $ygh$ with vertices $\gamma$, $\theta$, $\eta$, $\zeta$, $\epsilon$ and $\delta$ (respectively) form a set of tetrahedra in $P-R_{i}$. In the same way, one can identify a second 2-ball $Y$ in $\overline{R}_{i}$, which shares a series of links $\gamma\rightarrow\delta$ with $X$. The product of the triangles of $Y$ with vertex $y$ also form tetrahedra in $P-R_{i}$. 

\begin{figure}[htp] 
\centering{\input{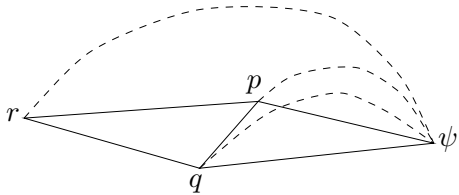}\par}
\caption{Example of a collapsed tetrahedron.}
\label{fig:collapsed}
\end{figure}

This process may be thought of as a mapping between two separate 2-balls of $\overline{R}_{i}$. One can continue to identify tetrahedra in this manner until almost every triangle of $\overline{R}_{i}$ is accounted for. Eventually, of course, the 2-balls will approach one another. This inevitably leads to collapsed tetrahedra and the breakdown of the manifold structure. In this case, one is forced to define a tetrahedron with only three vertices, which is impossible. This is illustrated in figure~\ref{fig:collapsed}, where a tetrahedron $pqr\psi$ is defined by the product of triangle $pqr$ and vertex $\psi$. This results in a collapsed tetrahedron $pq\psi$.

\noindent
These arguments have shown that $T$ is no longer a manifold if one insists that every triangle of $\overline{R}_{i}$ forms a tetrahedron in $P-R_{i}$ from the product of itself and another vertex of $\overline{R}_{i}$. 

\subsubsection{Additional Remarks}

After this work was completed, the author became aware that the concept of local constructivity was found to have relevance to the problem of enumerating the distinct triangulations of the 3-sphere~\cite{boulatov,dj3manifolds}. It was claimed that the existence of an exponential bound of the number of distinct 3-spheres may be proven analytically assuming that 3-spheres are locally constructive. It would appear that our result completes the proof. 

It was pointed out that the local constructivity of 3-spheres implies their algorithmic recognisability~\cite{dj3manifolds}. A manifold $\cal M$ is recognisable if there exists an algorithm which can prove that a second manifold $\cal M'$ is homeomorphic to $\cal M$. The algorithmic recognisability of the 3-sphere has already been proven~\cite{rubinstein,thompson}. In fact, it has also been proven that spheres of dimension greater than four\,\footnote{It is not known whether 4-spheres are recognisable.} are {\it unrecognisable}~\cite{novikov}. This result must imply that spheres of dimension greater than four cannot be shown to be locally constructive. One might expect our proof of local constructivity to generalise to higher dimensions. If this is the case, then one is forced to conclude that the proof is incomplete (or perhaps flawed). 

\vskip 5mm

\noindent
In order to complete the proof of equivalence one must show that every tetrahedron added to $R_{i}$ also satisfies the second condition. A tetrahedron $pqrs$ in $P-R_{i}$ may share one, two, three or four triangular faces with $\overline{R}_{i}$. Let us now treat each case individually to show that $T$$_{i}$ remains in the form of~(\ref{eq:t16}). 

\subsubsection{One Triangular Face Shared} \label{subsec:1tfs}

Let $pqr$ be a boundary triangle of $\overline{R}_{i}$. $pqs$, $prs$ and $qrs$ are internal triangles of $P-R_{i}$. In this case we perform a $(2,5)$ move on 5-simplices $abpqrs$ and $abpqrx$ as follows.
\begin{equation}
abpqr(s+x) \longrightarrow (abpq+abpr+abqr+apqr+bpqr)sx
\end{equation}
This results in the following transformation:
\begin{eqnarray}
T_{i} \longrightarrow T_{i+1} & = & abx(\overline{R}_{i}-pqr+pqs+prs+qrs) \nonumber \\
& & +x(a+b)(R_{i}+pqrs)+ab(P-R_{i}-pqrs)+Q \\
& = & abx\overline{R}_{i+1}+x(a+b)R_{i+1}+ab(P-R_{i+1})+Q 
\end{eqnarray}

\subsubsection{Two Triangular Faces Shared} \label{subsec:2tfs}

Let $pqr$ and $pqs$ be boundary triangles of $\overline{R}_{i}$. $prs$ and $qrs$ are internal triangles of $P-R_{i}$. In this case we perform a $(3,4)$ move on 5-simplices $abpqrx$, $abpqsx$ and $abpqrs$ as follows.
\begin{equation}
abrs(px+qx+pq) \longrightarrow (pqr+pqs+ars+brs)abx
\end{equation}
This results in the following transformation:
\begin{eqnarray}
T_{i} \longrightarrow T_{i+1} & = & abx(\overline{R}_{i}-pqr-pqs+prs+qrs) \nonumber \\
& & +x(a+b)(R_{i}+pqrs)+ab(P-R_{i}-pqrs)+Q \\
& = & abx\overline{R}_{i+1}+x(a+b)R_{i+1}+ab(P-R_{i+1})+Q 
\end{eqnarray}

\subsubsection{Three Triangular Faces Shared} \label{subsec:3tfs} 

Let $pqr$, $pqs$ and $prs$ be boundary triangles of $\overline{R}_{i}$. $qrs$ is an internal triangle of $P-R_{i}$. In this case we perform a $(4,3)$ move on 5-simplices $abpqrx$, $abpqsx$, $abprsx$ and $abpqrs$ as follows.
\begin{equation}
abp(qrx+qsx+rsx+qrs) \longrightarrow (ab+ap+bp)qrsx
\end{equation}
This results in the following transformation:
\begin{eqnarray}
T_{i} \longrightarrow T_{i+1} & = & abx(\overline{R}_{i}-pqr-pqs-prs+qrs) \nonumber \\
& & +x(a+b)(R_{i}+pqrs)+ab(P-R_{i}-pqrs)+Q \\
& = & abx\overline{R}_{i+1}+x(a+b)R_{i+1}+ab(P-R_{i+1})+Q 
\end{eqnarray}

\subsubsection{All Four Triangular Faces Shared} \label{subsec:4tfs} 

Now all four triangular faces of $pqrs$ are boundary triangles of $\overline{R}_{i}$. These correspond to the last move which transforms $T$$_{n-1}\longrightarrow T_{n}= T_{A}$. It is a $(5,2)$ move on the 5-simplices $abpqrx$, $abpqsx$, $abprsx$, $abqrsx$ and $abpqrs$ as follows.
\begin{equation}
ab(pqrx+pqsx+prsx+qrsx+pqrs) \longrightarrow (a+b)pqrsx
\end{equation}
This results in the following transformation:
\begin{eqnarray}
T_{i} \longrightarrow T_{i+1} & = & abx(\overline{R}_{i}-pqr-pqs-prs-qrs) \nonumber \\
& & +x(a+b)(R_{i}+pqrs)+ab(P-R_{i}-pqrs)+Q \\
& = & x(a+b)P+Q=T_{A}
\end{eqnarray}
since $P=R_{n-1}+pqrs$ and $\overline{R}_{n-1}$ is the boundary of a punctured 3-sphere given by $\overline{R}_{n-1}=pqr+pqs+prs+qrs$.

\vskip 5mm

\noindent
We have now shown that both conditions of local constructivity are satisfied. This proves that the Alexander and $(k,l)$ moves are equivalent in five dimensions. By showing that the Alexander and $(k,l)$ moves are equivalent we have also in effect proven that the $(k,l)$ moves are ergodic in the space of combinatorially equivalent simplicial manifolds $C$. $\Box$

\vskip 5mm

\noindent
Gross and Varsted were unable to fully generalise their arguments because local constructivity could not be proved for arbitrary dimension. Nevertheless, it was proved that the Alexander and $(k,l)$ moves are equivalent in all dimensions, assuming that spheres are locally constructive. Leading from comments made earlier, it seems that our proof of local constructivity certainly cannot be generalised to dimensions greater than four. In the context of ergodicity, we are interested in the local constructivity of $(d-2)$-spheres. Therefore, based on Novikov's result (that spheres of dimension greater than four are unrecognisable)~\cite{novikov}, one concludes that the proof of ergodicity cannot be generalised to spheres of dimensions greater than six. 

During the course of this work it was discovered that $(k,l)$ moves had already been proved to be ergodic in the space of combinatorially equivalent simplicial manifolds for arbitrary dimension~\cite{pachner1,pachner2}. One cannot extend the proof of ergodicity to the space of five dimensional homeomorphic triangulations, due to the negativity of the Hauptvermutung, which implies that $C\not\subseteq H$. In other words, there may exist some triangulations which cannot be reached by the Alexander or $(k,l)$ moves. The physical significance of this possibility is not clear. In fact, the Hauptvermutung is false in four dimensions and remains unsolved for the 4-sphere\,\footnote{This fact was pointed out to the author by W.~B.~R.~Lickorish~\cite{lick}.}. The result could have implications for four dimensional dynamical triangulations. 

\section{Non-computability} \label{sec:noncomp}

Any initial triangulation $T$ may be transformed into any other final triangulation $T'$ homeomorphic to $T$, via a {\it finite} series of $(k,l)$ moves~\cite{pachner1,pachner2}. At a practical level, it is hoped that the {\it number} of moves ${\cal N}_{(k,l)}$ required to transform $T$ into $T'$ is as small as possible. By this, we mean a number which modern computers can manipulate -- since a finite number can be unimaginably large. If $\cal N$$_{(k,l)}$ is not a computationally managable quantity then certain triangulations may effectively never be reached by a Markov chain. This would result in an {\it effective} breakdown in ergodicity. In this section we consider this concept in the context of five dimensional triangulations. 

\subsubsection{Unrecognisability of Manifolds} 

A manifold $\cal A$ is said to be {\it unrecognisable} if there exists no algorithm which can prove that a second manifold $\cal B$ is homeomorphic to $\cal A$. Unrecognisability is analogously defined for simplicial manifolds. This property of manifolds was shown to have relevance to dynamical triangulations by Ben-Av and Nabutovsky~\cite{nabutovsky}. They proved that the unrecognisability of a simplicial manifold $T$ implies that the number of moves ${\cal N}_{(k,l)}$ needed to transform one triangulation $T$ into a second $T'$ (both of volume $\leq N$) is not bounded by any recursive function of $N$. An example of a {\it very} fast growing but nevertheless {\it computable} (recursive) function $f$ is
\begin{equation}
f(N) = N^{N}\underbrace{\uparrow\uparrow\cdots\uparrow\uparrow}_{N^{N}} N^{N}\hspace{5mm}\mbox{for $N \in{\dbl N}$},
\end{equation}
using Knuth's up-arrow notation~\cite{knuth} where all powers associate to the right\,\footnote{$f(1)=1\uparrow1=1$, $f(2)=4\uparrow\uparrow\uparrow\uparrow4$ ({\it extremely} large).}.
\begin{eqnarray}
m\uparrow n & = & \underbrace{m\times m\times\cdots\times m\times m}_{n} = m^{n} \\
m\uparrow\uparrow n & = & \underbrace{m\uparrow m\uparrow\cdots\uparrow m\uparrow m}_{n} \\
m\uparrow\uparrow\uparrow n & = & \underbrace{m\uparrow\uparrow m\uparrow\uparrow\cdots\uparrow\uparrow m\uparrow\uparrow m}_{n} 
\end{eqnarray}

\subsubsection{Computational Ergodicity}

If ${\cal N}_{(k,l)}$ {\it is} bounded by a computable function then the moves are said to be {\it computationally ergodic}. Note that computational ergodicity of the $(k,l)$ depends on the topology of the simplicial manifold in question. The matter of computational ergodicity was first addressed in four dimensions by Nabutovsky and Ben-Av~\cite{nabutovsky} because certain 4-manifolds are known to be unrecognisable. The number of distinct triangulations (microcanonical partition function) $W(N_{4})$ of a simplicial 4-manifold appears to be bounded by a computable function of $N_{4}$, see chapter~\ref{chap:chap3}. With this in mind, one may infer that ${\cal N}_{(k,l)}$ is non-computable if $T$ can only be transformed into $T'$ (both of volume $\leq N_{4}$) via intermediate triangulations with volumes not bounded by computable functions of $N_{4}$. These intermediate triangulations may be interpreted as {\it barriers} in the space of triangulations.

Ambj{\o}rn and Jurkiewicz argued that such barriers should exist for all scales of volume. They devised an experiment to test for their existence in the triangulation space of $S^{4}$, whose recognisability is not known. This involved growing a minimal $S^{4}$ triangulation to a finite target volume and then shrinking it by increasing $\kappa_{4}$. If the volume eventually reduced to $N_{4}=6$ then this is taken as evidence supporting the non-existence of barriers in $S^{4}$. All independent triangulations tested were successfully shrunk to the minimal $S^{4}$. Ambj{\o}rn and Jurkiewicz reached the conclusion that either $S^{4}$ is recognisable, or, there are very few or no triangulations separated from the minimal $S^{4}$ by large barriers.

To test the significance of this result, de Bakker~\cite{bakkerbarriers} replicated the experiment for the 5-sphere, which is known to be unrecognisable~\cite{novikov}. Again, no evidence of barriers were found. This proved that the results presented by Ambj{\o}rn and Jurkiewicz tell us nothing about the recognisability of $S^{4}$. 

Given that $S^{5}$ is known to be unrecognisable, one wonders why no barriers were observed. Some possible explanation were suggested by de Bakker~\cite{bakkerbarriers}, which are also applicable to 4-spheres. Perhaps barriers are very large, even for small triangulations. Alternatively, the proportion of triangulations separated by large barriers in triangulations space may be very small. Consequently, the chance of observing such triangulations would also be very small. 

It seems that the whole issue of computational ergodicity is rather academic. Perhaps it is more relevant to consider whether ${\cal N}_{(k,l)}$ is computationally managable rather than computationally ergodic. Computationally managability refers to whether modern compters can perform ${\cal N}_{(k,l)}$ transformations over a reasonable time scale. It relates to the question of whether triangulations may be reached in practice. This is a more appropriate consideration since, as we have seen, even a recursive function can be extremely fast growing. How does this issue affect our study of dynamical triangulations in five dimensions? We have no evidence so far that barriers exist in the triangulation space of $S^{5}$. Even if 5-spheres were recognisable, there still may be triangulations separated by a prohibitively large number of moves. 

        

\input epsf.tex

\chapter{Exponential Bound} \label{chap:chap3}

\section{Canonical Partition Function} \label{sec:cpf}

\begin{figure}[ht]
\leavevmode
\hbox{\epsfxsize=2.7cm \epsfbox{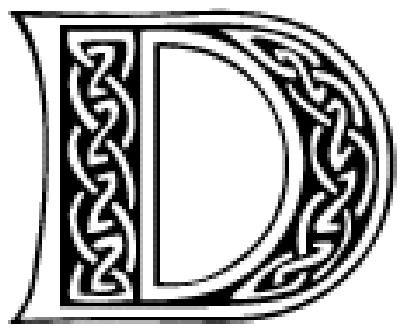}}
\parbox{13.2cm}{\baselineskip=16pt plus 0.01pt \vspace{-21mm}YNAMICAL triangulations is well-defined as a statistical model provided that the canonical partition function is an exponentially bounded function of the volume in the thermodynamic limit. The purpose of this chapter is to $\,$prove that this is $\,$indeed the case. $\,$We begin by $\,$reminding the reader}
\end{figure}
\vspace{-7.4mm}
that the grand canonical partition function of five dimensional dynamical triangulations is given by
\begin{equation}
Z(\kappa_{3},\kappa_{5}) = \sum_{T\,\in\,{\cal T}}\exp(\kappa_{3}N_{3}-\kappa_{5}N_{5}). \label{eq:xbcpf} 
\end{equation}
If $Z$ includes a summation over topology, then it is automatically divergent. In this thesis, the topology is fixed to that of 5-spheres -- though the following treatment may be applied to simplicial manifolds of any topology. Equation~(\ref{eq:xbcpf}) can be expressed equivalently as a sum over all volumes and a sum over all triangulations $T$ of a given volume $N_{5}$.
\begin{eqnarray}
Z(\kappa_{3},\kappa_{5}) & = & \sum_{N_{5}}\exp(-\kappa_{5}N_{5})\sum_{T\,\in\,{\cal T}(N_{5})}\exp(\kappa_{3}N_{3}) \\
& = & \sum_{N_{5}}\exp(-\kappa_{5}N_{5})\,Z_{c}(\kappa_{3},N_{5}) \label{eq:rewrite}
\end{eqnarray}
The canonical partition function $Z_{c}(\kappa_{3},N_{5})$ is a sum over all triangulations with fixed volume $N_{5}$. It is not possible to calculate $Z_{c}$ in practice because the five dimensional $(k,l)$ moves all change $N_{5}$. One must therefore allow the volume to fluctuate. Consequently, an extra term is added to the action in the grand canonical partition function.
\begin{equation}
Z'(\kappa_{3},N_{5}^{t},\gamma) = \sum_{T\,\in\,{\cal T}}\exp(\kappa_{3}N_{3}-\kappa_{5}N_{5}-\gamma(N_{5}-N_{5}^{t})^{2})
\end{equation}

\subsection{Entropy Function} \label{subsec:entfunc}

The microcanonical partition (entropy) function $W(N_{5})$ is obviously equal to $Z_{c}(\kappa_{3},N_{5})$ when $\kappa_{3}=0$. Explicitly, we have:
\begin{eqnarray}
{W}(N_{5}) & \equiv & \sum_{T\,\in\,{\cal T}(N_{5})}1 \\
& = & Z_{c}(0,N_{5})
\end{eqnarray}
Notice that in equation~(\ref{eq:rewrite}) the $\exp(-\kappa_{5}N_{5})$ factors tend to zero exponentially as $N_{5}\rightarrow\infty$. So, in order for $Z$ to be finite, the canonical partition function must be {\it exponentially bounded} from above as a function of volume.
\begin{eqnarray}
Z_{c}(\kappa_{3},N_{5}) & = & \sum_{T\,\in\,{\cal T}(N_{5})}\exp(\kappa_{3}N_{3}) \label{eq:bound1} \\
& \leq & \exp(\kappa_{5}^{c}N_{5}) \hspace{5mm} \mbox{as $N_{5}\rightarrow\infty$} \label{eq:bound}
\end{eqnarray}
Here $\kappa_{5}^{c}$ is the (finite) {\it critical simplex coupling}. If we now substitute 
\begin{equation}
Z_{c}(\kappa_{3},N_{5})=\exp(\kappa_{5}^{c}N_{5}) \label{eq:omegabound}
\end{equation}
into (\ref{eq:rewrite}), we get
\begin{equation}
Z = \sum_{N_{5}}\exp(-\kappa_{5}N_{5})\exp(\kappa_{5}^{c}N_{5}). \label{eq:divcon}
\end{equation}
It is now becomes clear that $Z$ is divergent if $\kappa_{5}^{c}\geq\kappa_{5}$ and convergent if $\kappa_{5}^{c}<\kappa_{5}$. From equations~(\ref{eq:bound1}) and (\ref{eq:bound}) we notice that $\kappa_{5}^{c}$ is a function of $\kappa_{3}$. In general, $\kappa_{5}^{c}$ is also a function of $N_{5}$, i.e.
\begin{equation}
\kappa_{5}^{c} = \kappa_{5}^{c}(\kappa_{3},N_{5}).
\end{equation}
This dependence on $N_{5}$ relates to power law subleading corrections to the exponential divergence. 

The partition function $Z$ is thus well-defined (finite) {\it only} if $\kappa_{5}^{c}$ is finite in the thermodynamic limit. The question we therefore must address is: does $\kappa_{5}^{c}$ have an asymptotic limit as $N_{5}\rightarrow\infty$? The existence of such an asymptote would imply an {\it exponential bound} on the number of distinct triangulations of a given topology as $N_{5}\rightarrow\infty$. In the next section we describe how this can be answered by numerical means. 

\section{Numerical Approach} \label{sec:numapp}

So far, we have shown that an exponential bound exists providing $\kappa_{5}^{c}$ has an asymptotic limit as $N_{5}\rightarrow\infty$. It is hoped that a numerical study will allow us to make a firm statement regarding this matter. Of course, numerical evidence could never constitute a proof of existence, or otherwise.

From equation~(\ref{eq:omegabound}) we see that the critical simplex coupling $\kappa_{5}^{c}(\kappa_{3},N_{5})$ is defined as
\begin{equation}
\kappa_{5}^{c}(\kappa_{3},N_{5}) \equiv \frac{\partial\ln Z_{c}(\kappa_{3},N_{5})}{\partial N_{5}}.
\end{equation}
Our Monte Carlo simulations evaluate the following sum.
\begin{eqnarray}
Z'(\kappa_{3},N_{5}^{t},\gamma) & = & \sum_{T\,\in\,{\cal T}}\exp(\kappa_{3}N_{3}-\kappa_{5}N_{5}-\gamma(N_{5}-N_{5}^{t})^{2}) \\
& = & \sum_{N_{5}}\exp(\ln Z_{c}(\kappa_{3},N_{5})-\kappa_{5}N_{5}-\gamma(N_{5}-N_{5}^{t})^{2}))
\end{eqnarray}
One can determine the most important contributions to $Z'(\kappa_{3},N_{5}^{t},\gamma)$ by partially differentiate with respect to $N_{5}$ and setting the result to zero.
\begin{equation}
0 = \sum_{N_{5}}(\kappa_{5}^{c}-\kappa_{5}-2\gamma(N_{5}-N_{5}^{t}))\exp(\ln Z_{c}(\kappa_{3},N_{5})-\kappa_{5}N_{5}-\gamma(N_{5}-N_{5}^{t})^{2}))
\end{equation}
Of course, $Z'(\kappa_{3},N_{5}^{t},\gamma)$ is not zero. This leaves us with the following relation.
\begin{equation}
\kappa_{5}^{c} = \kappa_{5}+2\gamma(N_{5}-N_{5}^{t})
\end{equation}
The saddle point approximation then relates the critical simplex coupling $\kappa_{5}^{c}$ to the mean volume $\langle N_{5}\rangle$.
\begin{equation}
\kappa_{5}^{c}(\kappa_{3},\langle N_{5}\rangle) = \kappa_{5}+2\gamma(\langle N_{5}\rangle-N_{5}^{t}) \label{eq:kappa5c}
\end{equation}

Here we describe the method of measuring $\kappa_{5}^{c}$ for a given target volume $N_{5}^{t}$. An initial arbitrary $\kappa_{5}$ is chosen at the beginning of the Monte Carlo simulation. Once the triangulation is fully grown, the simulation then passes into the thermalisation stage, see appendix~\ref{app:therm}. Then, $\kappa_{5}$ is tuned using the following iterative procedure.
\begin{equation}
\kappa_{5} \longrightarrow \kappa_{5}'=\kappa_{5}+2\gamma\,(N_{5}-N_{5}^{t}) \label{eq:tune}
\end{equation}
Throughout this thesis $\gamma$ is fixed to $\approx 0.001$. The simplex coupling $\kappa_{5}$ is updated after each tuning iteration. If $N_{5}>N_{5}^{t}$, then $\kappa_{5}$ is decreased, see equation~(\ref{eq:tune}). This means that the $(k,l)$ moves that increase $N_{5}$ have a lower change in action $\Delta S$, whereas moves that decrease $N_{5}$ have a higher $\Delta S$. This has the effect of increasing $N_{5}$. The opposite is true when $N_{5}<N_{5}^{t}$. The algorithm therefore tunes the system to the preset target volume.

Clearly, for a simplicial 5-manifold we have the inequality $N_{3}\gg N_{5}$. In fact, one may write $N_{3}=\alpha N_{5}$, where $\alpha$ is a function of $N_{5}$ and $\alpha>1$. Using this relation we may rewrite the canonical partition function as follows.
\begin{eqnarray}
Z_{c}(\kappa_{3},N_{5}) & = & \sum_{T\,\in\,{\cal T}(N_{5})}\exp(\kappa_{3}N_{3}) \\
& = & \exp(\alpha\kappa_{3}N_{5})\sum_{T\,\in\,{\cal T}(N_{5})}1
\end{eqnarray}
Therefore, in order to show that $Z_{c}(\kappa_{3},N_{5})$ is exponentially bounded we only need to prove this to be true for {\it one} value of $\kappa_{3}\geq 0$. For simplicity we choose to examine the divergence of $Z_{c}(0,N_{5})=W(N_{5})$, i.e. with $\kappa_{3}=0$. 

The pseudo-critical simplex coupling $\kappa_{5}^{c}$ was calculated using our measurements of $\langle N_{5} \rangle$, see equation (\ref{eq:kappa5c}). Calculation of the standard error in the mean volume $s_{m}(\langle N_{5} \rangle)$ allows us to evaluate the error in $\kappa_{5}^{c}$ using the following relation. 
\begin{equation}
\sigma(\kappa_{5}^{c}(\langle N_{5}\rangle)) = 2\gamma\,s_{m}(\langle N_{5}\rangle)
\end{equation}

\subsection{Exponential Bound for $d \leq 4$} \label{sec:dleq4}

The enumeration of distinct triangulations is a problem that was first addressed in the 1960s, though purely for mathematical motivations. It was Tutte who first proved that the number of distinct triangulations $W(N_{2})$ of a 2-manifold is exponentially bounded~\cite{tutte}. In fact, it has been shown that for a 2-manifold of genus $g$ we have
\begin{equation}
{W}(g,N_{2}) = \exp(\mu_{c}N_{2}){N_{2}}^{\gamma(g)-3}(1+\mbox{O}({N_{2}}^{-1})),
\end{equation}
where $\mu_{c}$ is the critical constant and $\gamma(g)=\frac{5}{2}g-\frac{1}{2}$~\cite{aventropy3d}. 

The generalisation of dynamical triangulations to dimensions greater than two naturally led to the question of whether such an exponential bound exists in these dimensions. This matter was first investigated numerically by Ambj{\o}rn and Varsted, who studied $S^{3}$ triangulations of volumes up to $N_{3}=14$k~\cite{aventropy3d}. It was found that their results were consistent with the existence of a bound. A second study using large volumes (up to $N_{3}=128$k) corroborated their findings~\cite{ckrentropy3d}. Attempts were subsequently made to construct an analytic proof -- with varying degrees of success~\cite{boulatov,dj3manifolds,mogami}. 

\vskip 5mm

\noindent
With no known formal proof in four dimensions, the exponential bound issue was also first considered numerically. Results produced by Catterall {\it et al.} (for volumes up to $N_{4}=32$k) seemed to indicate that $W(N_{4})$ was {\it not} exponentially bounded~\cite{ckrentropy4d}. This scenario corresponds to a factorial divergence of entropy
\begin{equation}
Z_{c}(\kappa_{0},N_{4}) \sim (N_{4}!)^{\delta},
\end{equation}
where $\delta$ is a positive constant\footnote{$\exp(x\delta \ln x) = x^{x\delta} \sim (x!)^{\delta}$}, estimated to be $\delta=0.026(5)$~\cite{ckrentropy4d}. Further numerical work was presented by Ambj{\o}rn and Jurkiewicz (with $N_{4} \leq 64$k), which contradicted the results of Catterall {\it et al.}~\cite{ajentropy4d}. The matter was resolved when a third study (with $N_{4} \leq 128$k) produced evidence consistent with the existence of an exponential bound~\cite{bmentropy4d}. More recently, certain advances have since been made towards an analytic proof~\cite{acmgeometry,bbcmcoverings,cmentropy,cmholonomy}. We now turn our attention to five dimensions. History has shown that analytic proofs are difficult to construct; so let us attack the problem by numerical means.

\section{Simulation Results} \label{sec:results}

One may establish the possible existence of an asymptotic limit, and hence an exponential bound, by first measuring $\kappa_{5}^{c}(\langle N_{5}\rangle)$ over a range of finite volumes $N_{5}$. The functional form of $\kappa_{5}^{c}(\langle N_{5}\rangle)$ may then be determined by fitting curves to the data using chi-squared hypothesis testing -- see appendix~\ref{app:curve} for details. Similar studies in four dimensions have shown that this approach can be very costly in terms of computational resources. Furthermore, the results require careful interpretation before any conclusion may be drawn. These facts influenced the planning of our strategy -- mainly in terms of the chosen target volumes. 

The most effective approach clearly entails striking a balance between the range $\cal R$ over which volumes are sampled, the number $\cal N$ of volumes we sample and the accuracy\footnote{In this sense `accuracy' is loosely interpreted as the inverse of the error in $\kappa_{5}^{c}$.} $\cal A$ with which we measure $\kappa_{5}^{c}$ for a given volume. It is reasonable to suppose that for a given amount of computational resources there exist optimal values of these variables that give the most reliable determination of $\kappa_{5}^{c}(\langle N_{5} \rangle)$. Calculating these optimal values is however rather tricky. 

It was first decided that $\cal N$ should be kept relatively small, thus avoiding the problems encountered by de Bakker~\cite{bakkersmitentropy}. Obviously, to have any hope of achieving our goal, we must have ${\cal N}\geq 3$. The accuracy $\cal A$ typically increases as ${\cal M}^{-\frac{1}{2}}$ where $\cal M$ is the number of independent measurements taken. Consequently, gains in accuracy will tend to diminish for longer Monte Carlo simulations. Our main concern is that $\cal A$ and $\cal R$ are chosen such that logarithmic divergence may be easily {\it distinguished} from power law convergence, see later. In other words, the accuracy should be sufficiently great that the form of $\kappa_{5}^{c}(\langle N_{5}\rangle)$ is compatible with {\it either} divergence {\it or} convergence, not both. 

After taking these issues into consideration, the number of volumes $\cal N$ was chosen to be five, ranging over an order of magnitude. Measurements of $N_{5}$ were taken over one million Monte Carlo sweeps for each target volume: 5k, 10k, 20k, 40k and 70k. The results are summarised in table~\ref{tab:xbresults}, with errors shown in parentheses.
\begin{table}[htp]
\begin{center}
\begin{tabular}{|c||c|c|c|c|c|} \hline
$N_{5}^{t}$ & 5\mbox{k} & 10\mbox{k} & 20\mbox{k} & 40\mbox{k} & 70\mbox{k} \\ \hline
$\langle N_{5} \rangle$ & 5000.1(2) & 10005.0(6) & 20020.0(1.0) & 40002.3(2) & 70011.8(1.6) \\ \hline
$\ln\langle N_{5} \rangle$ & 8.5172 & 9.2108 & 9.9045 & 10.5967 & 11.1564 \\ \hline
$\kappa_{5}^{c}(\langle N_{5} \rangle)$ & 0.8117(5) & 0.8277(6) & 0.8404(6) & 0.8498(4) & 0.8571(6) \\ \hline
\end{tabular}
\caption{Pseudo-critical simplex coupling $\kappa_{5}^{c}(\langle N_{5} \rangle)$ for target volumes $N_{5}^{t}$.}
\label{tab:xbresults} 
\end{center}
\end{table}

\subsection{Curve Fitting Analysis} \label{subsec:cfa}

\begin{figure}[ht]
\centerline{\epsfxsize=14cm \epsfbox{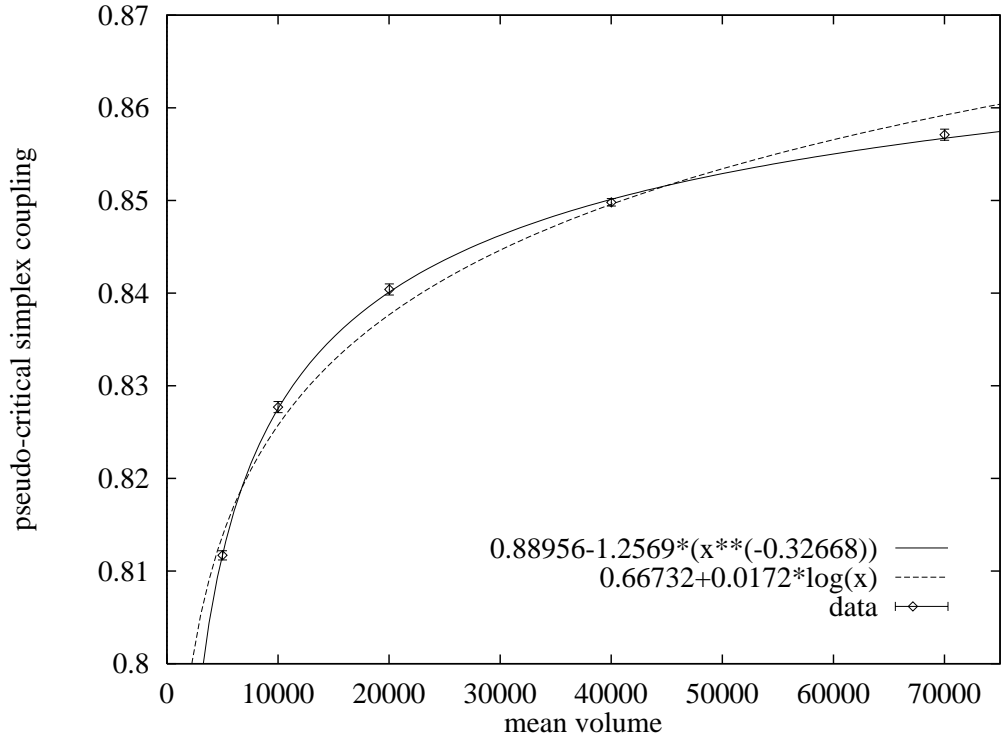}} 
\caption{Pseudo-critical simplex coupling $\kappa_{5}^{c}(\mbox{x})$ where $\mbox{x}=\langle N_{5} \rangle$, with best fit power law (solid) and logarithmic (dashed) curves. (In the key `log' actually means $\ln$.)}
\label{fig:basiclin}
\end{figure}

One may now fit curves to the data and hence determine whether or not $\kappa_{5}^{c}(\langle N_{5}\rangle)$ is finite as $N_{5}\rightarrow\infty$. We begin by fitting a power law curve of the form
\begin{equation}
\kappa_{5}^{c} = a-b\,\langle N_{5}\rangle{}^{-c}, \label{eq:powerform}
\end{equation}
which has an asymptote $\kappa_{5}^{c}\rightarrow a$ as $N_{5} \rightarrow \infty$, where $a$, $b$ and $c$ are positive. This curve is therefore consistent with the existence of an exponential bound. Conversely, a logarithmic curve of the form
\begin{equation}
\kappa_{5}^{c} = p+q \ln\langle N_{5}\rangle \label{eq:logform}
\end{equation}
does {\it not} have an asymptotic limit since $\kappa_{5}^{c}\rightarrow\infty$ as $N_{5}\rightarrow\infty$, where $q>0$. In other words, such a curve is incompatible with an exponential bound. Using the chi-squared method, the best fit curves were respectively found to be
\begin{eqnarray}
\kappa_{5}^{c} & = & 0.890(7)-1.3(4) \langle N_{5}\rangle{}^{-0.33(4)} \label{eq:power} \\ 
\chi^{2}  & \approx & 1.38 \hspace{3mm}\mbox{with 2 d.o.f.} \nonumber
\end{eqnarray}
and
\begin{eqnarray}
\kappa_{5}^{c} &  = & 0.667(4)+0.0172000(1) \ln \langle N_{5}\rangle \label{eq:ln} \\ 
\chi^{2} & \approx & 61.8 \hspace{3mm}\mbox{with 3 d.o.f.}, \nonumber
\end{eqnarray}
where the errors in the coefficients are shown in parentheses. These curves are plotted along with the numerical data of table~\ref{tab:xbresults} in figure~\ref{fig:basiclin}\,\footnote{The coefficients were not rounded off, in order to show the very best fit.}.

From figure~\ref{fig:basiclin} it is obvious that the power law fit is much better than the logarithmic fit. The power law curve passes almost {\it exactly} through all five data points. This may not be surprising given that the power law has fewer degrees-of-freedom. Nevertheless, the observed goodness-of-fit is proven by comparing their respective values of $\chi^{2}$ per degree-of-freedom. The power law fit has $\chi^{2}\approx 0.7$ per d.o.f., compared with $\chi^{2}\approx 21$ per d.o.f. for the logarithmic fit. (The errors in the measurements of $\langle N_{5}\rangle$ are very small, and are therefore neglected in the curve fitting analysis.)

\vskip 5mm

\begin{figure}[ht]
\centerline{\epsfxsize=14cm \epsfbox{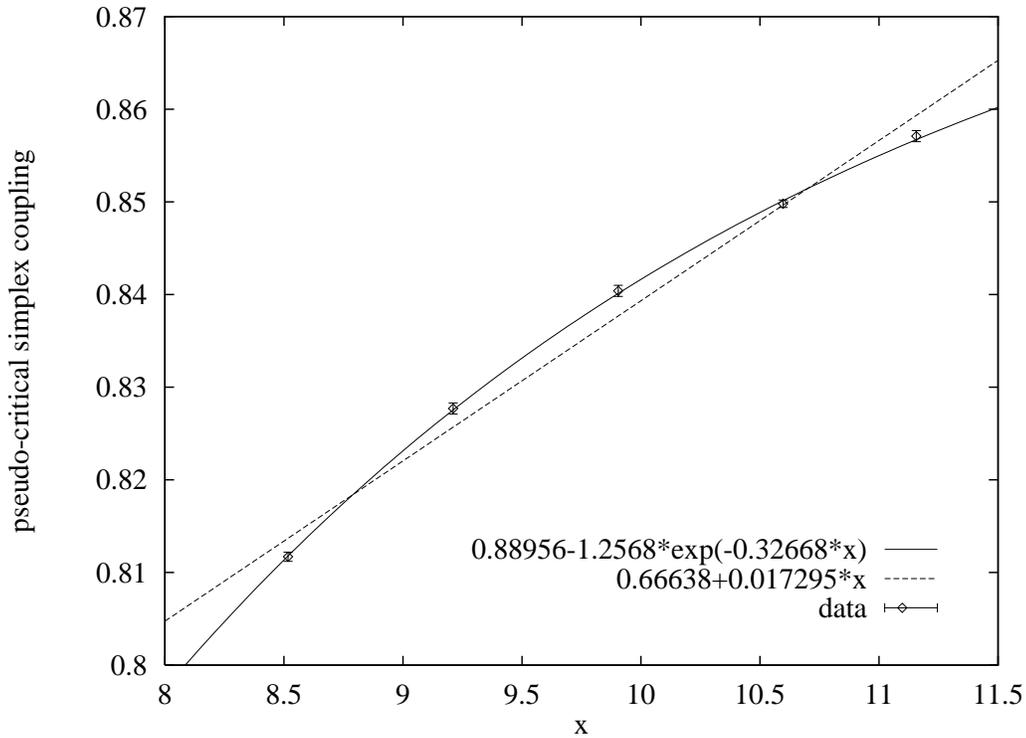}}
\caption{Pseudo-critical simplex coupling $\kappa_{5}^{c}(\mbox{x})$ where $\mbox{x}=\ln\langle N_{5}\rangle$, with best fit power law (solid) and logarithmic (dashed) curves.}
\label{fig:basiclog}
\end{figure}

\noindent
The goodness-of-fit of these curves can be made more apparent by plotting $\kappa_{5}^{c}(\langle N_{5} \rangle)$ against $\ln \langle N_{5} \rangle$. On such a graph a logarithmic divergence would appear as a straight line. This is done by making the following transformations.
\begin{eqnarray}
\label{eq:transform}
\mbox{x} & = & \ln \langle N_{5} \rangle \\
\kappa_{5}^{c} = a-b\,\langle N_{5} \rangle {}^{-c} & \longrightarrow & \kappa_{5}^{c} = a-b \exp(-c\mbox{x}) \label{eq:transpower} \\
\kappa_{5}^{c} = p+q \ln \langle N_{5} \rangle & \longrightarrow & \kappa_{5}^{c} = p+q\mbox{x} \label{eq:transln} 
\end{eqnarray}
Curves of the form (\ref{eq:transpower}) and (\ref{eq:transln}) were fitted to the numerical data. Not surprisingly, the coefficients $a, b, c, p$ and $q$ of (\ref{eq:transpower}) and (\ref{eq:transln}) were the same as those measured earlier for equations (\ref{eq:power}) and (\ref{eq:ln}). This exercise served as an useful check of the numerics. The best fit curves were found to be
\begin{eqnarray}
\kappa_{5}^{c} & = & 0.890(7)-1.3(3)\exp(-0.33(4)\mbox{x}) \label{eq:exppower} \\ 
\chi^{2}  & \approx & 1.40 \hspace{3mm} \mbox{with 2 d.o.f.} \nonumber
\end{eqnarray}
and
\begin{eqnarray}
\kappa_{5}^{c} & = & 0.662(2)+0.0173(2)\mbox{x} \label{eq:linln} \\ 
\chi^{2} & \approx & 61.6 \hspace{3mm} \mbox{with 3 d.o.f.}. \nonumber
\end{eqnarray}
These curves are plotted along with the data in figure~\ref{fig:basiclog}.
The best fit curves reinforce the result found earlier that the power law curves fit the data much better than the logarithmic curves. This body of evidence appears to be consistent with the existence of an exponential bound. 

\subsubsection{Three Parameter Logarithmic Fit}

\begin{figure}[htp]
\begin{center}
\leavevmode
(a){\hbox{\epsfxsize=14cm \epsfbox{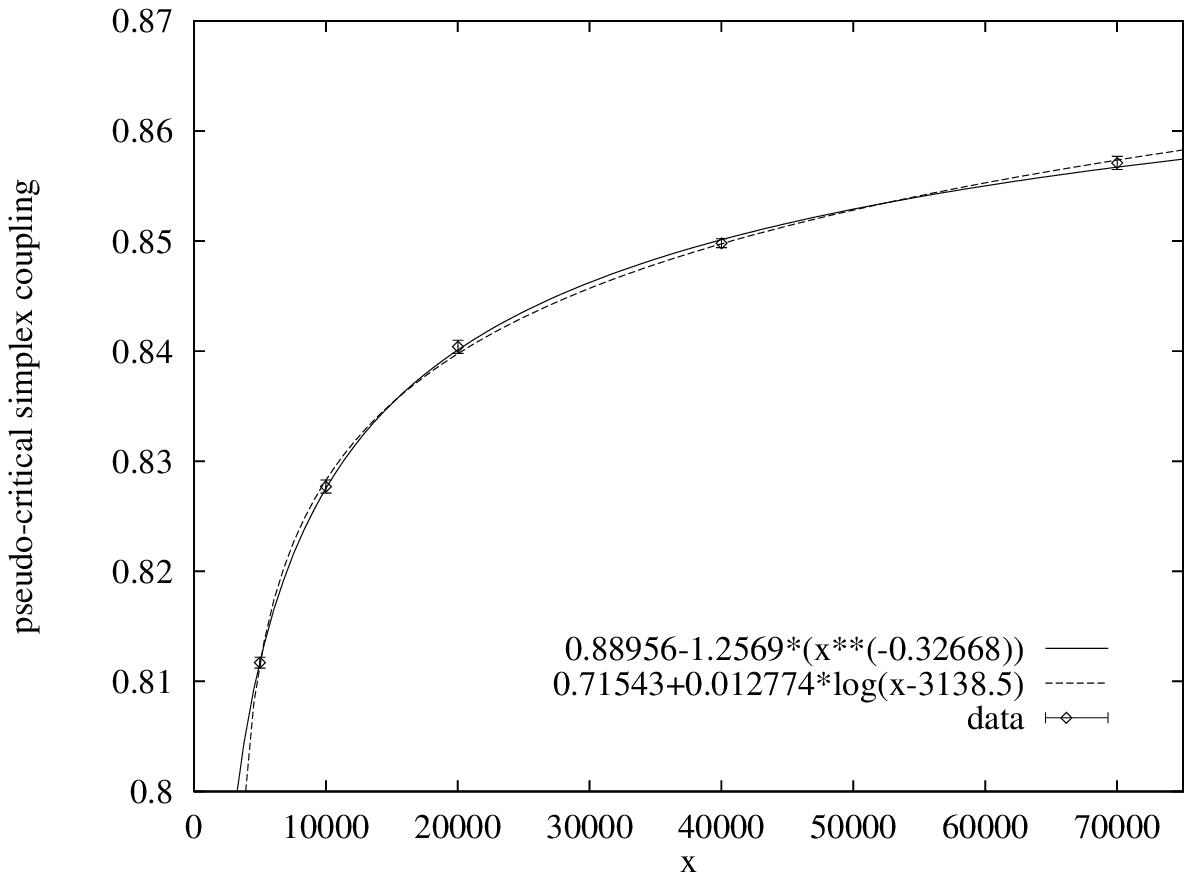}}}
(b){\hbox{\epsfxsize=14cm \epsfbox{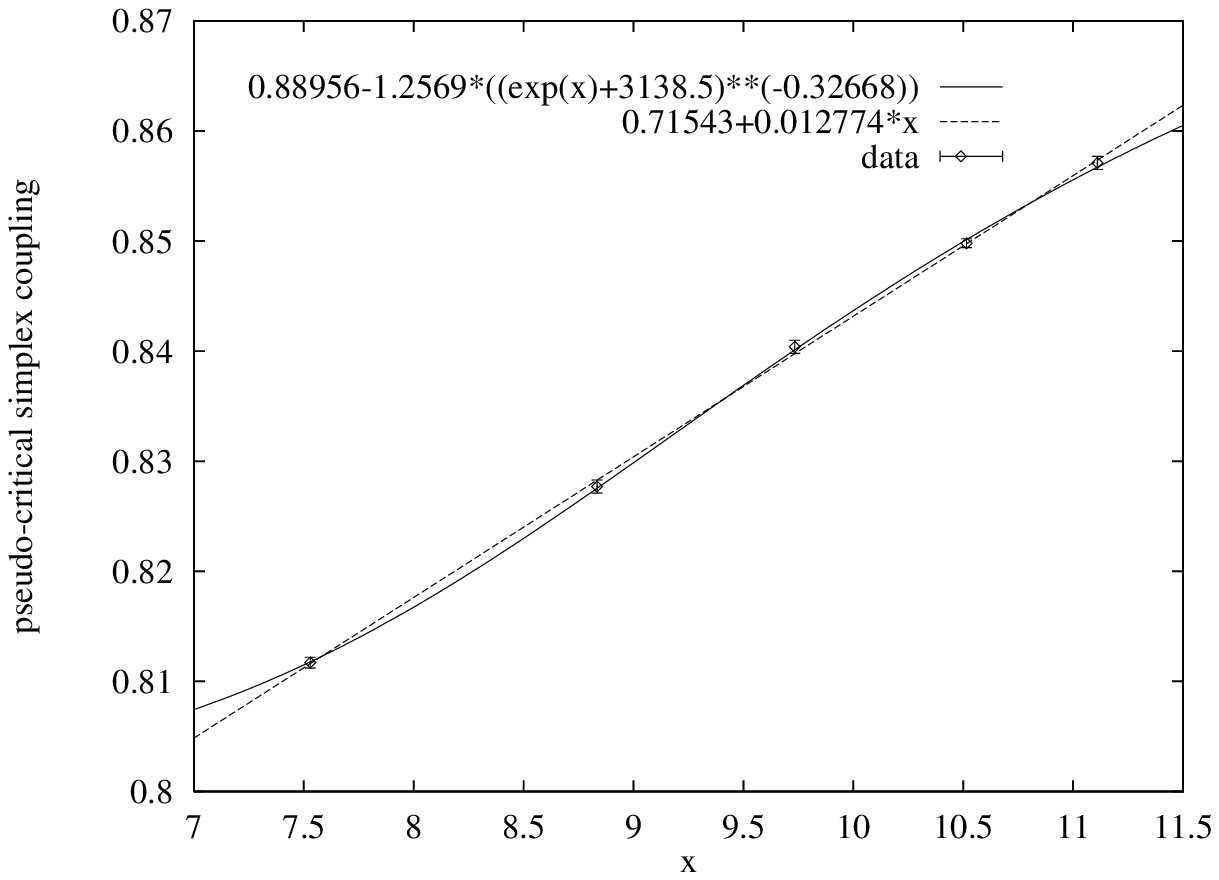}}}
\caption{(a) Pseudo-critical simplex coupling $\kappa_{5}^{c}(\mbox{x})$ where $\mbox{x}= \langle N_{5}\rangle$, with best fit power law (solid) and logarithmic (dashed) curves. (b) Pseudo-critical simplex coupling $\kappa_{5}^{c}(\mbox{x})$ where $\mbox{x}=\ln (\langle N_{5}\rangle-r)$, with best fit power law (solid) and logarithmic (dashed) curves.}
\label{fig:logshift}
\end{center}
\end{figure}

The logarithmic fit may be improved by introducing a third parameter in the form of a scale term $r$ to give
\begin{equation}
\kappa_{5}^{c} = p+q\ln(\langle N_{5}\rangle-r). \label{eq:lnshift}
\end{equation}
This functional form results in a much better fit of
\begin{eqnarray}
\kappa_{5}^{c} & = & 0.72(3)+0.013(2)\ln(\langle N_{5}\rangle-3139(12)) \label{eq:logshifteq} \\
\chi^{2} & \approx & 2.3 \hspace{3mm} \mbox{with 2 d.o.f.}. \nonumber
\end{eqnarray}
This new logarithmic fit is clearly much better than the previous fit of (\ref{eq:ln}). From figure~\ref{fig:logshift}(a), it is clear that fits (\ref{eq:power}) and (\ref{eq:logshifteq}) are equally good. 
Again, these curves are more easily compared when plotted against a log scale. This is done by making the following transformations
\begin{eqnarray}
\mbox{x} & = & \ln(\langle N_{5} \rangle-r) \\
\kappa_{5}^{c} = a-b \langle N_{5} \rangle {}^{-c} & \longrightarrow & \kappa_{5}^{c} = a-b\,(\exp(\mbox{x})+r)^{-c} \label{eq:transpowerlog} \\
\kappa_{5}^{c} = p+q \ln( \langle N_{5}\rangle-r) & \longrightarrow & \kappa_{5}^{c} = p+q\mbox{x} \label{eq:translnlog} 
\end{eqnarray}
Figure~\ref{fig:logshift}(b) also shows that both curves are very good fits to the data. 

In conclusion, both fits are compatible with the data. Therefore one cannot use these results to make a statement regarding the possible existence of an exponential bound in five dimensions. From figure~\ref{fig:logshift}(b) it is clear that the two curves begin to diverge steeply for $\mbox{x}$~\hbox{\begin{picture}(9,7)(0,2)\put(0,-2){\shortstack{$>$\\[-2pt]$\sim$}}\end{picture}}~11.5. This illustrates the fact that as $\cal R$ is increased, one fit will become progressively better than the other. It will become increasingly less likely that both curves give equally good fits.

\subsubsection{Smaller Triangulations}

\begin{figure}[ht]
\centerline{\epsfxsize=14cm \epsfbox{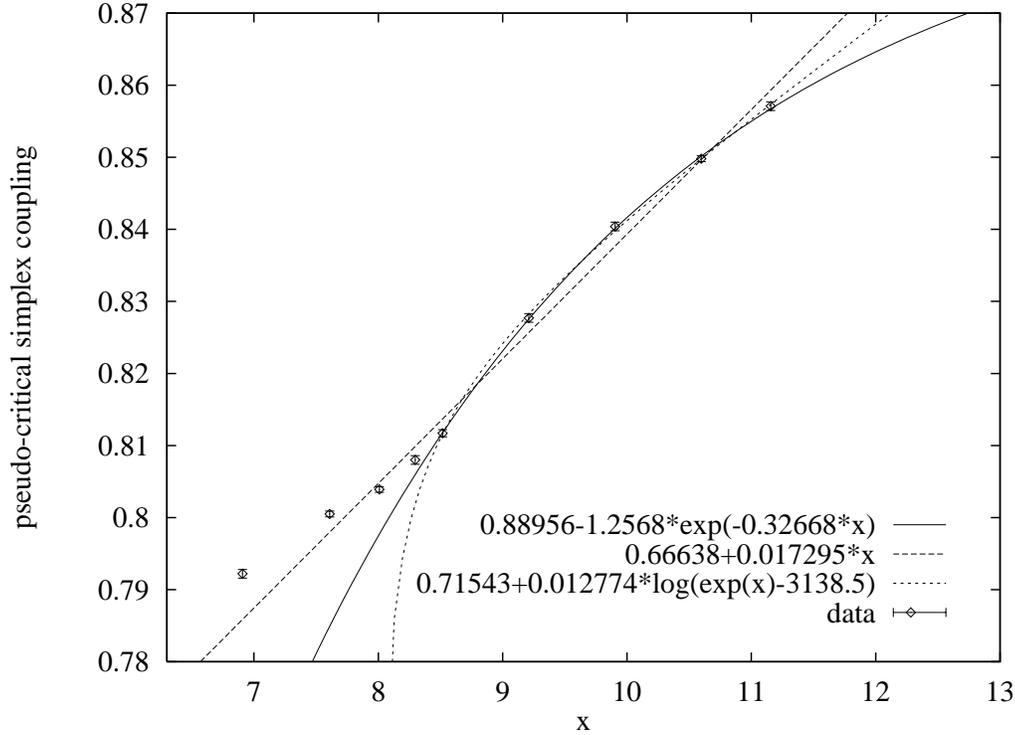}}
\caption{Pseudo-critical simplex coupling $\kappa_{5}^{c}(\mbox{x})$ where $\mbox{x}=\ln\langle N_{5}\rangle$, with best fit power law (solid), logarithmic (dashed) and scale term logarithmic (dotted) curves.}
\label{fig:all_log}
\end{figure}

Unfortunately, it was not really possible to measure $\kappa_{5}^{c}$ for volumes much greater than 70k; the limiting factor being computer power\,\footnote{For example, one million sweeps for $N_{5}^{t}=128$k takes at least 3 months, see appendix~\ref{app:perform}.}. Further measurements were made of $\kappa_{5}^{c}$ for $N_{5}^{t}<5$k, in order to check their compatibility with (\ref{eq:power}), (\ref{eq:logform}) and (\ref{eq:logshifteq}). Again, $N_{5}$ was measured over one million sweeps for target volumes 1k, 2k, 3k and 4k. The results are presented in figure~\ref{fig:all_log}.

It is plain to see that the new data is not even remotely consistent with any of the best fit curves. By differentiating (\ref{eq:powerform}), (\ref{eq:logform}) and (\ref{eq:lnshift}) twice with respect to volume we (respectively) get:
\begin{eqnarray}
\frac{\mbox{d}^{\!\,2}\kappa_{5}^{c}}{\mbox{d}N_{5}{}^{2}} & = & -bc(c+1){N_{5}}^{-(c+2)} \\
\frac{\mbox{d}^{\!\,2}\kappa_{5}^{c}}{\mbox{d}N_{5}{}^{2}} & = & -q{N_{5}}^{-2} \\
\frac{\mbox{d}^{\!\,2}\kappa_{5}^{c}}{\mbox{d}N_{5}{}^{2}} & = & -q\,(N_{5}-r)^{-2}
\end{eqnarray}
Obviously, these functions are negative for all $N_{5}$. If our numerical data is plotted against $N_{5}$ we see that the second derivative of $\kappa_{5}^{c}$ is positive near $N_{5}\approx 3$k ($\mbox{x}\approx 8$). This means that there exist no functions of the form (\ref{eq:powerform}), (\ref{eq:logform}) or (\ref{eq:lnshift}) that can accurately fit {\it all} the data\,\footnote{By `accurately' we mean: pass through each error bar.}. From this we can conclude that the data for $N_{5}^{t}<5$k is {\it not} described by the large $N_{5}$ limit of $\kappa_{5}^{c}(N_{5})$. In other words, $\kappa_{5}^{c}(N_{5})$ must have higher order corrections, thus giving either:
\begin{equation}
\kappa_{5}^{c} = a - b{{\langle N_{5}\rangle}^{-c}} + p{\langle N_{5}\rangle}^{q} + \cdots
\end{equation}
or
\begin{equation}
\kappa_{5}^{c} = s + t\ln\langle N_{5}\rangle + p{\langle N_{5}\rangle}^{q} + \cdots
\end{equation}
This shows that we must discard all data for $N_{5}<5$k from any curve fitting analysis. It is interesting that very similar effects were observed in four dimensions~\cite{ajentropy4d}. 

\section{Concluding Remarks} \label{sec:remarks}

In conclusion, it is not possible to determine whether $\kappa_{5}^{c}(N_{5})$ is finite as $N_{5}\rightarrow\infty$. Hence, one cannot make any firm statement regarding the existence of an exponential bound. The data is entirely consistent with convergent as well as divergent curves -- the power law fit is marginally better. The issue may only be resolved by measuring $\kappa_{5}^{c}$ for larger triangulations, as was the case in four dimensions~\cite{bmentropy4d}. As we increase the range $\cal R$ over which measurements are taken, it will become easier to differentiate between convergent and divergent data.

This whole approach raises an important question: how can we ever be sure that our data corresponds to the {\it true} large $N_{5}$ behaviour? In general, we expect subleading corrections to contribute significantly to $\kappa_{5}^{c}(N_{5})$ for small volumes -- but what are `small' volumes? This problem may be tackled (in principle) in the following way:
\begin{itemize}
\item First measure $\kappa_{5}^{c}$ over large $\cal R$ (orders of magnitude).
\item Divide the data into a small number of sets, say five.
\item Then determine the best fit coefficients for each set of data.
\item The final step involves comparing the sets of coefficients.
\end{itemize}
If the coefficients are the same (within errors) over a series of consecutive sets, then the fit is likely to represent the true large $N_{5}$ behaviour. Of course, the measurement of $\kappa_{5}^{c}$ over orders of magnitude is no easy task in practice.
        

\input epsf.tex

\chapter{Phase Structure} \label{chap:chap4}

\section{Action} \label{sec:action}

\begin{figure}[ht]
\leavevmode
\hbox{\epsfxsize=2.25cm \epsfbox{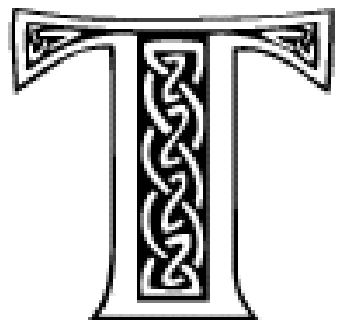}}
\parbox{13.65cm}{\baselineskip=16pt plus 0.01pt \vspace{-20mm}HE action normally used in dynamical triangulations is the discretised Einstein-Hilbert action $S[T,\kappa_{d-2}]$. Like its continuum counterpart $S[g_{\mu\nu},\Lambda,G]$, it has two terms $\!\,$derived from the physical notions of $\!\,$spacetime $\!$curvature and the cosmological constant $\Lambda$, which are linear in $N_{d-2}$ and $N_{d}$}
\end{figure}
\vspace{-6.75mm}
respectively. For the following analysis, it is convenient to group the $d+1$ elements of the $f$-vector into a set $\cal N$$_{\!d}$.
\begin{equation}
{\cal N}_{d} = \{ N_{0},\ldots,N_{d} \} \label{eq:simpvar}
\end{equation} 
In this thesis we do not restrict ourselves to the study of dynamical triangulations with the conventional Einstein-Hilbert action, which is derived for arbitrary dimension in section~\ref{subsec:discretisation}.
\begin{equation}
S[T,\kappa_{d-2}] = \kappa_{d}N_{d}-\kappa_{d-2}N_{d-2} \label{eq:ehdt}
\end{equation}
Instead, we consider the {\it most general} action $S_{d}^{g}$ that is linear in the members of $\cal N$$_{\!d}$ for a given dimension $d$. Let $S_{d}^{g}$ be a linear function of the members of a set $\cal N$$_{\!\:\!d}^{g}$, where
\begin{equation}
{\cal N}_{d}^{g} \subset {\cal N}_{d}. \label{eq:subset}
\end{equation}
One may then define a second subset ${\cal N}_{d}'$ of ${\cal N}_{d}$ such that
\begin{equation}
{\cal N}_{d}^{g} \cup {\cal N}_{d}' \subseteq {\cal N}_{d}. \label{eq:union}
\end{equation}
If the knowledge of $\cal N$$_{\!d}^{g}$ is sufficient to determine the members of $\cal N$$_{\!d}'$ using the Euler-Poincar\'{e} and Dehn-Sommerville relations, then $S_{d}^{g}$ is the most general action. Furthermore, the definition of $S_{d}^{g}$ stipulates that the number of members of $\cal N$$_{\!d}^{g}$ must be minimal and that $N_{d}\in{\cal N}_{d}^{g}$. 

In this case, the discretised Einstein-Hilbert action corresponds to a certain limit of the most general action. Each term in the action has an associated coupling constant. Therefore by taking this approach we, in effect, consider a larger space of parameters. However, as we shall see, this is not always the case. The question we must ask ourselves is: for a given dimension $d$, knowledge of which $\cal N$$_{\!d}^{g}$ allows us to calculate the members of $\cal N$$_{\!d}^{\prime}$? This question is now answered for dimensions two to five inclusively\,\footnote{This is done formally using Gaussian elimination. Basically, the number of terms of the action is $d+1-n$, where $n$ is the number of independent equations relating the $d+1$ variables.}.

For $d=2$, there are two linearly independent Euler-Poincar\'{e} and Dehn-Sommerville relations, (\ref{eq:dsep2d1}) and (\ref{eq:dsep2d2}), between three variables $N_{0}$, $N_{1}$ and $N_{2}$. This means that the most general action has only one term -- which is linear in $N_{2}$.
\begin{eqnarray}
N_{0} - N_{1} + N_{2} & = & 2 \label{eq:dsep2d1} \\
2 N_{1} - 3 N_{2}     & = & 0 \label{eq:dsep2d2}
\end{eqnarray}

For $d=3$, the four variables $N_{0}$, $N_{1}$, $N_{2}$ and $N_{3}$ are related by two linearly independent Euler-Poincar\'{e} and Dehn-Sommerville relations: (\ref{eq:dsep3d1}) and (\ref{eq:dsep3d2}).
\begin{eqnarray}
N_{0} - N_{1} + N_{2} - N_{3} & = & 0 \label{eq:dsep3d1}\\
N_{2} - 2 N_{3}               & = & 0 \label{eq:dsep3d2}
\end{eqnarray}
In this case, the most general action $S_{3}^{g}$ has {\it two} terms. It is important to note that $S_{3}^{g}$ cannot be a function of {\it any} two terms. For example, knowledge of $N_{2}$ and $N_{3}$ does not allow us to calculate $N_{0}$ and $N_{1}$.

For $d=4$, there are five variables $N_{0}$, $N_{1}$, $N_{2}$, $N_{3}$ and $N_{4}$ related by {\it three} linearly independent Euler-Poincar\'{e} and Dehn-Sommerville relations: (\ref{eq:dsep4d1}), (\ref{eq:dsep4d2}) and (\ref{eq:dsep4d3}). Again, this means that the most general action $S_{4}^{g}$ has two terms. Similarly, knowledge of $N_{3}$ and $N_{4}$ does not allow us to solve for the remaining variables.
\begin{eqnarray}
N_{0} - N_{1} + N_{2} - N_{3} + N_{4} & = & 2 \label{eq:dsep4d1} \\
2 N_{1} - 3 N_{2} + 4 N_{3} - 5 N_{4} & = & 0 \label{eq:dsep4d2} \\
2 N_{3} - 5 N_{4}                     & = & 0 \label{eq:dsep4d3}
\end{eqnarray}
It is a mere coincidence that the number of terms in $S_{3}^{g}$ and $S_{4}^{g}$ is the same as in the Einstein-Hilbert action $S[T,\kappa_{d-2}]$. It is important to remember that there is nothing physically significant about this fact. 

Usually, in practice, the action used in the study of dynamical triangulations is $S[T,\kappa_{d-2}]$, defined by equation~(\ref{eq:ehdt}). However, in certain cases a different, but equivalent, action $S[T,\kappa_{0}]$ is utilised in which the `curvature term' is linear in $N_{0}$.
\begin{equation}
S[T,\kappa_{0}] = \kappa_{d}N_{d}-\kappa_{0}N_{0} \label{eq:sk0tau}
\end{equation}
This is done purely for practical reasons; Monte Carlo simulation computer programs may be written for arbitrary dimension very efficiently in this way~\cite{catterallprogram} (see appendix~\ref{app:dtcp}). As we shall see, this is not generally possible for $d>4$.

\subsubsection{Five Dimensions}

The situation becomes a little more complicated in $d=5$ as we have six variables $N_{0}$, $N_{1}$, $N_{2}$, $N_{3}$, $N_{4}$ and $N_{5}$ constrained by three linearly independent Euler-Poincar\'{e} and Dehn-Sommerville relations: (\ref{eq:dsep5d1}), (\ref{eq:dsep5d2}) and (\ref{eq:dsep5d3}).
\begin{eqnarray}
N_{0}-N_{1}+N_{3}-3 N_{5} & = & 0 \label{eq:dsep5d1} \\
N_{2}-2N_{3}+5N_{5}       & = & 0 \label{eq:dsep5d2} \\
N_{4}-3N_{5}              & = & 0 \label{eq:dsep5d3} 
\end{eqnarray}
This means that the most general action $S_{5}^{g}$ has {\it three} terms\,\footnote{This fact was first pointed out by de Bakker~\cite{bakkerbarriers}.}. We are now faced with a situation where $S_{5}^{g}$ has {\it more} terms than $S[T,\kappa_{3}]$. It is important to note that $S[T,\kappa_{3}]$ cannot be rewritten in the form of (\ref{eq:sk0tau}), because it is impossible to express $N_{3}$ in terms of $N_{0}$ and $N_{5}$ alone. Again, the most general action cannot be expressed in terms of {\it any} set of three variables. Using equations~(\ref{eq:dsep5d1}), (\ref{eq:dsep5d2}) and (\ref{eq:dsep5d3}) we find that knowledge of $\cal N$$_{\!d}^{g}=\{N_{0},N_{3},N_{5}\}$ allows us to compute the values of $\cal N$$_{\!d}^{\prime}=\{N_{1},N_{2},N_{4}\}$. We therefore choose our most general action to be
\begin{equation}
S_{5}^{g}[T,\kappa_{0},\kappa_{3}] = \kappa_{5}N_{5}-\kappa_{3}N_{3}-\kappa_{0}N_{0}. \label{eq:5ds530}
\end{equation} 
Minor modifications were made to the Metropolis algorithm in order to accommodate this new action; details of which are given in appendix~\ref{app:dtcp}. Of course, an equally acceptable action would be
\begin{equation}
S_{5}^{g}[T,\kappa_{1},\kappa\,'_{\!3}] = \kappa\,'_{\!5}N_{5}-\kappa\,'_{\!3}N_{3}-\kappa_{1}N_{1}. \label{eq:5ds531}
\end{equation}
One can relate actions~(\ref{eq:5ds530}) and (\ref{eq:5ds531}) using equation~(\ref{eq:dsep5d1}), which gives the following transformations.
\begin{eqnarray}
\kappa\,'_{\!5} & = & \kappa_{5}-3\kappa_{0} \\
\kappa\,'_{\!3} & = & \kappa_{3}-\kappa_{0} \\
\kappa_{1}      & = & \kappa_{0} 
\end{eqnarray}
Clearly, the five dimensional Einstein-Hilbert action $S[T,\kappa_{3}]$ corresponds to the $\kappa_{0}=0$ limit of $S_{5}^{g}[T,\kappa_{0},\kappa_{3}]$. On that note, let us now clarify our reasons for considering these general actions.

\vskip 5mm

\noindent
The principle reason for researching five dimensional dynamical triangulations is based on the realisation that the most general space of parameters is {\it three dimensional}. Rather than attaching any great significance to the discretised Einstein-Hilbert action, we consider the most general action that is linear in the components of the $f$-vector. In doing so, we are treating dynamical triangulations as a statistical model of geometry in its own right without worrying too much about `gravity'.

It is suspected that this larger space of couplings may disclose rich phase structure, which could conceivably lead to interesting physics. It is hoped, at the very least, that this perspective may uncover new facets of dynamical triangulations, which could teach us more about the theory in general and its potential application to quantum gravity. For example, this viewpoint may shed new light on the phenomenon of singular vertices (see chapter~\ref{chap:chap6}). These are known to be an important feature of dynamical triangulations, and may even drive the phase transition in four dimensions. 

\subsection{Space of Parameters} 

\begin{figure}[htp] 
\centering{\input{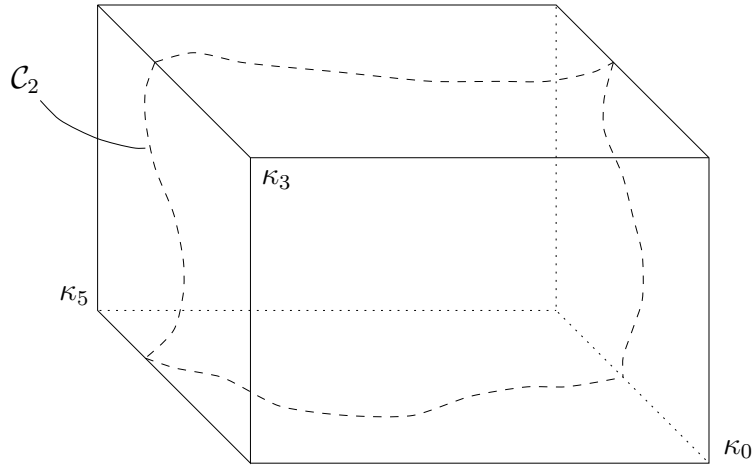}\par}
\caption{Two dimensional coupling constant surface ${\cal C}_{2}$.}
\label{fig:paraspace}
\end{figure}

So far, we have established that $S_{5}^{g}$ has three terms, and hence three parameters. Only two of these parameters are freely independent; the simplex coupling $\kappa_{5}$ is effectively fixed by the chosen target volume (see chapter~\ref{chap:chap3}). This means that the space of coupling constants is a two dimensional surface ${\cal C}_{2}$ defined at $\kappa_{5}=\kappa_{5}^{c}(\kappa_{0},\kappa_{3},N_{5})$, as shown schematically in figure~\ref{fig:paraspace}. Points on ${\cal C}_{2}$ are identified by the coordinates $(\kappa_{0},\kappa_{3})$. 

Our aim is to search for non-trivial phase structure in ${\cal C}_{2}$. As ${\cal C}_{2}$ is two dimensional, a thorough exploration is likely to be very costly in terms of computer power and time. For this reason, all numerical work requires careful forward planning to make the best possible use of available resources. We shall begin our investigation by considering the $\kappa_{3}=0$ and $\kappa_{0}=0$ limits of $S_{5}^{g}[T,\kappa_{0},\kappa_{3}]$. It is hoped that these preliminary simulations will allow us to identify phase transitions and characterise the associated phases. The results are presented and analysed in section~\ref{sec:prelim}.

If any phase transitions are found, then the next natural step is to determine their order, either by finite size scaling analysis or other means (see section~\ref{sec:fss}). Our main concern is to determine whether a transition is continuous or otherwise, as we are ultimately interested in the existence of physical continuum limits. The final stage of the numerical work involves exploring regions of ${\cal C}_{2}$ where $\kappa_{0}\neq0$ and $\kappa_{3}\neq0$ (see section~\ref{sec:2dcoupling}). These simulations will be limited to strategically chosen regions of ${\cal C}_{2}$. Before going any further, let us first review some relevant background theory of phase transitions and critical phenomena.

\subsection{Critical Phenomena} \label{sec:critphenom}

Dynamical triangulations can only give us a non-perturbative formulation of quantum gravity if it has a physical continuum limit. Such a limit is reached by reducing the lattice spacing (link length) $a$ to zero. This can only be done if the system is at a {\it critical point}.

One can define the connected correlation function $\Gamma_{ij}$ between two lattice spin variables $\sigma_{i}$ and $\sigma_{j}$ (at lattice sites $i$ and $j$ respectively) as
\begin{equation}
\Gamma_{ij} = \langle\sigma_{i}\,\sigma_{j}\rangle-\langle\sigma_{i}\rangle\langle\sigma_{j}\rangle,
\end{equation}
where $\langle\sigma_{i}\rangle$ represents the expectation value of $\sigma$ at site $i$. In general, the correlation function $\Gamma_{ij}$ is a function of $x$, the distance separating sites $i$ and $j$ in lattice units. It is a measure of the fluctuations in $\sigma$ across the lattice. Away from any critical point we find that correlations between lattice variables fall off exponentially for large $x$ as
\begin{equation}
\Gamma_{ij}(x) \sim x^{-\lambda} \exp \left( -\frac{x}{\xi} \right),
\end{equation}
where $\lambda$ is a positive number and $\xi$ is the {\it correlation length} (in lattice units). At high temperatures $\sigma_{i}$ become uncorrelated as $x \rightarrow \infty$ since the mean value of $\sigma_{i}$ is zero. Typically, at low temperatures we expect the lattice variables to be perfectly correlated. However, the {\it fluctuations} in $\sigma_{i}$ are zero.

A continuum limit exists only if the ensemble is invariant under conformal transformations. This occurs if there are fluctuations in the spin variables at {\it all\,} length scales (up to the size of the lattice). For this to be the case $\xi$ must diverge. Correlation lengths only diverge at continuous phase transitions. At critical points the microscopic details of the lattice become irrelevant and $\Gamma_{ij}(x)$ falls off according to a power law function
\begin{equation}
\Gamma_{ij}(x) \sim x^{-\lambda'},
\end{equation}
where $\lambda'=d-2+\eta$ and $d$ is the dimensionality of the system and $\eta$ is a {\it critical exponent}. In this instance, the renormalised mass $m_{r}$ is related to $\xi$ by
\begin{equation}
m_{r} = (\xi a)^{-1}. \label{eq:cont}
\end{equation}
Therefore, in order for $m_{r}$ to be finite we require $\xi \rightarrow \infty$ as $a \rightarrow 0$. 

\section{Preliminary Monte Carlo Simulations} \label{sec:prelim}

Our primary objective is to map the phase structure of ${\cal C}_{2}$. This is done by measuring quantities that are sensitive to phase transitions, i.e. derivatives of the free energy $F$. 
\begin{equation}
F \propto \log Z
\end{equation}
Past studies have shown that it can also be useful to measure geometrical quantities such as geodesic distances. Monte Carlo simulations are run over a wide range of $\kappa_{0}$ and $\kappa_{3}$ for triangulations with target volumes of $N_{5}^{t}=10$k. Measurements of $N_{0}$, $N_{3}$, $N_{5}$ and the average geodesic distance $\overline{g}$ (see section~\ref{subsec:geodesic}) were taken at regular intervals over $10^{5}$ sweeps. This range was progressively widened until kinematic upper and lower bounds were reached. Knowledge of these bounds aid the characterisation of phases and hence leads to a more complete picture of ${\cal C}_{2}$. Finally, the resolution is increased by running more simulations closer to the presumed phase transitions, thus identifying their location with ever greater precision. 

\subsection{Derivatives of the Free Energy} \label{subsec:derivatives}

\begin{figure}[htp]
\begin{center}
\leavevmode
(a){\hbox{\epsfxsize=14cm \epsfbox{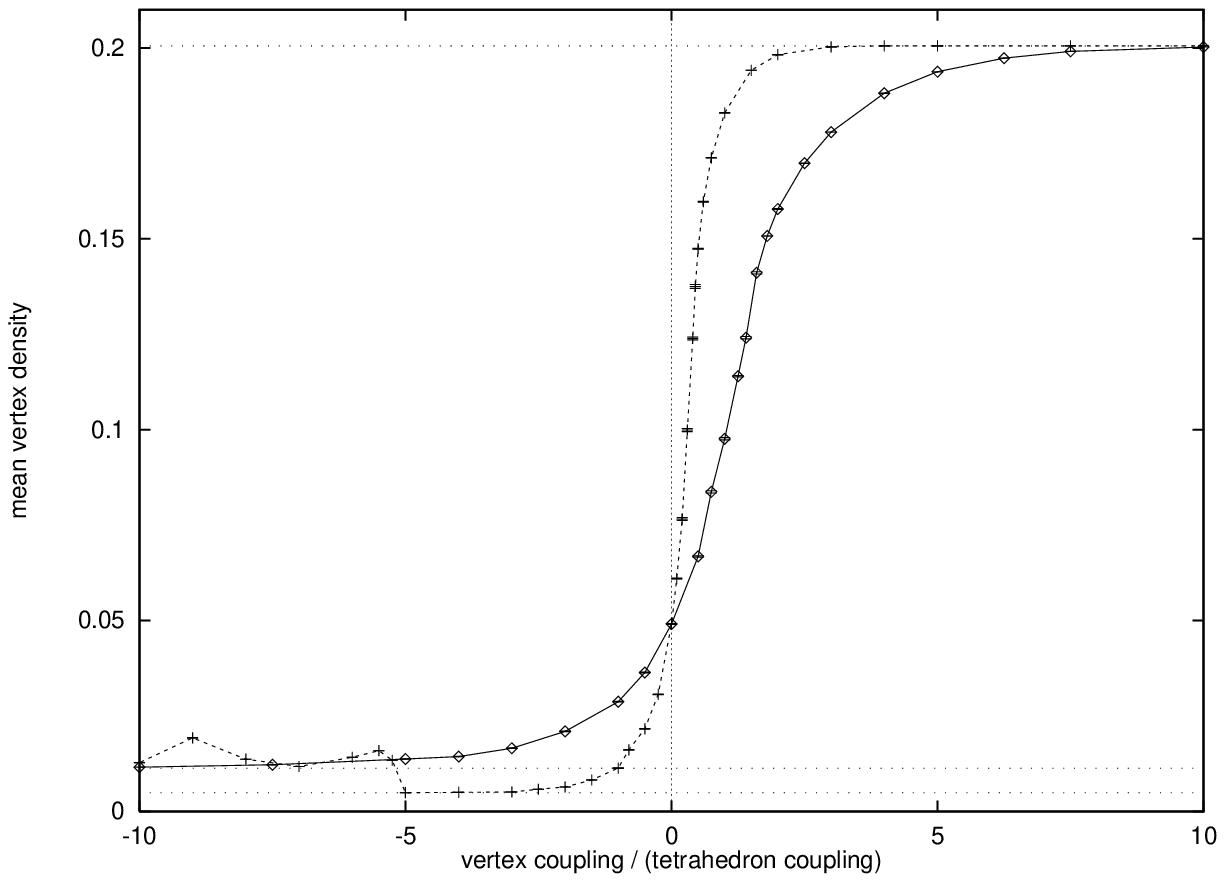}}}
(b){\hbox{\epsfxsize=14cm \epsfbox{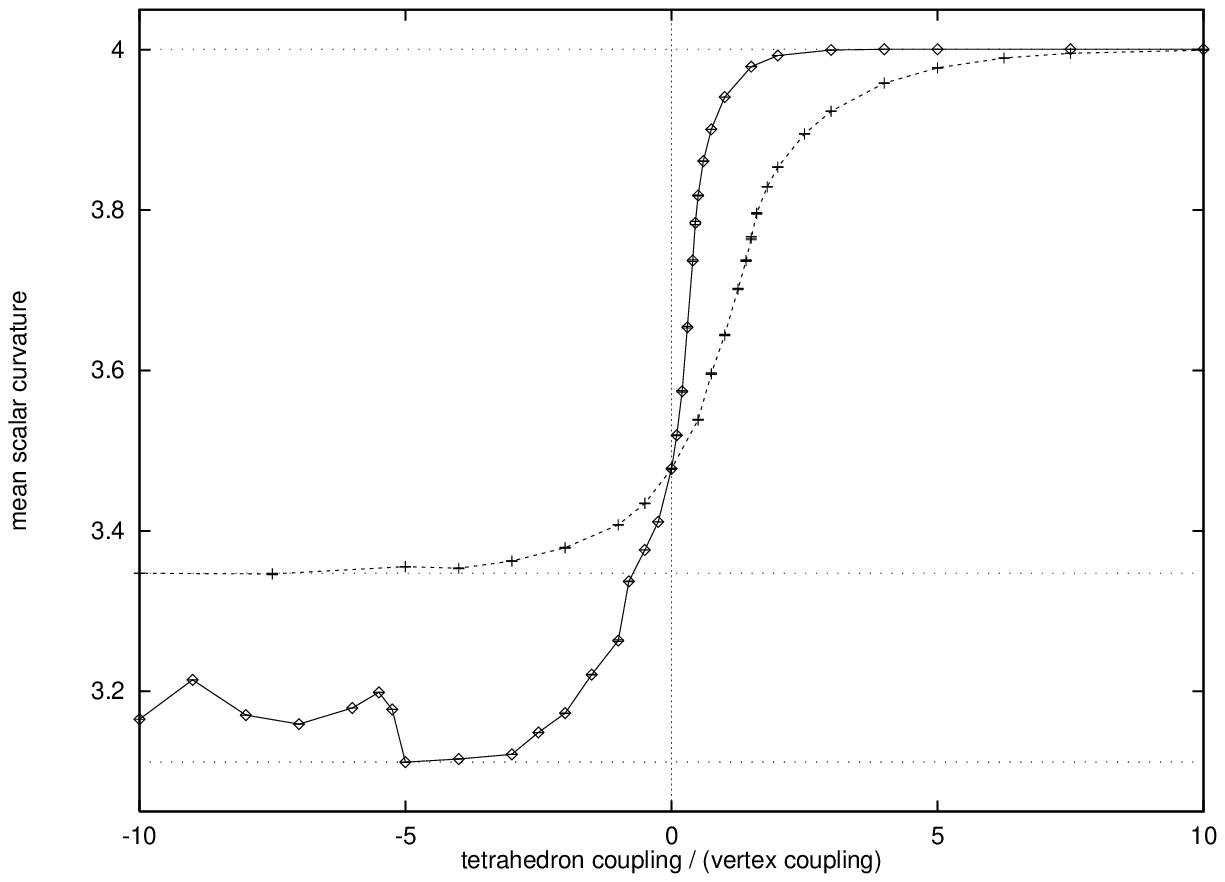}}}
\caption{(a) Mean vertex density $\langle\rho_{0}\rangle$ versus the vertex coupling $\kappa_{0}$ (solid) and the tetrahedron $\kappa_{3}$ (dashed). (b) Mean scalar curvature $\langle R\;\!\rangle$ versus $\kappa_{3}$ (solid) and $\kappa_{0}$ (dashed).}
\label{fig:zoomoutR}
\end{center}
\end{figure}

Let us begin by considering {\it first} derivatives of the free energy. Along the $\kappa_{0}$ axis of ${\cal C}_{2}$, the first derivative of the free energy with respect to $\kappa_{0}$ is the first cumulant of the distribution of $N_{0}$. It is proportional to the vertex density $\rho_{0}$. A discontinuity in $\rho_{0}$ as a function of $\kappa_{0}$ is indicative of a first order phase transition.
\begin{equation}
\rho_{0} = \frac{N_{0}}{N_{5}}
\end{equation}
Figure~\ref{fig:zoomoutR}(a) shows two plots of the mean vertex density $\langle\rho_{0}\rangle$, in which the solid (dashed) curve shows the relationship between $\langle\rho_{0}\rangle$ and $\kappa_{0}$ ($\kappa_{3}$) respectively. 
(Note that the curves intersect at $\kappa_{0}=\kappa_{3}=0$, as one would expect.) From this graph, it is immediately apparent that both plots highlight two distinct phases. Basically, in each case we find a weak coupling phase with large $\rho_{0}$ and a strong coupling phase with small $\rho_{0}$. 

One can plainly see that both curves have the same asymptote of $\rho_{0}\approx0.200$ for $\kappa_{0}\rightarrow\infty$ and $\kappa_{3}\rightarrow\infty$, which is marked by a dotted line. It appears that this limit represents the kinematic upper bound of $\rho_{0}$. This observation clearly opens up the possibility that the $\kappa_{3}=0$ and $\kappa_{0}=0$ limits share a {\it common} weak coupling phase.

In contrast, the strong coupling phases are distinct. The $\kappa_{0}\rightarrow -\infty$ limit has an asymptote of $\rho_{0}\approx0.010$, which one might assume to represent the kinematic lower bound of $\rho_{0}$.  In comparison, there does not appear to be an obvious asymptotic limit for $\kappa_{3}\rightarrow -\infty$. Indeed, $\rho_{0}$ fluctuates significantly as a function of $\kappa_{3}$, but only for $\kappa_{5}<3\kappa_{3}$ for $\kappa_{3}~\hbox{\begin{picture}(9,7)(0,2)\put(0,-2){\shortstack{$<$\\[-2pt]$\sim$}}\end{picture}}-5$. From these results it is evident that $\rho_{0}\approx0.010$ is not the true kinematic lower bound. This fact is somewhat surprising. A dotted line indicates a possible lower bound at $\rho_{0}\approx0.005$. Time series plots of $N_{0}$ have shown that all ensembles are thermalised. It seems likely that the ensembles for $\kappa_{5}<3\kappa_{3}$ for $\kappa_{3}~\hbox{\begin{picture}(9,7)(0,2)\put(0,-2){\shortstack{$<$\\[-2pt]$\sim$}}\end{picture}}-5$ have become trapped in metastable states, and that an asymptotic limit {\it does} exist for $\rho_{0}\approx0.005$. 

\vskip 5mm

\noindent
One physically tangible observable is the scalar curvature $R$, which is effectively the mean hinge local volume. It is proportional to the first derivative of the free energy along the $\kappa_{3}$ axis.
\begin{equation}
R = \frac{N_{3}}{N_{5}} 
\end{equation}
Figure~\ref{fig:zoomoutR}(b) shows two plots of the mean scalar curvature $\langle R\,\rangle$, in which the solid (dashed) curve shows the relationship between $\langle R\,\rangle$ and $\kappa_{3}$ ($\kappa_{0}$) respectively. By comparison with figure~\ref{fig:zoomoutR}(a), we find that both pairs of plots are qualitatively very similar. This illustrates the intimate (though indirect) connections between certain members of ${\cal N}_{d}$. The mean scalar curvature tends to an asymptotic value of $R\approx4.00$ in the weak coupling limit, which appears to represent the kinematic upper bound of $R$. One also finds an asymptote of $R\approx3.35$ for $\kappa_{0}\rightarrow -\infty$. A dotted line highlights a possible kinematic lower bound of $R\approx3.11$. Closer inspection reveals a small kink near $\kappa_{3}\approx-1$. The possible significance of this will be investigated later.

So far, we have already detected non-trivial phase structure in ${\cal C}_{2}$. Furthermore, we have uncovered interesting similarities and differences between phases along the $\kappa_{0}$ and $\kappa_{3}$ axes of ${\cal C}_{2}$. These are now investigated further by looking at second derivatives of the free energy.

\vskip 5mm

\begin{figure}[htp]
\begin{center}
\leavevmode
(a){\hbox{\epsfxsize=14cm \epsfbox{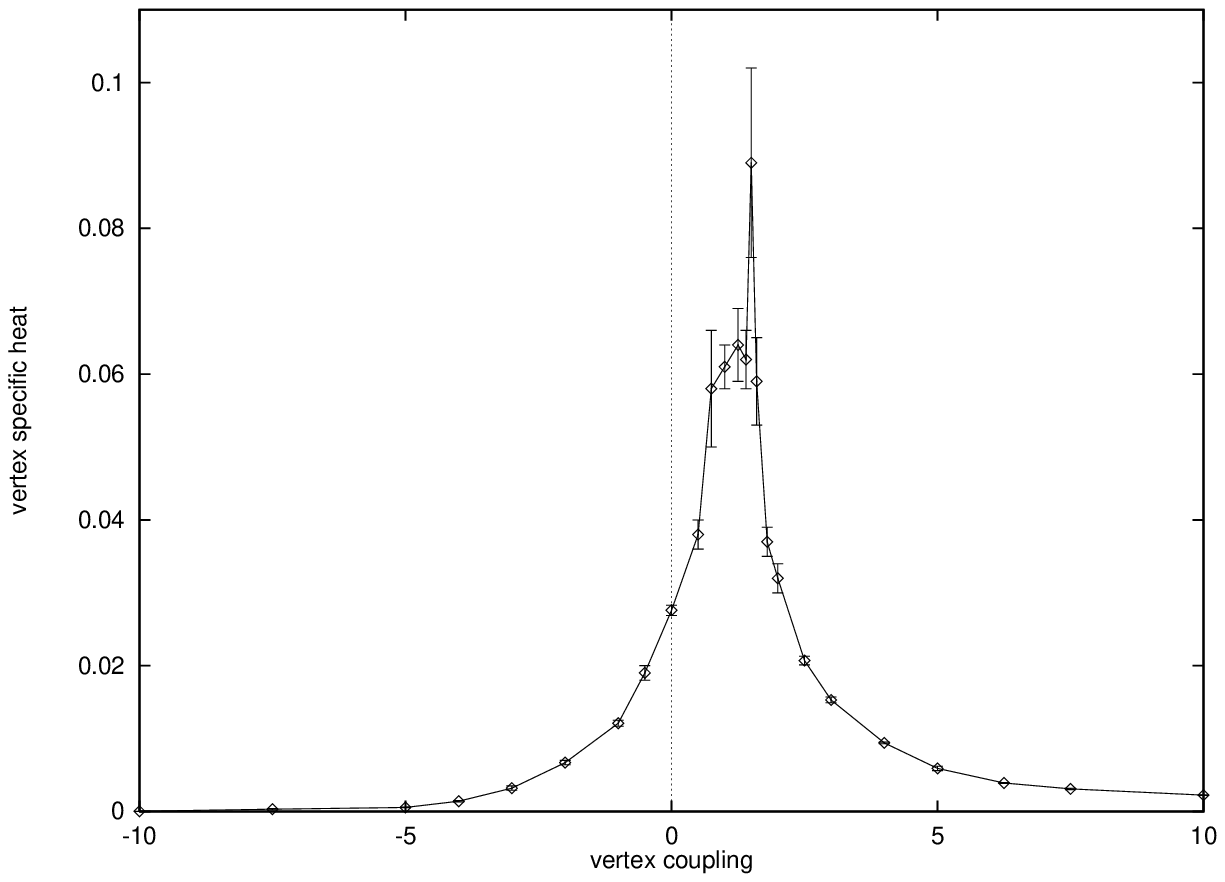}}}
(b){\hbox{\epsfxsize=14cm \epsfbox{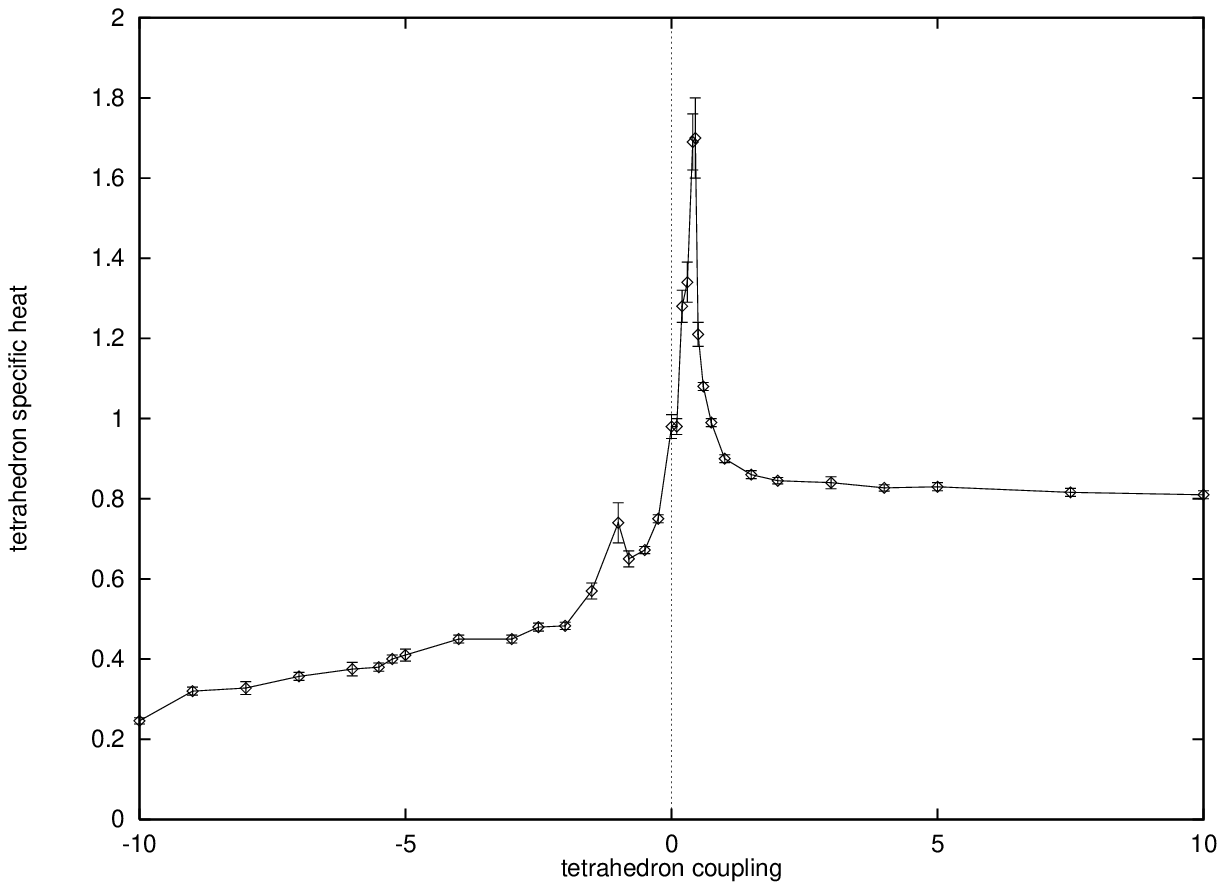}}}
\caption{(a) Vertex specific heat $c_{2}(\kappa_{0},N_{5})$ versus the vertex coupling $\kappa_{0}$. (b) Tetrahedron specific heat $c_{2}(\kappa_{3},N_{5})$ versus the tetrahedron coupling $\kappa_{3}$.}
\label{fig:zoomoutN0susN3sus}
\end{center}
\end{figure}

\noindent
Phase transitions are normally identified by divergences in normalised second cumulants, which are essentially second derivatives of the free energy. Let us first consider the $\kappa_{3}=0$ limit. The normalised second cumulant of $c_{2}(\kappa_{0},N_{5})$ is a measure of the fluctuations in $N_{0}$ and may be interpreted as the vertex specific heat. 
\begin{equation}
c_{2}(\kappa_{0},N_{5}) = \frac{1}{\langle N_{5}\rangle}\frac{\partial\,^{2}F(\kappa_{0},N_{5})}{\partial{\kappa_{0}}^{2}} = \frac{\langle {N_{0}}^{2}\rangle -\langle N_{0}\,\rangle ^{2}}{\langle N_{5}\rangle}
\end{equation}
Figure~\ref{fig:zoomoutN0susN3sus}(a) is a plot of $c_{2}(\kappa_{0},N_{5})$ versus $\kappa_{0}$. The sharp peak at $\kappa_{0}=1.5$ indicates the presence of a phase transition in this vicinity. Analogously, a divergence in the tetrahedron specific heat $c_{2}(\kappa_{3},N_{5})$ signifies the existence of a phase transition along the $\kappa_{3}$ axis. 
\begin{equation}
c_{2}(\kappa_{3},N_{5}) = \frac{1}{\langle N_{5} \rangle}\frac{\partial\,^{2}F(\kappa_{3},N_{5})}{\partial{\kappa_{3}}^{2}} = \frac{\langle {N_{3}}^{2}\rangle -\langle N_{3}\,\rangle ^{2}}{\langle N_{5}\rangle}
\end{equation}
Indeed, a peak is observed at $\kappa_{3}=0.45$ (see figure~\ref{fig:zoomoutN0susN3sus}(b)). It is interesting to see a small peak in $c_{2}(\kappa_{3},N_{5})$ near $\kappa_{3}=-1$, which coincides with the observed kink in $\langle R(\kappa_{3})\rangle$ (see figure~\ref{fig:zoomoutR}(b)). This may indicate the presence of a second phase transition. From visual inspection, the peaks in $c_{2}(\kappa_{0},N_{5})$ and $c_{2}(\kappa_{3},N_{5})$ appear to coincide with the points of inflexion of $\langle \rho_{0}(\kappa_{0})\rangle$ and $\langle R(\kappa_{3})\rangle$, as one would expect. Collectively, these results prove that five dimensional dynamical triangulations has a non-trivial phase structure. This was not entirely unexpected given the situation for $d\leq 4$. 

\subsection{Geodesic Distances} \label{subsec:geodesic}

Triangulated lattices may be viewed as discretisations of continuum manifolds. We have already established that the concept of curvature may be extended to triangulations. In fact, geodesics also have a natural definition on such discretised spaces. 

\begin{figure}[htp] 
\centering{\input{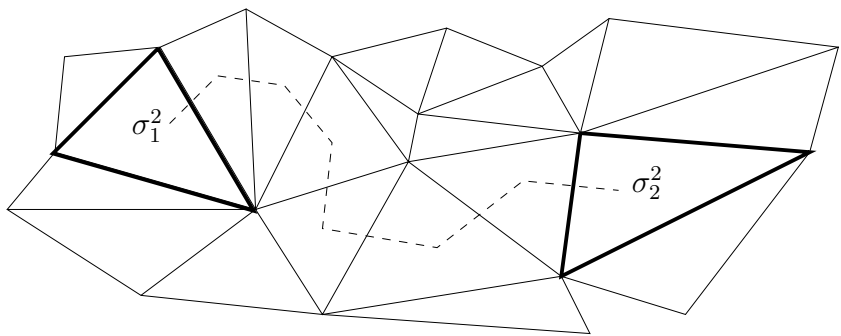}\par}
\caption{Geodesic between two 2-simplices $\sigma_{1}^{2}$ and $\sigma_{2}^{2}$.}
\label{fig:geodesic}
\end{figure}

A geodesic may be defined as the shortest path connecting two points on a manifold. For our purposes, this definition requires some modification since we deal with {\it simplicial} manifolds. We understand a geodesic between two arbitrary simplices $\sigma_{1}^{d}$ and $\sigma_{2}^{d}$ to mean the minimum number of simplices that link $\sigma_{1}^{d}$ and $\sigma_{2}^{d}$ along faces. (Alternatively, one could define a geodesic as the shortest distance between two vertices along links. Both definitions would give the same qualitative results.) Figure~\ref{fig:geodesic} illustrates the geodesic (dotted line) between simplices $\sigma_{1}^{2}$ and $\sigma_{2}^{2}$ of a two dimensional simplicial manifold\,\footnote{Unlike geodesics defined on continuum manifolds, `discrete' geodesics are not necessarily unique. In other words, there may exist more than one path of the same length connecting two simplices (or vertices).}. In this particular case the geodesic distance is 7 units.

By measuring the {\it average geodesic distance} $\overline{g}$ one can learn more about the geometric of aspects dynamical triangulations. For a given Monte Carlo configuration, one can evaluate $\overline{g}$ by calculating the geodesic distances between many random pairs of simplices. Ideally, one ought to average over all pairs. However, this approach can become very costly in terms of computer resources since the number of pairs grows factorially with volume. Fortunately, reasonable estimates of $\overline{g}$ may be had by sampling just a small proportion of the total number of pairs. This is done by randomly selecting a predetermined number of `origin simplices'. Then, for each origin simplex the algorithm calculates the geodesic distance to all other simplices, thus giving an average value $\overline{g}$. 

\vskip 5mm

\begin{figure}[htp]
\begin{center}
\leavevmode
(a){\hbox{\epsfxsize=14cm \epsfbox{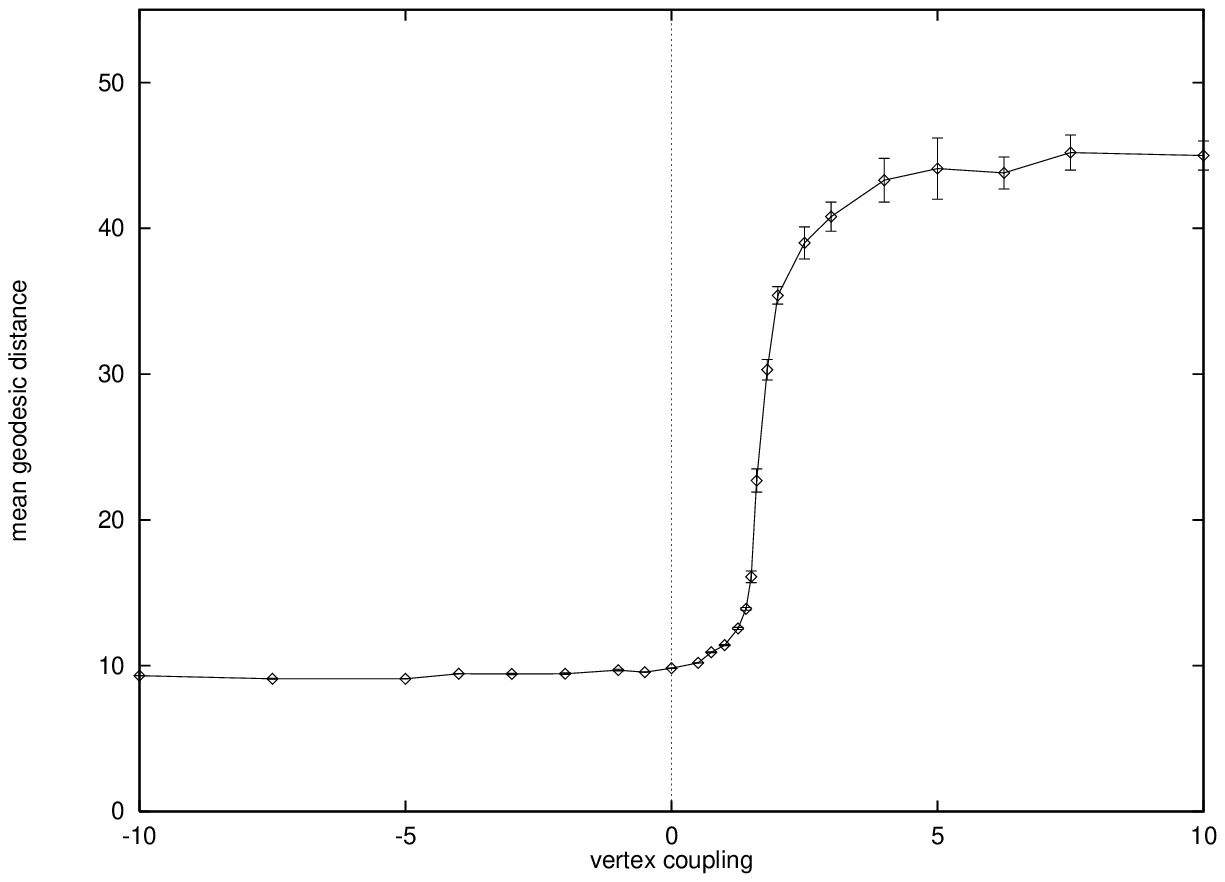}}}
(b){\hbox{\epsfxsize=14cm \epsfbox{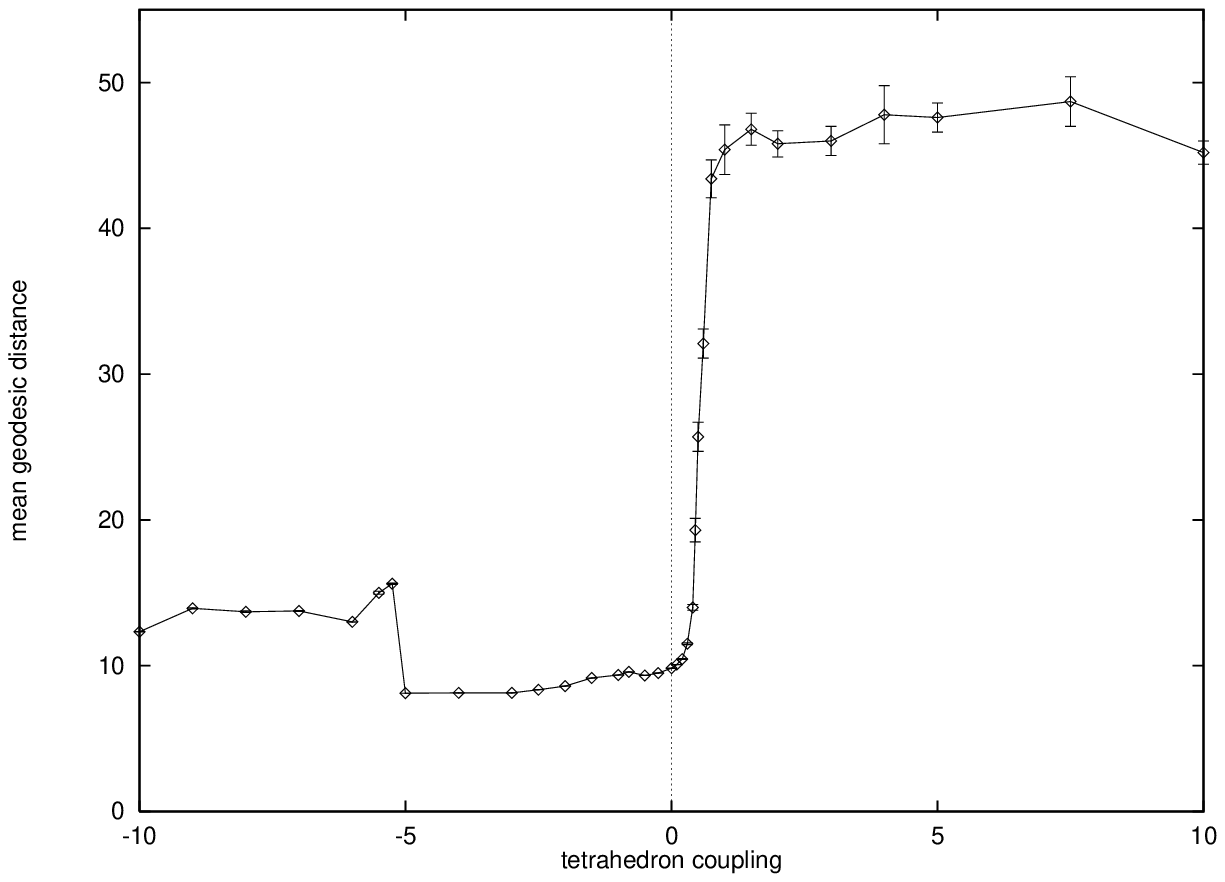}}}
\caption{Mean geodesic distance $\langle\,\overline{g}\,\rangle$ versus (a) the vertex coupling $\kappa_{0}$ and (b) the tetrahedron coupling $\kappa_{3}$.}
\label{fig:zoomoutd}
\end{center}
\end{figure}

\noindent
Figures~\ref{fig:zoomoutd}(a) and~\ref{fig:zoomoutd}(b) show plots of the mean (value of the average) geodesic distance $\langle\,\overline{g}\,\rangle$ versus $\kappa_{0}$ and $\kappa_{3}$ respectively. 
In both cases we find that the weak coupling triangulations are elongated, with average geodesic distances of about 45 units. This is consistent with the existence of a common weak coupling phase. In figure~\ref{fig:zoomoutd}(a), the $\kappa_{0}\rightarrow -\infty$ asymptote corresponds to triangulations with $\overline{g}\approx9$. For $\kappa_{3}\rightarrow -\infty$, we observe fluctuations commensurate with those in $\rho_{0}$ and $R$. In light of these findings, we shall adopt the nomenclature prevalent in four dimensions, in which the strong and weak coupling phases are commonly referred to as {\it crumpled} and {\it elongated} phases respectively. 

So far, we have discovered that the mean geodesic distance is as much as four times greater in the elongated phase than in the crumpled phases. When this fact is coupled with the large fluctuations in $\overline{g}$, we are naturally led to suspect that the dominant triangulations of the elongated phase may be {\it branched polymers}. These tree-like structures (with variable branching order) have been observed in two, three and four dimensional dynamical triangulations. This matter will be resolved later in chapter~\ref{chap:chap5}.

\vskip 5mm 

\noindent
Previous studies have shown that the fluctuations in $\overline{g}$ are highly sensitive to phase transitions~\cite{ckr4dphase}. The normalised second cumulant $c_{2}(\overline{g},N_{5})$ of the distribution of $\overline{g}$ is given by
\begin{equation}
c_{2}(\overline{g},N_{5}) = \frac{\langle\,\overline{g}\,^{2}\,\rangle -\langle\,\overline{g}\,\rangle ^{2}}{\langle N_{5}\rangle}.
\end{equation}
It was found that $c_{2}(\overline{g},N_{5})$ is sensitive to both phase transitions. In the crumpled phases $c_{2}(\overline{g},N_{5})$ is very small, typically of order $10^{-6}$. Near the transitions we find that $c_{2}(\overline{g},N_{5})$ increases very sharply by orders of magnitude and fluctuates greatly in the elongated phase. Hence, $c_{2}(\overline{g},N_{5})$ behaves as an order parameter. 

The large fluctuations in $\overline{g}$ are most probably due to long autocorrelation times (compared to the measurement intervals). The branched nature of these triangulations means that many $(k,l)$ moves must be performed to generate statistically independent configurations. The local update algorithm can be made more efficient by utilising a technique known as {\it baby universe surgery}~\cite{jainbaby,ambjornbaby}. In general, this involves identifying a branch $\cal B$ of a triangulation $T$ and cutting it off along an $S^{d-1}$ submanifold of volume $N_{d-1}$. Another portion of the manifold $\cal P$ homeomorphic to $B^{d}$ is selected at random. If the $S^{d-1}$ boundary of $\cal P$ also has volume $N_{d-1}$, then $\cal P$ and $\cal B$ may be exchanged and reconnected. This procedure is a non-local move and generally decreases the autocorrelation time. These non-local moves were not employed in our simulations.

\section{Finite Size Scaling} \label{sec:fss}

So far we have discovered clear signals indicating the existence of two phase transitions along the $\kappa_{0}$ and $\kappa_{3}$ axes of ${\cal C}_{2}$. Our next goal is to ascertain the order of these phase transitions. More specifically, we hope to determine whether the transitions are of first order or continuous, using finite size scaling methods. This is achieved in practice by looking at how second cumulants scale over a range of target volumes $N_{5}^{t}$. 

As we shall see, this type of numerical work requires substantial amounts of computational resources, which obviously increase with volume. For this reason, the target volumes to be studied were chosen carefully, with a number of factors taken into consideration. These include the available CPU (central processing unit) time, computer performance\,\footnote{Computer performance tests show that Monte Carlo simulation times grow faster than linearly with volume. Details are given in appendix~\ref{app:perform}.} and possible finite size effects. It was first decided that only three target volumes would be studied, which is the bare minimum. Past studies of three and four dimensional dynamical triangulations were also influencing factors, for they served as useful reference points. The target volumes were finally chosen to be 10k, 20k and 30k. 

\vskip 5mm

\noindent
In general, the critical couplings are functions of volume, i.e. $\kappa_{0}^{c}=\kappa_{0}^{c}(N_{5})$ and $\kappa_{3}^{c}=\kappa_{3}^{c}(N_{5})$. Presumably, a prerequisite condition for well-defined thermodynamic limits is that $\kappa_{0}^{c}(N_{5})$ and $\kappa_{3}^{c}(N_{5})$ converge as $N_{5} \rightarrow \infty$~\cite{ckr4dphase}. Unfortunately, given that we have only three data sets, it is not possible to determine whether this is the case. As an example, consider the function $\kappa_{0}^{c}(N_{5})$. If $\kappa_{0}^{c}(N_{5})$ is a monotonic function, then one can always find perfect fits for the two following functional forms (where $a$, $b$ and $c$ are positive constants).
\begin{eqnarray}
\mbox{divergent:} \hspace{5mm} \kappa_{0}^{c} & = & a + b{N_{5}}^{c} \label{eq:divk0} \\
\mbox{convergent:} \hspace{5mm} \kappa_{0}^{c} & = & a - b{N_{5}}^{-c} \label{eq:conk0} 
\end{eqnarray}
In other words, it is impossible to differentiate between the divergent and convergent functions. At this stage, one can only {\it assume} that $\kappa_{0}^{c}(\infty)$ and $\kappa_{3}^{c}(\infty)$ are finite. To produce evidence that supports or contradicts this assumption would require an extensive numerical effort, similar in scale to that presented in chapter~\ref{chap:chap3}. It is therefore not in our interests to pursue this line of research at the present time. 

\subsection{Second Cumulant Scaling} \label{subsec:scs}

It is assumed that the normalised second cumulant $c_{2}(\kappa_{0},N_{5})$ scales with volume at the phase transition as 
\begin{equation}
c_{2}(\kappa_{0},N_{5}) \sim {N_{5}}^{\alpha/\nu d_{H}}, \label{eq:joseph} 
\end{equation}
where $\alpha$ and $\nu$ are critical indices and $d_{H}$ is the internal Hausdorff (fractal) dimension\,\footnote{An analogous relation exists for $c_{2}(\kappa_{3},N_{5})$.}. These quantities are related by Josephson's scaling relation.
\begin{equation}
\alpha = 2 - \nu d_{H}
\end{equation}
The product $\nu d_{H}$ has the bound $\nu d_{H}\geq 1$. A first order phase transition is characterised by having $\nu d_{H}=1$. Hence, using Josephson's relation we see that in this instance $c_{2}(\kappa_{0},N_{5})$ would diverge linearly with volume. The phase transition is continuous if $0<\alpha <1$ and $1<\nu d_{H} <2$. Relation~(\ref{eq:joseph}) may be rewritten simply as
\begin{equation}
c_{2}(\kappa_{0},N_{5}) \sim {N_{5}}^{\Delta},
\end{equation}
where $\Delta=\alpha /\nu d_{H}$. A continuous phase transition therefore has $0<\Delta <1$. Our task is to measure the normalised second cumulants at the phase transitions for each target volume and hence determine the scaling behaviour.

The accuracy with which one can measure second cumulants at phase transitions depends on two factors. Apart from the usual statistical error, an added error is incurred from the uncertainty in the location of the phase transition itself. This drawback is exacerbated if the location is determined {\it from} measurements of second cumulants. Fortunately, measurements of $c_{2}(\overline{g},N_{5})$, which behaves as an order parameter, allow us to locate transitions with good accuracy. Table~\ref{tab:location} shows the locations of the phase transitions for each target volume $N_{5}^{t}$, with estimates of the uncertainty given in parentheses.
\begin{table}[htp]
\begin{center}
\begin{tabular}{|c||c|c|} \hline
$N_{5}^{t}$ & $\kappa_{0}^{c}(N_{5})$ & $\kappa_{3}^{c}(N_{5})$ \\ \hline
10k & 1.5375(5) & 0.454(1) \\ \hline
20k & 1.595(1)  & 0.469(2) \\ \hline
30k & 1.640(1)  & 0.480(1) \\ \hline
\end{tabular}
\caption{Locations of the phase transitions.}
\label{tab:location}
\end{center}
\end{table}

\vskip 5mm

\noindent
Extensive Monte Carlo simulations were then run near the phase transitions, in an effort to minimise statistical errors. These ranged up to 8 million sweeps for $N_{5}^{t}=10$k and up to 2 million sweeps for $N_{5}^{t}=30$k. It was found that $c_{2}(\overline{g},N_{5})$ no longer behaves as an order parameter for sufficiently high resolutions of the coupling constants. The scaling of $c_{2}(\overline{g},N_{5})$ was found to be similar for both the $\kappa_{3}=0$ and $\kappa_{0}=0$ limits. Basically, in the crumpled phases $c_{2}(\overline{g},N_{5})$ is small and decreases with volume, and the phase transitions are characterised by an acute increase in $c_{2}(\overline{g},N_{5})$ which becomes progressively sharper with volume. 

The vertex and tetrahedron specific heats were measured close to the phase transitions. At this level of resolution, distinct peaks in $c_{2}(\kappa_{0},N_{5})$ and $c_{2}(\kappa_{3},N_{5})$ are only observed for $N_{5}^{t}=30$k. In fact, for $N_{5}^{t}=10$k, peaks in the specific heats are hardly perceptible. The results are shown in table~\ref{tab:chi0chi3}. Graphical presentation of the results cannot be justified due to the poor statistics. The errors, shown in parentheses, were estimated using the binning technique to counter the effect of correlated data (see appendix~\ref{app:binning}).
\begin{table}[htp]
\begin{center}
\begin{tabular}{|c||c|c|} \hline
$N_{5}^{t}$ & $c_{2}(\kappa_{0}^{c},N_{5})$ & $c_{2}(\kappa_{3}^{c},N_{5})$ \\ \hline
10k & 0.09(1)  & 1.81(6) \\ \hline
20k & 0.105(7) & 1.75(10) \\ \hline
30k & 0.14(2)  & 2.0(3) \\ \hline
\end{tabular}
\caption{Scaling of the vertex and tetrahedron specific heats.}
\label{tab:chi0chi3}
\end{center}
\end{table}
It is clear that one cannot make any firm deductions regarding $\Delta$ using these results. 

\vskip 5mm

\noindent
This approach to determining the order of the phase transitions has obviously not proven to be very effective. It seems that this type of undertaking is simply too ambitious given the computational hardware available. In retrospect, these difficulties should have been expected. Fortunately, there exist certain numerical algorithms which can improve the efficiency of such studies. For example, histogramming methods have been successfully implemented by other researchers in four dimensions~\cite{ckr4dphase,focusing,basfirst}, did not improve matters in our case. This may have been due to an insufficient quantity of raw data. 

Analogous studies in four dimensions have shown that finite size scaling results require careful interpretation. A study by Catterall {\it et al.} using target volumes of up to $N_{4}^{t}=8$k found $\Delta$ to be 0.259(7), which corresponds to a continuous phase transition~\cite{ckr4dphase}. Further numerical simulations with target volumes up to $N_{4}^{t}=32$k, determined $\Delta$ to be 0.81(4)~\cite{basfirst}. This result opened up the possibility that $\Delta\rightarrow 1$ for large $N_{4}$. It is entirely possible that a similar effect could exist in five dimensions. In conclusion, this approach is fraught with difficulties mainly relating to the sheer scale of the problem. We shall now consider an alternative way of determining the order of phase transitions.

\subsection{Metastability} \label{subsec:meta}

In this section we exploit the fact that first order phase transitions are characterised by metastability. It is possible to observe the co-existence of multiple phases in thermal equilibrium from Monte Carlo time series plots. Given sufficient data, one may then plot histograms of the data, in which metastability would manifest itself as multiple peaks in the distribution. These tests may reveal the existence of first order phase transitions. Of course, the absence of metastable effects does not necessarily mean that a transition is not of first order. One could envisage cases in which the latent heat is undetectably small\,\footnote{This was found to be the case for $N_{5}^{t}=10$k and $N_{5}^{t}=20$k.}. Since latent heat increases with volume, we shall only consider the time series plots of our largest target volumes ($N_{5}^{t}=30$k).

\begin{figure}[htp]
\begin{center}
\leavevmode
(a){\hbox{\epsfxsize=14cm \epsfbox{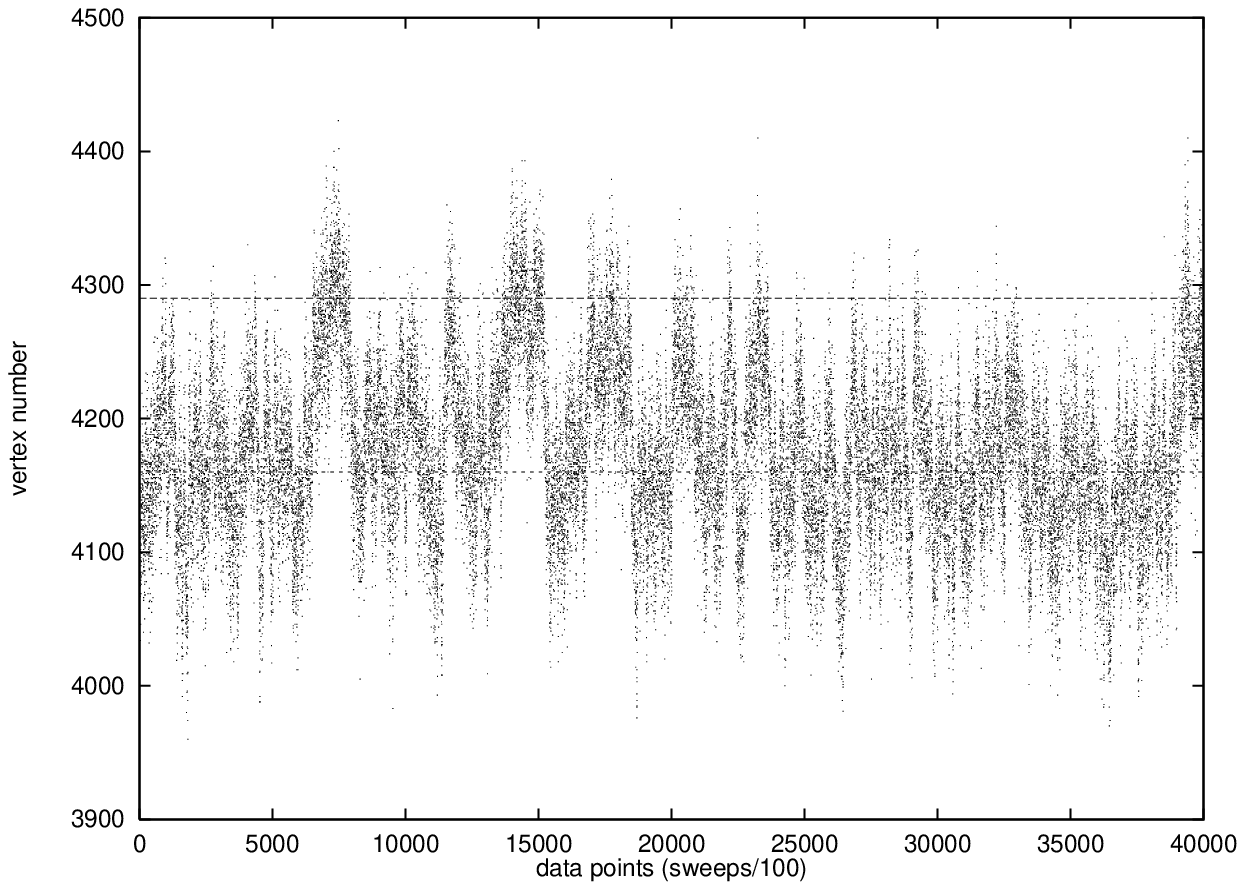}}}
\vskip 10mm
(b){\hbox{\epsfxsize=14cm \epsfbox{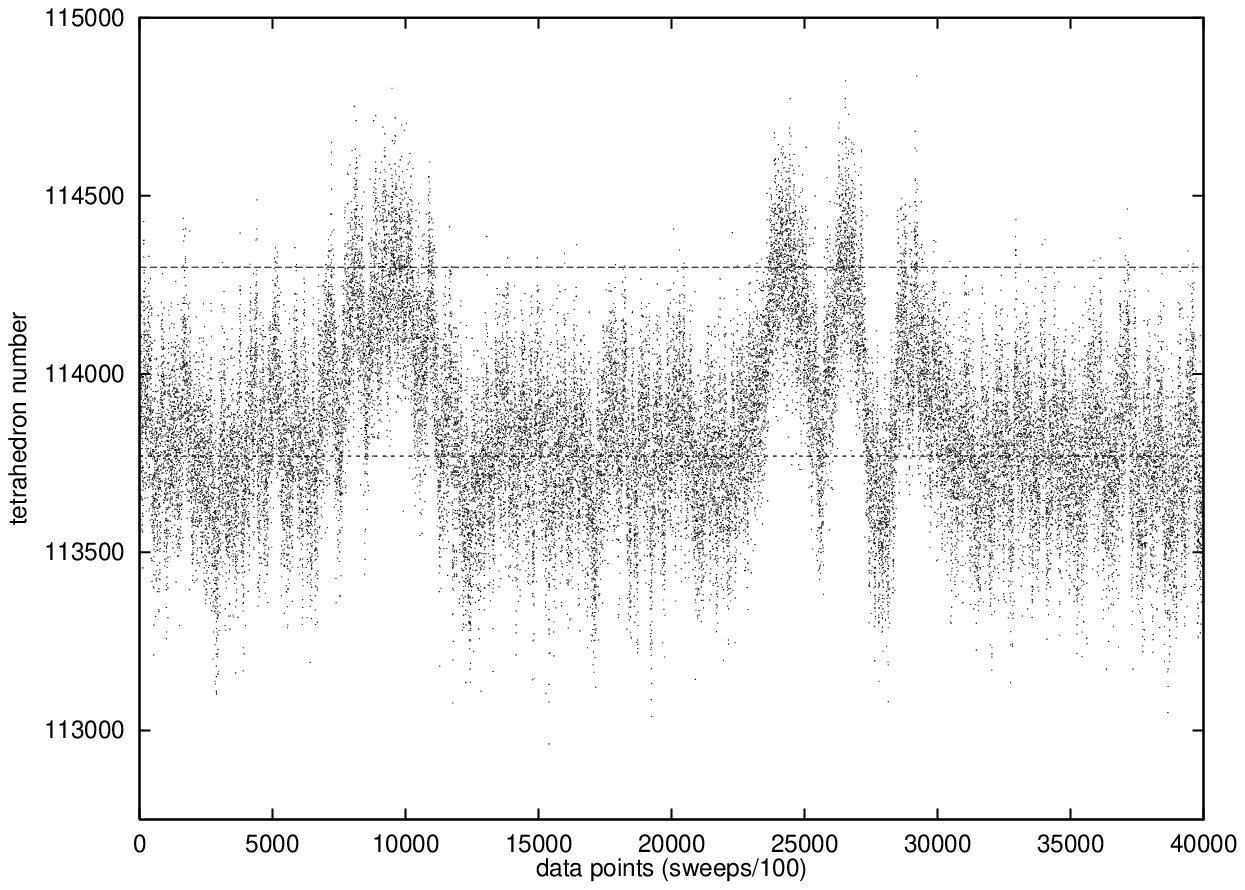}}}
\caption{Monte Carlo time series plots of (a) the vertex number $N_{0}$ for $\kappa_{0}=1.640$ and (b) the tetrahedron number $N_{3}$ for $\kappa_{3}=0.480$.}
\label{fig:bistable}
\end{center}
\end{figure}

Figure~\ref{fig:bistable}(a) is a time series plot of $N_{0}$ taken over 4 million sweeps for $\kappa_{0}=1.640$. 
(It becomes immediately apparent that the data is highly correlated -- this is evident from the vertical streaking effect. However, this is not too important for our current consideration.) From closer examination, it appears that the ensemble flips between two bistable states, which are highlighted by horizontal lines. This effect is not very clear by any means and hence may not be genuine. If this interpretation is correct then the $N_{0}=4160$ and $N_{0}=4290$ dashed horizontal lines correspond to the crumpled and elongated phases respectively.

Figure~\ref{fig:bistable}(b) is a time series plot of $N_{3}$ taken over 4 million sweeps for $\kappa_{3}=0.480$. In this case, we also find evidence of bistability -- though much stronger. Here the crumpled and elongated phases are highlighted by the $N_{3}=113770$ and $N_{3}=114300$ dashed horizontal lines respectively. 

In both cases, we find that the ensemble appears more settled in the crumpled phase. From this observation, one may cautiously deduce that the exact location of the phase transition are at marginally greater values of $\kappa_{0}$ and $\kappa_{3}$. Unfortunately, it was not possible to extend this study due to time constraints. The CPU (central processing unit) time required to complete these time series runs is of the order of 7 weeks -- even for our most powerful computer. In conclusion, these preliminary results appear to be consistent with first order phase transitions. However, the study must be extended to larger volumes before our claims can be accepted with confidence. At this point we conclude our investigation of phase transitions and concentrate on further exploration of ${\cal C}_{2}$.

\section{Two Dimensional Coupling Space ${\cal C}_{2}$} \label{sec:2dcoupling} 

Until now, we have only considered the $\kappa_{3}=0$ and $\kappa_{0}=0$ limits of ${\cal C}_{2}$, where $S_{5}^{g}[T,\kappa_{0},\kappa_{3}]$ effectively has only two terms. The next stage in our investigation involves probing regions of ${\cal C}_{2}$ where $\kappa_{0}$ {\it and} $\kappa_{3}$ are non-zero. Given the difficulties encountered earlier, it is clearly not feasible to undertake any form of finite size scaling analysis in ${\cal C}_{2}$. Our objective is simply to detect other non-trivial phase structure. 

It is tempting to speculate that there may exist a line of phase transitions (phase boundary) passing through $(1.5375,0)$ and $(0,0.454)$ in ${\cal C}_{2}$. One may then consider whether the phase boundary extends indefinitely or ends in a critical point, as in the phase diagram of water. Such question are very difficult, if not impossible, to answer predictively. It is not even clear whether such a boundary would be straight or curved. The preliminary simulations of section~\ref{sec:prelim} showed marked differences between the strong coupling phases of the $\kappa_{3}=0$ limit and the $\kappa_{0}=0$ limit. This may hint at the existence of other structure in the region of ${\cal C}_{2}$ where $\kappa_{0}<0$ and $\kappa_{3}<0$, which would explain the small peak in $c_{2}(\kappa_{3},N_{5})$ found near $\kappa_{3}=-1$. Let us now attempt to resolve some of these issues.

\vskip 5mm

\begin{figure}[htp]
\begin{center}
\leavevmode
(a){\hbox{\epsfxsize=14cm \epsfbox{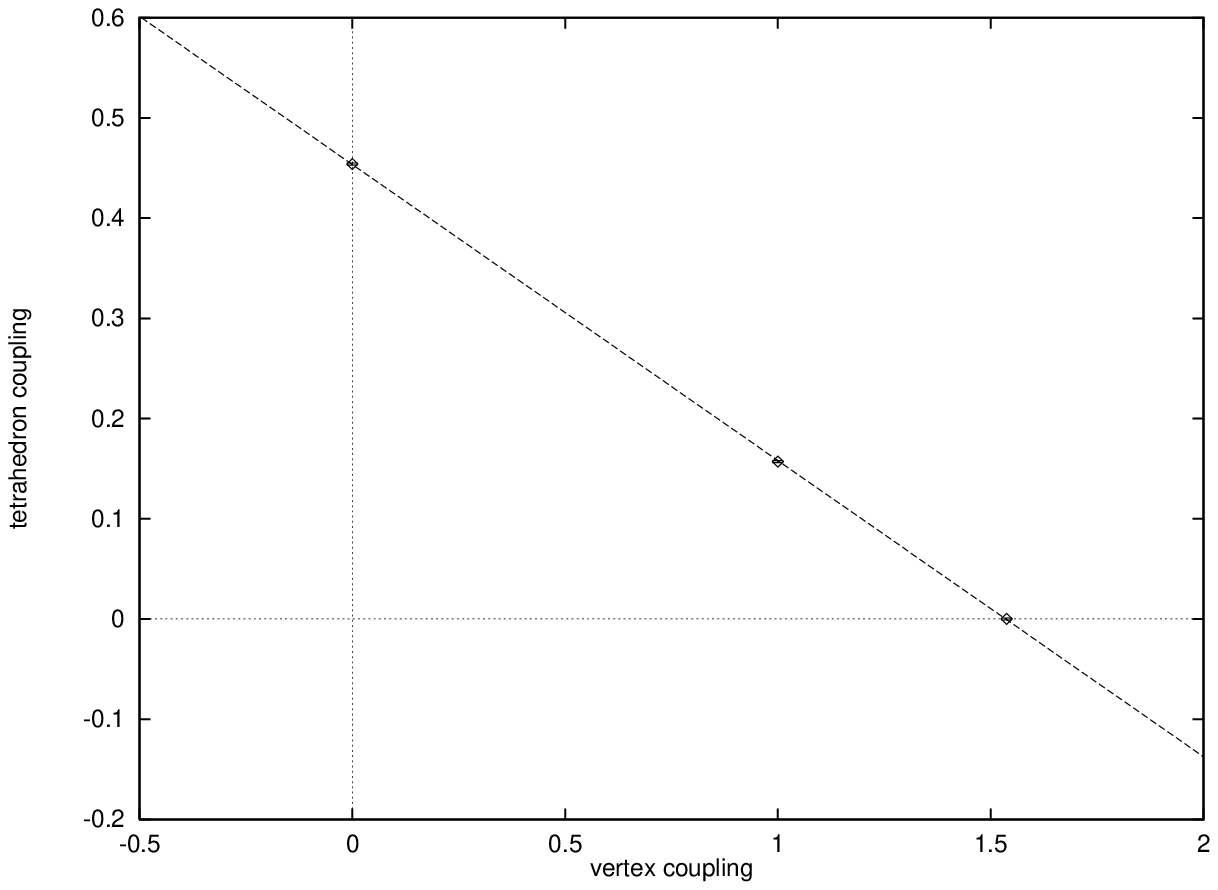}}}
(b){\hbox{\epsfxsize=14cm \epsfbox{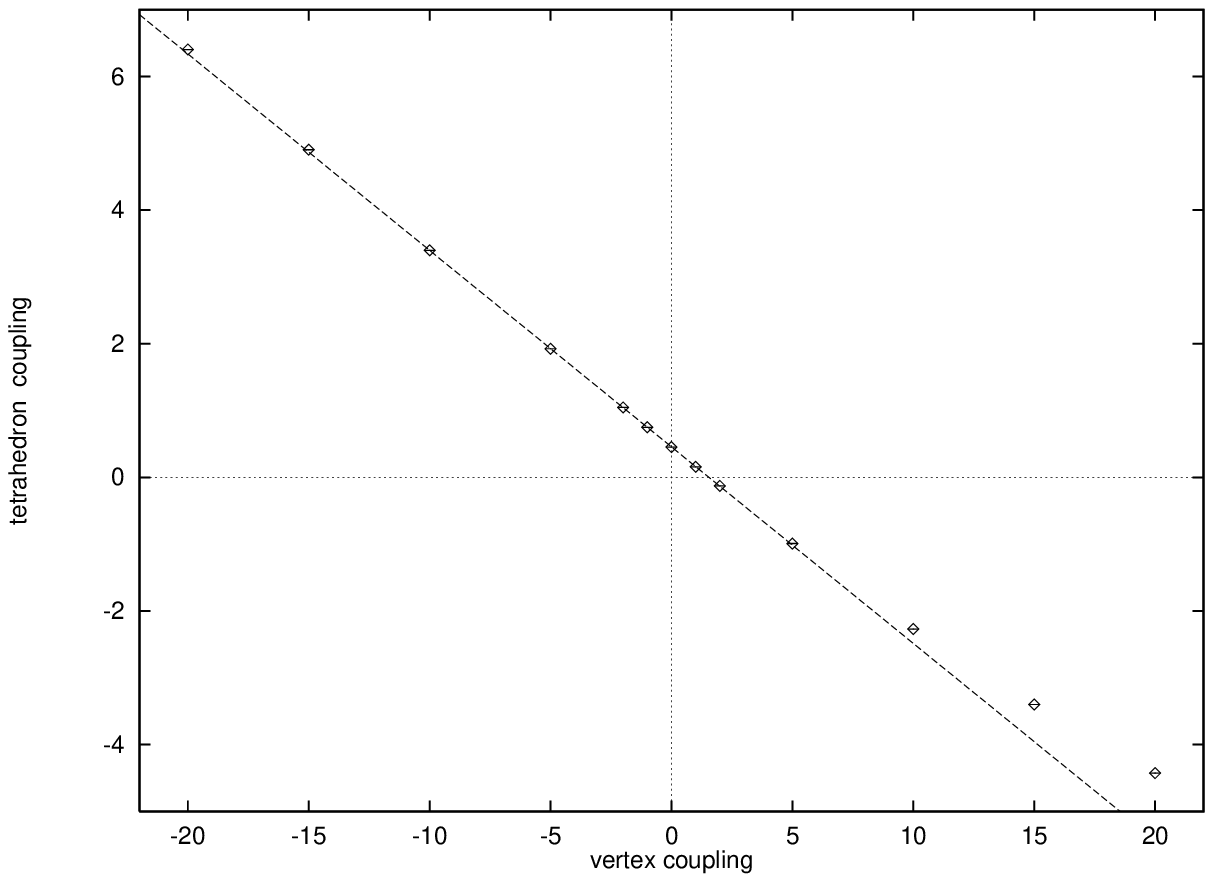}}}
\caption{(a) Possible line of phase transitions. (b) Phase boundary in ${\cal C}_{2}$.}
\label{fig:c2space}
\end{center}
\end{figure}

\noindent
We begin by exploring ${\cal C}_{2}$ for triangulations with target volumes of $N_{5}^{t}=10$k. This was done by first fixing $\kappa_{0}$ and then tuning $\kappa_{3}$ to phase transitions. The first series of simulations were run for $\kappa_{0}=1.0$, from which a transition was detected at $\kappa_{3}=0.157(1)$. This result effectively proves the existence of a line of phase transitions in ${\cal C}_{2}$ (see figure~\ref{fig:c2space}(a)). The equation of the best fit line was found to be
\begin{equation}
\kappa_{3} = -0.2955(1)\kappa_{0}+0.4536(1),
\end{equation}
with $\chi^{2}\approx1.9 \mbox{ d.o.f.}^{-1}$. The good fit indicates that the phase boundary is a straight line.
The phase boundary was then traced in the same manner at regular intervals ranging from $\kappa_{0}=-20$ to $\kappa_{0}=20$. In each case, $\kappa_{3}$ was tuned until the location of the phase transitions were known to within at least $\pm\,0.001$. The results are shown in figure~\ref{fig:c2space}(b).

\subsection{Phase Boundary} \label{subsec:pb}

From figure~\ref{fig:c2space}(b), it is evident that the phase boundary does not form a straight line in ${\cal C}_{2}$. Analysis has shown that the boundary {\it is} straight in the approximate range of $-2<\kappa_{0}<2$. However, significant deviations are observed for large positive values of $\kappa_{0}$. Careful observation also reveals minor deviations for large negative values of $\kappa_{0}$. If we ignore these deviant points then the best fit line is given by
\begin{equation}
\kappa_{3} = -0.2943(3)\kappa_{0}+0.4564(4),
\end{equation}
with $\chi^{2}\approx23 \mbox{ d.o.f.}^{-1}$. This line is drawn in figure~\ref{fig:c2space}(b). The possibility of an ending to the phase boundary remains, as yet, an open question. The curved nature of the phase boundary is not currently understood. 

\vskip 5mm

\noindent
Interestingly, it was noticed that the phase transition appears `sharper' for large positive values of $\kappa_{0}$. By this we mean that for a given resolution of $\kappa_{3}$, the first derivatives of the free energy appear to be more like discontinuous functions of $\kappa_{3}$. We illustrate this point using the $\kappa_{0}=-20$ and $\kappa_{0}=+20$ simulations as examples (see table~\ref{tab:sharp}).
\begin{table}[htp]
\begin{center}
\begin{tabular}{|c|c||c|c|c|} \hline
$\kappa_{0}$ & $\kappa_{3}$ & $\langle\rho_{0}\rangle$ & $\langle R\,\rangle$ & $\langle\,\overline{g}\,\rangle$ \\ \hline
$+20.0$ & $-4.425$ & 0.19185(3) & 3.97006(8) & 40.2(13) \\ \hline
$+20.0$ & $-4.426$ & 0.01285(1) & 3.6867(7)  & 10.44(1) \\ \hline
$-20.0$ & $+6.405$ & 0.1443(6)  & 3.812(2)   & 21.7(11) \\ \hline
$-20.0$ & $+6.404$ & 0.1440(6)  & 3.811(2)   & 21.6(14) \\ \hline
$-20.0$ & $+6.403$ & 0.1431(6)  & 3.808(2)   & 19.2(7) \\ \hline
\end{tabular}
\caption{Measurements taken near phase transitions.}
\label{tab:sharp}
\end{center}
\end{table}

\noindent
This may indicate that the phase boundary is more strongly first order for large positive values of $\kappa_{0}$. Further exploration of ${\cal C}_{2}$ is temporarily postponed. This chapter is ended with another look at the strong coupling phases.

\section{Strong Coupling Regimes} \label{sec:strong}

Triangulations in the crumpled phase of the $\kappa_{3}=0$ limit rapidly thermalise to a particular domain of the space of triangulations $\cal T$, which is asymptotically fixed as $\kappa_{0}\rightarrow -\infty$. Time series plots of $N_{0}$ have confirmed this fact. Naturally, one assumes that this domain corresponds to the kinematic lower bound of $N_{0}$. Let us explain the basis for this assumption. 

In the $\kappa_{3}=0$ limit, the action is a function of $N_{0}$ and $N_{5}$. When $\kappa_{0}$ is large, triangulations with large $N_{0}$ will dominate the partition function -- due to their increased statistical weight. Obviously, for a given $N_{5}$, one can identify a subset of $\cal T$ in which $N_{0}$ and $\rho_{0}$ are {\it maximal}. As $\kappa_{0}\rightarrow\infty$ we expect this subset to saturate the partition function. Conversely, one can identify another subset of $\cal T$ in which $N_{0}$ and $\rho_{0}$ are {\it minimal}. One expects this subset to saturate the partition function as $\kappa_{0}\rightarrow -\infty$. However, certain measurements of $\rho_{0}$ for $\kappa_{3}<\kappa_{3}^{c}$ have shown that this is not the case (see figure~\ref{fig:zoomoutR}(a)). This observation raises the question: why does the $\kappa_{0}\rightarrow -\infty$ limit of $\rho_{0}$ not correspond to the true kinematic lower bound? 

\vskip 5mm

\noindent
In constrast, the crumpled phase of the $\kappa_{0}=0$ limit is more complicated --  in the sense that there does not appear to be a single asymptotic domain of $\cal T$ to which the ensembles equilibrate. In actual fact, there is evidence of a kinematic lower bound for $N_{3}$ and $R$, but triangulations for $\kappa_{3}\rightarrow -\infty$ do not coincide with the bound as one would expect. What could be the reason for this? This effect may be due to ensembles becoming trapped in metastable states. 

One expects triangulations with small $N_{3}$ to dominate for $\kappa_{3}<\kappa_{3}^{c}$, and those with minimal $N_{3}$ to saturate the partition function as $\kappa_{3}\rightarrow -\infty$. Our observations may be explained by the existence of a number of domains in $\cal T$ having minimal $N_{3}$. In other words, the free energy surface has a number of vacua. Time series plots of $N_{3}$ 
seem to be consistent with this theory -- we find that during thermalisation $N_{3}$ often reaches apparently stable equilibria followed by sudden drops. These may translate to the ensemble migrating from one minimum of the free energy to another via random fluctuations. It is possible to test for the existence of such metastable states by simple experiments. This is done by thermalising a set of different starting configurations. If the final state varies, then this is indicative of metastable states. This avenue of research was not explored, due to time constraints.

\vskip 5mm

\begin{figure}[ht]
\centerline{\epsfxsize=14cm \epsfbox{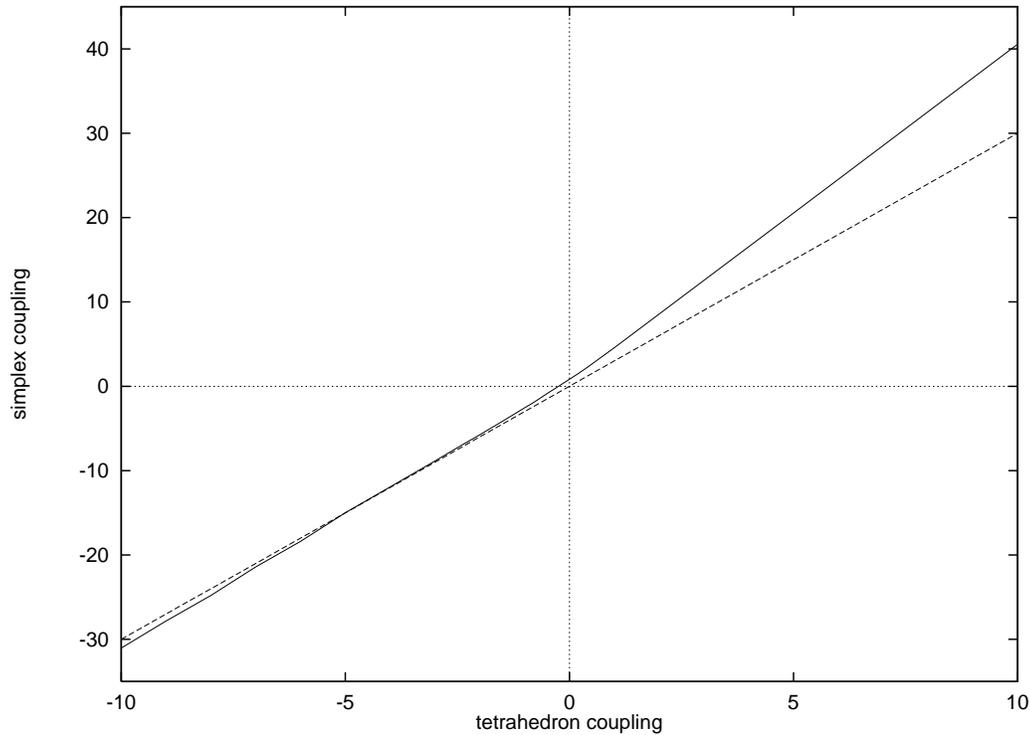}}
\caption{Tuned simplex coupling $\kappa_{5}$ versus tetrahedron coupling $\kappa_{3}$.}
\label{fig:k5k3}
\end{figure}

\noindent
Using equation~(\ref{eq:deltani}) one can write down an expression for the change in $S_{5}^{g}[T,\kappa_{0},\kappa_{3}]$ for any five dimensional $(k,l)$ move. In the $\kappa_{3}=0$ limit, it is written as
\begin{equation}
\Delta S_{5}^{g}[T,\kappa_{0},0]_{(k,l)} = \kappa_{5}(l-k)-\kappa_{0}({}^{l}\mbox{C}_{6}-{}^{k}\mbox{C}_{6}) = -\Delta S_{5}^{g}[T,\kappa_{0},0]_{(l,k)}.
\end{equation}
Numerical simulations have shown that $\Delta S_{5}^{g}[T,\kappa_{0},0]_{(k,l)}$ is positive for $k=1,2,3$ and hence negative for $k=4,5,6$. This is true for all $\kappa_{0}$. In the $\kappa_{0}=0$ limit, the analogous relation reads
\begin{equation}
\Delta S_{5}^{g}[T,0,\kappa_{3}]_{(k,l)} = \kappa_{5}(l-k)-\kappa_{3}({}^{l}\mbox{C}_{3}-{}^{k}\mbox{C}_{3}) = -\Delta S_{5}^{g}[T,0,\kappa_{3}]_{(l,k)}. \label{eq:deltastrongk3}
\end{equation}
In this case we find that for $\kappa_{3}=-10$, the change in action $\Delta S_{5}^{g}[T,0,\kappa_{3}]_{(k,l)}$ is positive for $k=1,2,4$ and hence negative for $k=3,5,6$. Of course, when $\kappa_{0}=\kappa_{3}=0$ the change in action is positive for $k=1,2,3$. This means that $\Delta S_{5}^{g}[T,0,\kappa_{3}]_{(3,4)}$ changes sign at some point in the range $-10<\kappa_{3}<0$. By substituting $k=3$ and $l=4$ into~(\ref{eq:deltastrongk3}) we get
\begin{equation}
\Delta S_{5}^{g}[T,0,\kappa_{3}]_{(3,4)} = \kappa_{5}-3\kappa_{3} = -\Delta S_{5}^{g}[T,0,\kappa_{3}]_{(4,3)}. 
\end{equation}
In other words, the sign change occurs when $\kappa_{5}=3\kappa_{3}$. Figure~\ref{fig:k5k3} plots the tuned simplex coupling $\kappa_{5}$ versus $\kappa_{3}$ (for $\kappa_{0}=0$). 
The straight line corresponds to $\kappa_{5}=3\kappa_{3}$. Clearly, we see that $\kappa_{5}<3\kappa_{3}$ for $\kappa_{3}~\hbox{\begin{picture}(9,7)(0,2)\put(0,-2){\shortstack{$<$\\[-2pt]$\sim$}}\end{picture}}-5$. It is exactly this region where we observe the strange metastable effects. In conclusion, it appears that the sign change of $\Delta S_{5}^{g}[T,0,\kappa_{3}]_{(3,4)}$ results in a destabilisation effect, which seems to prevent the ensemble from reaching true equilibrium. 

Notice from figure~\ref{fig:k5k3} that $\kappa_{5}=3\kappa_{3}$ near $\kappa_{3}=-5$. Recall that $\kappa_{5}$ is calculated via an iterative tuning procedure during the thermalisation stage. The change in action due to the $(3,4)$ move for $\kappa_{3}=-5$ is zero when $\kappa_{5}=15$. In this case, $\kappa_{5}$ was tuned to $-14.992139$. Clearly, these values satisfy $\kappa_{5}>3\kappa_{3}$. This accounts for the apparent complete thermalisation. This simulation was repeated as an experiment. This time $\kappa_{5}$ was tuned to $-15.138613$. As expected, the ensemble became trapped in metastable states. The critical simplex coupling $\kappa_{5}$ obviously plays a crucial role in determining whether the ensemble reaches true equilibrium. This type of effect will always emerge whenever the overall change in action of a $(k,l)$ move is close to zero.

        
\input epsf.tex

\chapter{Fractal Geometry and Stacked Spheres} \label{chap:chap5}

\section{Spatial Distributions} \label{sec:spatdist}

\begin{figure}[ht]
\leavevmode
\hbox{\epsfxsize=1.6cm \epsfbox{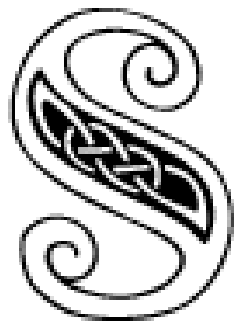}}
\parbox{14.25cm}{\baselineskip=16pt plus 0.01pt \vspace{-20.5mm}O far, our extensive series of Monte Carlo simulations have revealed the existence of a phase boundary and at least two phases in $\cal C$$_{2}$. The weak coupling phase corresponds to elongated triangulations with large scalar curvature and vertex density. In contrast, the strong coupling regimes of the $\kappa_{3}=0$ and $\kappa_{0}=0$ limits}
\end{figure}
\vspace{-6mm}
appear to be distinct. The main purpose of this section is to characterise these phases in geometrical terms. This goal may be achieved by visualising the the $S^{5}$ triangulations at a global scale; in other words, as structures embedded in $\dbl R$$^{6}$. Specifically, this entails measuring the number of simplices $N$ at a given geodesic distance $r$ from an arbitrary origin simplex\,\footnote{Details of the algorithm are provided in appendix~\ref{app:sdf}}. The resulting {\it spatial distribution} $N(r)$ is normalised and averaged over all possible origin simplices and over many Monte Carlo configurations. Therefore, the mean of the distribution is, by definition, located at $r=\langle\,\overline{g}\,\rangle$, the mean geodesic distance. From such information one can gain knowledge of the spatial distribution of simplices and, as we shall see later, learn more about the geometrical and fractal nature of the triangulations. This will give us a more complete picture of the phase structure as a whole.

\vskip 5mm

\begin{figure}[htp]
\begin{center}
\leavevmode
(a){\hbox{\epsfxsize=14cm \epsfbox{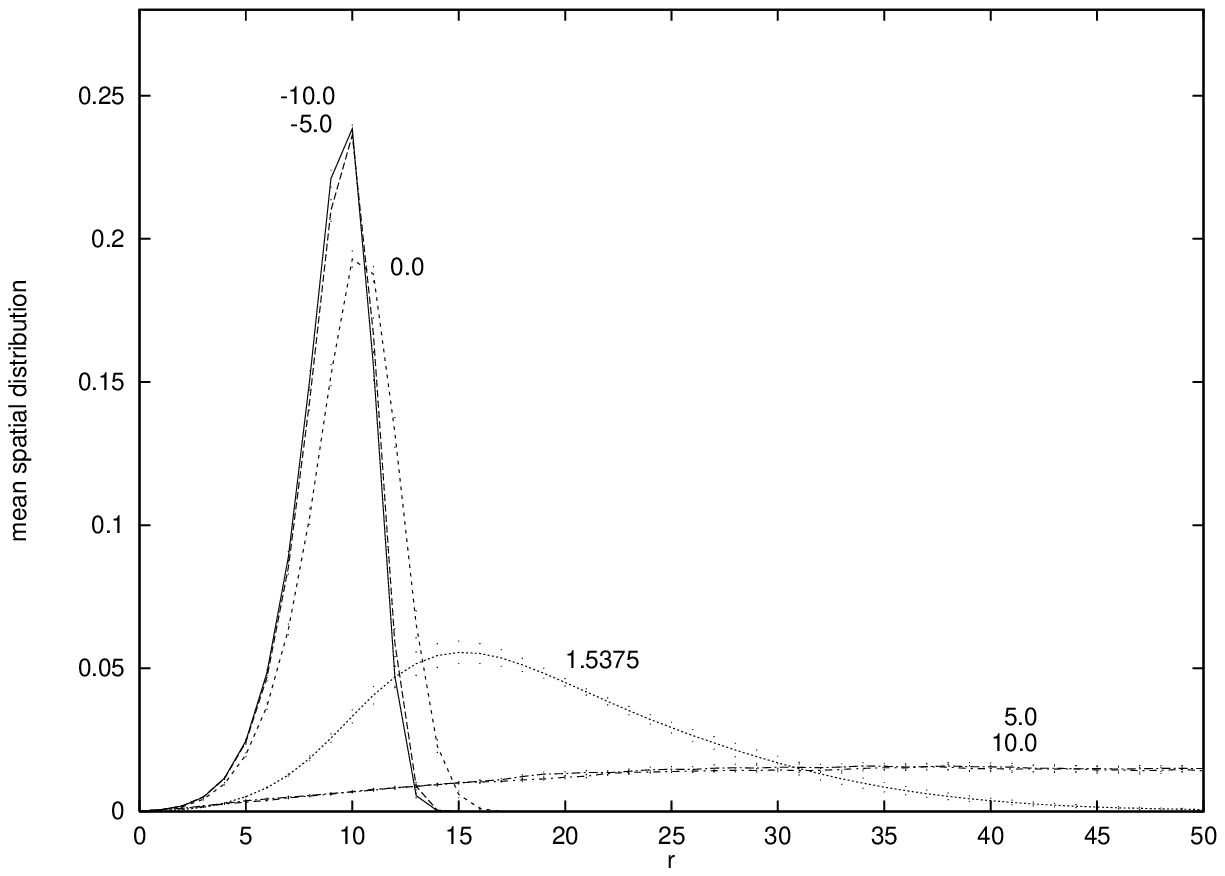}}}
(b){\hbox{\epsfxsize=14cm \epsfbox{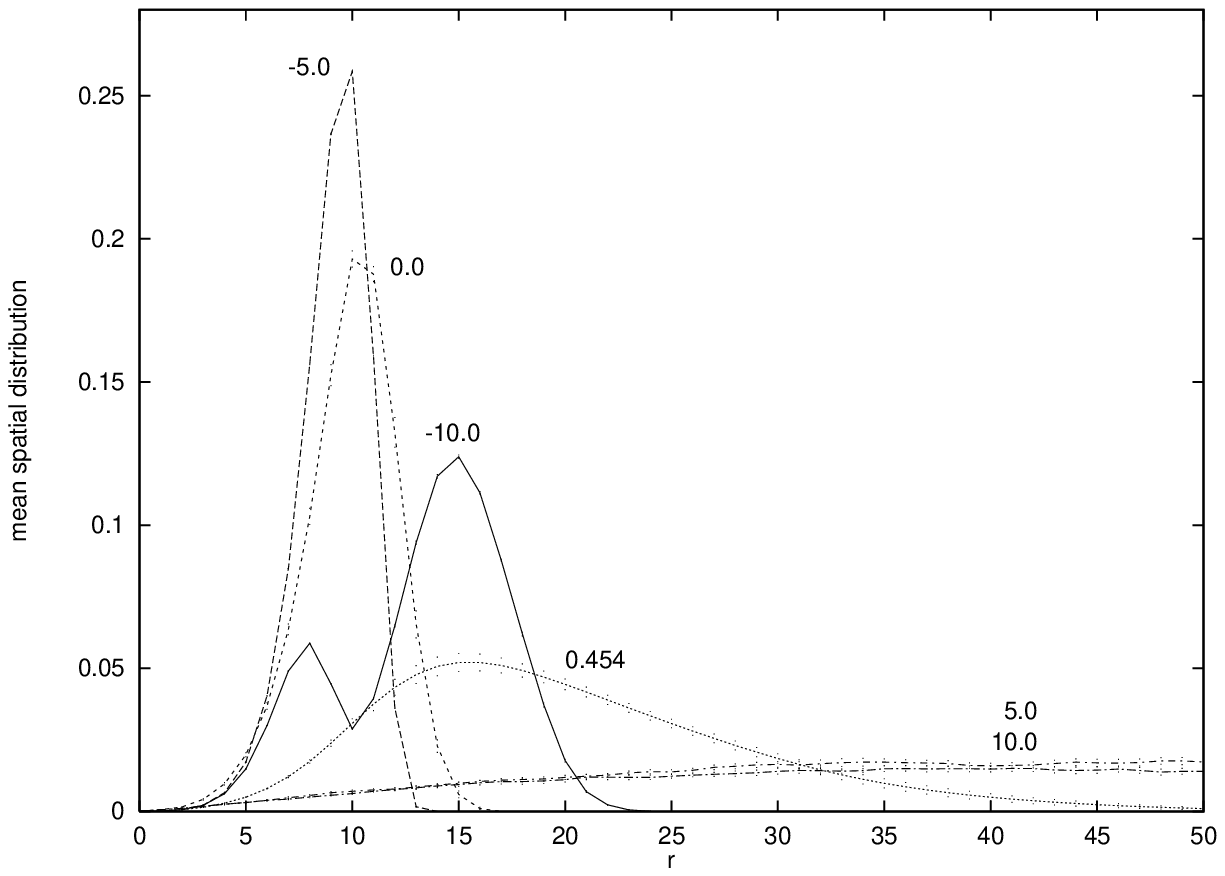}}}
\caption{Mean spatial distribution $\langle N(r)\rangle$ for (a) the $\kappa_{3}=0$ limit and (b) the $\kappa_{0}=0$ limit, for triangulations with target volumes $N_{5}^{t}=10$k.}
\label{fig:nr10k}
\end{center}
\end{figure}

\noindent
The first step of this project is to establish how $N(r)$ varies with $\kappa_{0}$ and $\kappa_{3}$. Measurements of $N(r)$ were taken at regular intervals over $10^{5}$ Monte Carlo sweeps, for target volumes of $N_{5}^{t}=10$k. Simulations were run over a wide range of couplings, which included points near the phase transitions: $\kappa_{0}=1.5375$ and $\kappa_{3}=0.454$. Figures~\ref{fig:nr10k}(a) and \ref{fig:nr10k}(b) show plots of the mean spatial distribution $\langle N(r)\rangle$ for the $\kappa_{3}=0$ and $\kappa_{0}=0$ limits respectively, where each curve is labelled by its corresponding coupling. 
Both graphs are plotted with the same scale to allow comparisons to be made. Measurements of $N(r)$ for $\kappa_{0}=\kappa_{3}=0$ serve as a point of reference. The error estimates\,\footnote{The errors were estimated using the Jackknife method, which is explained in appendix~\ref{app:jack}.} (indicated by dots) are considered reliable as they are compatible with the fluctuations in $N(r)$. 

\subsection{Results and Interpretation}

From figures~\ref{fig:nr10k}(a) and \ref{fig:nr10k}(b), we find that the weak coupling spatial distributions are virtually indistinguishable. This reaffirms the fact that the weak coupling regimes of the $\kappa_{3}=0$ and $\kappa_{0}=0$ limits correspond to the same elongated phase. (Some simplices of these triangulations are separated by geodesic distances of up to 140 units.) One can plainly see that the spatial distributions near the phase transitions are also very similar. This could be consistent with the phase transitions being of the same order, as originally suspected. Perhaps one should not read too much into this observation as there are hidden errors derived from the uncertainty in the location of the phase transitions. 

Let us now consider the spatial distributions in the strong coupling regimes. It is here that we find striking differences between the two sets of distributions. The $\kappa_{0}\ll\kappa_{0}^{c}$ plots of figure~\ref{fig:nr10k}(a) are consistent with the existence of an asymptotic set of crumpled triangulations, and correlate perfectly with measurements of $\langle\rho_{0}(\kappa_{0})\rangle$ and $\langle R(\kappa_{0})\rangle$ from section~\ref{subsec:derivatives}. These results highlight the distinction between the two strong coupling regimes. From figure~\ref{fig:nr10k}(b) it is clear that $\kappa_{3}=-5.0$ plot corresponds to highly crumpled triangulations. This is also in agreement with earlier results. The most surprising result of this study relates to the $\kappa_{3}=-10.0$ plot, which has {\it two} peaks! This spatial distribution of simplices accounts for the unexpected increases observed in $\rho_{0}$, $R$ and $\overline{g}$ for $\kappa_{3}~\hbox{\begin{picture}(9,7)(0,2)\put(0,-2){\shortstack{$<$\\[-2pt]$\sim$}}\end{picture}}-5$. These represent the possible metastable configurations discussed in section~\ref{sec:strong}.

One may wonder what the corresponding triangulations look like. Unfortunately, it is not possible to uniquely determine the geometry of the triangulations from knowledge of $N(r)$ alone. Despite this fact, one can deduce certain features from basic arguments. First of all, the double peak tells us that the geometry is not even remotely spherical. One can only conclude that the geometry is comprised of a number of `baby universes' connected by thin necks. The last and perhaps most powerful deduction is that the number of these necks is small -- probably just one or two. This becomes clear when we consider the spatial distribution of the elongated phase which are believed to be branched polymers. As these triangulations have a large number of necks, the mean spatial distribution is smoothed out, resulting in a single peak. 

\section{Fractal Structure} \label{sec:fractal}

This section deals with the fractal nature of five dimensional triangulations. Our primary aim is to prove that the elongated phase is characterised by branched polymer triangulations. This is done by evaluating the {\it internal Hausdorff dimension} $d_{H}$ as a function of $r$. In general, the internal Hausdorff dimension of a geometrical structure is defined in terms of the volume $V$ contained within a radius $r$.
\begin{equation}
V \propto r^{\,d_{H}} \label{eq:hausdorff}
\end{equation}
This definition may be naturally extended to triangulations, by considering the number of simplices contained within a given geodesic distance $r$ of an arbitrary origin simplex. Now, the internal Hausdorff dimension of a triangulation is a function of $r$, which is given by
\begin{equation}
V(r) = \alpha\,r^{\,d_{H}(r)}, \label{eq:df}
\end{equation}
where $\alpha$ is a constant and $V(r)$ is the volume of a $d$-ball with geodesic radius $r$. This expression is valid only for a limited range of $r$, since we are dealing with finite triangulations. Of course, $N(r)$ is the volume of a shell at a geodesic radius $r$. Therefore, $V(r)$ is given by the sum over $N(r)$ from 0 to $r$. By averaging over all origin simplices, one obtains the following expression.
\begin{equation}
\langle V(r)\rangle = \sum_{i=0}^{r}\langle N(i\;\!)\rangle \label{eq:vr}
\end{equation}
This is calculated in practice using the results of section~\ref{sec:spatdist}. By taking logs and differentiating (\ref{eq:hausdorff}) with respect to $\ln r$, we obtain a formal definition of the internal Hausdorff dimension.
\begin{equation}
d_{H} = \frac{\mbox{d}\ln V}{\mbox{d}\ln r} \label{eq:dhr}
\end{equation}
This may also be extended to (discrete) triangulations by replacing derivatives with differences. Using (\ref{eq:vr}) one can define an expression for the mean internal Hausdorff dimension function.
\begin{equation}
\langle d_{H}(r)\rangle = \frac{\ln\langle V(r)\rangle -\ln\langle V(r-1)\rangle}{\ln r -\ln(r-1)} \label{eq:dhdiff}
\end{equation}
Using equation (\ref{eq:vr}) one can then easily express $d_{H}(r)$ in terms of $N(r)$.

\subsection{Branched Polymers}

\begin{figure}[htp]
\begin{center}
\leavevmode
(a){\hbox{\epsfxsize=14cm \epsfbox{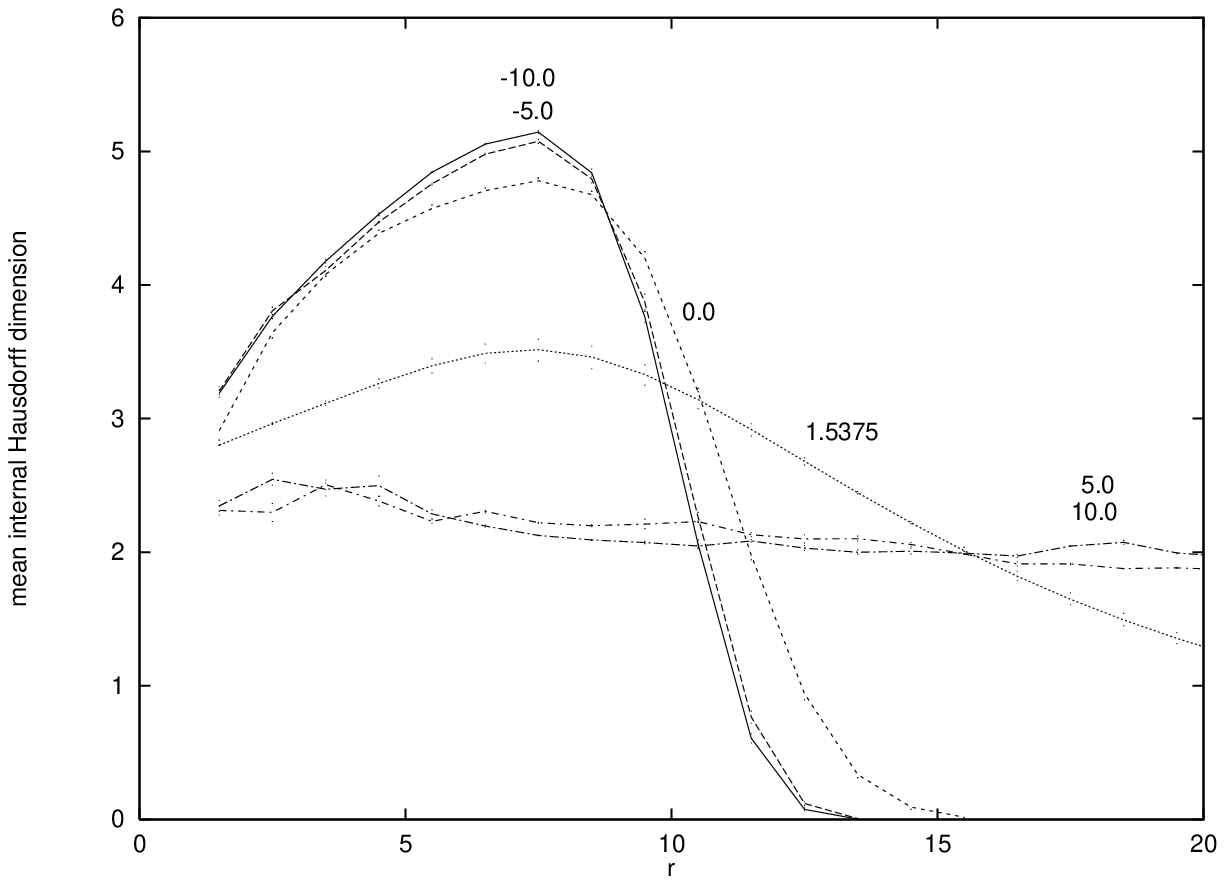}}}
(b){\hbox{\epsfxsize=14cm \epsfbox{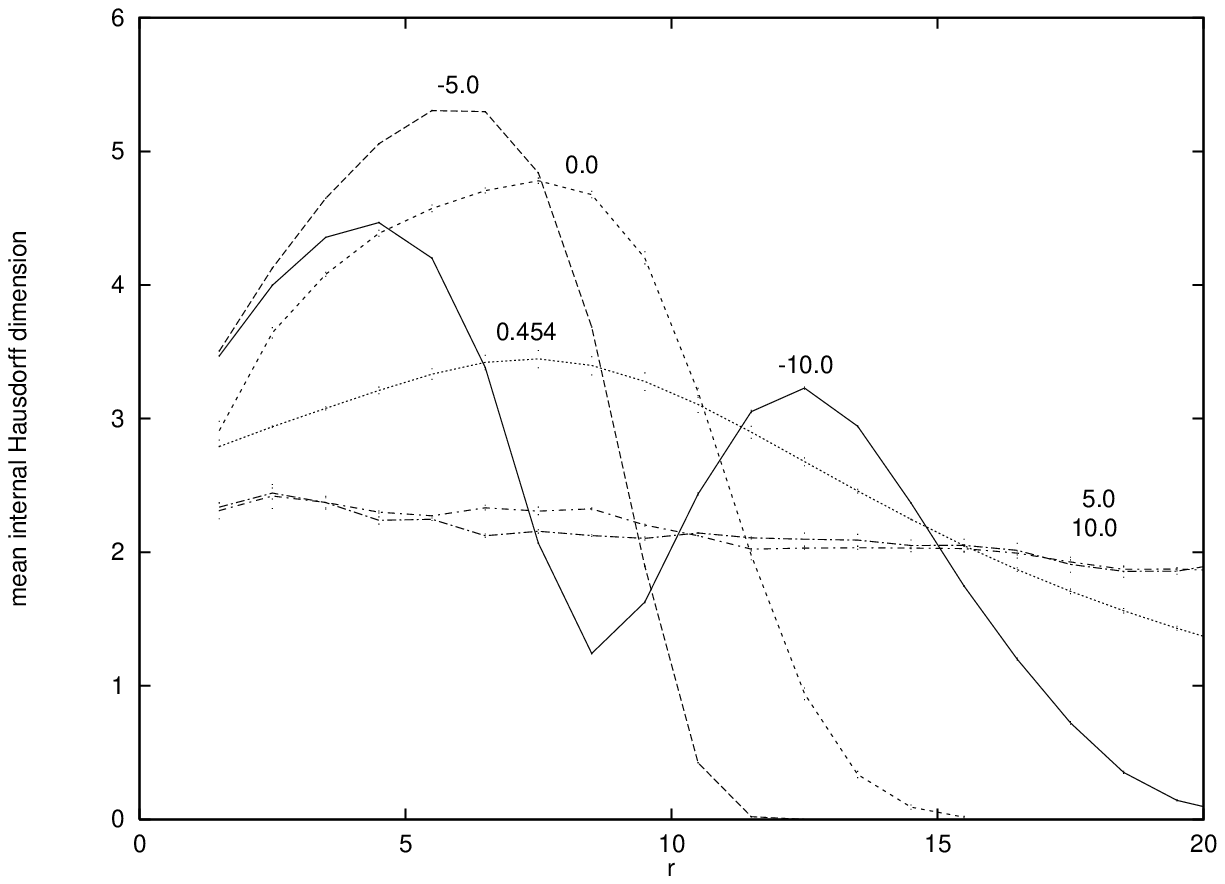}}}
\caption{Mean internal Hausdorff dimension $\langle d_{H}(r)\rangle$ for (a) the $\kappa_{3}=0$ limit and (b) the $\kappa_{0}=0$ limit, for triangulations with target volumes $N_{5}^{t}=10$k..}
\label{fig:dh10k}
\end{center}
\end{figure}

Figures~\ref{fig:dh10k}(a) and \ref{fig:dh10k}(b) show plots of $\langle d_{H}(r)\rangle$ for the $\kappa_{3}=0$ and $\kappa_{0}=0$ limits respectively\,\footnote{The method of error estimation is discussed in appendix~\ref{app:sdf}.}. 
The internal Hausdorff dimension of the triangulations is given by the maximum of $d_{H}(r)$, which widens into a plateau as the target volume is increased. From these figures one can plainly see that the fractal dimension of the weak coupling triangulations is $d_{H}\approx2$. This result proves that the weak coupling triangulations are branched polymers. The strong coupling triangulations are characterised by large internal Hausdorff dimension. It is important to realise that these results are subject to finite size effects; one can learn more by measuring $d_{H}(r)$ as a function of volume. 

Further Monte Carlo simulations were run in order to establish the scaling behaviour of $d_{H}(r)$. The study was repeated for triangulations of target volumes of $N_{5}^{t}=20$k and $N_{5}^{t}=30$k. It was found that $d_{H}\approx 2$ in the weak coupling phase, regardless of volume. From now on we shall refer to the common weak coupling phase as the {\it branched polymer phase}. In the strong coupling regime of the $\kappa_{3}=0$ limit, we find that $d_{H}$ increases approximately linearly with volume. This would be consistent with $d_{H}\rightarrow\infty$ in the thermodynamic limit, as conjectured by Ambj{\o}rn~\cite{quantgeo}. Double peaks in $N(r)$ were also detected for $\kappa_{3}\ll\kappa_{3}^{c}$. Finally, the internal Hausdorff dimensions measured near the phase transitions also increase with volume. However, these results are not considered reliable due to relatively poor statistics and hidden errors due to the uncertainty in the location of the phase transitions. 

\section{Stacked Spheres} \label{sec:stack}

For a given space of triangulations $\cal T$ with fixed volume, it is clear that one can identify a subspace of $\cal T$ in which $N_{0}$ is minimal. These triangulations correspond to the lower kinematic bound of $N_{0}$. Conversely, there also exists a subspace of $\cal T$ in which $N_{0}$ is maximal. These triangulations of the upper kinematic bound are known as {\it stacked spheres}. This section deals with the analytic study of these structures. The name given to these structures is quite apt when one considers how they are constructed. Minimal $d$-spheres are composed of $d+2$ vertices and $d+2$ simplices pairwise connected along their faces. Stacked $d$-spheres are generated from minimal $d$-spheres by performing {\it only} vertex insertion moves. In a sense, new simplices are `stacked onto' the sphere. It is the fact that only $(1,d+1)$ moves are used that allows us to study stacked spheres by analytical means. 

\vskip 5mm

\noindent
It is a trivial matter to calculate the scalar curvature $R$ of a stacked sphere. Consider a minimal 5-sphere, which has $N_{3}=35$ and $N_{5}=7$,\,\footnote{The number of $i$-(sub)simplices of a minimal $d$-sphere is given by ${}^{d+2}$C$_{i+1}$.}. Using equation~(\ref{eq:deltani}) one can show that $\Delta N_{3}=+10$ and $\Delta N_{5}=+5$ for the $(1,6)$ move. These results give us the following relations for $N_{3}$ and $N_{5}$ after $n$ vertex insertion moves, where $n\in{\dbl N}$.
\begin{eqnarray}
N_{3} & = & 35 + 20n \label{eq:ssN3} \\
N_{5} & = & 7 +5n \label{eq:ssN5}
\end{eqnarray}
By eliminating $n$ and dividing by $N_{5}$ we get the following equation for the scalar curvature, expressed in terms of $N_{5}$.
\begin{equation}
R = \frac{N_3}{N_5} = 4 + \frac{7}{N_5} \label{eq:Rstacked}
\end{equation}
Clearly, we find that $R\rightarrow 4$ in the thermodynamic limit. One must note that (\ref{eq:Rstacked}) does not hold for all volumes -- only those that satisfy (\ref{eq:ssN5}). Substitution of $N_{5}=10,002$ into (\ref{eq:Rstacked}) gives $R\approx 4.0007$. This value of the scalar curvature matches exactly the weak coupling data presented in figure~\ref{fig:zoomoutR}(b). This proves that the partition function in this limit is saturated by stacked spheres. It is important to bear in mind that these structures are {\it branched polymer} stacked spheres. These are constructed by performing vertex insertion moves on {\it randomly chosen} simplices. One can generate {\it singular stacked spheres} if the moves are performed on simplices in a specific pattern. This resulting triangulation has a vertex with a very large local volume. Such structures clearly do not resemble branched polymers. 

\section{Walkup's Theorem} \label{sec:walkup}

It is the fact that stacked spheres represent a kinematic bound that allows for the existence of inequalities relating $N_{1}$ and $N_{0}$, at least in three and four dimensions. Collectively the inequalities are called {\it Walkup's theorem}~\cite{walkup}. These results have been central to certain advances made in the understanding of the weak coupling limit of dynamical triangulations~\cite{gabrielli,gionti}.
\begin{eqnarray}
d=3 : \hspace{10mm} N_{1} & \geq & 4 N_{0} - 10 \label{eq:3dwalk} \\ 
d=4 : \hspace{10mm} N_{1} & \geq & 5 N_{0} - 15 \label{eq:4dwalk}
\end{eqnarray}
Inequalities (\ref{eq:3dwalk}) and (\ref{eq:4dwalk}) are true for all spheres and hold with equality only for stacked spheres\,\footnote{It is worth noting that there exist four dimensional triangulations which satisfy the lower bounds of inequality (\ref{eq:4dwalk}) that are {\it not} stacked 4-spheres. This is because in even dimensions there are $(k,l)$ moves that transform triangulations without changing the $f$-vector. In other words, it is a necessary but insufficient condition for stacked 4-spheres to satisfy the lower bounds of Walkup's theorem in four dimensions. This does not happen in odd dimensions.}. 

\subsection{Generalisation of Walkup's Theorem}

This chapter is concluded with a simple proof of Walkup's theorem by considering the effect of $(k,l)$ moves on stacked spheres. In fact, the proof is a generalisation of Walkup's theorem to arbitrary dimension. In one and two dimensions $N_{1}$ is uniquely defined for any given $N_{0}$.
\begin{eqnarray}
d=1 : \hspace{10mm}N_{1} & = & N_{0} \label{eq:1dwalk}\\
d=2 : \hspace{10mm} N_{1} & = & 3 N_{0} - 6 \label{eq:2dwalk}
\end{eqnarray}
The following analysis applies to all dimensions $d>0$. From chapter~\ref{chap:chap1} we know that a minimal $d$-sphere is composed of ${}^{d+2}$C$_{i+1}$ $i$-(sub)simplices. 
\begin{eqnarray}
N_{0} & = & {}^{d+2}\mbox{C}_{1} = d+2 \\
N_{1} & = & {}^{d+2}\mbox{C}_{2}  
\end{eqnarray}
Starting from a minimal $d$-sphere one can construct a stacked $d$-sphere by performing a series of $n$ vertex insertion moves. Again, using equation~(\ref{eq:deltani}) one can calculate $\Delta N_{0}$ and $\Delta N_{1}$ for a $(1,d+1)$ move.
\begin{eqnarray}
\Delta N_{0} & = & +1 \\
\Delta N_{1} & = & +(d+1-{}^{1}\mbox{C}_{d}) 
\end{eqnarray}
Using these equations one can write very simple expressions for $N_{0}$ and $N_{1}$ in terms of just $d$ and $n$.
\begin{eqnarray}
N_{0} & = & d+2+n\Delta N_{0} = d+2+n \\
N_{1} & = & {}^{d+2}\mbox{C}_{2}+n\Delta N_{1}={}^{d+2}\mbox{C}_{2}+n(d+1-{}^{1}\mbox{C}_{d}) 
\end{eqnarray}
By eliminating $n$ one can express $N_{1}$ in terms of only $N_{0}$ and $d$. The following equation is true for any stacked $d$-sphere.
\begin{equation}
N_{1} = {}^{d+2}\mbox{C}_{2} + (N_{0} - d-2)(d+1-{}^{1}\mbox{C}_{d}) \label{eq:eqwalk}
\end{equation}
Equation (\ref{eq:eqwalk}) clearly reduces to (\ref{eq:1dwalk}) and (\ref{eq:2dwalk}) when $d=1$ and $d=2$ respectively. It is conjectured that the following inequality holds for any $d$-sphere.  
\begin{equation}
N_{1} \geq {}^{d+2}\mbox{C}_{2} + (N_{0} - d-2)(d+1-{}^{1}\mbox{C}_{d}) \label{eq:ineqwalk}
\end{equation}
This inequality is offered as a generalised form of Walkup's theorem. It clearly reduces to Walkup's theorem when $d=3$ and $d=4$. This is taken as evidence supporting the validity of our conjecture. The following sections form a proof of this result.

\subsection{Proof Strategy} \label{subsec:proofstrategy}

The strategy for proving our generalisation of Walkup's theorem is as follows. We begin with a triangulation that satisfies (\ref{eq:eqwalk}), i.e. a stacked sphere. It is known that a stacked sphere may be transformed into any other combinatorially equivalent (piecewise linear homeomorphic) triangulation using a finite series of $(k,l)$ moves~\cite{pachner1,pachner2}. If one can show that the effect of each $(k,l)$ move is consistent with (\ref{eq:ineqwalk}), then one may be able to conclude that the lower bound holds for all $d$-spheres. 

\vskip 5mm

\noindent
From now on we only consider $d>1$. Our conjecture reads
\begin{equation}
N_{1} \geq {}^{d+2}\mbox{C}_{2}+(N_{0}-d-2)(d+1). \label{eq:ineqwalkd>1}
\end{equation}
Inequality (\ref{eq:ineqwalkd>1}) may be shown to be false if there exist $(k,l)$ moves that have $\delta < d+1$, where $\delta$ is defined as follows.
\begin{equation}
\delta = \frac{\Delta N_{1}}{\Delta N_{0}} = \frac{{}^{d+2-k}\mbox{C}_{d}-{}^{k}\mbox{C}_{d}}{{}^{d+2-k}\mbox{C}_{d+1}-{}^{k}\mbox{C}_{d+1}} 
\end{equation}
For $k=1$ (vertex insertion moves) and $k=d+1$ (vertex deletion moves) we see that $\delta=+(d+1)$ which means that the moves {\it are} consistent with inequality (\ref{eq:ineqwalkd>1}). Now $\Delta N_{0}=0$ for all moves with $1<k<d+1$. Therefore the only moves that are {\it inconsistent} with the inequality are those with $1<k<d+1$ {\it and} $\Delta N_{1}<0$. Using $\Delta N_{1}={}^{d+2-k}$C$_{d}-{}^{k}$C$_{d}$ we see that $\Delta N_{1}<0$ only when $k=d,d+1$. Therefore $\delta<d+1$ only for the $(d,2)$ move. Such a move could {\it potentially} be used to generate triangulations that do not satisfy (\ref{eq:ineqwalkd>1}). This situation presents us with two possibilities. Either
\begin{itemize}
\item (a) our conjecture is wrong, or
\item (b) $\cal N$$_{(2,d)} \geq \cal N$$_{(d,2)}$,
\end{itemize}
where $\cal N$$_{(k,l)}$ is the total number of $(k,l)$ moves required to generate an arbitrary triangulation from a stacked sphere. Option (a) seems unlikely because (\ref{eq:ineqwalkd>1}) is known to reduce to Walkup's theorem for $d=3$ and $d=4$. In the next section we hope to convince the reader that (b) is true.

\subsection{Clusters and $(d,2)$ Moves}

\begin{figure}[ht] 
\centerline{\epsfxsize=16cm \epsfbox{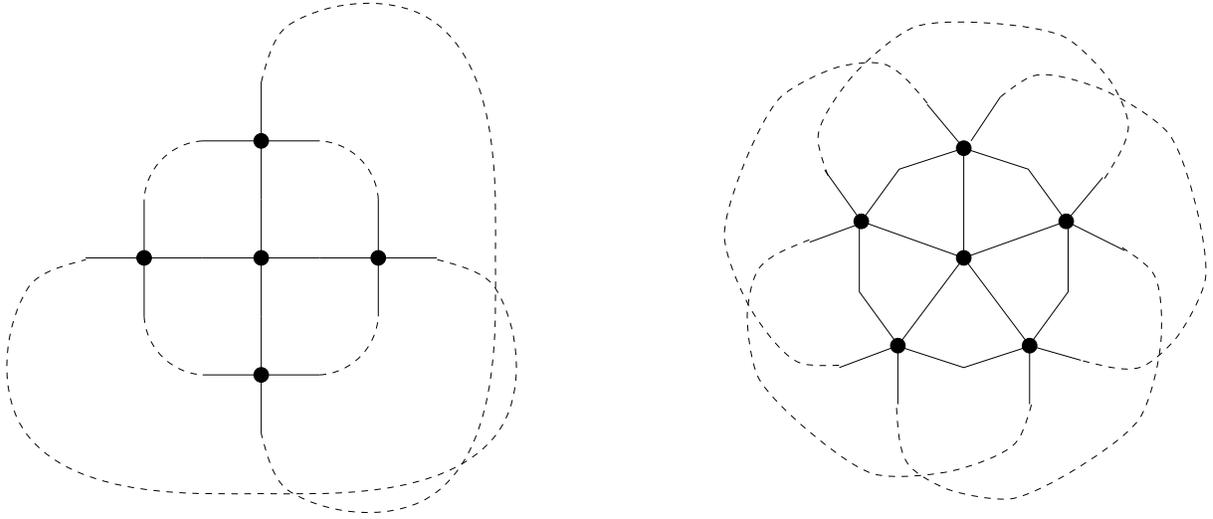}}
\caption{Graph of a minimal 3-sphere (left) and a minimal 4-sphere (right).}
\label{fig:dual34seed}
\end{figure}

From here on, it is necessary to visualise simplices in their dual form -- details of which are to be found in chapter~\ref{chap:chap2}. The reader is reminded that simplices and faces are represented by nodes and edges respectively. In $d$-dimensions each node has $d+1$ edges connecting it with $d+1$ other nodes such that pairs of nodes are connected by {\it only one edge}. This condition imposes that simplices are pairwise connected. In this chapter, these ideas are extended to construct entire triangulations. Their dual versions are called {\it graphs}. Let us begin by considering the graphs of minimal 3-spheres and minimal 4-spheres. These are shown in figure~\ref{fig:dual34seed},\,\footnote{The graphs are drawn symmetrically merely for clarity.}. 

\begin{figure}[htp] 
\centering{\input{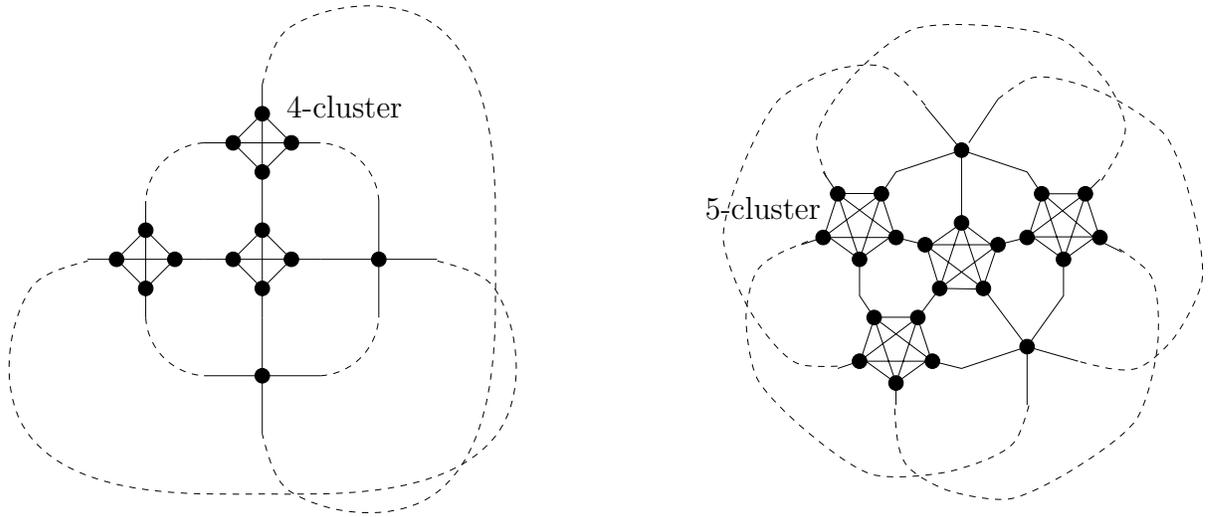}\par}
\caption{Graph of a stacked 3-sphere (left) and a stacked 4-sphere (right).}
\label{fig:34cluster}
\end{figure}

\noindent
Before going any further, we must introduce the concept of a {\it cluster}. An $n$-cluster in $d$-dimensions is a set of $n$ nodes pairwise interconnected by ${}^{n}$C$_{2}$ edges. Each node in the cluster has $d-n+2$ {\it free edges} linked to neighbouring clusters. In general, an $n$-cluster contains ${}^{n}$C$_{i}$ $i$-subclusters, for $i=2,\ldots,n-1$. An $i$-cluster may be regarded as the dual of a $(d+1-i)$-(sub)simplex star. The action of a $(k,l)$ move in the dual picture is therefore the replacement of a $k$-cluster with an $l$-cluster. For stacked $d$-spheres, $d$-clusters can only exist as subclusters of $(d+1)$-clusters, because only $(1,d+1)$ moves are allowed. This is evident from figure~\ref{fig:34cluster} which shows graphs of stacked 3-spheres and stacked 4-spheres, cf. figure~\ref{fig:dual34seed}. Their constituent 4-clusters and 5-clusters are labelled.

\vskip 5mm

\begin{figure}[htp] 
\centering{\input{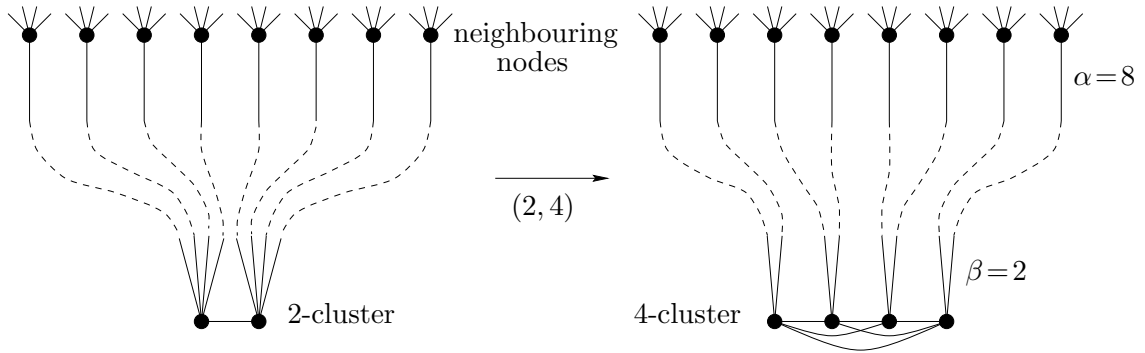}\par}
\caption{Dual representation of the $(2,4)$ move for a stacked 4-sphere.}
\label{fig:2dmove}
\end{figure}

\noindent
In the dual picture, a $(2,d)$ transformation may be considered as the replacement of a 2-cluster with a $d$-cluster. A 2-cluster is the structure formed by two connected nodes. Each node has $d$ free edges since the total number of edges is $d+1$. The $2d$ free edges are pairwise connected to $\alpha=2d$ neighbouring nodes. Now, each node of a $d$-cluster has $\beta=2$ free edges. The fact that $\alpha\not<\beta d$ means that the nodes of the $d$-cluster {\it can} be pairwise connected with the neighbouring nodes\,\footnote{This argument only applies to $d>2$ since $(2,d)$ moves are not defined for $d<3$.}. In other words, it is always possible to perform a $(2,d)$ move whilst maintaining the manifold structure of the triangulation. Figure~\ref{fig:2dmove} illustrates this argument schematically for the four dimensional case.

\begin{figure}[htp] 
\centering{\input{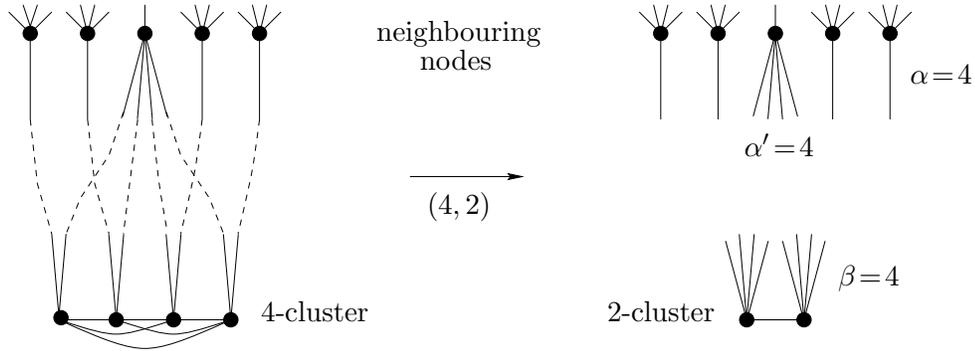}\par}
\caption{Dual representation of the $(4,2)$ move for a stacked 4-sphere.}
\label{fig:d2move}
\end{figure}

Let us now consider $(d,2)$ transformations. (These moves cannot be performed on minimal $d$-spheres as $N_{d}\geq d+2$ for $d$-spheres.) Each node of a $(d+1)$-cluster has one free edge connected to another node outside the cluster. By removing $d$ nodes of a $(d+1)$-cluster we are left with one neighbouring node with $\alpha'=d$ free edges and $d$ other free edges from $d$ neighbouring nodes. These are to be connected with the 2 nodes of the replacement 2-cluster. Again, figure~\ref{fig:d2move} is a schematic illustration of the four dimensional case.
Clearly, one cannot pairwise connect the nodes with this edge configuration. Forced connection results in multiple edges, which signifies the breakdown of the manifold structure. This is because $\alpha<\beta d$. In other words, $(d,2)$ moves cannot be performed on stacked $d$-spheres\,\footnote{In fact, it can be shown that this is the case for all moves with $l<k<d+1$.}. Logically, we conclude that a $(d,2)$ move can only be executed as the inverse of a $(2,d)$ move. Therefore $d$-clusters only exist as subclusters of $(d+1)$-clusters or as the end product of $(2,d)$ transformations. For these reasons we conclude that ${\cal N}_{(2,d)}\geq{\cal N}_{(d,2)}$. This fact proves that our generalisation of Walkup's theorem is correct. $\Box$

\vskip 5mm

\noindent
The dual formulation of $(k,l)$ moves were (apparently) first employed by Godfrey and Gross~\cite{godfreygross} to aid their visualisation in dimensions greater than two. It seems, however, that this is first time the dual graphs have been used to visualise entire triangulations and solve problems in this manner -- although their use was recognised by Gabrielli~\cite{gabrielli}. This work has demonstrated the power of simple graph theory concepts to produce analytic results in dynamical triangulations. The actual generalisation of Walkup's theorem is not in itself a significant result. It is hoped that more complex applications of graph theory may lead to new analytic results.

        

\input epsf.tex

\chapter{Singular Structures} \label{chap:chap6}

\section{Background} \label{sec:background}

\begin{figure}[ht]
\leavevmode
\hbox{\epsfxsize=2.7cm \epsfbox{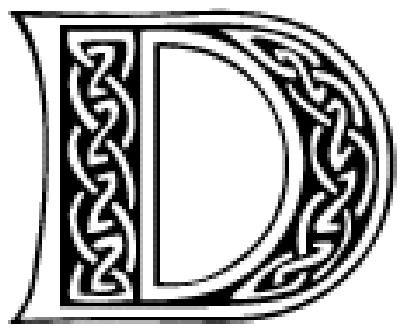}}
\parbox{13.21cm}{\baselineskip=16pt plus 0.01pt \vspace{-20.5mm}URING 1995 a new feature of dynamical triangulations was discovered by Hotta {\it et al.}~\cite{hinptp}. This development centred on measurements of the vertex local volume distribution $\cal N$$(n_{0})$ of four dimensional triangulations. They found that $\,$triangulations $\,$in the strong coupling phase are $\,$characterised}
\end{figure}
\vspace{-5.5mm}
by a small number of vertices with unusually large local volumes. Here we present a short historical review of the discovery and subsequent developments.

\begin{figure}[ht] 
\centering{\input{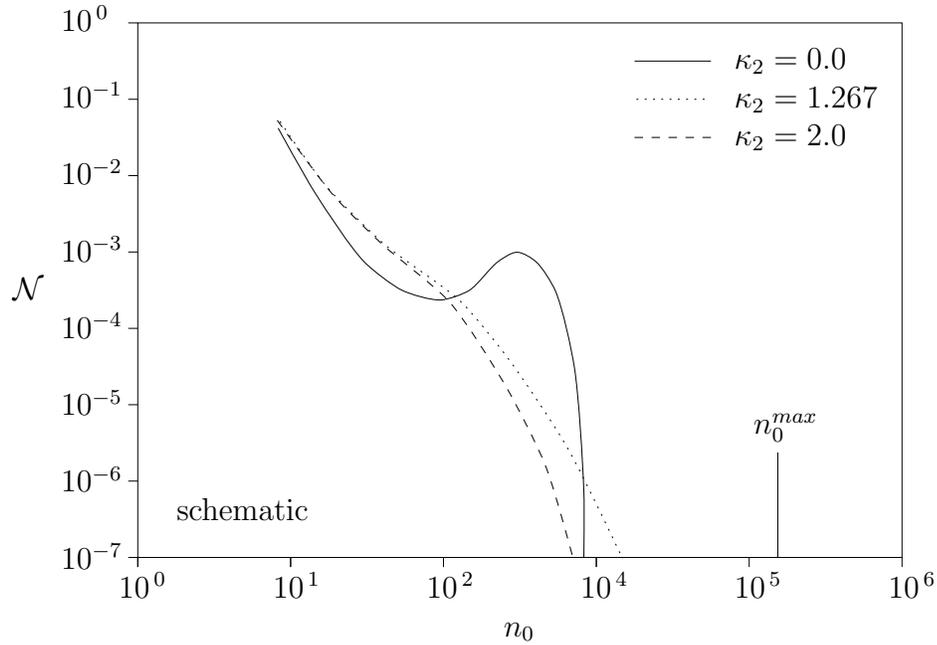}\par}
\caption{Vertex local volume distribution $\cal N$$(n_{0})$ for four dimensional triangulations with target volumes $N_{4}^{t}=32$k. The singular vertices have local volume $n_{0}^{max}$ of order $10^{4}$.}
\label{fig:singular}
\end{figure}

Figure~\ref{fig:singular} is a schematic normalised log-log graph of $\cal N$$(n_{0})$ for four dimensional triangulations with target volumes of $N_{4}^{t}=32$k, drawn from results presented by Hotta {\it et al.}~\cite{hinptp}. 
It shows three plots which correspond to the crumpled phase ($\kappa_{2}=0.0$), the branched polymer phase ($\kappa_{2}=2.0$) and to a point near the phase transition ($\kappa_{2}=1.267$). For $\kappa_{2}\geq\kappa_{2}^{c}$, the distribution is as one may na\"{\i}vely expect: the vast majority of vertices have small\,\footnote{It is easy to show that $n_{0}>d+1$ for simplicial $d$-manifolds.} local volumes ($n_{0}\sim 10^{1}$) and $\cal N$$(n_{0})$ rapidly tails off to zero for $n_{0}\sim10^{3}$. In contrast, the distribution of the strong coupling phase is markedly different; the main distinction being the existence of a small number of vertices with {\it very} large vertex local volumes (about a third of the triangulation volume). It was confirmed that this peak in $\cal N$$(n_{0})$ is representative of two vertices with approximately equal local volumes $n_{0}^{max}$~\cite{hinptp}.

Hotta {\it et al.} also looked at how $\cal N$$(n_{0})$ scales with volume~\cite{hinptp}. In the weak coupling phase, there was found to be no significant dependence of $\cal N$$(n_{0})$ on volume. However, for $\kappa_{2}<\kappa_{2}^{c}$, it was found that $n_{0}^{max}$ increases {\it linearly} with volume. As a consequence, these vertices were dubbed {\it singular vertices} since their local volumes seemed to diverge in the thermodynamic limit. This was perceived to be a potential problem in reaching a continuum limit. For this reason, Hotta {\it et al.} constructed a new variant of the discretised Einstein-Hilbert action $S[T,\kappa_{2}]$ by introducing an additional term which inhibited the formation of singular vertices. The new action is written as follows.
\begin{equation}
S[T,\kappa_{2},\lambda] = S[T,\kappa_{2}]+\frac{\lambda}{2}\sum_{\sigma^{0}\:\!\in\,T}(n(\sigma^{0})-k)\,^{2}
\end{equation}
The extra term is a summation over all $N_{0}$ vertices $\sigma^{0}$ of triangulation $T$. Here, $\lambda$ is a coupling parameter and $k$ is a constant, arbitrarily chosen to be 5, since $n_{0}\geq5$ in four dimensions. Monte Carlo simulations were performed in order to establish the effect of $\lambda$ on the singular vertices. It was found that the singular vertices vanish when $\lambda$ \hbox{\begin{picture}(9,7)(0,2)\put(0,-2){\shortstack{$ > $\\[-2pt]$\sim$}}\end{picture}} $10^{-3}$. In fact, the phase transition also disappears and the dominant triangulations are branched polymers for all values of $\kappa_{2}$. For $\lambda$~\hbox{\begin{picture}(9,7)(0,2)\put(0,-2){\shortstack{$<$\\[-2pt]$\sim$}}\end{picture}}~$10^{-6}$, the additional term has negligible effect and there is no suppression of singular vertices. From this investigation, it became evident that singular vertices are an important feature of four dimensional dynamical triangulations, and may even drive the phase transition~\cite{hinptp}.

In a later paper by Hotta {\it et al.} (with the same title), it was claimed that no observations were made of {\it links} with unusually large local volumes~\cite{hinnpb}. Furthermore, it was found that $d$-dimensional triangulations (for $\kappa_{d-2}=0$) have $d-2$ singular vertices, at least for $3<d<7$. It was thought that perhaps these singular triangulations may be the result of the ensemble becoming trapped in metastable states. This possibility was explored by running Monte Carlo simulations with three very different starting configurations. It was found that in each case the ensemble thermalised to the same set of singular triangulations. This was taken as evidence to suggest that singular vertices are {\it not} a result of incomplete thermalisation, but a genuine feature of the model~\cite{hinnpb}. It was also discovered that toroidal triangulations in the strong coupling phase are {\it also} characterised by singular vertices. In fact, the $\cal N$$(n_{0})$ distribution for $T^{4}$ was found to be virtually identical to that for $S^{4}$ (at least for large volumes). On the basis of this fact, one could tentatively speculate that the existence of singular vertices might be independent of topology. 

\subsection{Further Developments} \label{subsec:furdev}

Further research was carried out by Catterall {\it et al.} on singular vertices~\cite{ckrtsing}, which we shall now review. Monte Carlo simulations were run in four and five dimensions using $S[T,\kappa_{0}]$ with $\kappa_{0}=0$. This choice of coupling ensured that all triangulations carried equal weight in the partition function$\;\!$\footnote{In five dimensions, $S[T,\kappa_{0}]$ is {\it only} equivalent to the discretised Einstein-Hilbert action $S[T,\kappa_{3}]$ when $\kappa_{0}=0$ and $\kappa_{3}=0$.}. On the basis of their numerical results it was conjectured that such $d$-dimensional triangulations are characterised by a single singular $(d-3)$-subsimplex. This {\it primary} singular subsimplex is the product of ${}^{d-2}$C$_{i+1}$ {\it secondary} singular $i$-subsimplices, for $i=0,\ldots,d-4$. These results {\it appear} to be in contradiction with those of Hotta {\it et al.}, who claim that no singular links were observed in four dimensions~\cite{hinnpb}. However, this is probably not the case because Hotta {\it et al.} describe a link as singular if it has an unusually large local volume. In this thesis -- following the convention set by Catterall {\it et al.} -- a singular link is defined as the product of two singular vertices, regardless of its local volume. Driven by numerical results, it was conjectured that the local volume of primary singular subsimplices increase with volume as 
\begin{equation}
n_{d-3}^{max} \propto {N_{d}}^{\frac{2}{3}}.
\end{equation}
In addition, it was conjectured that {\it all} secondary singular $i$-subsimplex local volumes scale {\it linearly} with volume, for $i=0,\ldots,d-2$. All multiple singular structures of a given dimension were confirmed to have approximately equal local volumes. 

\subsubsection{Dual-Spheres and Local Entropy}

These observations were explained heuristically in terms of entropy. Consider an $i$-subsimplex $\sigma^{i}$ of a $d$-dimensional triangulation $T$. The star $St(\sigma^{i})$ of $\sigma^{i}$ is homeomorphic to $B^{d}$, because $T$ is a simplicial $d$-manifold. Since $\sigma^{i}$ has $i+1$ vertices, each simplex of $St(\sigma^{i})$ has $d-i$ vertices that are {\it not} in common with $\sigma^{i}$. These structures are a set of $(d-i-1)$-subsimplices which form an $S^{d-i-1}$ submanifold of $T$. Moreover, this structure is a submanifold of the $S^{d-1}$ boundary of $St(\sigma^{i})$. The $S^{d-i-1}$ submanifold is called the {\it dual-sphere} of $\sigma^{i}$. The local volume $n(\sigma^{i})$ of $\sigma^{i}$ is therefore equal to the volume of its dual-sphere. To a given $i$-subsimplex $\sigma^{i}$, one may associate a {\it local entropy} $\omega(\sigma_{s}^{i})$. This is equal to the entropy of its dual-sphere, which is denoted by $\Omega\:\!(S^{d-i-1})$.

Let us now consider some examples, beginning with $i=d-2$. In this case, the dual-sphere of a $(d-2)$-subsimplex $\sigma^{d-2}$ is a 1-sphere. Clearly, the $S^{1}$ dual-sphere has a unique triangulation for a given volume (up to automorphisms). This means that the local entropy $\omega\:\!(\sigma^{d-2})$ does {\it not} increase with volume. Consequently, the formation of {\it singular} $(d-2)$-subsimplices are not entropically favoured. Now consider the $S^{2}$ dual-sphere of a $(d-3)$-subsimplex $\sigma^{d-3}$. The entropy of a 2-sphere is known to grow exponentially with volume~\cite{tutte}. This fact implies that the local entropy of $\sigma^{d-3}$ also increases exponentially. This accounts for the existence of {\it singular} $(d-3)$-subsimplices. For singular subsimplices of lower dimension ($i<d-3$), the corresponding dual spheres have dimension $d>2$. The entropy of these structures are known to increase (at least) exponentially with volume. Hence, for the same reasons, this explains the existence of singular $i$-subsimplices, for $i=0,\ldots,d-4$. The problem with this picture is that it cannot account for the apparent non-existence of singular vertices in three dimensions. According to the reasoning of Catterall {\it et al.}, vertices (in three dimensional triangulations) with large local volumes should be entropically favoured since their dual-sphere are 2-spheres. This point seems to be a problem and is recognised as being so~\cite{renkcomm}.

This brings us to the question: why should there be only {\it one} singular primary subsimplex? Catterall {\it et al.} answer this important question with the following argument. Consider a triangulation with a number of randomly distributed singular vertices. Each vertex strives to maximise its local volume. This can be done if the vertices tend to {\it share} their local volume. In other words, the dual-sphere of a singular vertex will itself contain other singular vertices. Eventually some singular vertices will dominate by `absorbing' others. This coalescence continues until only $d-2$ vertices are left. These maximise their local volumes by sharing simplices and hence form a primary singular subsimplex.

It is quite easy to determine the intersection of these dual-spheres. Consider the four dimensional case, where two $S^{3}$ dual-spheres whose intersection forms a 2-sphere. This can be shown using simple geometry. Of course the situation becomes more complicated in higher dimensions. In five dimensions we have three 4-spheres dual to the singular vertices. Again, simple algebra shows that each pair of 4-spheres intersects at a 3-sphere dual to a singular link. In fact, one can show that all three 4-spheres intersect at a 2-sphere dual to the singular triangle. In higher dimensions, one can follow the same pattern to find that the intersection submanifold is always $S^{2}$, the dual-sphere of the primary singular subsimplex.

\section{Three Dimensions} \label{sec:threedimensions}

According to the entropy arguments proposed by Catterall {\it et al.}, one would expect to observe singular vertices in three dimensions; since the dual-sphere of a vertex in three dimensions is a 2-sphere, whose entropy is known to grow exponentially with volume~\cite{tutte}. None have ever been detected~\cite{ckrtsing}. The purpose of this section is to resolve this apparent discordance. Let us begin by repeating the work of Catterall {\it et al.}, as their numerical results were not published in explicit form. 

\begin{figure}[ht]
\centerline{\epsfxsize=14cm \epsfbox{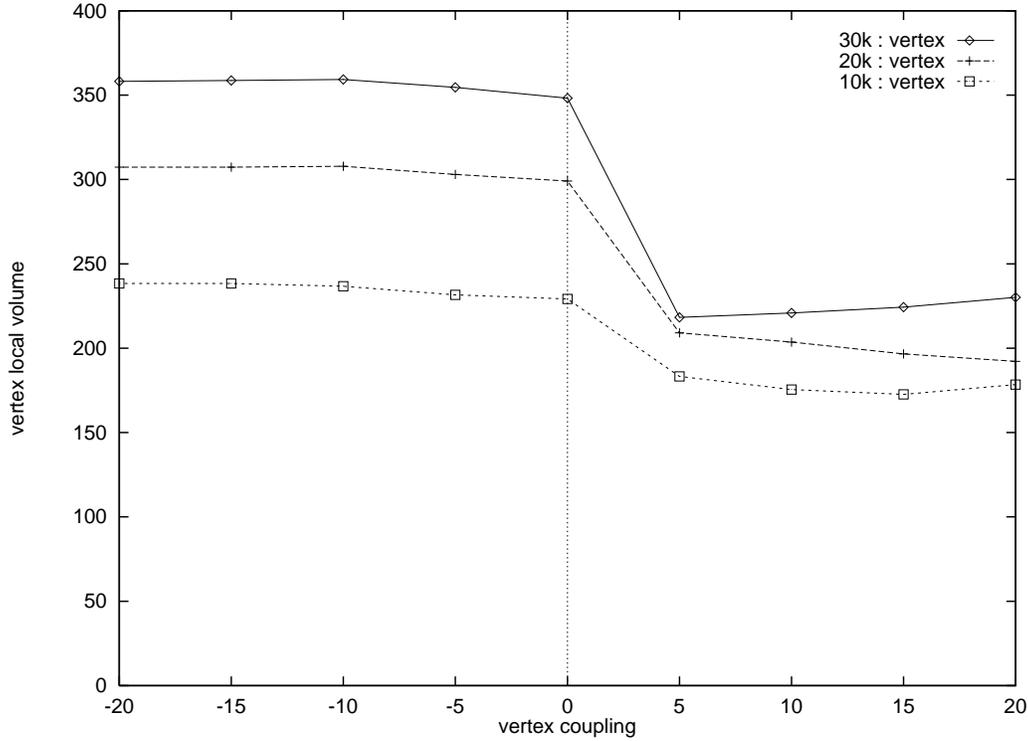}}
\caption{Maximum vertex local volume $\langle n_{0}^{max}\rangle$ versus $\kappa_{0}$ for three dimensional triangulations. (The lines serve only to guide the eye.)}
\label{fig:3dvertices}
\end{figure}

A new function incorporated into the standard dynamical triangulations code allowed for the measurement of the vertex local volume distribution ${\cal N}(n_{0})$ and hence detection of any singular vertices that may exist (for $d<6$). The algorithm is described in appendix~\ref{app:singcode}. Monte Carlo simulations were run for target volumes up to $N_{5}^{t}=30$k and over a wide range of $\kappa_{0}$. In each case, the maximum vertex local volume $n_{0}^{max}$ was measured $10^{3}$ times over a period of $10^{5}$ sweeps, yielding a mean value $\langle n_{0}^{max} \rangle$. The results presented in figure~\ref{fig:3dvertices} clearly show that singular vertices are not found in three dimensions. 

\subsection{Non-existence of Singular Vertices} \label{subsec:nonexist}

In this section, our aim is to construct a plausible explanation of {\it why} singular vertices are not prevalent in three dimensional triangulations. The following arguments are developed from those put forward by Catterall {\it et al.}~\cite{ckrtsing}.

\vskip 5mm

\noindent
Consider a simplicial 3-manifold $T$. Now identify an arbitrary $B^{3}$ submanifold of $T$ with volume $N_{3}$. The 3-ball submanifold has a 2-sphere boundary with volume $N_{2}$. Let the space of triangulations of the 3-ball be symbolised by $\cal T$$_{3\mbox{-}ball}$. There exists a special subspace of ${\cal T}_{3\mbox{-}ball}$ whose constituent triangulations have $N_{2}=N_{3}$, at least for $N_{3}>4$. The triangulations of this subspace, denoted by ${\cal T}_{sing,0}$, are described as {\it vertex-singular}, since every tetrahedron of the 3-ball is common to a single vertex. In other words, we have ${\cal T}_{\,sing,0}\subset{\cal T}_{\,3\mbox{-}ball}$. Furthermore, it is clear that the 2-sphere boundary is the dual-sphere of a vertex iff $N_{2}=N_{3}$. One can now define the subspace of {\it non-vertex-singular} triangulations $\cal T$$_{\!non,0}$ as follows.
\begin{equation}
{\cal T}_{\,sing,0}\cup{\cal T}_{\,non,0} \subseteq {\cal T}_{\,3\mbox{-}ball}
\end{equation}
This space of 3-balls are obviously triangulated such that $N_{2}\neq N_{3}$. 

Now one must consider how the entropy of these structures grow with volume. The entropy of 3-balls is given by $\Omega({\cal T}_{3\mbox{-}ball})=\Omega(B^{3})$. The entropy of vertex-singular triangulations effectively grows as that of 2-spheres because $N_{2}=N_{3}$; therefore we have $\Omega({\cal T}_{sing,0})=\Omega(S^{2})$. The entropy associated with these spaces of triangulations can be equated to give
\begin{equation}
\Omega\:\!({\cal T}_{\,sing,0})+\Omega\:\!({\cal T}_{\,non,0}) = \Omega\:\!({\cal T}_{\,3\mbox{-}ball}), 
\end{equation}
which may be rewritten as 
\begin{equation}
\Omega\:\!(S^{2}) + \Omega\:\!({\cal T}_{\,non,0}) = \Omega\:\!(B^{3}). \label{eq:3domvert}
\end{equation}
The observed non-existence of singular vertices for $\kappa_{0}=0$ would therefore be due to 
\begin{equation}
\Omega\:\!({\cal T}_{non,0}) \gg \Omega\:\!({\cal T}_{sing,0}).
\end{equation}
Using equation~(\ref{eq:3domvert}) one can re-express this inequality as
\begin{equation}
\Omega\:\!(B^{3}) \gg 2\:\!\Omega\:\!(S^{2}). \label{eq:3dentropy}
\end{equation}
For a given simplicial $d$-manifold of fixed volume $N_{d}$, there exists an upper kinematic bound to the local volume of a vertex, which is a function of $N_{d}$. We shall call this upper bound $n_{0}^{kin}$. The non-existence of singular vertices may then be explained by comparing the entropy of 3-balls and 2-spheres with volume $n_{0}^{kin}$ according to inequality (\ref{eq:3dentropy}). Let us illustrate these ideas with an example. Consider a three dimensional triangulation of volume $N_{3}=1000$. In this case $n_{0}^{kin}$ could be, say, 400. That is to say: it is geometrically impossible for any vertex to be common to more than 400 simplices. If the entropy of an arbitrary 3-ball of volume $N_{3}=400$ is much greater than twice the entropy of a 2-sphere of volume $N_{2}=400$, then vertex-singular triangulations (with volume $N_{3}=1000$) will not contribute significantly to the partition function. Obviously, in order to extend this idea to the thermodynamic limit, one must consider the {\it growth rates} of entropy.

Although it may be possible to prove inequality~(\ref{eq:3dentropy}) in principle, it is likely to be very difficult\,\footnote{Depending on the growth rates of $\Omega\:\!(B^{3})$ and $\Omega\:\!(S^{2})$; inequality~(\ref{eq:3dentropy}) may not be true {\it for all} $N_{3}$. As physicists, however, we are only really interested in the thermodynamic limit.}. We therefore offer the result as a conjecture. One can, however, show that 
\begin{equation}
\Omega\:\!(B^{3}) \gg \Omega\:\!(S^{2}). \label{eq:b3s2}
\end{equation}
This is seen by realising that the boundary of a 3-ball is a 2-sphere. In this case the space of triangulations of $S^{2}$ may be regarded as a special subspace of the space of triangulations of $B^{3}$, cf. ${\cal T}_{\,sing,0}\subset{\cal T}_{\,3\mbox{-}ball}$. From this fact one can deduce (\ref{eq:b3s2}). This inequality may be trivially extended to arbitrary dimension $d$ giving
\begin{equation}
\Omega\:\!(B^{d}) \gg \Omega\:\!(S^{d-1}). 
\end{equation}
By combining inequalities~(\ref{eq:3dentropy}) and (\ref{eq:b3s2}), we find that singular vertices can only exist in three dimensions if
\begin{equation}
2 \Omega\:\!(S^{2}) > \Omega\:\!(B^{3}) > \Omega\:\!(S^{2}). \label{eq:3ddoubineq}
\end{equation}
In order for (\ref{eq:3ddoubineq}) to hold in the thermodynamic limit, the entropy functions of $S^{2}$ and $B^{3}$ must grow at the same rate in this limit. This seems unlikely. 

\vskip 5mm

\noindent
Our new way of looking at the problem involves {\it comparing} the entropy of the singular vertices with those of `ordinary' triangulations of $B^{3}$, and determining which is the greater. This is in contrast to the approach taken by Catterall {\it et al.} who only considered the entropy of the dual-spheres. These arguments are valid for $\kappa_{0}=0$, since all triangulations are equally weighted in the partition function. For $\kappa_{0}\neq 0$, one must consider the effect of the action. Figure~\ref{fig:3dvertices} shows that vertices in the weak coupling phase have slightly lower local volumes, because the action forces extra vertices into the triangulations.

\section{Four Dimensions} \label{sec:fourdim}

The next step in our investigation of singular structures takes us to four dimensions. All studies published to date have concentrated on $\kappa_{0}=0$ or regions near the phase transition. For a more complete picture, it was decided to measure the vertex local volume distribution ${\cal N}(n_{0})$ across a much wider range of couplings, including negative values of $\kappa_{0}$. For the time being we shall focus on the special case of $\kappa_{0}=0$, and attempt to reproduce the findings of previous studies. Measurements were recorded of the two largest vertex local volumes and the corresponding link local volume. If the two vertices did not form a link in the triangulation, then the link local volume was zero by definition. For each Monte Carlo simulation measurements were taken at $10^{2}$ sweep intervals over a period of $10^{5}$ sweeps, for target volumes up to $N_{4}^{t}=40\mbox{k}$. 

\subsubsection{Singular Vertices}

\begin{figure}[htp]
\begin{center}
\leavevmode
(a){\hbox{\epsfxsize=14cm \epsfbox{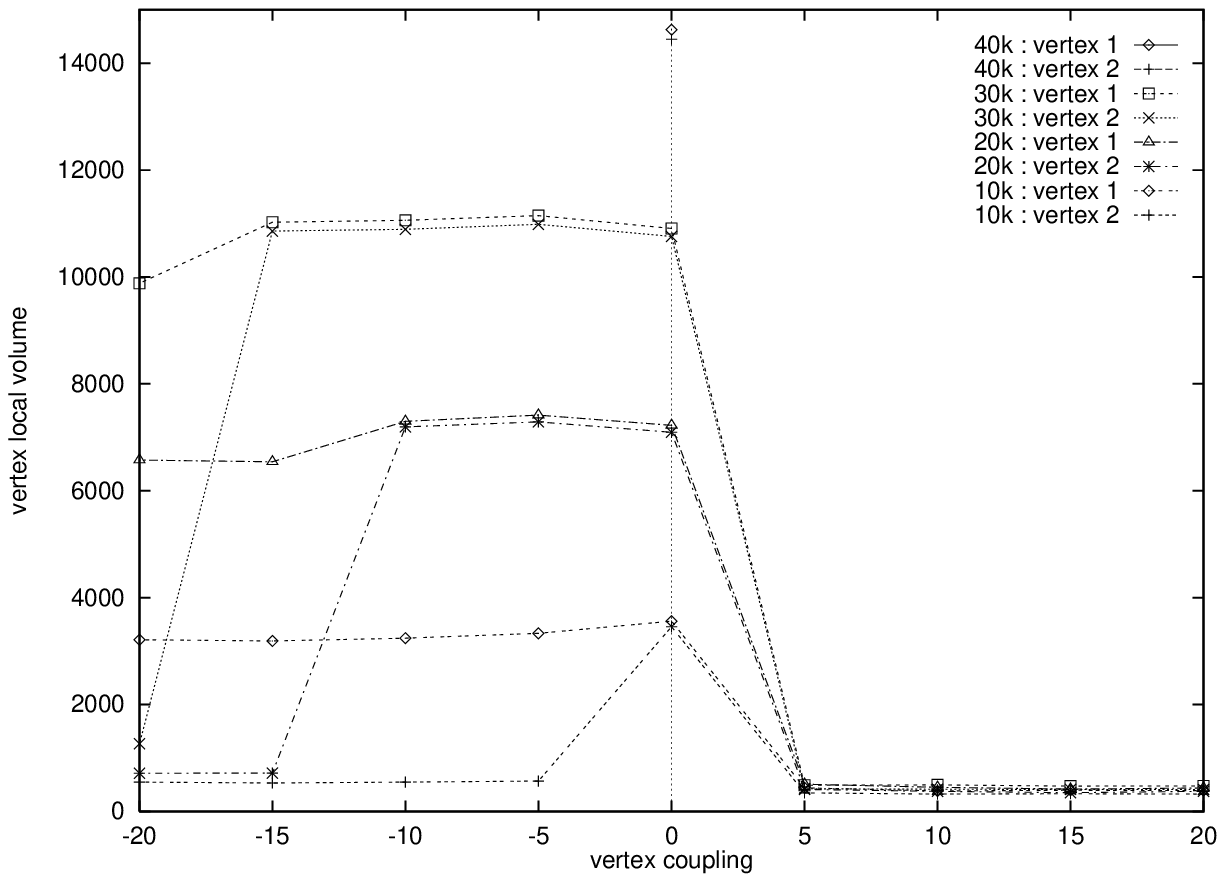}}}
(b){\hbox{\epsfxsize=14cm \epsfbox{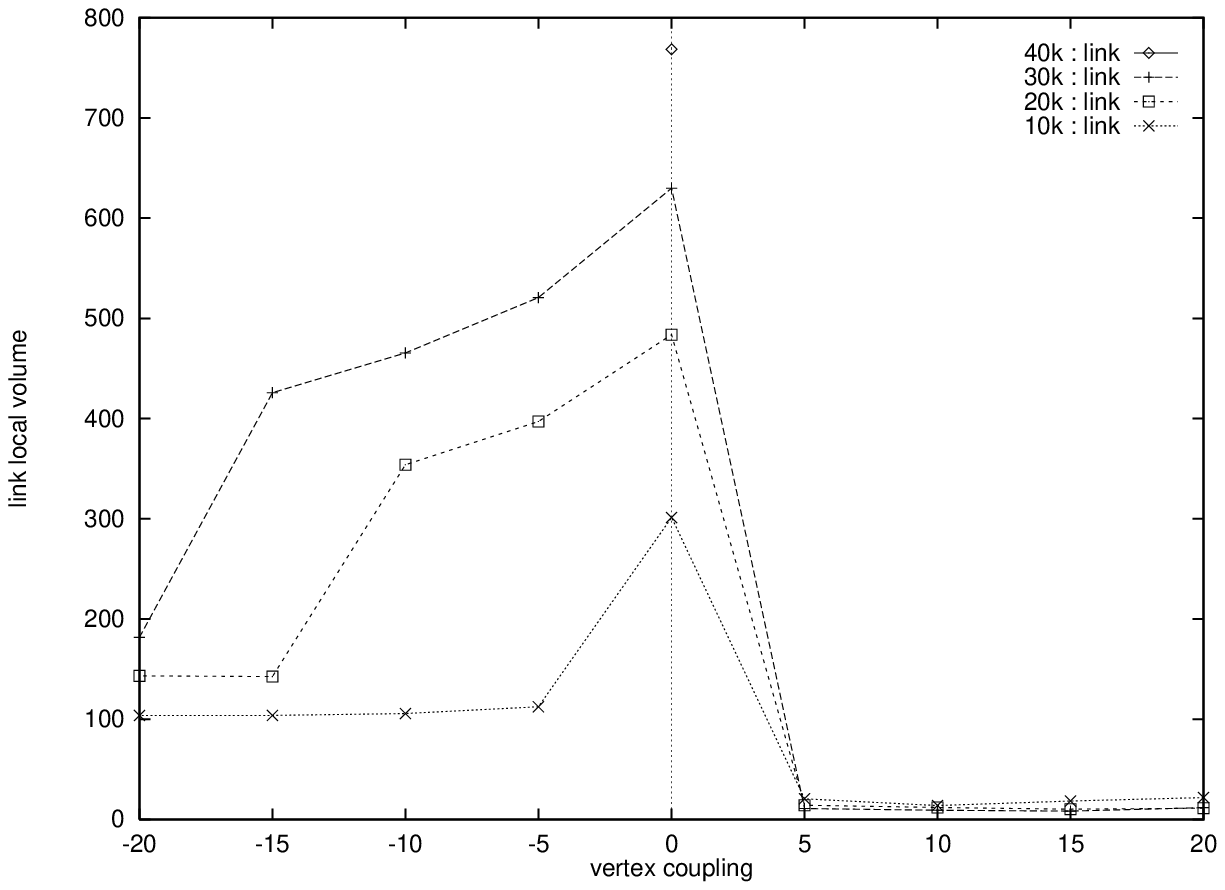}}}
\caption{(a) Local volume of singular vertices versus vertex coupling $\kappa_{0}$. (b) Local volume of corresponding singular link versus $\kappa_{0}$. (The lines serve only to guide the eye.)}
\label{fig:4dsingular}
\end{center}
\end{figure}

As expected, two singular vertices were detected for each target volume with $\kappa_{0}=0$. These are labelled as $\sigma^{0}_{s1}$ and $\sigma^{0}_{s2}$. They were found to have {\it approximately}\, equal local volumes, which consistently differed slightly. For this reason, the two quantities were distinguished by calculating two separate averages. The largest and second largest vertex local volumes are denoted by $n(\sigma^{0}_{s1})$ and $n(\sigma^{0}_{s2})$ respectively. Figure~\ref{fig:4dsingular}(a) plots the local volume of the two singular vertices versus $\kappa_{0}$ for each target volume.
The $\kappa_{0}=0$ data was fitted to the following curves using the chi-squared method, which is described in appendix~\ref{app:curve}.
\begin{eqnarray}
n(\sigma_{s1}^{0}) & = & a_{1}+b_{1}{N_{4}}^{c_{1}}\label{eq:4dvertfit1} \\
n(\sigma_{s2}^{0}) & = & a_{2}+b_{2}{N_{4}}^{c_{2}}\label{eq:4dvertfit2}
\end{eqnarray}
The best fit curves (both with $\chi^{2}\approx 5~\mbox{d.o.f.}^{-1}$) have coefficients $c_{1}=1.017(2)$ and $c_{2}=1.006(6)$. Again, as expected, these results are consistent with the conjecture that the local volumes of secondary singular subsimplices diverge linearly with triangulation volume. Furthermore, the two singular vertices were found to be common to approximately one third of the triangulation volume, since $b_{1}=0.307(6)$ and $b_{2}=0.34(2)$. This is also in perfect agreement with the work of Hotta {\it et al.}~\cite{hinptp,hinnpb} and Catterall {\it et al.}~\cite{ckrtsing}. 

One should bear in mind that (\ref{eq:4dvertfit1}) and (\ref{eq:4dvertfit2}) are valid only for large volumes, since the relationship between vertex local volumes and $N_{4}$ is discrete in nature. In fact, $n_{0}$ is not even defined for $N_{4}<6$. Consequently, one cannot assume that the large $N_{4}$ behaviour is consistent with a curve which intercepts the origin. This fact was accounted for by including constants $a_{1}$ and $a_{2}$ in the fitted curves.

\subsubsection{Singular Link}

Let us now consider the second conjecture made by Catterall {\it et al.} which states that the local volume of primary singular subsimplices diverge with triangulation volume according to a two-thirds power law~\cite{ckrtsing}. Figure~\ref{fig:4dsingular}(b) plots the local volume of the singular link (formed by the product of singular vertices $\sigma_{s1}^{0}$ and $\sigma_{s2}^{0}$) versus $\kappa_{0}$ for each target volume. If the link local volume $n_{s}^{1}$ is zero, then this implies that the two singular vertices are not connected. From the Monte Carlo data it became apparent that $\sigma^{0}_{s1}$ and $\sigma^{0}_{s2}$ {\it always} form a link $\sigma_{s}^{1}$ with local volume $n(\sigma_{s}^{1})$. In other words, $n(\sigma_{s}^{1})$ was always {\it non-zero}. By fitting the $\kappa_{0}=0$ results to the following curve
\begin{equation}
n(\sigma_{s}^{1}) = p+q{N_{4}}^{r},
\end{equation}
we find that $r=0.66(1)$ with $\chi^{2}\approx 17~\mbox{d.o.f.}^{-1}$. This is perfectly consistent with the conjectured two-thirds power law divergence. 

It was claimed by Catterall {\it et al.} that it is possible to understand the origin of this power law divergence in terms of simple geometry~\cite{ckrtsing}. In four dimensions the dual-sphere of a vertex is a 3-sphere, and the dual-sphere of a link is the 2-sphere intersection of two 3-spheres. By assuming that simplices are evenly distributed one may equate the volumes of simplicial and smooth spheres. It is then claimed that the two-thirds power is derived from the fact that the volume of smooth 2-spheres and 3-spheres scale with the square and the cube of the radius respectively~\cite{ckrtsing}. When considering singular structures in five dimensions, we indeed encounter three overlapping 3-spheres which intersect at a 2-sphere. However, each 3-sphere is the intersection of two 4-spheres. Presumably one could apply the same arguments to the 4-spheres. Given that the local volume of the 4-spheres diverge linealy with triangulations volume, one would conclude by the same arguments that the local volume of the 3-sphere diverges according to a three-quarters power law. It seems that this leads to a contradiction since singular links in five dimensions are known to diverge linearly with volume (see figure~5 in~\cite{ckrtsing}). 

\subsubsection{Additional Remarks}

So far we have detected the presence of singular links in four dimensional triangulations for $\kappa_{0}=0$, and reproduced known results. Close scrutiny of Monte Carlo time series plots has revealed that, occasionally, the local volume of one of the singular vertices dramatically reduces to `normal levels' (but always remains the second largest local volume). However, the local volume of the link never reduces to zero. In other words, the two vertices with the largest local volumes are always connected by a link. The most astonishing thing about this phenomenon is that the transformation from singular link to singular vertex and back again occurs very quickly (typically within about 200 Monte Carlo sweeps). Furthermore, these events are very rare -- with an occurrence rate of only about 1 in $10^{3}$ measurements. This type of phenomenon was apparently first reported by Catterall {\it et al.} following an investigation into the possible existence of `volume barriers' in the space of triangulations~\cite{ckrtsing}. It was found that the ensemble `tunnels' between triangulations characterised by a singular link (so-called super-crumpled states), a single singular vertex or those with no singular vertices. These occurrences were only observed for small triangulations $N_{4}^{t}=4$k.

Our time series plots show that the ensemble is generally very stable in the space of triangulations with singular links. These rare tunneling events may be a finite size effect. However, our results are consistent with a tunneling rate that is independent of the volume, at least up to $N_{4}^{t}=40$k. This effect has also been observed for $\kappa_{0}<0$. In fact, the tunneling rate is marginally higher deeper into the crumpled phase. The reason behind these events is not currently well-understood. This brings us to our last comment of the section. 

These tunneling events may be related to another phenomenon concerning singular structures discovered deep in the crumpled phase of four dimensional dynamical triangulations. From figure~\ref{fig:4dsingular}(a) it is evident that triangulations deep within the crumpled phase have only one singular vertex, and hence no singular link. Figure~\ref{fig:4dsingular}(b) also illustrates this fact -- though less convincingly. Interestingly, this effect was predicted by Bia{\l}as {\it et al.}~\cite{bbptmother}. Perhaps these effects deserve closer attention.

\subsection{Generalisation of Entropy Arguments} \label{subsec:genentarg}

The non-existence of singular vertices in three dimensions was earlier accounted for in terms of entropy. In the simple three dimensional case one only considers the entropy contributions of a single subsimplex. In higher dimensions, primary singular subsimplices are formed by the product of secondary singular subsimplices. This complication requires some modification of our (new) arguments. In this section, we generalise our arguments to arbitrary dimension.

Consider an arbitrary $B^{d}$ submanifold of a $d$-dimensional triangulation. Let the space of triangulations of this structure be represented by ${\cal T}_{d\mbox{-}ball}$. For simplicial $d$-manifolds, the star of a $(d-3)$-subsimplex is homeomorphic to $B^{d}$, which has an $S^{d-1}$ boundary. Now, the dual-sphere of a $(d-3)$-subsimplex is an $S^{2}$ submanifold of the $S^{d-1}$ boundary.
Such $d$-balls are described as {\it $(d-3)$-subsimplex-singular} because each simplex of the $d$-ball is common to the $(d-3)$-subsimplex. The corresponding space of triangulations is symbolised by ${\cal T}_{sing,d-3}$. From our definitions of ${\cal T}_{d\mbox{-}ball}$ and ${\cal T}_{sing,d-3}$ it is clear that ${\cal T}_{\,sing,d-3}\subset{\cal T}_{\,d\mbox{-}ball}$. One can define the subspace of {\it non-$(d-3)$-subsimplex-singular} triangulations $\cal T$$_{\!non,d-3}$ such that the following expression is true.
\begin{equation}
{\cal T}_{\,sing,d-3}\cup{\cal T}_{\,non,d-3} \subseteq {\cal T}_{\,d\mbox{-}ball}.
\end{equation}
One must now consider the entropy of these structures as a function of volume. The entropy associated with these spaces of triangulations can be equated to give
\begin{equation}
\Omega\:\!({\cal T}_{\,sing,d-3})+\Omega\:\!({\cal T}_{\,non,d-3}) = \Omega\:\!({\cal T}_{\,d\mbox{-}ball}). \label{eq:bahfah}
\end{equation}
It was claimed by Catterall {\it et al.} that the existence of singular subsimplices can be accounted for by considering the entropy of the dual-spheres individually. In this thesis, we propose that it makes more sense to consider the entropy associated with the {\it entire} singular structure, whose neighbourhood is homeomorphic to a $d$-ball. One then {\it compares} the entropy (or growth rates of entropy) of the singular structure $d$-ball and an arbitrary $d$-ball of the same volume. These ideas may be clarified with an example.

Consider the {\it purely hypothetical} case of two singular vertices in three dimensions whose product forms a singular link. The vertices have $S^{2}$ dual-spheres which intersect at a 1-sphere dual to the link. Our previous understanding of the existence of singular structures accounted for their existence by quantifying the entropy of the various singular subsimplices individually. Instead, one ought to consider the entropy of the entire singular structure -- that is to be treated as a single object. Previous thought would have it that the entropy associated with the link is effectively that of the 1-sphere. In fact, we propose that the entropy $\Omega$ associated with the structure is given by
\begin{equation}
\Omega = 2\Omega(S^{2})-\Omega(S^{1}),
\end{equation}
since the entropy contribution of the 1-sphere intersection is overcounted. 

By generalising this argument to higher dimensions, one can show that the entropy associated with a primary singular subsimplex in $d$-dimensions is given by
\begin{equation}
\Omega\:\!({\cal T}_{sing,d-3}) = \sum_{i=0}^{d-3} {}^{d-2}\mbox{C}_{i+1}\,(-1)^{i}\,\Omega(S^{d-i-1}). \label{eq:priment}
\end{equation}
The existence of the singular structure depends on
\begin{equation}
\Omega({\cal T}_{sing,d-3}) \gg \Omega({\cal T}_{non,d-3}) \label{eq:entgg}
\end{equation}
By substituting (\ref{eq:bahfah}) and (\ref{eq:priment}) in (\ref{eq:entgg}) and using the fact that $\Omega({\cal T}_{d\mbox{-}ball})=\Omega(B^{d})$, we get the following expression.
\begin{equation}
2 \sum_{i=0}^{d-3} {}^{d-2}\mbox{C}_{i+1}\,(-1)^{i}\,\Omega\:\!(S^{d-i-1}) \gg \Omega(B^{d}) \label{eq:entcond}
\end{equation}
Clearly, this inequality applies to $d$-balls of the same volume $N_{d}$. Using the same arguments explained earlier, one can show that this is true when
\begin{equation}
N_{d}(B^{d}) = \sum_{i=0}^{d-3}{}^{d-2}\mbox{C}_{i+1}(-1)^{i}N_{i}(S^{i}), \label{eq:arbvol}
\end{equation}
where $N_{i}(S^{i})$ is the volume of an $S^{i}$ dual-sphere. (One assumes that multiple dual-spheres of the same dimension have equal volumes.) In other words, inequality (\ref{eq:entcond}) is valid only for structures which satisfy (\ref{eq:arbvol}).

For example, by substituting $d=3$ in (\ref{eq:entcond}) we see that singular vertices exist if the following inequality is satisfied, cf. inequality (\ref{eq:3dentropy}).
\begin{equation}
2\Omega(S^{2}) \gg \Omega(B^{3})
\end{equation}
In the same way one can show that the existence of singular structures in four dimensions would be due to 
\begin{equation}
4\Omega(S^{3})-2\Omega(S^{2}) \gg \Omega(B^{4}), \label{eq:foursing}
\end{equation}
providing the volumes of the dual-spheres satisfy
\begin{equation}
N_{4}(B^{4}) = 2N_{3}(S^{2})-N_{2}(S^{2}).
\end{equation}
Unfortunately, it is very difficult to show that (\ref{eq:foursing}) is true. Obviously, in the thermodynamic limit one considers the growth rates of entropy and so (\ref{eq:arbvol}) does not apply.

\section{Five Dimensions} \label{sec:singfive}

Finally, let us now turn our attention to singular structures in five dimensional triangulations. These have already been investigated to a certain extent by Catterall {\it et al.}~\cite{ckrtsing}. Here, our objective is to add to their work in the light of the newly discovered phase structure of five dimensional dynamical triangulations. When $\kappa_{0}=\kappa_{3}=0$, one expects to observe three singular vertices and three singular links which collectively form a singular triangle. Our algorithm is therefore designed to measure the three largest vertex local volumes of a given triangulation. It then calculates the local volumes of the three links formed by pairs of the three vertices. Lastly, it determines the local volume of the triangle formed by the product of all three vertices. Monte Carlo simulations were run for the $\kappa_{3}=0$ and $\kappa_{0}=0$ limits over a wide range of couplings, with target volumes up to $N_{5}^{t}=40$k. 

\subsection{Vertex Local Volume Distributions} \label{subsec:vlvd}

Let us begin by taking snapshots of the vertex local volume distribution ${\cal N}(n_{0})$ at various points in ${\cal C}_{2}$, which are identified by coordinates $(\kappa_{0},\kappa_{3})$. These distributions (taken from triangulations with $N_{5}^{t}=30$k) are neither normalised or averaged over a number of triangulations. Their features are more discernible when plotted on a logarithmic scale. In each plot a horizontal line indicates the triangulation volume and a vertical line shows the mean vertex local volume $\overline{n}_{0}$, defined as follows.
\begin{equation}
\overline{n}_{0} = \frac{1}{N_{0}}\sum_{\sigma^{0}\in\,T}n(\sigma^{0}) = \frac{6N_{5}}{N_{0}}
\end{equation}

\begin{figure}[htp]
\begin{center}
\leavevmode
(a){\hbox{\epsfxsize=14cm \epsfbox{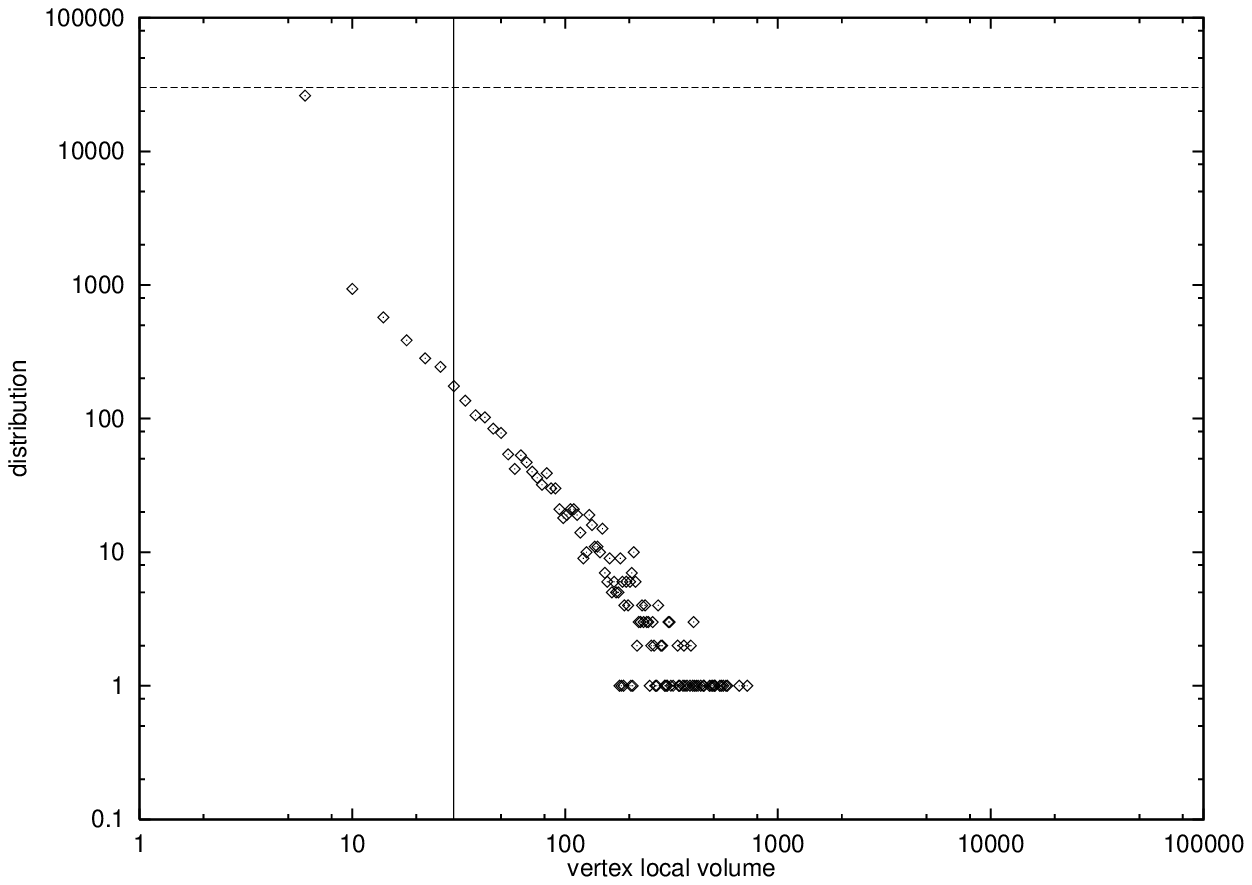}}}
(b){\hbox{\epsfxsize=14cm \epsfbox{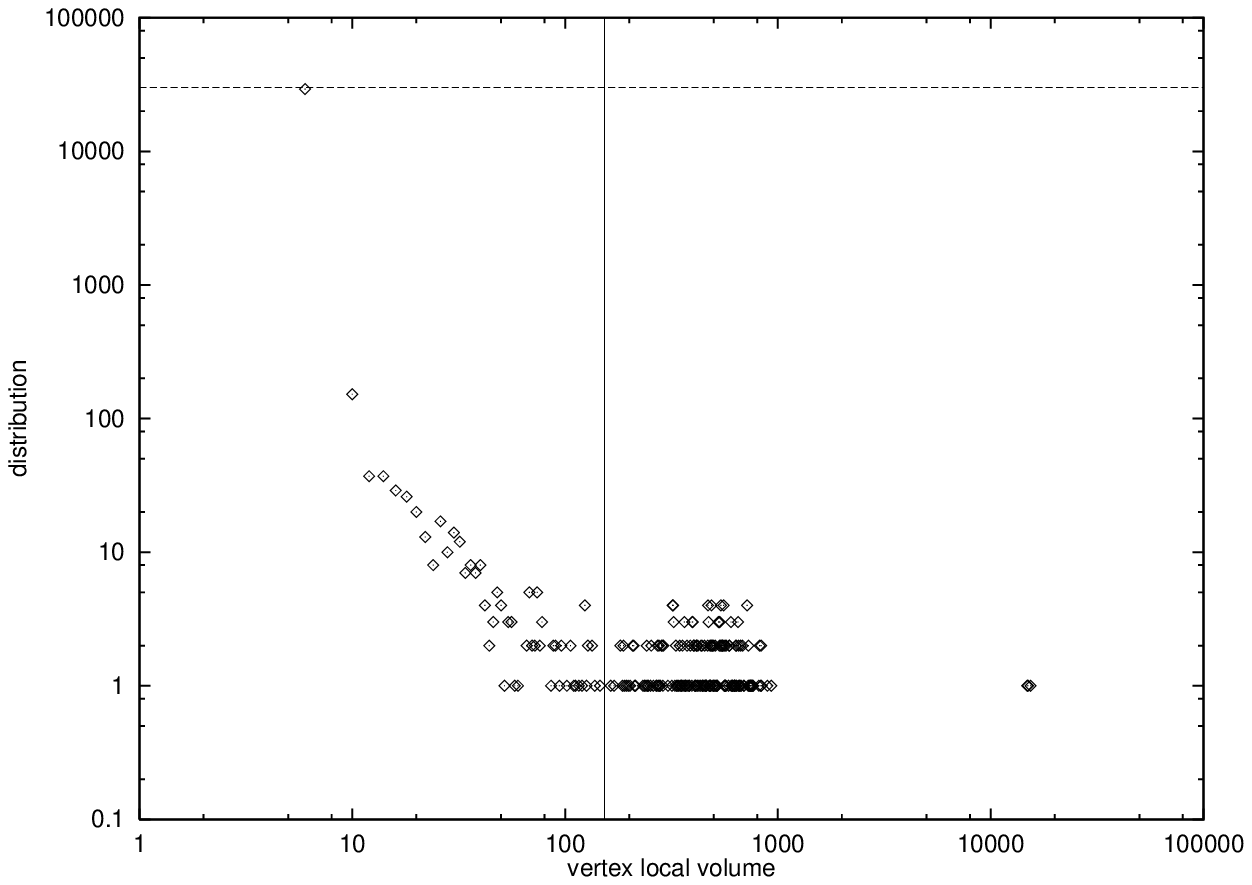}}}
\caption{Vertex local volume distributions for points (a) $(0,5)$ and (b) $(0,0)$ in ${\cal C}_{2}$.}
\label{fig:fivesingular1}
\end{center}
\end{figure}

The distribution plotted in figure~\ref{fig:fivesingular1}(a) was recorded at point $(0,5)$ in ${\cal C}_{2}$, which is deep within the branched polymer phase. 
It is immediately evident that the vast majority of vertices have a local volume of six. This is the minimum for five dimensional triangulations, i.e. $n_{0}\geq 6$. The distribution has $\overline{n}_{0}\approx 30$ and tails off inversely proportional to $n_{0}$. The largest vertex local volumes are of order $10^{2}$. Not surprisingly, this is qualitatively very similar to the analogous distribution of the four dimensional branched polymer phase, schematically drawn in figure~\ref{fig:singular}. The absence of singular vertices in this phase is accounted for by the fact that large $\kappa_{3}$ (and/or large $\kappa_{0}$) forces extra vertices into the triangulations. 

As expected, the distribution for $(0,0)$ is radically different. The main distinction being the existence of a small number of vertices with very large local volumes (in excess of $10^{4}$). These are plainly seen in figure~\ref{fig:fivesingular1}(b). Analysis of the raw data confirms that these points represent {\it three} singular vertices. Other differences between the $(0,0)$ and $(0,5)$ distributions include far fewer vertices with $n_{0}\sim 10^{1}$. This feature also mirrors the four dimensional distribution (see figure~\ref{fig:singular}). 

\begin{figure}[htp]
\begin{center}
\leavevmode
(a){\hbox{\epsfxsize=14cm \epsfbox{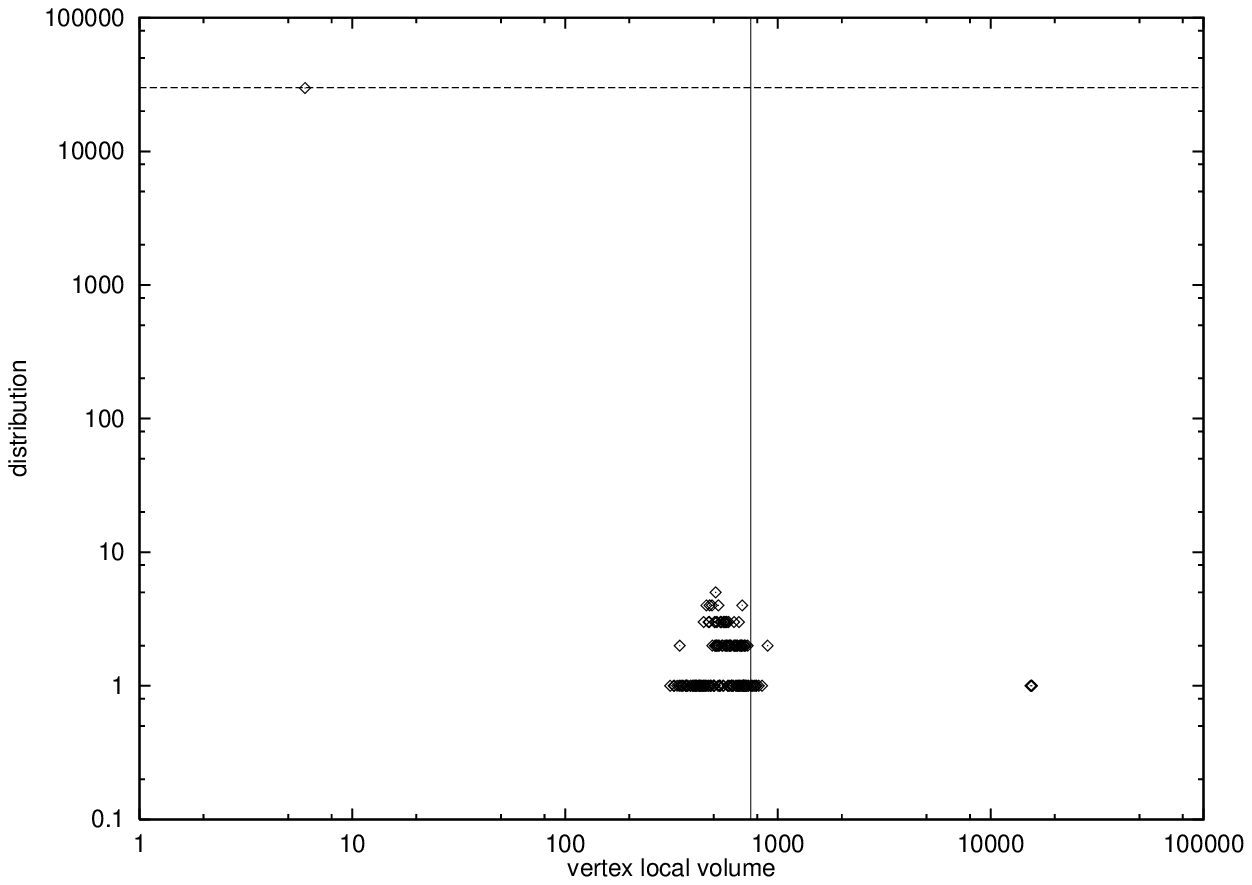}}}
(b){\hbox{\epsfxsize=14cm \epsfbox{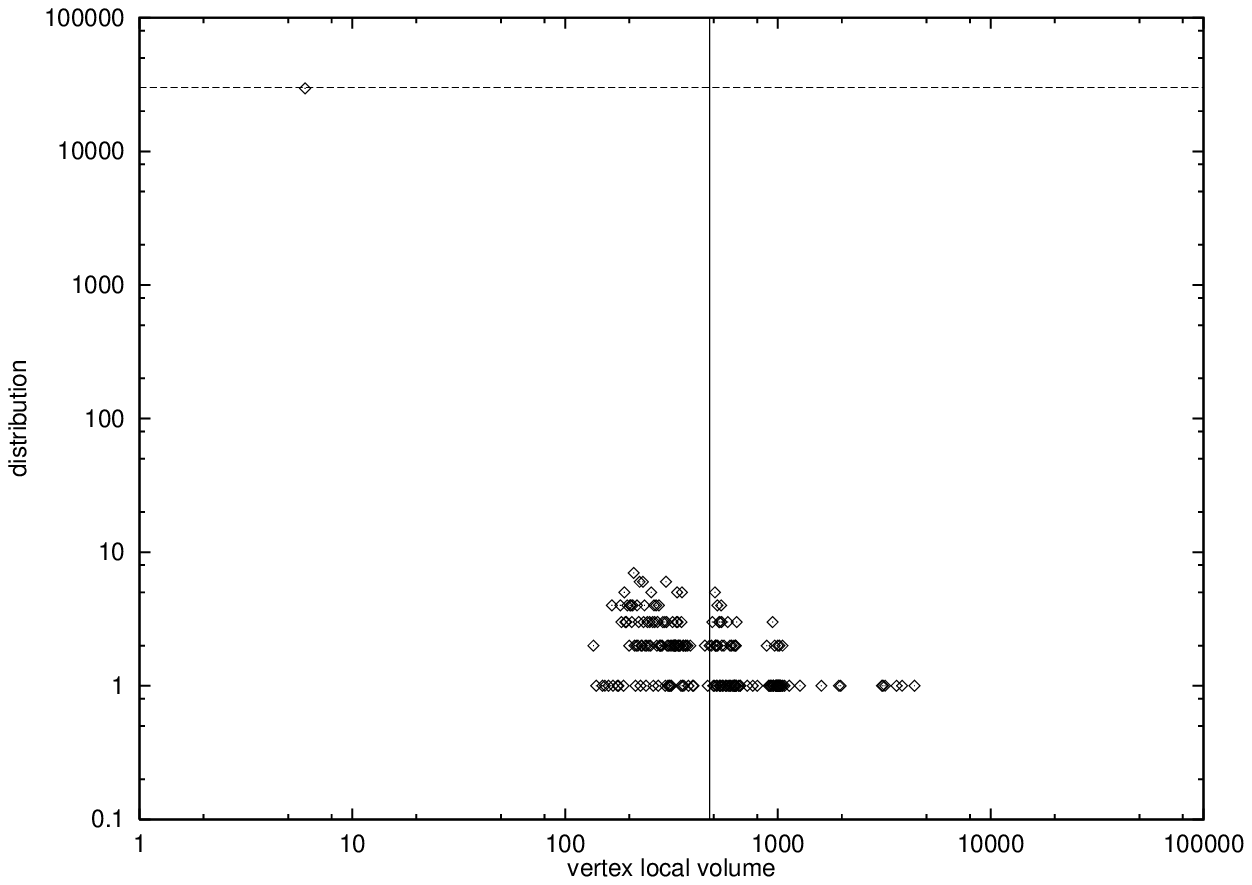}}}
\caption{Vertex local volume distributions for points (a) $(-10,0)$ and (b) $(0,-5)$ in ${\cal C}_{2}$.}
\label{fig:fivesingular2}
\end{center}
\end{figure}

Figure~\ref{fig:fivesingular2}(a) corresponds to a triangulation with $(-10,0)$ that is representative of the strong coupling limit of $\kappa_{0}\rightarrow -\infty$ (see figure~\ref{fig:zoomoutR}(a)).
By comparing this figure with the previous two, one can see how the transition from the branched polymer phase to the crumpled phase affects the distribution of vertex local volumes. There are two main changes that occur. The first is obviously the appearance of three singular vertices, and the second is the disappearance of vertices with small local volumes of order $6<n_{0}<10^{1}$. This results in an overall increase in $\overline{n}_{0}$, shown by the vertical lines. In fact, using the ideas presented in section~\ref{sec:stack} one can show that (in the thermodynamic limit) the mean vertex local volume is bounded from below by $\overline{n}_{0}\geq 30$, or in general $\overline{n}_{0}\geq d(d+1)$.

Our last distribution snapshot, shown in figure~\ref{fig:fivesingular2}(b), was taken for $(0,-5)$. This point might be located in a new phase, whose possible existence was bought to light in section~\ref{sec:prelim}. The distribution is quite unlike the others. As with the previous three cases, we find that virtually all vertices have $n_{0}=6$. In this phase there are no vertices with very large local volumes, though a small number have $n_{0}\sim 10^{3}$. More surprisingly perhaps, there are {\it no}\, vertices with $6<n_{0}<100$. Clearly, this distribution is fundamentally different to those of the branched polymer phase and the strong coupling phase of the $\kappa_{3}=0$ limit. The origin of this distribution is not currently understood, though the key to its understanding may be the realisation that the five dimensional Einstein-Hilbert action cannot be expressed in terms of $N_{5}$ and $N_{0}$. In this strong coupling regime, the ensemble seeks to minimise $N_{3}$ {\it irrespectively} of $N_{0}$.

\subsection{Primary and Secondary Singular Subsimplices} \label{subsec:secsingsub}

\begin{figure}[ht] 
\centering{\input{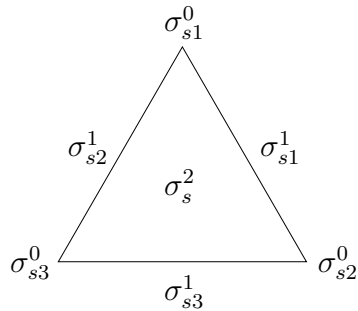}\par}
\caption{Labelling of primary and secondary singular subsimplices.}
\label{fig:singtri}
\end{figure}

The local volumes of secondary singular subsimplices are conjectured to diverge linearly with volume~\cite{ckrtsing}. In five dimensions, this translates to the linear divergence of vertex and link local volumes with $N_{5}$. As in four dimensions, the local volumes of the three vertices are only approximately equal. These are denoted by $n(\sigma_{s1}^{0})$, $n(\sigma_{s2}^{0})$ and $n(\sigma_{s3}^{0})$, such that $n(\sigma_{s1}^{0})>n(\sigma_{s2}^{0})>n(\sigma_{s3}^{0})$. The link local volumes are given by $n(\sigma_{s1}^{1})$, $n(\sigma_{s2}^{1})$ and $n(\sigma_{s3}^{1})$. These structures form a triangle with local volume $n(\sigma_{s}^{2})$ in the arrangement shown in figure~\ref{fig:singtri}.

\begin{figure}[htp]
\begin{center}
\leavevmode
(a){\hbox{\epsfxsize=14cm \epsfbox{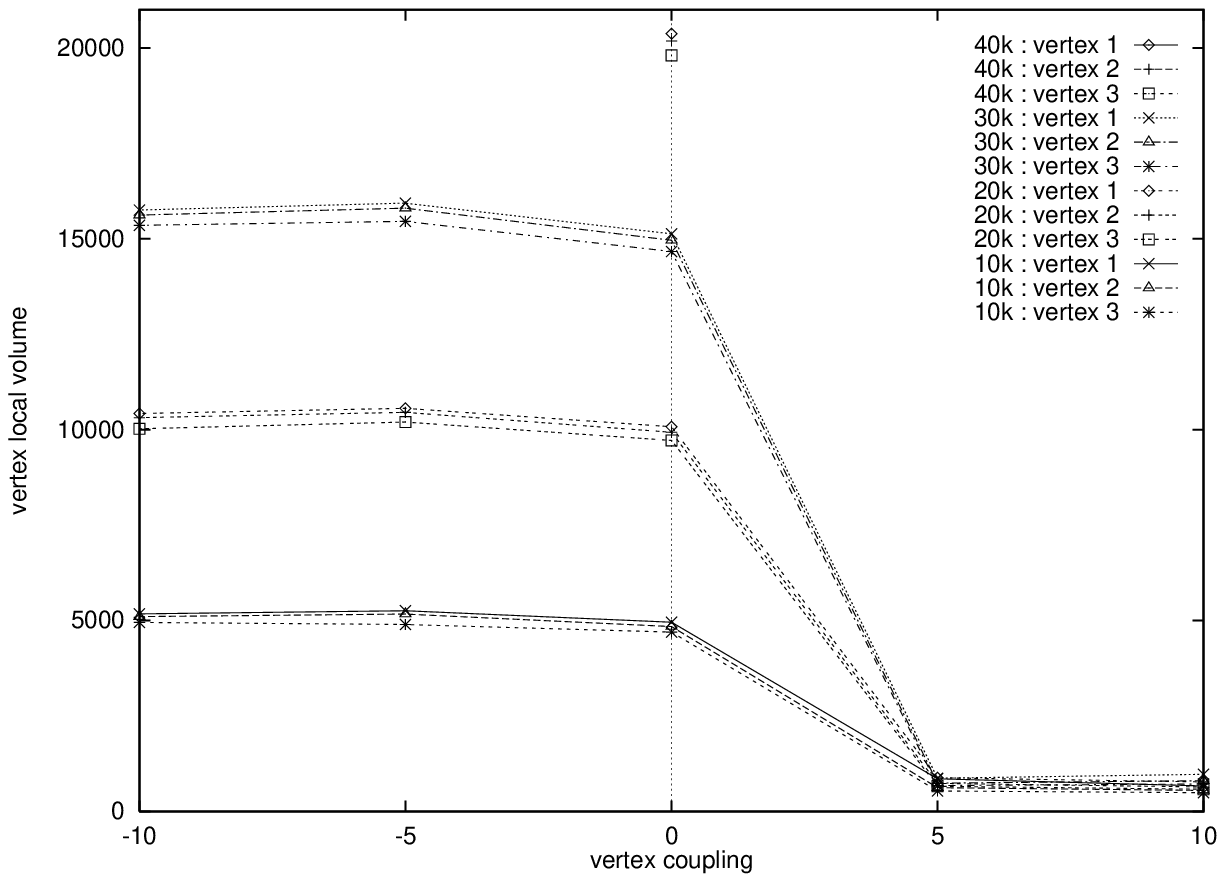}}}
(b){\hbox{\epsfxsize=14cm \epsfbox{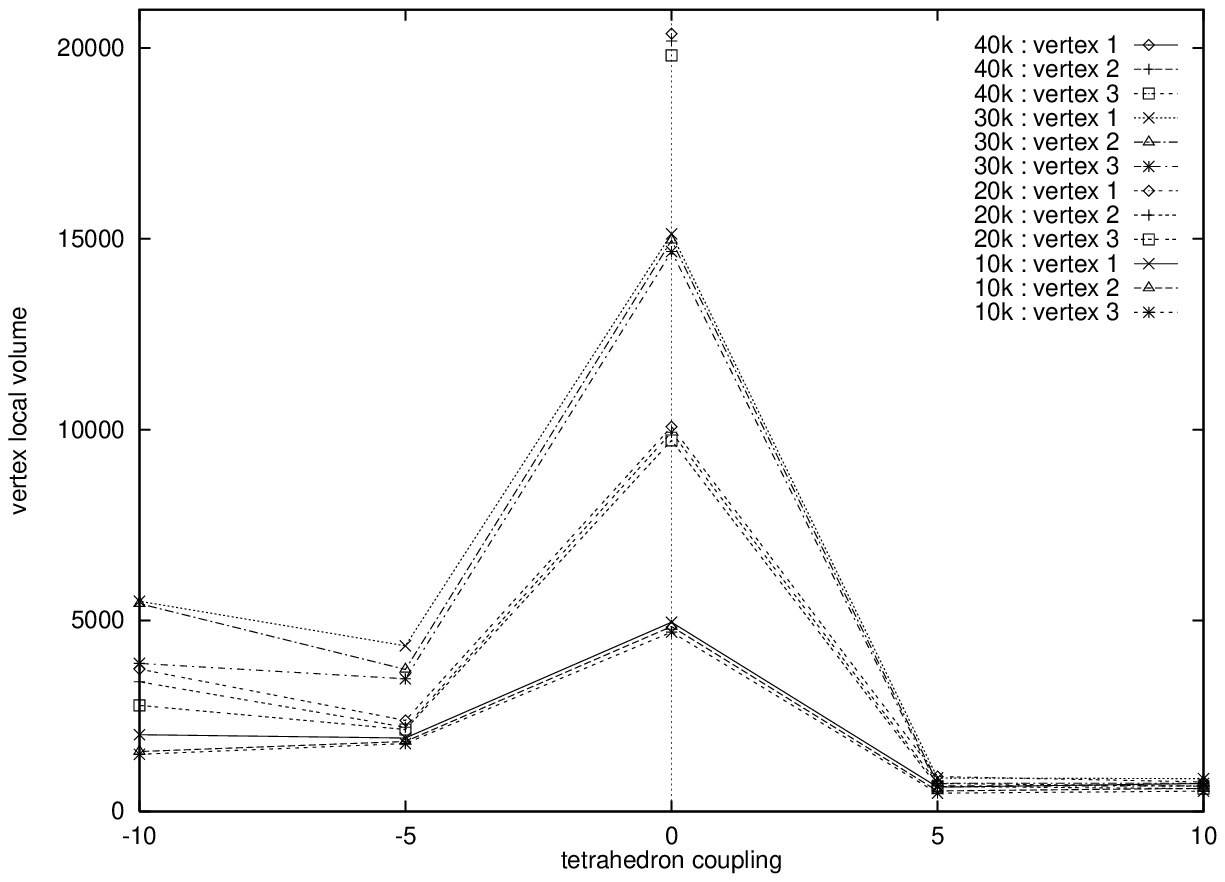}}}
\caption{Scaling plots of the three largest vertex local volumes versus (a) the vertex coupling $\kappa_{0}$ and (b) the tetrahedron coupling $\kappa_{3}$. (Lines serve only to guide the eye.)}
\label{fig:5dvertices}
\end{center}
\end{figure}

\begin{figure}[htp]
\begin{center}
\leavevmode
(a){\hbox{\epsfxsize=14cm \epsfbox{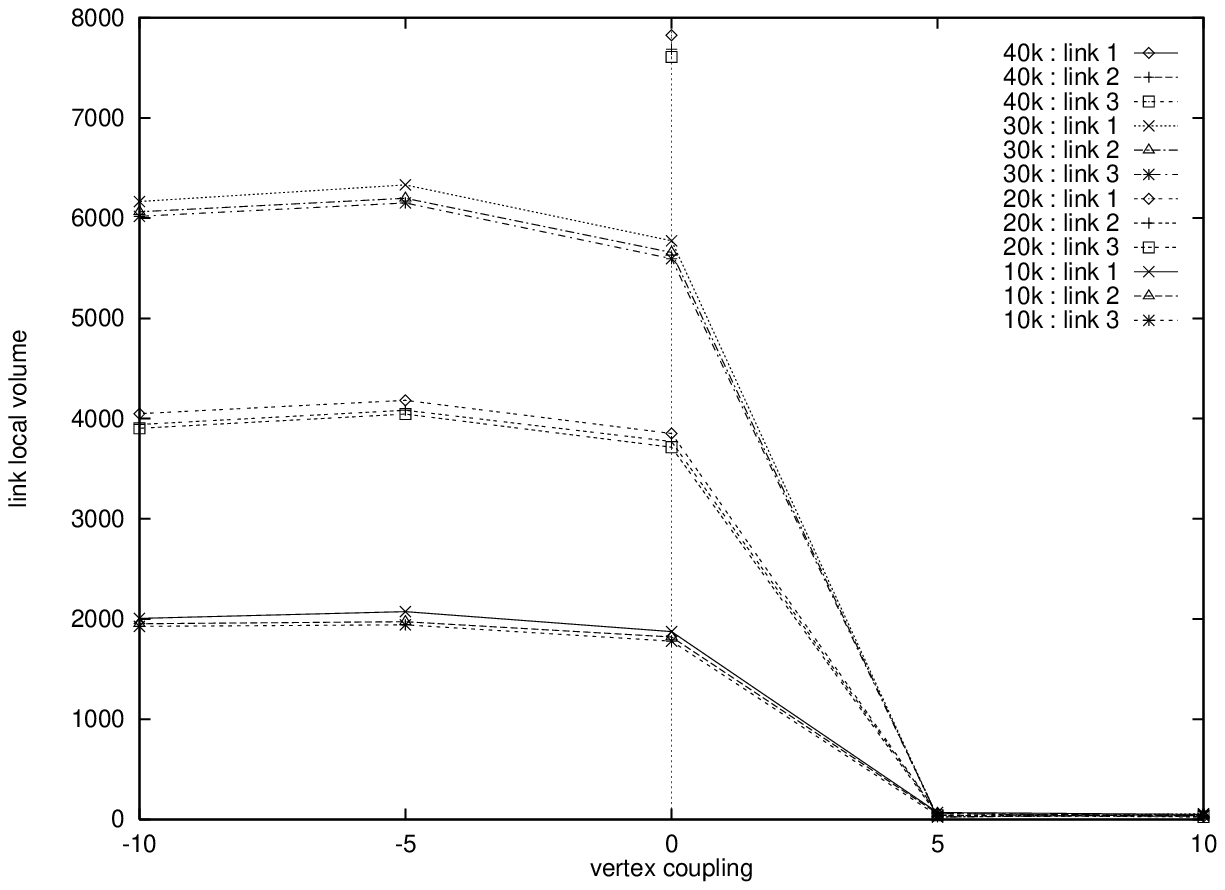}}}
(b){\hbox{\epsfxsize=14cm \epsfbox{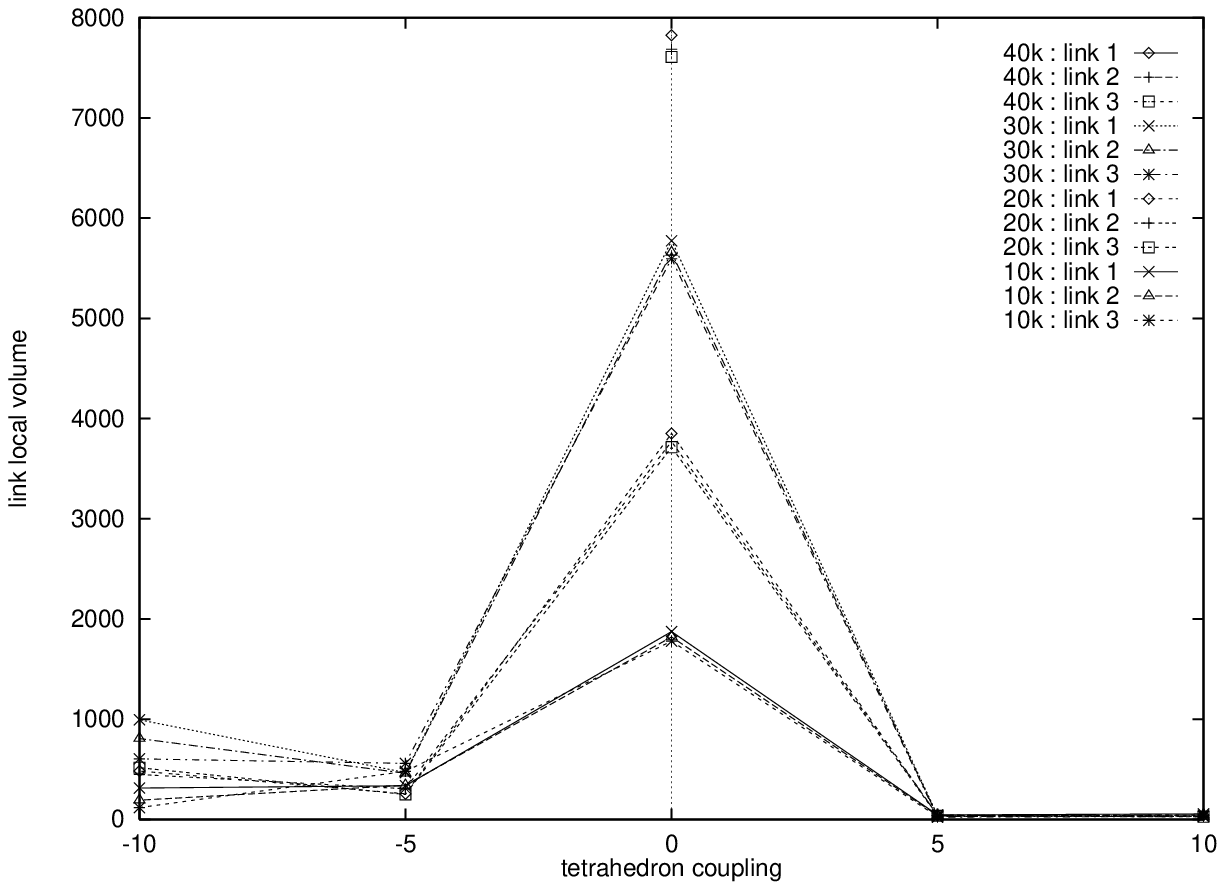}}}
\caption{Scaling plots of the three largest link local volumes versus (a) the vertex coupling $\kappa_{0}$ and (b) the tetrahedron coupling $\kappa_{3}$. (Lines serve only to guide the eye.)}
\label{fig:5dlinks}
\end{center}
\end{figure}

Figures~\ref{fig:5dvertices}(a) and~\ref{fig:5dvertices}(b) plot the mean values of $n(\sigma_{s1}^{0})$, $n(\sigma_{s2}^{0})$ and $n(\sigma_{s3}^{0})$ versus $\kappa_{0}$ and $\kappa_{3}$ respectively, for each target volume. 
On each graph, a key labels the vertex local volumes of each target volume in a logical manner. Figures~\ref{fig:5dlinks}(a) and~\ref{fig:5dlinks}(b) plot the mean values of $n(\sigma_{s1}^{1})$, $n(\sigma_{s2}^{1})$ and $n(\sigma_{s3}^{1})$ versus $\kappa_{0}$ and $\kappa_{3}$ respectively, for each target volume.
Again, a key labels the link local volumes in each graph. The $\kappa_{0}=\kappa_{3}=0$ data was fitted to the following sets of curves.
\begin{eqnarray}
n(\sigma_{s1}^{0}) & \!\!=\, a_{1}+b_{1}{N_{5}}^{c_{1}} \hspace{20mm} n(\sigma_{s1}^{1}) & \!\!=\, f_{1}+g_{1}{N_{5}}^{h_{1}} \\
n(\sigma_{s2}^{0}) & \!\!=\, a_{2}+b_{2}{N_{5}}^{c_{2}} \hspace{20mm} n(\sigma_{s2}^{1}) & \!\!=\, f_{2}+g_{2}{N_{5}}^{h_{2}} \\
n(\sigma_{s3}^{0}) & \!\!=\, a_{3}+b_{3}{N_{5}}^{c_{3}} \hspace{20mm} n(\sigma_{s3}^{1}) & \!\!=\, f_{3}+g_{3}{N_{5}}^{h_{3}}
\end{eqnarray}
The exponent coefficients of the best fit curves are given in table~\ref{tab:expcoeff} along with an indication of the `goodness-of-fit' in terms of $\chi^{2}$ per degree of freedom. As expected, these results are consistent with linear divergence. 
\begin{table}[htp]
\begin{center}
\begin{tabular}{|c||c|c||c|c|} \hline
$i$ & $c_{i}$ & $\chi^{2}~\mbox{d.o.f.}^{-1}$ & $h_{i}$ & $\chi^{2}~\mbox{d.o.f.}^{-1}$ \\ \hline
1 & 1.021(2) & 229 & 1.029(3) & 228 \\ \hline
2 & 1.022(2) & 237 & 1.01(1) & 11.3 \\ \hline
3 & 1.01(1) & 2.7 & 1.01(1) & 10.1 \\ \hline
\end{tabular}
\caption{Best fit exponent coefficients for secondary singular subsimplices.}
\label{tab:expcoeff} 
\end{center}
\end{table}

\begin{figure}[htp]
\begin{center}
\leavevmode
(a){\hbox{\epsfxsize=14cm \epsfbox{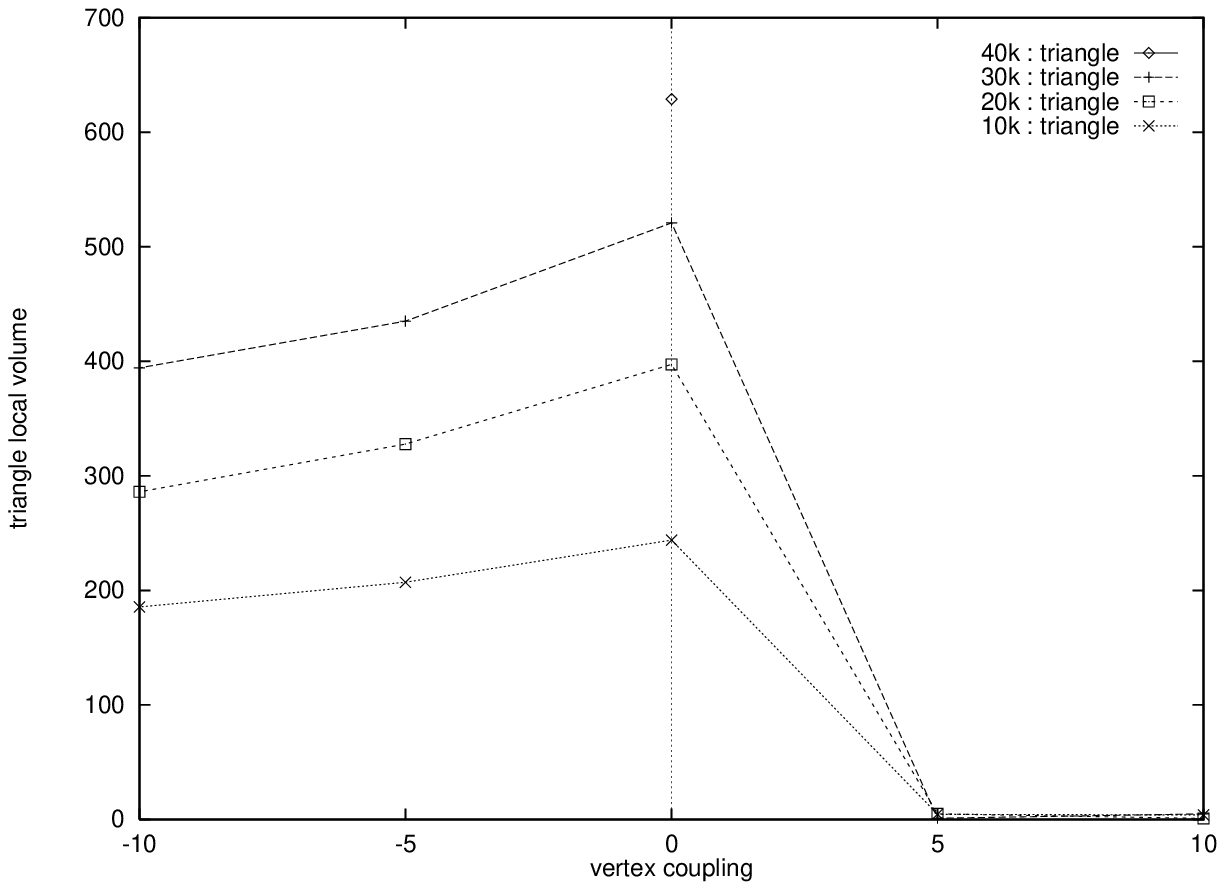}}}
(b){\hbox{\epsfxsize=14cm \epsfbox{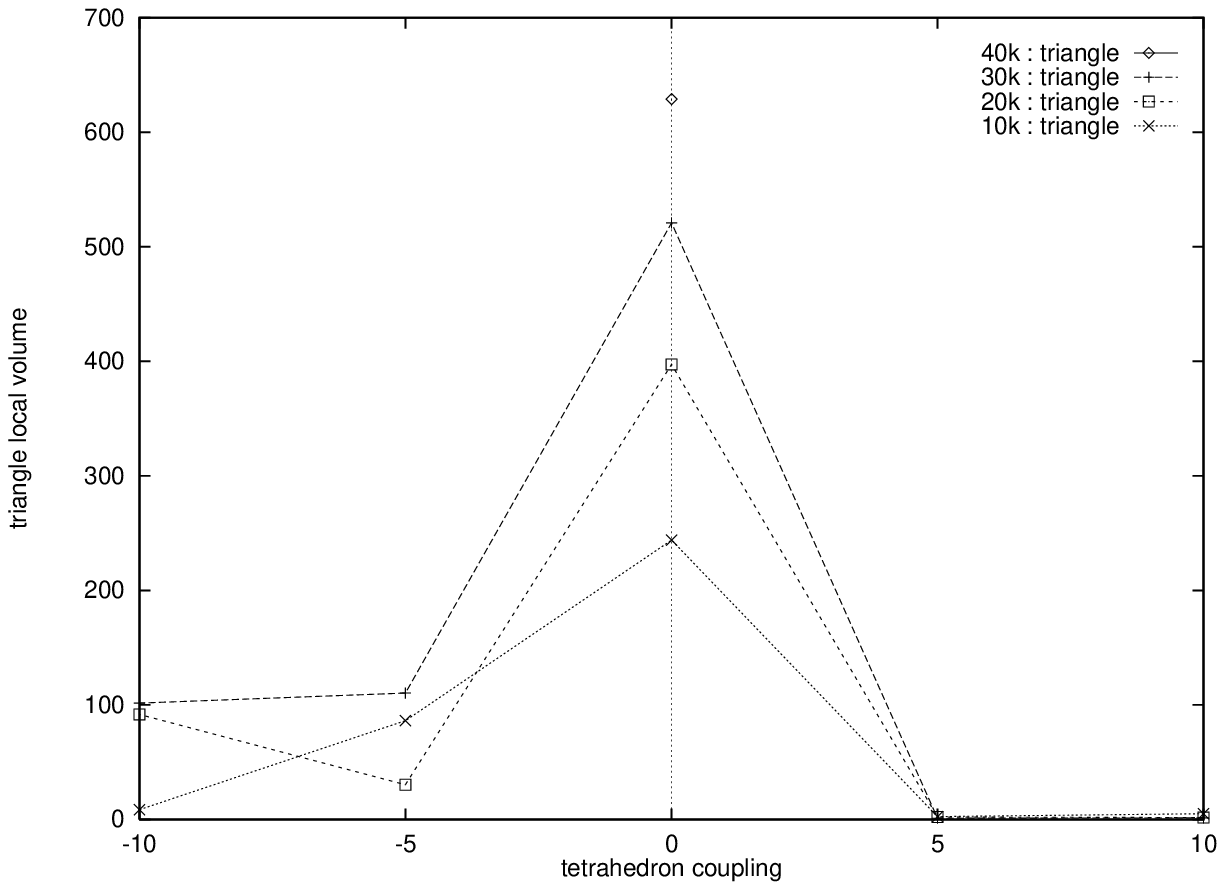}}}
\caption{Scaling plots of the largest triangle local volume versus (a) the vertex coupling $\kappa_{0}$ and (b) the tetrahedron coupling $\kappa_{3}$. (Lines serve only to guide the eye.)}
\label{fig:5dtriangle}
\end{center}
\end{figure}

Let us now consider the primary singular subsimplex, which in five dimensions is a triangle. The algorithm measures the local volume of the product of the three singular vertices, whether they form a triangle or not. Figures~\ref{fig:5dtriangle}(a) and~\ref{fig:5dtriangle}(b) plot the mean value of $n(\sigma_{s}^{2})$ versus $\kappa_{0}$ and $\kappa_{3}$ respectively, for each target volume. 
In the same manner, we fit the $\kappa_{0}=\kappa_{3}=0$ data to the following curve.
\begin{equation}
n(\sigma_{s}^{2}) = p+q{N_{5}}^{r} \label{eq:5dtri1} 
\end{equation}
The best fit curve was found to have $r=0.63(2)$ with $\chi^{2}\approx 3.2$ d.o.f.$^{-1}$. These results are reasonably consistent with the conjectured two-thirds power law divergence. 

\subsubsection{New Results}

Our study has shown that the vertex local volume distribution ${\cal N}(n_{0})$ of five dimensional triangulations is independent of volume. There is a direct correspondence between the branched polymer phases of four and five dimensional dynamical triangulations. This is not so surprising. For large $\kappa_{0}$ and/or $\kappa_{3}$, the action promotes vertex insertion moves. Consequently, the number of vertices of the triangulations grows which in effect bars the existence of singular structures.

One can draw parallels between singular structures in four dimensions and the $\kappa_{3}=0$ limit of five dimensions. It was interesting to find that triangulations in the crumpled phase are not always characterised by singular triangles. Monte Carlo time series plots of $n(\sigma_{s1}^{0})$, $n(\sigma_{s2}^{0})$ and $n(\sigma_{s3}^{0})$ for $\kappa_{0}=0$ showed that occasionally the ensemble tunnels between states with singular triangles and singular links. Like the four dimensional case, the tunneling events occured very quickly (within 200 sweeps). However, in five dimensions the tunneling rate was of order 1 in 100. It was found that the three largest vertex local volumes always form a triangle.

Deeper into the crumpled phase ensembles were found to thermalise towards triangulations with singular {\it links} for $\kappa_{0}\ll 0$. Analogous effects have been observed in four dimensions and may represent metastable states in the space of triangulations~\cite{ckrtsing}. A simple experiment was devised in order to test this was the case in five dimensions. The ensembles were originally thermalised from cold configurations, i.e. branched polymers. The simulations were repeated by thermalising from the $\kappa_{0}=0$ triangulations. It was found that the singular triangles remained intact and stable for all $\kappa_{0}<0$. Despite this, the ensemble continued to tunnel between singular triangle and singular link states in the same manner. These are the results which are shown in figures~\ref{fig:5dvertices}(a),~\ref{fig:5dlinks}(a) and~\ref{fig:5dtriangle}(a). In conclusion, this evidence suggests that triangulations with singular links (in five dimensions) represent metastable configurations. 

Close examination of the results reveals that $n(\sigma_{s1}^{0})$, $n(\sigma_{s2}^{0})$ and $n(\sigma_{s3}^{0})$ increase slightly for $\kappa_{0}<0$. This is presumably attributable to the action which tends to reduce $N_{0}$. It appears that the local volumes of secondary singular subsimplices diverge linearly with volume throughout the crumpled phase. It is difficult to confirm the two-thirds power law scaling in this limit, since we have data for only three target volumes. Interestingly, the singular triangle local volume decreases (possibly to an asymptote) as $\kappa_{0}\rightarrow -\infty$. Triangulations with $\kappa_{3}\ll0$ in the $\kappa_{0}=0$ limit have no singular vertices. This fact was discovered earlier in section~\ref{subsec:vlvd}. Unfortunately, it is not clear from our limited study how (or whether) the vertex local volume distribution depends on triangulation volume. 

\subsection{Entropy}

\begin{figure}[ht] 
\centering{\input{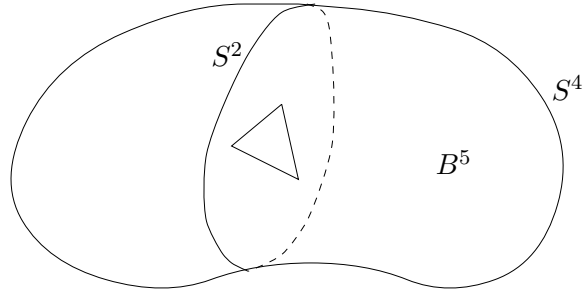}\par}
\caption{Schematic diagram of a triangle-singular $B^{5}$ with $S^{4}$ boundary. The $S^{2}$ dual-sphere of the triangle in the $B^{5}$ is a submanifold of the $S^{4}$ boundary.}
\label{fig:fivedual}
\end{figure}

As an explicit example of our generalised entropy arguments, we shall now apply these  ideas to singular structures in five dimensional triangulations. Consider an arbitrary $B^{5}$ submanifold of a five dimensional triangulation. The boundary of a 5-ball is a 4-sphere. Let the space of triangulations of this structure be represented by ${\cal T}_{5\mbox{-}ball}$. For simplicial 5-manifolds, the star of a triangle is homeomorphic to a 5-ball, and its dual-sphere is a 2-sphere. Clearly, the dual-sphere is a submanifold of the 4-sphere boundary of the star. This is illustrated schematically in figure~\ref{fig:fivedual}.
Such 5-balls are described as {\it triangle-singular}. The corresponding space of triangulations is symbolised by ${\cal T}_{sing,2}$. Clearly, from our definitions of ${\cal T}_{5\mbox{-}ball}$ and ${\cal T}_{sing,2}$ we find that ${\cal T}_{sing,2} \subset {\cal T}_{5\mbox{-}ball}$. One can then define the subset of {\it non-triangle-singular} triangulations ${\cal T}_{non,2}$ such that
\begin{equation}
{\cal T}_{sing,2} \cup {\cal T}_{non,2} \subseteq {\cal T}_{5\mbox{-}ball}. 
\end{equation}
The entropy of an arbitrary 5-ball may be represented by $\Omega(B^{5})$. By substituting $d=5$ into equation~(\ref{eq:priment}) one can show that the entropy associated with the singular structure as a whole is given by
\begin{equation}
\Omega\:\!({\cal T}_{sing,2}) = 3\:\!\Omega\:\!(S^{4}) - 3\:\!\Omega\:\!(S^{3}) + \Omega\:\!(S^{2}). \label{eq:5dsingent}
\end{equation}
The entropy of these spaces of triangulations may be equated as follows.
\begin{equation}
\Omega\:\!({\cal T}_{sing,2}) + \Omega\:\!({\cal T}_{non,2}) = \Omega({\cal T}_{5\mbox{-}ball}) \label{eq:5dent}
\end{equation}
Therefore, the existence of the observed singular structures in five dimensions would be due to 
\begin{equation}
\Omega\:\!({\cal T}_{sing,2}) \gg \Omega\:\!({\cal T}_{non,2}). \label{eq:5dentineq}
\end{equation}
By substituting equations~(\ref{eq:5dsingent}) and (\ref{eq:5dent}) into inequality~(\ref{eq:5dentineq}) and using the fact that $\Omega({\cal T}_{5\mbox{-}ball})=\Omega(B^{5})$ one can derive the following expression.
\begin{equation}
6\:\!\Omega\:\!(S^{4}) - 6\:\!\Omega\:\!(S^{3}) + 2\:\!\Omega\:\!(S^{2}) \gg \Omega\:\!(B^{5}) \label{eq:fiveentropy}
\end{equation}
Again, such an expression is very difficult, if not impossible, to verify. One cannot even qualitatively substantiate this relation. Obviously, we are interested in comparing the entropy of two 5-balls {\it of the same volume}. Now, the triangle-singular 5-ball is comprised of three 4-spheres, three 3-spheres and one 2-sphere. Let their volumes be given by $N_{5}(B^{5})$, $N_{4}(S^{4})$, $N_{3}(S^{3})$ and $N_{2}(S^{2})$ respectively. Using (\ref{eq:arbvol}) one can show that inequality~(\ref{eq:fiveentropy}) is only valid when the volumes satisfy the following equation (assuming that the 4-spheres and 3-spheres have equal volume).
\begin{equation}
N_{5}(B^{5}) = 3N_{4}(S^{4})-3N_{3}(S^{3})+N_{2}(S^{2})
\end{equation}

\subsection{New Phase Structure}

In this chapter, it has become clear that the distinction between the two strong coupling regimes (first brought to light in chapter~\ref{chap:chap4}) is directly related to the phenomenon of singular structures. Our study has shown that the crumpled phase of the $\kappa_{3}=0$ limit is characterised by the existence of singular triangles. This phase is known to include the point $(0,0)$ in ${\cal C}_{2}$. One may then trivially deduce that the $\kappa_{0}=0$ limit also corresponds to the same phase at $\kappa_{3}=0$ and perhaps its vicinity. Having observed vertex local volume distributions for $\kappa_{3}<0$ that are fundamentally distinct from those of $\kappa_{3}=0$, one may wonder whether they represent different phases in ${\cal C}_{2}$. The main purpose of this section is to explore this possibility.

Recall that evidence of a phase transition near $\kappa_{3}=-1.0$ was discovered in section~\ref{sec:prelim}. It seems very likely that this point in ${\cal C}_{2}$ separates two phases highlighted clearly in figures~\ref{fig:5dvertices}(b) and~\ref{fig:5dlinks}(b). To avoid confusion, the location of the new phase transtion is denoted by $\kappa_{3}^{\ast}$. For clarity and convenience, the phases are named as follows: the {\it crumpled} phase is located in the region $\kappa_{3}^{\ast}<\kappa_{3}<\kappa_{3}^{c}$ and the {\it compact}\,\footnote{This name reflects the fact that triangulations of the compact phase have smaller mean geodesic distances than those of the crumpled phase, as seen in figure~\ref{fig:zoomoutd}(b).} phase is found at $\kappa_{3}<\kappa_{3}^{\ast}$. 

\begin{figure}[htp]
\begin{center}
\leavevmode
(a){\hbox{\epsfxsize=14cm \epsfbox{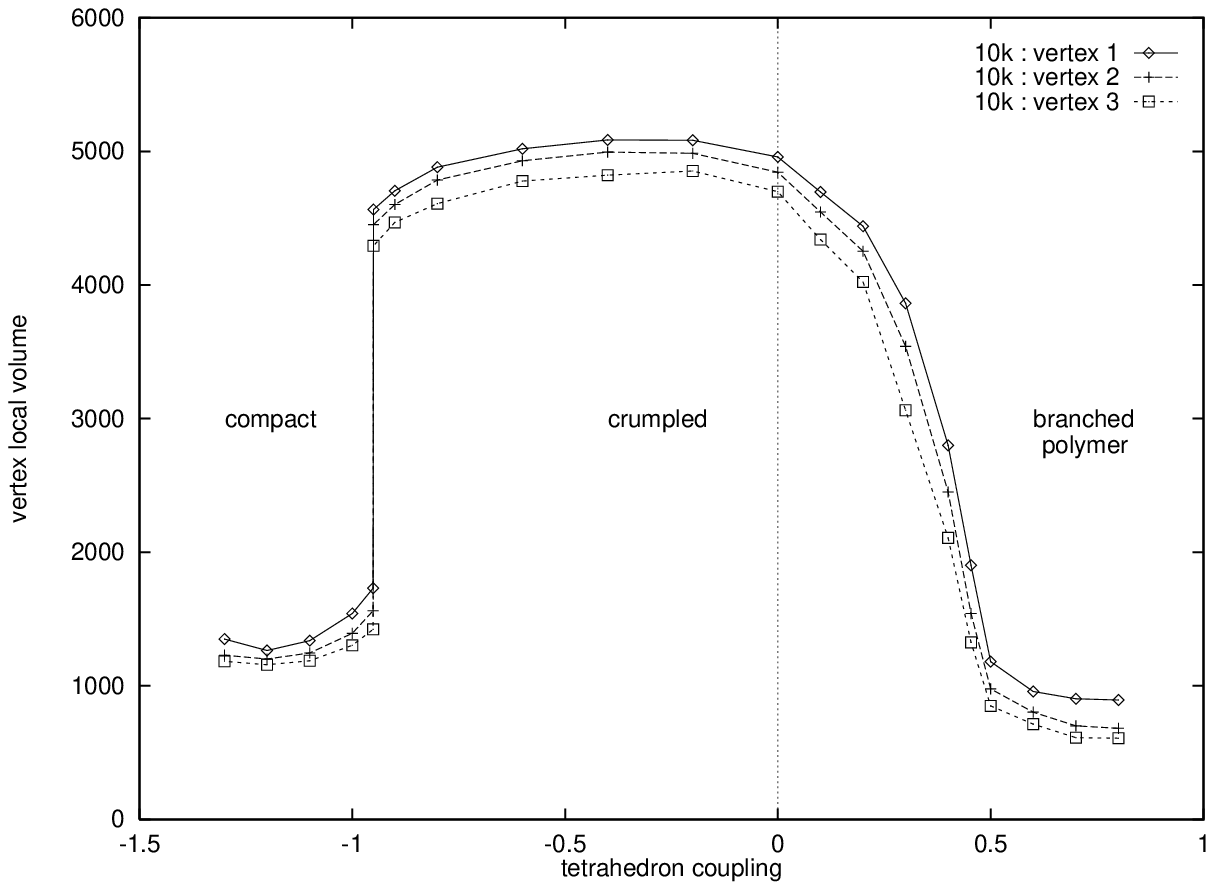}}}
(b){\hbox{\epsfxsize=14cm \epsfbox{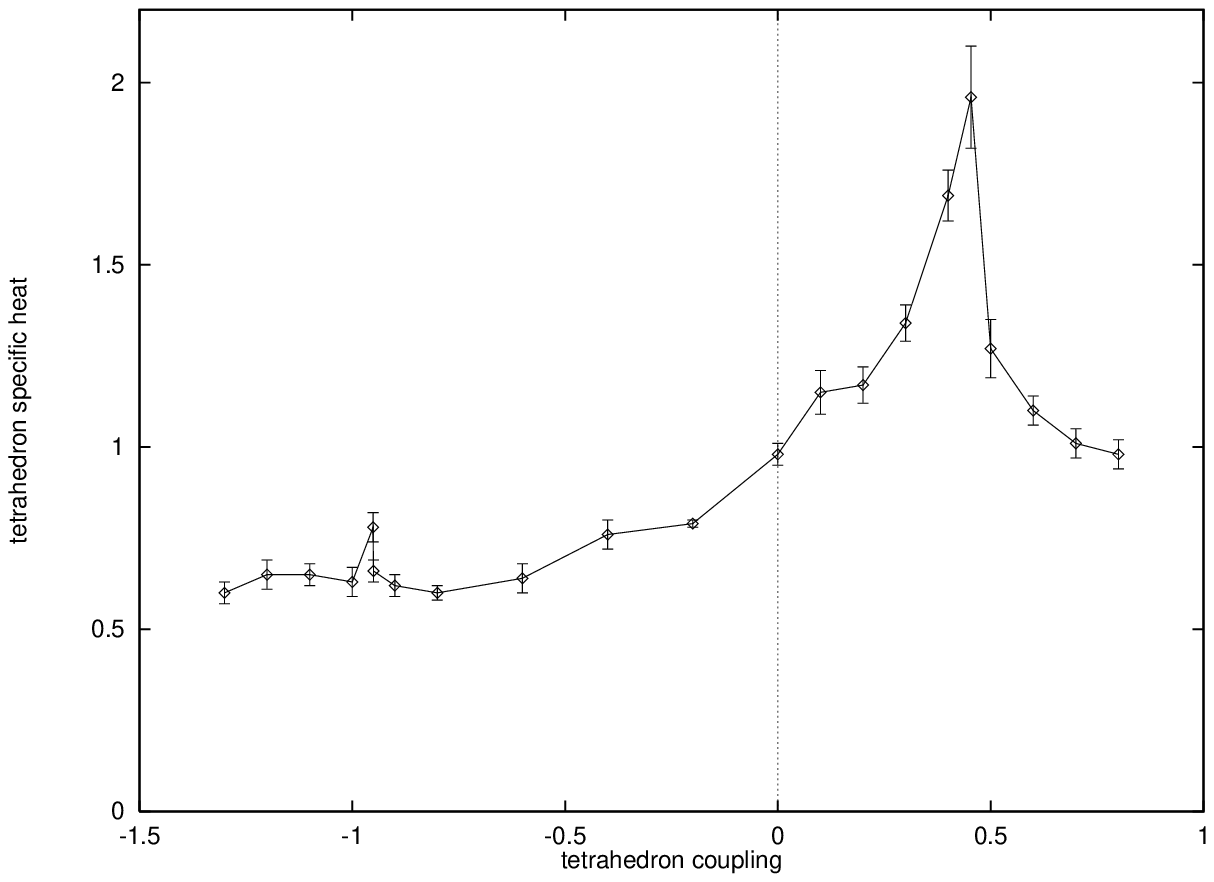}}}
\caption{\label{(a)} Vertex local volume versus the tetrahedron coupling $\kappa_{3}$ for $\kappa_{0}=0$. (b) Tetrahedron specific heat $\chi_{3}$ versus $\kappa_{3}$.}
\label{fig:5dk3zoom}
\end{center}
\end{figure}

Our specific goal is therefore to confirm that a phase transition does separate these phases. This is done by measuring the tetrahedron specific heat $\chi_{3}(\kappa_{3})$ and the local vertex distribution for higher resolutions of $\kappa_{3}$. These simulations were run for triangulations with target volume $N_{5}^{t}=10$k. Figures~\ref{fig:5dk3zoom}(a) and~\ref{fig:5dk3zoom}(b) (on page~\pageref{(a)}) plot the three largest vertex local volumes and $\chi_{3}$ versus $\kappa_{3}$ respectively. 
At $\kappa_{3}=0$, triangulations are characterised by three singular vertices. As $\kappa_{3}$ is increased, the singular vertices are progressively suppressed. By comparing both figures, their disappearance is obviously marked by the phase transition $\kappa_{3}^{c}\approx 0.454(1)$. As $\kappa_{3}$ is {\it decreased} we find that the singular vertices vanish near $\kappa_{3}^{\ast}\approx 0.951$. Again, this correlates with a small peak in $\chi_{3}$. It would appear that the transition is strongly first order. 

\vskip 5mm

\begin{figure}[htp]
\centerline{\epsfxsize=14cm \epsfbox{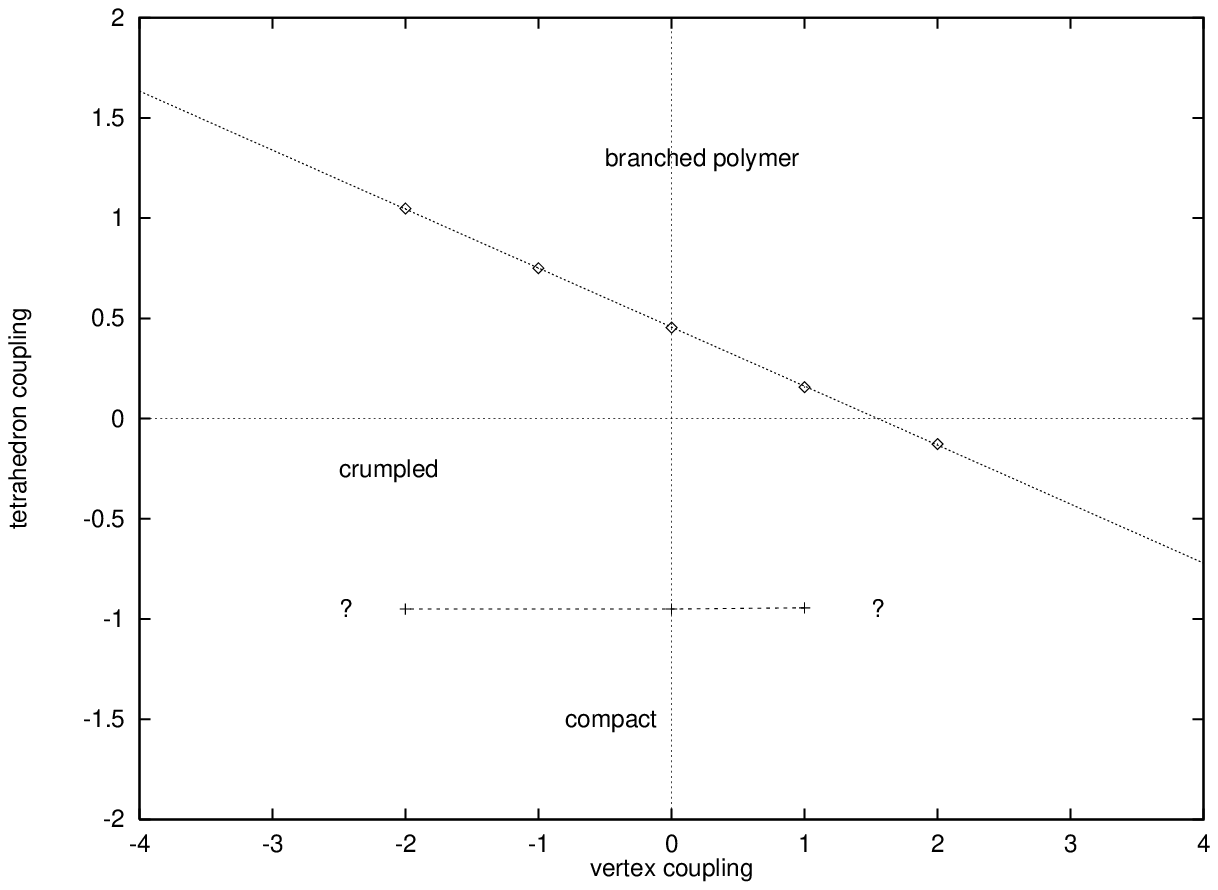}}
\caption{\label{Known} Phase structure of ${\cal C}_{2}$.}
\label{fig:c2}
\end{figure}

\noindent
The new phase transition was traced for $\kappa_{0}>0$ and for $\kappa_{0}<0$. Again, this was done by fixing $\kappa_{0}$ and tuning $\kappa_{3}$ to the transition. The transition was traced from $\kappa_{0}=-2$ to $\kappa_{0}=1$ (see figure~\ref{fig:c2} on page~\pageref{Known}).
The known phase are labelled accordingly. The question marks indicate regions in ${\cal C}_{2}$ which are yet to be explored. These results confirm the existence of a second line of phase transitions. Unfortunately, it was not possible to reconnoitre ${\cal C}_{2}$ any further due to time constraints. 
        
\input epsf.tex

\chapter{Summary} \label{chap:chap7}

\begin{figure}[ht]
\leavevmode
\hbox{\epsfxsize=2.52cm \epsfbox{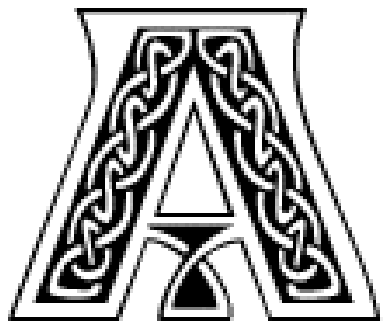}}
\parbox{13.35cm}{\baselineskip=16pt plus 0.01pt \vspace{-19mm}LTHOUGH dynamical triangulations has been actively researched for over ten years, physicists are yet to fully understand its possible links with a consistent theory of quantum gravity in four dimensions. It appears that the na\"{\i}ve generalisation of two dimensional dynamical triangulations does}
\end{figure}
\vspace{-5mm}
not yield a physical continuum limit. Despite this fact, research into dynamical triangulations for $d>2$ has continued, albeit in new directions. One such direction relates to the implementation of modified actions. This is usually done either by including matter fields or by adding measure terms. The net effect is the addition of an extra coupling constant. One then hopes that a physical continuum limit may be reached at a point in the expanded space of couplings.

In this thesis, the space of couplings is expanded by considering the most general action that is linear in components of the $f$-vector. These actions happen to be equivalent to the discretised Einstein-Hilbert actions for $d<5$. The purpose of this work is to discover new aspects of dynamical triangulations by viewing the model from a fresh slant. This approach will teach us more about dynamical triangulations in general. Let us now summarise the results of the preceding five chapters.

\vskip 5mm

\noindent
This thesis is naturally divided into two parts. The first part aims to give credibility to the second. A fundamental requirement of Monte Carlo simulations is that local moves are ergodic in the space of configurations. In chapter~\ref{chap:chap2}, the five dimensional $(k,l)$ moves were proven to be ergodic in the space of combinatorially equivalent triangulations. Unfortunately, the negativity of the Hauptvermutung in five dimensions means that the proof cannot be extended (in general) to the space of homeomorphic simplicial 5-manifolds. It is not known (to the author) whether the Hauptvermutung is true for the 5-sphere. The physical implications of its negativity are not well-understood.

Beyond this formal proof, one must contend with the issues of computational ergodicity and the recognisability of manifolds. In the context of triangulations, the unrecognisability of simplicial manifolds $T$ and $T'$ implies that the number of $(k,l)$ moves needed to transform $T$ into $T'$ is not bounded by a recursive function of their volume. Of course, even recursive functions can grow extremely quickly. Therefore, it is argued that, on a practical level, it is more relevant to consider whether the number of $(k,l)$ moves required to transform $T$ into $T'$ is computationally manageable. This relates to whether a series of $(k,l)$ moves can be performed by modern computers in a reasonable time scale (i.e. days or weeks). 

Chapter~\ref{chap:chap3} aimed to show that the canonical partition function of five dimensional dynamical triangulations is exponentially bounded from above as a function of volume. Of course, numerical evidence can never constitute a proof. No attempt was made to construct an analytic proof -- mainly because of the obvious complexity of the problem. Unfortunately, the outcome was inconclusive since the results were found to be equally compatible with exponential and factorial divergence. The chapter was concluded by highlighting the problems associated with the numerical approach, and suggesting a possible solution.

\vskip 5mm

\noindent
The second part of the thesis deals with the more interesting aspects of dynamical triangulations, at least from a physicist's point of view. In chapter~\ref{chap:chap4}, Monte Carlo simulations reveal the existence of a phase boundary in ${\cal C}_{2}$, the two dimensional space of coupling constants. The weak coupling phase is characterised by elongated triangulations of high scalar curvature and vertex density. The triangulations of the strong coupling regimes are generally crumpled with low scalar curvature and vertex density. The phase transitions were identified by sharp peaks in the specific heats. The results also hinted at the existence of a third phase in the strong coupling regime of the $\kappa_{0}=0$ limit. 

Traditional finite size scaling methods failed to determine the order of the phase transition at points $(\kappa_{0}^{c},0)$ and $(0,\kappa_{3}^{c})$. This was probably due to a combination of insufficient statistics and strong finite size effects. However, time series plots showed some evidence of bistability, which may be indicative of first order phase transitions: the signal was somewhat clearer for $(0,\kappa_{3}^{c})$. Further simulations at large volumes are required for verification. The phase boundary was traced over a large range of couplings. It forms a straight line for small $\kappa_{0}$ and $\kappa_{3}$, but is slightly curved in regions away from the origin. The reason for this is unknown. Interestingly, it was found that the phase boundary appears to be more strongly first order as $\kappa_{0}\rightarrow\infty$.

In chapter~\ref{chap:chap5}, the weak coupling phase triangulations were found to have $d_{H}\approx 2$, irrespective of volume. This discovery confirmed their branched polymer nature. In fact, it was found that the partition function is saturated by (branched polymer) stacked spheres for large $\kappa_{0}$ and/or $\kappa_{3}$. Finally, a generalisation of Walkup's theorem to arbitrary dimension was proved from elementary arguments and simple graph theory concepts. This was achieved by constructing dual graphs of whole triangulations. The real result of this work is not so much the generalised inequality itself, but the methods used to prove it.

The thesis was brought to a close with a broad study of singular structures in chapter~\ref{chap:chap6}. Their existence is accounted for in terms of new entropy arguments, based on those proposed by Catterall {\it et al.}~\cite{ckrtsing}. They appear to explain the non-existence of singular vertices in three dimensions. Monte Carlo simulations were run for four dimensions over a range of volumes and couplings. Time series plots revealed that the singular link occasionally transforms into a singular vertex and back again very quickly (within about 200 sweeps). This phenomenon was observed for various values of $\kappa_{0}$ in the crumpled phase, including $\kappa_{0}=0$. It was found that triangulations deep within the crumpled phase are characterised by a single singular vertex. This confirmed the prediction made by Bia{\l}as {\it et al.}~\cite{bbptmother}. 

The very last section dealt with singular structures in five dimensions. It was found that singular structures play an important role in the $\kappa_{3}=0$ limit for $\kappa_{0}<\kappa_{0}$. Perhaps the most interesting outcome of this research was the discovery of a new `compact' phase in the strong coupling regime of the $\kappa_{0}=0$ limit, where the action is the discretised Einstein-Hilbert action. It was found that triangulations endowed with singular structures dominate the partition function in the range $\kappa_{3}^{\ast}<\kappa_{3}<\kappa_{3}^{c}$, where $\kappa_{3}^{\ast}$ is the location of the newly discovered phase transition. Its existence was hinted at by measurements of $c_{2}(\kappa_{3},N_{5})$ in section~\ref{sec:prelim}. The compact phase was found to be characterised by a new distribution of vertex local volumes, which is shown in figure~\ref{fig:fivesingular2}(b). Finally, the account ended with an in-depth study of the new phase transition, which appears to be strongly first order. The phase transition was then traced in ${\cal C}_{2}$ for $\kappa_{0}\neq0$.

\vskip 5mm

\noindent
Let us now look at how these results relate to dynamical triangulations in general. Our study of stacked spheres in chapter~\ref{chap:chap5} allows us to make the following generalisation. Consider $d$-dimensional dynamical triangulations, in which the action is {\it linear} in any components of the $f$-vector -- one of which must be $N_{d}$. One can {\it always} identify a `weak coupling phase' which is characterised by branched polymer triangulations. Let us substantiate this claim. One can generalise the ideas of section~\ref{sec:walkup} to show that branched polymer stacked spheres of a given volume $N_{d}$ correspond to the upper kinematic bound of $N_{i}$, for $i=0,\ldots,d-1$. In other words, regardless of the form of the action, one can always set the couplings either to $+\infty$ or $-\infty$ such that the partition function is saturated by branched polymer stacked spheres. Using equation~(\ref{eq:deltani}), it is quite simple to prove the following upper kinematic bound, which holds in the thermodynamic limit for $d>1$ and $i=0,\ldots,d$.
\begin{equation}
\frac{N_{i}}{N_{d}} \longrightarrow \frac{(d+1)(d-1)!}{i\,!\,(d-i+1)!} 
\end{equation}
The substitution of $d=4$ and $i=2$ yields $R \rightarrow 2.5$ as expected. Of course, when $i=0$ we are left with the upper kinematic bound of the vertex density.
\begin{equation}
\frac{N_{0}}{N_{d}} \longrightarrow \frac{1}{d}
\end{equation}

Singular structures are known to play an important role in dynamical triangulations, at least for $d>3$. Based on our knowledge of dynamical triangulations in four and five dimensions, one may speculate that a phase transition separates the weak coupling limit and zero coupling in all cases where the action is linear in components of the $f$-vector. For an action with $n$ independent couplings, one may expect a $(n-1)$-dimensional surface of phase transitions to exist in the space of couplings. This is exactly what we have found in five dimensions.

One can make further speculations as a result of this work. Na\"{\i}vely, dynamical triangulations could give us `quantum gravity' in any dimension we like. If this was the case, then one might expect the phase structure of four dimensions to be similar to that of the $\kappa_{0}=0$ limit of five dimensions, assuming that the Einstein-Hilbert action is special in some sense. Our research has shown that this is not the case. In fact, the phase structure of the $\kappa_{3}=0$ limit is more similar to the four dimensional case -- in terms of singular structures. From this, one may speculate that the four dimensional phase structure is as it is because the action is proportional to $N_{d}$ and $N_{0}$, {\it not} because it is the discretised Einstein-Hilbert action. 

\vskip 5mm

\noindent
Finally, let us end with some suggestions for future research. First of all, as mentioned earlier, the negativity of the Hauptvermutung implies that the $(k,l)$ moves may not be ergodic in the space of homeomorphic simplicial manifolds. It is not clear whether any unreachable triangulations are physically significant; that is, whether they should be included in the measure. This matter seems to be quite important and ought to be resolved. 

The existence of an exponential bound remains an outstanding problem in five dimensions. In the absence of an analytic proof, one must rely on numerical evidence. This issue might be `resolved' simply by extending the investigation of chapter~\ref{chap:chap3} to larger triangulations. However, there is no guarantee that the outcome would be conclusive.

Although much of ${\cal C}_{2}$ is now mapped, there are a number of issues which are yet to be addressed. Perhaps the first task would be to establish the order of the phase transitions at $(\kappa_{0}^{c},0)$ and $(0,\kappa_{3}^{c})$ with certainty. This may be achieved by seeking evidence of metastability, particularly over a range of volumes. One may then consider whether the whole boundary is of the same order. Other prospects include the possible existence of critical points. Perhaps the best course of action is to further trace the second phase boundary, which separates the crumpled and compact phases. The identification of a critical point in ${\cal C}_{2}$ is a real possibility, and ought to be pursued. If contact can be made with the continuum, then one must consider whether the theory has anything to do with `gravity'. The answer would be very interesting, regardless of the outcome. 

Our generalisation of Walkup's theorem demonstrates the power of simple graph theory concepts when applied to the dual construction of triangulations. Perhaps, this connection between dynamical triangulations and regular non-planar graphs can be exploited further. It would also be interesting to see if (or how) our results relate to recent analytic advances~\cite{weak,crumpledcritical}. Lastly, it would be nice to explain the nature of the vertex local volume distribution of the compact phase. This may help us understand more about the transition between the crumpled and compact phases and may reveal connections with other work. Perhaps this could be achieved in terms of the `Balls-in-Boxes' model~\cite{balls}.


        \appendix
        
\input epsf.tex

\chapter{Computer Programs} \label{app:appa}

\section{Dynamical Triangulations Computer Program} \label{app:dtcp}

\begin{figure}[ht]
\leavevmode
\hbox{\epsfxsize=1.3cm \epsfbox{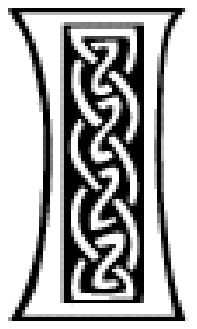}}
\parbox{14.55cm}{\baselineskip=16pt plus 0.01pt \vspace{-19mm}N this section we present a condensed overview of the main computer program used in our research of dynamical triangulations. In particular, we outline the general algorithm and mechanics of the Monte Carlo program, which was written in C $\,$by $\,$Simon Catterall $\,$for arbitrary $\,$dimension~\cite{catterallprogram}. $\,$This program, which $\,$we}
\end{figure}
\vspace{-5.8mm}
refer to as {\tt main}, was written in C primarily to utilise its {\it dynamic memory allocation} capability\,\footnote{From here on, any technical C terms will be printed in {\it italics}. Programs, functions and variables will be printed in {\tt lower case typed text} and global constants are shown in {\tt UPPER CASE TYPED TEXT}. For full details of the C programming language the reader is referred to a popular standard text by Kelley and Pohl~\cite{kelleypohl}.}. It also allowed for the effective use of {\it pointers}. These features are by no means mandatory as equivalent programs have been written in FORTRAN~\cite{bilkeburda}. One of the most difficult problems in creating such a program is finding a good way of abstractly representing the simplices and their connectivity. Let us now outline the framework of {\tt main} with this issue in mind.

Each simplex of a $d$-dimensional simplicial manifold may be represented by $d+1$ labels for each of its vertices and some form of identification of its $d+1$ neighbouring simplices. This knowledge is sufficient to reconstruct any triangulation. In our program, each simplex is abstractly represented by a set of variables known as a {\it structure}, which has an associated pointer. These pointers are grouped together into an array called {\tt simplex}\_{\tt point}. Each structure contains (amongst others) two main variables: an array of $d+1$ integers which label the vertices of the corresponding simplex and an array of $d+1$ pointers which point to the $d+1$ neighbouring structures (simplices). In {\tt main} these arrays are called {\tt vertex} and {\tt neighbour} respectively. These constructs enable the program to simulate triangulations of any dimension. This use of pointers allows us to `connect' neighbouring simplices in a very intuitive manner. 

\subsection{Update Algorithm}

Recall that a $(k,l)$ move in $d$-dimensions may be regarded as the replacement of an $i$-(sub)simplex with its dual $(d-i)$-(sub)simplex (see section~\ref{subsec:algform}). In this context we have $i=d+1-k$ and hence $d-i=d+1-l$ because $k+l=d+2$. This realisation of the $(k,l)$ moves allows for their definition for arbitrary dimension since a move is uniquely defined for a given $i$ and $d$, where $i=0,\ldots,d$. Here we describe the four main stages in the update procedure as they occur in order.

\begin{itemize}

\item In order to maintain ergodicity it is imperative that the moves are performed on random (sub)simplices. This is done by first selecting, at random, a pointer from {\tt simplex}\_{\tt point} which corresponds to a simplex, followed by a move of type {\tt sub}, where {\tt sub} $=i=d+1-k$. The (sub)simplex to be transformed is chosen by randomly selecting $d+2-k$ of the $d+1$ vertices of the simplex. 

\item Before executing the transformation the algorithm must first check that the chosen (sub)simplex is `legal'. An $i$-(sub)simplex is considered legal if its local volume is $d-i+1$. For example, one cannot perform a $(3,1)$ move on a vertex if its local volume is four. If the (sub)simplex is illegal then the algorithm restarts. Notice that this check is unnecessary if $i=d$ since vertex insertion moves are always legal.

\item Once a legal (sub)simplex has been identified, the corresponding transformation will be accepted by the Metropolis algorithm with a given probability -- determined by the change in the action $\Delta S$. If $\Delta S$ is zero or negative then the move is accepted. If $\Delta S$ is positive then the probability of acceptance is $\exp(-\Delta S)$. The original Metropolis algorithm of {\tt main} calculated $\Delta S[T,\kappa_{0}]$. Some minor alterations were made in order to study five dimensional dynamical triangulations with our new action $S_{5}^{g}[T,\kappa_{0},\kappa_{3}]$.

\item If the move is accepted by Metropolis then the algorithm checks that the move is geometrically `allowed'. This condition insists that the star of each vertex is homeomorphic to a $d$-ball and hence the triangulation retains its manifold structure. Transformations that do not satisfy the condition are therefore discarded. The importance of this point was stressed by Schleich and Witt~\cite{schleichwitt}.

\end{itemize}
If a candidate move passes through each step then the transformation will be executed. Having explained the update process of {\tt main}, let us now describe each phase of the Monte Carlo simulation life cycle. 

\subsection{Growth} \label{app:growth}

The C program written by Catterall is capable of generating simplicial manifolds of spherical topology for arbitrary dimension~\cite{catterallprogram}. This is made possible by realising that such structures can be grown from minimal $d$-spheres using $(k,l)$ moves. The initial configuration is equivalent to the surface of a $(d+1)$-simplex and is generated by a function named {\tt initial}\_{\tt config} which is called by {\tt main}. This can then be grown to any chosen target volume $N_{d}^{t}$ within technological bounds. This is done most efficiently by only performing vertex insertion moves that bypass the Metropolis algorithm. This process generates stacked $d$-spheres (see section~\ref{sec:stack}). 

Clearly, the triangulation target volume is limited by RAM (random access memory)\,\footnote{According to Catterall, the memory required for a $d$-dimensional triangulation of volume $N_{d}$ is of the order of $8(d+6)N_{d}$ bytes~\cite{catterallprogram}. For example, a five dimensional triangulation of volume $N_{5}=10$k requires approximately 0.9MB of memory, which is a modest amount for current technology.}. However, the numerical performance of Monte Carlo simulations is limited more by CPU power (see section~\ref{app:perform}). Once the required target volume has been attained, the algorithm then passes into the thermalisation stage. 

\subsection{Thermalisation} \label{app:therm}

In general, the initial fully grown triangulation will not be in thermal equilibrium. It is known from our investigations in chapter~\ref{chap:chap5} that stacked spheres saturate the weak coupling limit. If order to study the model at stronger coupling then one must allow a finite Monte Carlo time for the initial triangulations to thermlise. In these simulations Monte Carlo time is measured in units of sweeps. A sweep is defined as $N_{d}^{t}$ legal {\it attempted} moves. (Alternatively, one could equally define a sweep as $N_{d}^{t}$ completed updates.) The number of thermalisation sweeps {\tt THERM} is defined as a global constant. 

During this stage the simplex coupling $\kappa_{d}$ is tuned at regular intervals of {\tt TUNE}\_{\tt COUPLING} sweeps using the following iterative procedure
\begin{equation}
\kappa_{d} \longrightarrow {\kappa_{d}}'=\kappa_{d}+2\,\!\gamma\,\!(N_{d}-N_{d}^{t}), \label{eq:tune}
\end{equation}
where ${\kappa_{d}}'$ is the updated simplex coupling and $\gamma$ is related to the volume fluctuation parameter {\tt DV} by
\begin{equation}
{\tt DV} = \gamma^{-\frac{1}{2}}.
\end{equation}
It has been verified that thermal averages are independent of $\gamma$ for small $\gamma$, but have errors that grow as $\gamma\rightarrow0$~\cite{catterallprogram}. In our simulations $\gamma$ was chosen to be $\gamma\approx 0.0005$, corresponding to ${\tt DV}\approx32$. It is important that {\tt DV} is not set too low, as one may lose ergodicity. Ergodicity demands that all of the $d+1$ $(k,l)$ moves are implemented, which means that {\tt DV} {\it must} be greater than $d$. 

At intervals of {\tt CHECKPOINT} sweeps an output file called {\tt dump} is created. It contains all the necessary information needed to regenerate a triangulation of a given volume and coupling. One can set {\tt main} to read-in a {\tt dump} file so that triangulations do not have to be grown and thermalised for each simulation. Once the simplex coupling is tuned and the ensemble is fully thermalised {\tt main} then passes into the measurement stage.

\subsection{Measurement} \label{app:measure}

At this point the {\tt main} begins a series of {\tt SWEEPS} sweeps during which measurements of certain physical observables are made at regular intervals of {\tt GAP} sweeps. These include the triangulation volume $N_{d}$, the total number of vertices $N_{0}$ and the average geodesic distance $\overline{g}$ (see section~\ref{subsec:geodesic}). The number of origin simplices is controlled by {\tt HITS}. All measurements are sent to an output file called {\tt data}. As with the thermalisation stage, the triangulation connectivity is sent to {\tt dump} at regular intervals of {\tt CHECKPOINT} sweeps. 

It was necessary to measure $N_{3}$ in order to calculate the scalar curvature $R$ and tetrahedron specific heat $\chi_{3}$. This minor problem was solved by introducing a mechanism of monitoring the $f$-vector. A $d+1$ dimensional array of integers called {\tt Tot}\_{\tt up} was created, whose elements correspond to the components of the $f$-vector. Prior to the growing stage the elements of the array were set to those of a minimal 5-sphere.
\begin{equation}
{\tt Tot}\_{\tt up[i]} ={}^{d+2}\mbox{C}_{i+1} = {}^{7}\mbox{C}_{i+1} 
\end{equation}
Then for each $(k,l)$ move that was accepted, the elements of {\tt Tot}\_{\tt up} were updated accordingly using equation~(\ref{eq:deltani}). Table~\ref{tab:deltas} shows $\Delta N_{i}$ for each of the six $(k,l)$ moves. On completion of {\tt SWEEPS} sweeps, a final {\tt dump} file is created.
\begin{table}[htp]
\begin{center}
\begin{tabular}{|c|c||c|c|c|c|c|c|} \hline
$(k,l)$ & {\tt sub} & $\Delta N_{0}$ & $\Delta N_{1}$ & $\Delta N_{2}$ & $\Delta N_{3}$ & $\Delta N_{4}$ & $\Delta N_{5}$ \\ \hline
$(1,6)$ & 5 & +1   & +6   & +15   & +20   & +15   & +5 \\ \hline
$(2,5)$ & 4 & 0    & +1   & +5    & +10   & +9    & +3 \\ \hline
$(3,4)$ & 3 & 0    & 0    & +1    & +3    & +3    & +1 \\ \hline
$(4,3)$ & 2 & 0    & 0    & $-1$  & $-3$  & $-3$  & $-1$ \\ \hline
$(5,2)$ & 1 & 0    & $-1$ & $-5$  & $-10$ & $-9$  & $-3$ \\ \hline
$(6,1)$ & 0 & $-1$ & $-6$ & $-15$ & $-20$ &$ -15$ & $-5$ \\ \hline
\end{tabular}
\caption{$\Delta N_{i}$ for the five dimensional $(k,l)$ moves.}
\label{tab:deltas}
\end{center}
\end{table}

\section{Computer Performance Tests} \label{app:perform}

\begin{figure}[ht]
\centerline{\epsfxsize=14cm \epsfbox{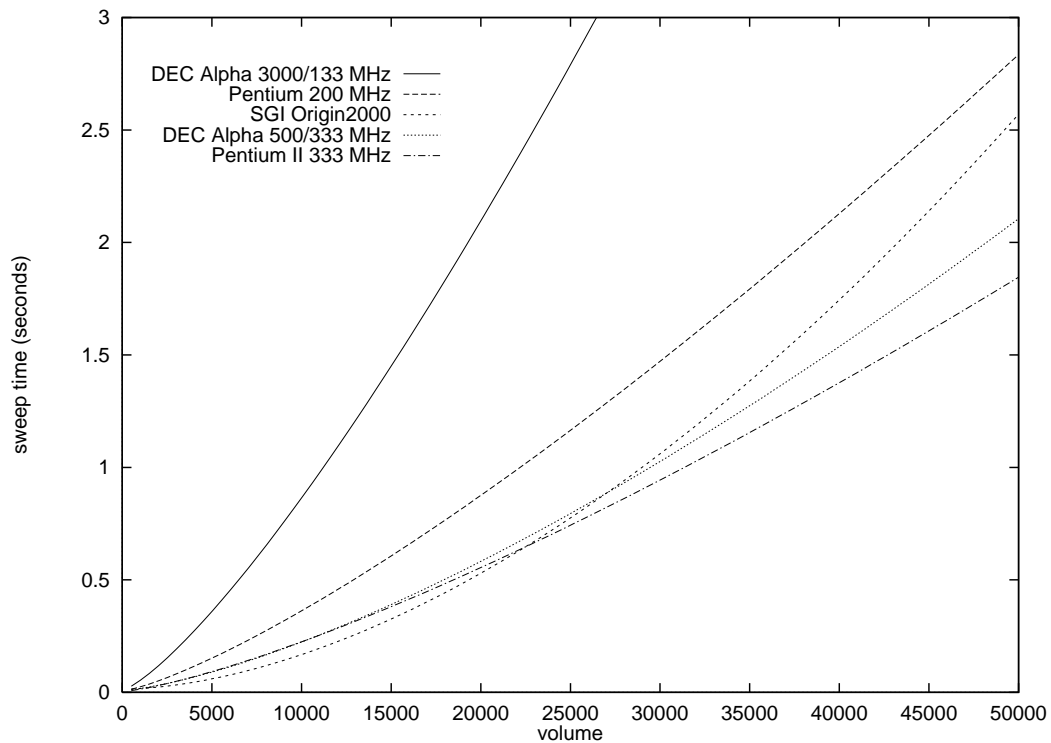}} 
\caption{Sweep time (in seconds) versus triangulation volume $N_{5}$.}
\label{fig:perform}
\end{figure}

In this section we give an indication of the scale of computational resources and time required for this type of work. During the course of this research Monte Carlo simulations were run on a variety of computers, which included:
\begin{itemize}
\item DEC Alpha 3000/133 MHz and 500/333 MHz workstations (Digital UNIX).
\item PCs: Pentium 200 MHz, 233 MHz and Pentium II 333 MHz (Linux).
\item SGI Origin2000 supercomputer -- non-parallel (IRIX). 
\end{itemize}
Tests were run on each computer to quantify their power when applied to our particular needs. The test involve measuring the CPU time needed to complete $10^{3}$ sweeps, with $\kappa_{3}=\kappa_{0}=0$ over a range of target volumes\,\footnote{In each case the code was compiled using {\tt cc -O3 ...}.}. Figure~\ref{fig:perform} shows the best fit curves of the resulting data. This is a plot of time per sweep (in seconds) versus triangulation volume $N_{5}$. A key identifies each curve with its corresponding computer. 

\noindent
It is clear that DEC Alpha 3000 machines are considerably slower than the other machines tested -- in fact, Pentium 200 MHz PCs are over twice as fast. The newer DEC Alpha 500/333 MHz and Pentium II 333 MHz are faster still. Interestingly, we see that the single processor Origin2000 is the fastest for small volumes, but slows very rapidly for larger volumes. It seems that this system is best suited for parallelised code. From these results we see that, for example, a Pentium 200 MHz PC can complete $10^{6}$ sweeps in about 100 hours. In conclusion, the most efficient means (in terms of cost as well as time) of running these simulations is by using a set of the most powerful PCs available (currently Pentium III processors). 

\section{Additional Code} \label{app:addcode}

Some additional code was incorporated into {\tt main} in order to study spatial distributions, and singular structures in various dimensions. These function were written by the author.

\subsubsection{Spatial Distributions} \label{app:sdf}

The spatial distribution of simplices $n(r)$ was evaluated by a C function called {\tt spatial}. It was called by {\tt main} at regular intervals of {\tt SPATIAL}\_{\tt GAP} sweeps. Memory is allocated in {\tt main} for a two dimensional array of {\it floating point numbers} called {\tt simplex}\_{\tt data}. The first dimension corresponds to the geodesic distance $r$. Its size is set to the maximum $r$ expected, typically $10^{2}$. The second dimension relates to the number of measurements of $n(r)$ taken during the Monte Carlo simulation. Its size will therefore be
\begin{equation}
\frac{\tt SWEEPS}{{\tt SPATIAL}\_{\tt GAP}}.
\end{equation}

Each time {\tt spatial} is called by {\tt main} it allocates memory for an array of integers {\tt spatial}\_{\tt tally} whose elements are initialised to zero. The $r^{\rm th}$ element counts the number of simplices separated by a geodesic distance of $r$. Then for each element of {\tt simplex}\_{\tt point} acting as an origin simplex $O_{s}$, {\tt spatial} calls {\tt geodesic} which returns the geodesic distance between $O_{s}$ and every other simplex. By using {\it every} simplex as $O_{s}$ one obtains an `exact' measurement of $n(r)$ rather than an estimate\,\footnote{This can become costly (in computational terms) for large volumes. Fortunately, it is not a problem for the scale of volumes studied here.}. Once this loop is completed, {\tt spatial}\_{\tt tally} is then normalised and passed back to {\tt simplex}\_{\tt data} in {\tt main}. As the volume is not fixed, {\tt spatial}\_{\tt tally} was normalised for each configuration {\it then} averaged over configurations, rather than vice versa.

When the measurement stage is finished, {\tt main} calls {\tt mean}\_{\tt spatial} which calculates the mean spatial distribution $\langle n(r)\rangle$ and estimates the errors $\sigma(\langle n(r)\rangle)$ using the Jackknife method (see appendix~\ref{app:jack}). This is the reason why the distributions must be stored in {\tt simplex}\_{\tt data}. Clearly, such storage of data can become a problem for intensive simulations. This algorithm is not very efficient.

\vskip 5mm

\noindent
A separate C program used measurements of $\langle n(r)\rangle$ and $\sigma(\langle n(r)\rangle)$ to calculate the normalised volume $\langle V(r) \rangle$ contained within a geodesic radius $r$ of an arbitrary origin simplex. The error in $\langle V(r)\rangle$ was calculated (rather crudely) by defining the upper and lower limits as follows.
\begin{eqnarray}
\langle V(r)\rangle^{+} & = & \sum_{i=0}^{r}\left(\langle n(i)\rangle+\sigma(\langle n(i)\rangle)\right) \\
\langle V(r)\rangle^{-} & = & \sum_{i=0}^{r}\left(\langle n(i)\rangle-\sigma(\langle n(i)\rangle)\right) 
\end{eqnarray}
This method may not give accurate estimates of uncertainty, but does have the virtue of being conservative. The mean fractal dimension function $\langle d_{H}(r)\rangle$ was calculated using equation (\ref{eq:dhdiff}) and the uncertainty in this quantity was estimated using equations (\ref{eq:dfeplus}) and (\ref{eq:dfeminus}).
\begin{eqnarray}
\langle d_{H}(r)\rangle^{+} & = & \frac{\ln\langle V(r+1) \rangle^{+}-\ln\langle V(r) \rangle^{+}}{\ln(r+1)-\ln(r)} \label{eq:dfeplus} \\
\langle d_{H}(r)\rangle^{-} & = & \frac{\ln\langle V(r+1) \rangle^{-}-\ln\langle V(r) \rangle^{-}}{\ln(r+1)-\ln(r)} \label{eq:dfeminus}
\end{eqnarray}

\subsubsection{Singular Structure} \label{app:singcode}

A new function of {\tt main} named {\tt vertex}\_{\tt order} was used to study singular structures for dimensions $d<6$. It is called at regular intervals of {\tt SPATIAL}\_{\tt GAP} sweeps. {\tt vertex}\_{\tt order} measures the local volume of each vertex by exploiting the fact that each {\it structure} stores an array of $d+1$ integers called {\tt vertex}, which label the vertices of the corresponding simplex. The algorithm may be divided into the following stages.

\begin{itemize}

\item Memory is allocated for an array of integers called {\tt vertex}\_{\tt tally}, whose elements are vertex integer labels. A loop is then run over all simplices and {\tt vertex}\_{\tt tally} counts the vertex labels. This process reveals the number of simplices common to a given vertex, i.e. its local volume. A check then confirms that the number of non-zero elements of {\tt vertex}\_{\tt tally} equals $N_{0}$.

\item A second array of integers called {\tt singular} is then generated which counts vertices according to their local volume. For example, the tenth element gives the number of vertices with a local volume of 10. The elements of {\tt singular} are then sent to an output file. 

\item From chapter~\ref{chap:chap6}, we know that the primary singular subsimplex is the product of $d-2$ vertices for $d>3$. The function then identifies the labels of the $d-2$ largest vertex local volumes and calculates the local volume of the corresponding subsimplex.

\item Given that a primary singular subsimplex has $d-2$ vertices, the function then calculates the local volume of each of the ${}^{d-2}$C$_{i+1}$ secondary singular $i$-subsimplices for $i=1, \ldots, d-3$. {\tt vertex}\_{\tt order} works out every combination of subsimplices and finds the local volume for each case. 

\end{itemize}
Once the measurement stage is completed, the output files are then analysed. Specifically, a separate program calculates the mean local volume of each structure and the standard error in the mean (see appendix~\ref{app:appb}). 

        
\input epsf.tex

\chapter{Error Analysis and Curve Fitting} \label{app:appb}

\section{Independent Samples} \label{app:indep}

\begin{figure}[ht]
\leavevmode
\hbox{\epsfxsize=1.95cm \epsfbox{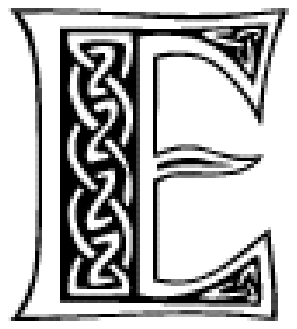}}
\parbox{13.9cm}{\baselineskip=16pt plus 0.01pt \vspace{-20mm}FFECTIVE error analysis is a vital part of Monte Carlo simulation methods. It allows us to establish the uncertainty associated with a given measurement. Let us first consider the simple case where we have $n$ independent (random) samples of $\,$an observable $X$, $\,$which $\,$form a normal $\,$distribution. $\,$In terms $\,$of}
\end{figure}
\vspace{-5mm}
Monte Carlo simulations, independent samples are those that are not correlated. Given a set of $n$ measurements of $X$, the mean $\overline{X}$ is, of course, simply defined as 
\begin{equation}
\overline{X}=\frac{1}{n} \sum_{i=1}^{n} X_{i}.
\end{equation}
(Since we are dealing with finite systems, it is important to remember that $\overline{X}$ is an estimate of the expectation value of $X$.) The standard deviation $\sigma$ and standard error in the mean $s_{m}$ of the sample are given by
\begin{equation}
\sigma^{2} = \frac{1}{n} \sum_{i=1}^{n}(X_{i}-\overline{X})^{2} \label{eq:isd}
\end{equation}
and
\begin{equation}
s_{m} = \frac{1}{\sqrt{n-1}}\,\sigma \approx \frac{1}{\sqrt{n}}\,\sigma \hspace{5mm} \mbox{for large $n$}. \label{eq:isem}
\end{equation}

One may envisage a scenario in which measurements are not independent, i.e. correlated. Under these circumstances, one cannot obtain reliable estimates of errors using equations (\ref{eq:isd}) and (\ref{eq:isem}). Such problems may be avoided by allowing the Monte Carlo ensemble sufficient `time' to decorrelate between successive measurements. This characteristic time is known as the {\it autocorrelation time} $\tau$ and can vary for different observables. Unfortunately, calculation of $\tau$ is a non-trivial matter and can be a clumsy process. In this thesis, the problem of correlated data was dealt with in another way -- by using a technique known as {\it binning}.

\section{Binning} \label{app:binning}

Suppose that our $n$ measurements of $X$ are correlated. Binning involves dividing the $n$ data sets into $b$ `bins' of equal size and calculating the average of each. If the size of each bin is greater than the autocorrelation time, then the averages can be considered as being independent. Therefore, in essence, binning is a way of generating independent data from correlated data~\cite{montvaymunster}. 

The mean $Y_{i}$ of bin $i$ is given by
\begin{equation}
Y_{i}=\frac{b}{n} \sum_{j=1}^{\frac{n}{b}}X_{\frac{n}{b}(i-1)+j} \hspace{5mm}\mbox{for $i=1, \ldots, b$.} 
\end{equation}
One can then calculate $\overline{Y}$, which is equal to $\overline{X}$.
\begin{eqnarray}
\overline{Y} & = & \frac{1}{b}\sum_{i=1}^{b}Y_{i} \\
& = & \frac{1}{n}\sum_{i=1}^{b}\sum_{j=1}^{\frac{n}{b}}X_{\frac{n}{b}(i-1)+j} \\
& = & \overline{X}
\end{eqnarray}
By treating the set of `measurements' $Y_{i}$ as independent data, one can define the standard deviation and standard error in the mean $s_{m}$ as
\begin{equation}
{\sigma}^{2} = \frac{1}{b}\sum_{i=1}^{b}(Y_{i}-\overline{Y})^{2} 
\end{equation}
and
\begin{equation}
s_{m} = \frac{1}{\sqrt{b-1}}\,\sigma \approx \frac{1}{\sqrt{b}}\,\sigma \hspace{5mm} \mbox{for large $b$}. 
\end{equation}
Clearly, when $b=n$, binning has no effect and we expect inaccurate estimates of the errors in $\overline{X}$. As $b$ is decreased, $Y_{i}$ become progressively decorrelated. Consequently, the error in $\overline{Y}$ (and hence $\overline{X}$) increases and thus becomes more accurate. When $b\approx\tau$, $Y_{i}$ are effectively independent and the resulting error estimates are accurate. This method is effective providing the sample size is much greater than the autocorrelation time.

\section{Jackknife} \label{app:jack}

As the name suggests, the jackknife is a rather crude (yet reliable) general purpose method of estimating errors~\cite{montvaymunster}. It is most commonly applied to samples with unknown distributions. The technique involves dividing a sample of $n$ independent measurements of $X$ into, say, $c$ blocks. One then calculates the $c$ averages $Y_{i}$ of the sample by leaving out each block in turn. 
\begin{equation}
Y_{i} = \frac{1}{c-1}\left(\sum_{j=1}^{\frac{n}{c}(i-1)}X_{j}+\sum_{j=\frac{n}{c}i}^{n}X_{j}\right)\hspace{5mm}\mbox{for $i=1,\ldots,c$.} 
\end{equation}
The mean of the jackknife estimators $\overline{Y}$ is given by
\begin{eqnarray}
\overline{Y} & = & \frac{1}{c}\sum_{i=1}^{c}Y_{i} \\
& = & \frac{1}{c}\sum_{i=1}^{c}\frac{1}{c-1}\left(\sum_{j=1}^{\frac{n}{c}(i-1)}X_{j}+\sum_{j=\frac{n}{c}i}^{n}X_{j}\right) \\
& = & \overline{X},
\end{eqnarray}
and the standard error in the mean $s_{m}$ is 
\begin{equation}
s_{m} = \sqrt{\frac{c-1}{c}\sum_{i=1}^{c}(Y_{i}-\overline{Y})^{2}}.
\end{equation}

\section{Curve Fitting} \label{app:curve}

The testing of observations against theoretical predictions is a fundamental aspect of scientific research. Here we are concerned with `goodness-of-fit' hypothesis testing -- specifically {\it chi-squared tests}~\cite{siegel}. For example, given a set of observed data, chi-squared tests allows us to quantify how well the data fits a hypothesised theoretical prediction. To be more precise, chi-squared tests tell us how confidently we may accept or reject our hypothesis.

Consider a set of $n$ observed data sets in the form $(x_{i},y_{i},\sigma_{i})$ and a theoretical prediction $y^{\ast}=f(x)$. Chi-squared tests involve calculating the $\chi^{2}$ statistic
\begin{equation}
\chi^{2} = \sum_{i=1}^{n} \left( \frac{y^{\ast}-y_{i}}{\sigma_{i}} \right)^{2}, 
\end{equation}
where $y^{\ast}$ is the predicted value of $y$ for a given $x$. Clearly, if we have a perfect fit then $\chi^{2}=0$. In order to judge the fit correctly, one must specify the number of degrees of freedom $k$ (or d.o.f.), which is defined as
\begin{equation}
k = n - p, 
\end{equation}
where $p$ is the number of parameters (coefficients) of $y^{\ast}=f(x)$. Once the functional form of $y^{\ast}$ is chosen, a computer program can then find the best fit by tuning the coefficients such that $\chi^{2}$ is minimised. (Typically the program outputs the best fit coefficients, their errors and the resulting $\chi^{2}$.) This fit is usually called the {\it null hypothesis} $H_{0}$.

For a given number of degrees of freedom, the chi-squared distribution is defined as
\begin{equation}
{\cal F}_{k}(x) = \frac{1}{2^{\frac{k}{2}}\Gamma(k/2)}x^{\frac{k}{2}-1}e^{-\frac{x}{2}}, \hspace{5mm}\mbox{for $x>0$}
\end{equation}
where $x$ represents $\chi^{2}$. The final step involves using the appropriate distribution to determine whether the null hypothesis can be accepted or rejected. Let us assume that for a certain null hypothesis $H_{0}$ the chi-squared statistic is $\chi^{2}=u$. In this case, the probability $Q(\chi^{2}|k)$ of getting $\chi^{2}\geq u$ if $H_{0}$ is true is given by
\begin{equation}
Q(\chi^{2}|k) = \frac{1}{2^{\frac{k}{2}}\Gamma(k/2)}\int_{u}^{\infty}x^{\frac{k}{2}-1}e^{-\frac{x}{2}}\mbox{d}x.
\end{equation}
For example, if $Q(\chi^{2}|k)$ is small, then it is considered unlikely that the discrepancies between observation and prediction are chance fluctuations and hence, $H_{0}$ may be rejected. 
	
	\newpage
        \addcontentsline{toc}{chapter}{Bibliography}
        \bibliographystyle{unsrt} 
        \bibliography{} 

\end{document}